%% file: Quantum.tex
\documentclass[a4paper,twocolumn,11pt,accepted=2025-07-21]{quantumarticle}
\pdfoutput=1
\usepackage[utf8]{inputenc}
\usepackage[english]{babel}
\usepackage[T1]{fontenc}
\usepackage[bbgreekl]{mathbbol}
\usepackage{amsmath,dsfont,mathtools,amssymb}
\usepackage{stmaryrd}
\usepackage{amsthm}
\usepackage{physics}
\usepackage{amsfonts}
\usepackage{bm}
\usepackage{enumitem}
\usepackage{orcidlink}
\usepackage{chngcntr}

\usepackage{layouts}
\usepackage{nowidow}
\usepackage{tikz}
\usepackage{multirow}
\usetikzlibrary{patterns}
\usetikzlibrary{quantikz2}
\usepackage{graphicx}
\usepackage{pgf}
\usetikzlibrary{positioning, calc}
\usepackage{caption}
\usepackage{subcaption}
\usepackage{dcolumn}
\usepackage{bm}
\usepackage{stmaryrd}
\usepackage[bbgreekl]{mathbbol}
\usepackage{amsmath,dsfont,mathtools}
\usepackage{amssymb}
\usepackage[normalem]{ulem}
\usepackage{color}
\usepackage{stmaryrd}
\usepackage{adjustbox}
\usepackage{physics}
\usepackage{amsfonts}
\usepackage{soul} 
\usepackage{todonotes}
\usepackage{overpic}
\usepackage{authblk}
\usepackage{amsthm}
\usepackage{comment}
\usepackage{tabularx}
\usepackage{algorithm}
\usepackage[noend]{algpseudocode}
\usepackage{thmtools}
\usepackage{thm-restate}
\usepackage{hyperref}
\usepackage{xcolor}
\usepackage{url}
\usepackage{orcidlink}
\usepackage{etoolbox}

\renewcommand{\norm}[1]{\left\lVert#1\right\rVert}

\newcommand{\figref}[2][]{\hyperref[#2]{\ref*{#2}#1}}
\newcommand{\R}{\mathbb{R}}

\newcommand{\bzero}{\boldsymbol{0}}

\newcommand{\przero}{\ket{0}\!\!\bra{0}}

\newcommand{\sumrw}{\sum_{\substack{ i_1,i'_1\in R(\omega) \\j_1,j'_1\in R(\omega)}}}
\newcommand{\sumrwb}{\sum_{\substack{ I,I'\in R(\omega) \\J,J'\in R(\omega)}}}
\newcommand{\E}{\mathbb{E}}

\newcommand{\hela}[1]{\textcolor{black}{#1}}
\newcommand{\leo}[1]{\textcolor{black}{#1}}
\definecolor{border_green_encoding}{HTML}{3C552D}
\definecolor{green_encoding}{HTML}{B8E986}
\definecolor{border_blue_ansatz}{HTML}{154239}
\definecolor{blue_ansatz}{HTML}{7FBFB2}
\definecolor{blue1}{HTML}{548ea1}
\definecolor{blue2}{HTML}{7f88bf}
\definecolor{blue3}{HTML}{7fbfb2}
\definecolor{grayplot}{HTML}{d6d6d6}

\newtheorem{theorem}{Theorem}
\newtheorem{corollary}{Corollary}

\newtheorem{lemma}{Lemma}
\newtheorem{definition}{Definition}
\newcommand{\noj}{\bar{\jmath}}
\newcommand{\Var}{\operatorname{\mathds{V}\!ar}}

\definecolor{customblue}{HTML}{17178B}

\usepackage{titletoc}

\usepackage{csquotes}
\MakeOuterQuote{"}

\usepackage{tabularx}
\usepackage{algorithm}
\usepackage[noend]{algpseudocode}

\usepackage{thmtools}
\usepackage{thm-restate}

\usepackage{hyperref}
\usepackage[capitalise]{cleveref}
\usepackage{xcolor}
\definecolor{quantumviolet}{HTML}{53257F}
\definecolor{navy}{RGB}{47,60,126}

\hypersetup{
    colorlinks=true,
    linkcolor=navy,
    citecolor=quantumviolet,
    urlcolor=quantumviolet,
    breaklinks=true,
}

\makeatletter
\newcommand{\IfRestatedTF}[2]{\ifthmt@thisistheone #2\else #1\fi}
\makeatother

\begin{document}

\title{Constrained and Vanishing Expressivity of Quantum Fourier Models}

\author{Hela Mhiri\orcidlink{0009-0001-3997-2646}}
\affiliation{Sorbonne Université, CNRS, LIP6, 75005 Paris, France}
\affiliation{ENSTA Paris, Institut Polytechnique de Paris, France}
\email{hela.mhiri@outlook.fr}
\author{Léo Monbroussou\orcidlink{0009-0005-4898-8155}}
\affiliation{Sorbonne Université, CNRS, LIP6, 75005 Paris, France}
\affiliation{CEMIS, Direction Technique, Naval Group, 83190 Ollioules, France}
\author{Mario Herrero-González\orcidlink{0000-0002-0506-7453}}
\affiliation{Sorbonne Université, CNRS, LIP6, 75005 Paris, France}
\affiliation{Quantum Software Lab, School of Informatics, University of Edinburgh, United Kingdom}
\author{Slimane Thabet\orcidlink{0000-0002-6291-3927}}
\affiliation{Sorbonne Université, CNRS, LIP6, 75005 Paris, France}
\affiliation{PASQAL SAS, 7 avenue Léonard de Vinci, 91300 Massy, France}
\author{Elham Kashefi\orcidlink{0000-0001-6280-9604}}
\affiliation{Sorbonne Université, CNRS, LIP6, 75005 Paris, France}
\affiliation{Quantum Software Lab, School of Informatics, University of Edinburgh, United Kingdom}
\author{Jonas Landman\orcidlink{0000-0002-2039-5308}}
\affiliation{Quantum Software Lab, School of Informatics, University of Edinburgh, United Kingdom}
\affiliation{QC Ware, Palo Alto, USA and Paris, France}

\maketitle

\begin{abstract}  
In this work, we highlight an unforeseen behavior of the expressivity of Parameterized Quantum Circuits (PQCs) for machine learning. 
A large class of these models, seen as Fourier series whose frequencies are derived from the encoding gates, were thought to have their Fourier coefficients mostly determined by the trainable gates.
Here, we demonstrate a new correlation between the Fourier coefficients of the quantum model and its encoding gates. In addition, we display a phenomenon of vanishing expressivity in certain settings, where some Fourier coefficients vanish exponentially as the number of qubits grows. These two behaviors imply novel forms of constraints which limit the expressivity of PQCs, and therefore imply a new inductive bias for quantum models. The key concept in this work is the notion of a frequency 
redundancy in the Fourier series spectrum, which determines its importance. Those theoretical behaviors are observed in numerical simulations.
\end{abstract}


\section{Introduction}

Quantum Machine Learning (QML) is an important field of study as an application of quantum computing \cite{wiebe_quantum_2015}. While many fault-tolerant algorithms propose significant advantages over their classical counterparts, many technological milestones must be reached for implementation. Variational quantum algorithms \cite{cerezo_variational_2021} are promising candidates for near term QML methods. Indeed, a popular approach consists of training Parametrized Quantum Circuits (PQCs) as neural networks. To do so, the classical data must be embedded in the Hilbert space through encoding gates, whereas trainable gates that depend on internal parameters will be optimized during the hybrid training procedure.

\vspace{-.1cm}
\begin{figure*}[tp]
    \centering    
    \input{Figures/Figure_Introduction_Merge.tikz}
    \caption{Parameterized quantum models with classical input $x$ and parameter vector $\theta$, a) can be seen as Fourier series in the classical input with frequencies $\omega \in \Omega$ and associated Fourier coefficients $c_\omega(\theta)$. b) illustrates the relation between the frequencies redundancies $|R(\omega)|$, i.e. the number of times a frequency appears in the spectrum, and their Fourier coefficients variance $\Var[c_\omega(\theta)]$. This connection constrains the expressivity of quantum Fourier models.}    \label{fig:Figure_Introduction}
\end{figure*}
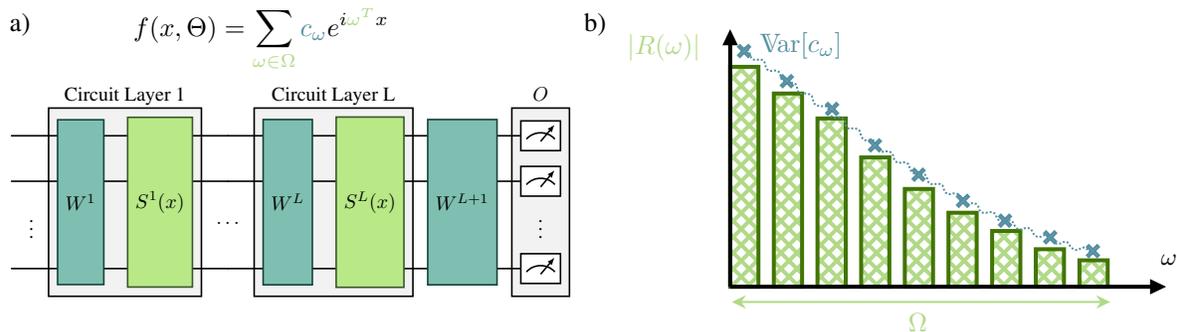

Many studies have been conducted to understand the potential and limitations of quantum models. Multiple
works focus on the trainability of such models and highlight the \hela{issues of local minima and barren plateaus} \cite{Anschuetz_2022,arrasmith_equivalence_2022,larocca_theory_2023,larocca_diagnosing_2022,mcclean_barren_2018,zoufal_generative_2021}.

On the other hand, another fundamental question concerns the expressivity of these models, namely which hypothesis class the quantum model is exploring.
From the seminal paper \cite{schuld_effect_2021}, we know that if one considers an encoding scheme where the classical input is encoded as the time evolution of some Hamiltonian,
the quantum model generated by the PQC can be described as a Fourier series in the classical input. The spectrum is determined by the encoding layers while the Fourier coefficients are mainly controlled by the trainable layers (See Section \ref{sec:Framework} for more details). 

In this work, we highlight a new connection between the Fourier coefficients and the encoding gates as illustrated in Fig.\ref{fig:Figure_Introduction}. This connection is made through the new concept of frequency \emph{redundancy}, which captures the number of times a frequency appears in the spectrum (see Definition \ref{def:Redundancy} for details).

The concept of frequency redundancy emerged from previous works \cite{landman_classically_2022, schreiber_classical_2023} attempting to find a classical approximation to any quantum Fourier model. Indeed, one could use the same frequencies in the spectrum, or a sample from them, to train a classical Fourier series that is guaranteed to have the same or better performance than the quantum model on generic machine learning tasks. If this question remains open in general \cite{jerbi_quantum_2023,sweke_potential_2023}, it was observed in \cite{landman_classically_2022} that some frequencies were more important to include in the Fourier series, as their Fourier coefficients had always a greater contribution. 

In this paper, we formally establish that some Fourier coefficients \hela{are less constrained} than others depending on the choice of the encoding Hamiltonians. Using the notion of redundancy, we show that the encoding strategy leads to an \textit{inductive bias} in the quantum model, as the Fourier coefficients with high redundancies will have a greater impact.
To prove the connection between a frequency redundancy and its importance in the Fourier model, we show that the variance of any Fourier coefficient is \hela{highly dependent on} the redundancy of its frequency. We further use this result to establish the \textit{vanishing expressivity} phenomenon, whereby some or all Fourier coefficients can suffer from exponentially vanishing variance  as the number of qubits increases. \hela{This vanishing expressivity phenomenon can be interpreted as the effective model expressivity. Namely, it implies that while using an exponentially big spectrum may seem at first as a trivial strategy to increase the model's expressivity, one should examine more carefully the \textit{effective} expressivity of the model.} Moreover, we establish another constraint on the 2-norm of the Fourier coefficients vector that holds for any quantum Fourier Model.

The manuscript is structured as follows. In Section \ref{sec:Framework}, we introduce the framework we consider in the study of quantum models from PQCs. We also define relevant mathematical tools and concepts required for our study. In Section \ref{sec:Main_Results}, we present our main results showing the inductive bias in the quantum model that arises from spectrum properties. 
In Section \ref{sec:Discussion}, we discuss in more details the impact of the proven Fourier model constraints on the model behavior.  Finally, we present numerical simulations in Section \ref{sec:Simulations} supporting our theoretical results and we conclude in Section \ref{sec:Conclusion}.

\subsection*{Related Work}

In \cite{schuld_effect_2021}, it has been shown that a large class of parameterized quantum models with classical data encoding can be expressed as Fourier series. Moreover, by considering reuploading schemes, the spectrum size can be efficiently increased leading to quantum models that serve as universal function approximators \cite{perez-salinas_data_2020,schuld_effect_2021}. Hence, this framework arises as a powerful tool to further assess the performance and potential of such quantum models. Specifically, much focus has been accorded to the study of the quantum spectrum and its characteristics \cite{schuld_effect_2021,peters_generalization_2023,shin_exponential_2022} by exploring different choices of the encoding Hamiltonians.
Besides, in \cite{jaderberg_let_2023}, the author proposed to learn the frequencies of the quantum model that fit best the learning task at hand.  
On the other hand, a variety of works used this Fourier representation to propose
dequantization schemes of  quantum models.
\hela{In \cite{landman_classically_2022}, an efficient construction of a classical model based solely on the circuit description was proposed. The classical model was expected to have similar performances to the quantum model on the training data. However, it has been pointed out in \cite{jerbi_quantum_2023,sweke_potential_2023} that both models may not necessarily converge to the same solution, leading to different generalization performances. This statement has been further examined in \cite{jerbi_shadows_2023} and in \cite{thabet_2024} where the authors show that there exist quantum models that cannot be dequantized by any efficient method.} Thus, it is now highly important to understand the inductive bias of quantum Fourier models that differentiate them from their classical ``dequantizers''.

In this work, we focus on studying the bias in the quantum model Fourier coefficients and how it is linked to the structure of the quantum circuit (i.e. encoding strategy, observable, etc.). In this line of work, \cite{peters_generalization_2023} related the phenomenon of benign overfitting in quantum interpolating Fourier models to the frequency distribution and state preparation showing that the encoding strategy provides an inductive bias that impacts directly the generalization performance. Moreover, \cite{caro_encoding-dependent_2021} provided encoding-dependent generalization bounds for such models.
Nonetheless, these works provide generic results about quantum Fourier models performance
without taking into account the parameterized Fourier coefficients structure in finer detail. In a more recent work \cite{barthe_gradients_2023}, the authors 
explored encoding dependent concentration of the Fourier coefficients in the quantum reuploading scheme under the Haar measure assumption proving a phenomenon of vanishing expressivity for high frequencies in the limit of an infinite depth circuit. In addition, \cite{xiong_fundamental_2023} explored exponential concentration sources in quantum Reservoir computing by  using the Fourier representation. 
In this work, we further explore the degree to which the Fourier coefficients concentrate around their mean under different assumptions on the \hela{trainable unitaries} of the circuit and show that the encoding strategy constraints the Fourier coefficients' variance for arbitrary \hela{trainable unitaries}, limiting their theoretical expressivity.

\section{Framework}\label{sec:Framework}

In this Section, we present the framework used throughout this work. We first describe the considered circuit structure and then recall how to define the Fourier representation of the associated quantum model in Section \ref{subsec:Fourier_Model}. \hela{Moreover}, we introduce the notion of frequency redundancy and how it could be tuned through the choice of the encoding strategy \hela{(i.e. encoding Hamiltonians)}. Finally, in Section \ref{subsec:Mathematical_Framework}, we  discuss key metrics for quantifying the expressivity of quantum models.

\subsection{Quantum Fourier Model}\label{subsec:Fourier_Model}

We consider a standard supervised learning task,
where a parameterized function $f$, called \textit{model},
must be optimized to match targets in a finite
dataset.
 We define \textit{quantum models} on \textit{n} qubits as the family of parameterized functions $f : \mathcal{X} \times \Theta \rightarrow \mathbb{R}$ obtained by measuring the expectation value of some hermitian observable $O$, such that
 \begin{equation}\label{Eq:quantum_Model}
    f(x,\theta) = \bra{0} U(x,\theta)^\dagger O U(x,\theta) \ket{0} \;,
 \end{equation}
  where $U(x,\theta)$ is a $2^n$-dimensional unitary , $\theta \in \Theta$ is the vector of trainable parameters and $x= (x_1,\dots,x_D) \in \mathcal{X} \subset \mathbb{R}^D$ is the classical data vector \hela{with $D$ components}.

We consider a circuit unitary  composed of alternating \textit{encoding} and \textit{trainable} layers as depicted in Fig.$\ref{fig:Figure_Introduction}$ of the form 
\begin{equation}\label{eq:circuit_ansatz}
    U(x,\theta) = W^{L+1}(\theta)\left[\ \prod_{l=1}^L S^l(x)W^l(\theta)\right]\;,
\end{equation}
where $L$ is the total number of circuit layers (i.e. a circuit layer is made of an encoding layer and a trainable layer), $W^l(\theta)$s are formed by trainable gates depending on the parameter vector $\theta$, which is optimized during training. \hela{On the other hand,} $S^l(x)$s only depend on input data values \hela{and are fixed during the training}. \hela{This scheme where the classical data is encoded using $L$ alternated layers was originally proposed in \cite{perez-salinas_data_2020} and the associated model of the form in Eq.\eqref{Eq:quantum_Model} is denoted by quantum reuploading model.}

In the remainder of this work, we adopt the \textit{Hamiltonian encoding} strategy where the classical input components are encoded as the time evolution of some Hamiltonians \hela{$H_l^{(k)}$ such that the encoding unitaries are of the form} $S^l(x)= \prod_{k=1}^D e^{-ix_kH_l^{(k)}}$. From the seminal work \cite{schuld_effect_2021}, we know that if one considers the Hamiltonian encoding strategy, then the quantum model generated by the circuit described in Eq.(\ref{eq:circuit_ansatz})xxx can be written as a truncated Fourier Series with a spectrum $\Omega$ depending on the eigenvalues of the encoding Hamiltonians. The associated Fourier coefficients depend mainly on the trainable unitaries. Under these assumptions, we call the obtained model a quantum Fourier model (QFM), which is expressed as follows

\begin{equation}\label{Eq:quantum_Fourier_Model}
    f(x,\theta) = \sum_{\omega \in \Omega} c_\omega(\theta) e^{i \omega^T x}\;.    
\end{equation}

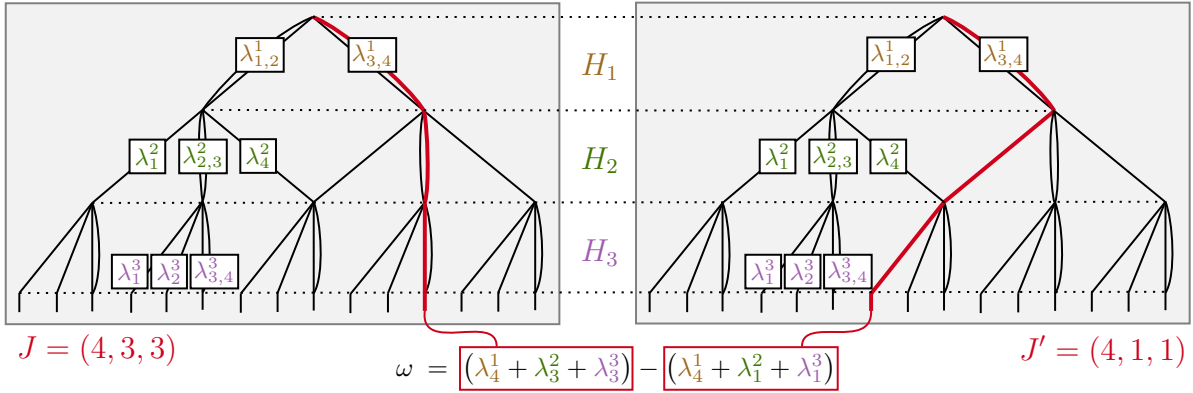
\begin{figure*}[tp]
    \centering
    \input{Figures/Quantum_Spectrum_Trees.tikz}
    \caption{\textbf{Quantum spectrum tree.}   The frequencies of a quantum Fourier model are derived from the eigenvalues of the encoding Hamiltonians. Each path in the quantum spectrum tree represents a different choice of eigenvalues $\lambda_{j_\ell}^\ell$ from Hamiltonians $H_\ell$s. Some edges are duplicated, expressing that some eigenvalues are degenerate for some Hamiltonians (not at scale). Each frequency $\omega$ in the model comes from the difference of two paths in the tree, as shown in the example in red.
    For a quantum model acting on $n=2$ qubits with $L=3$ circuit layers and encoding Hamiltonians $H_1, H_2 \text{ and } H_3$, we give in red a pair of paths in the tree $J=(4,3,3), J'=(4,1,1)$ generating the frequency $\omega = (\lambda_4^1 + \lambda_3^2+ \lambda_3^3)-(\lambda_4^1 + \lambda_1^2+ \lambda_1^3)$.
    }    \label{fig:quantum_Spectrum_Trees}
\end{figure*}

The above equation tends to imply that $c_{\omega}(\theta)$ is solely determined by the parameterized unitaries $W^l(\theta)$s. However, in this work, we show that the dependence of the Fourier coefficients on the encoding gates is more subtle. \hela{First, we begin by highlighting} the relation between the frequencies and the encoding Hamiltonian's eigenvalues. \hela{To do so,} we expand here Eq.\eqref{Eq:quantum_Model} in the case of one-dimensional input vectors ($D=1$). \hela{ The generalization of the Fourier expansion to higher dimensional input is detailed in Appendix \ref{app:high_dim}}. 
We denote by $d=2^n$ the dimension of the Hilbert space (with $n$ the number of qubits) and we assume without loss of generality \footnote{One can simply consider that $S^l(x) = P D P^{-1}$ and \emph{inject} $P$ in the expression of $W^l$ and $P^{-1}$ in $W^{l+1}$.} that $S^l(x) = \text{diag}(\lambda^l_1, \dots, \lambda^l_d)$. We also  drop the explicit dependence on $\theta$ in $W^{l}(\theta)$ for ease of notation and obtain
\color{black}
\begin{equation}\label{Eq:quantum_Fourier_Model_developped}
    \begin{aligned}
   c_{J,J'} &= \bra{0}\prod_{l=1}^L W^{l \dagger}\ketbra{j'_l}{j'_l} \widetilde{O} \prod_{l=L}^1 \ketbra{j'_l}{j'_l}W^l \ket{0}\;,\\
   f(x,\theta) &= \sum_{J, J' \in \llbracket 1,d \rrbracket^L} c_{J,J'}  e^{-ix \left(\sum_{l=1}^L \lambda^l_{j_l} -  \lambda^{l}_{j'_{l}}\right)}\;,
    \end{aligned}
\end{equation}
\color{black}
where we introduce the shorthand $\widetilde{O} = W^{L+1 \dagger} O W^{L+1}$. We note that $J=(j_1, \dots, j_L)$ is a multi-index where each component $j_l \in [|1,d|]$ refers to the choice of the $j^{th}$ eigenvalue of the Hamiltonian $H_l$ ($J$ maps to a path in the tree from Fig.\ref{fig:quantum_Spectrum_Trees}).

From Eq.\eqref{Eq:quantum_Fourier_Model_developped}, we see that the spectrum $\Omega$ can be constructed from the eigenvalues of the encoding Hamiltonians in each layer as follows
\begin{equation}
\label{Eq:Spectrum_def}
    \Omega = \left\{ \sum_{j_l \in J} \lambda^l_{j_l} - \sum_{j'_l \in J'} \lambda^{l}_{j'_{l}} \middle| (J,J')  \in \llbracket 1,d \rrbracket^{2L}  \right\}\;.
\end{equation}

We note here that the spectrum $\Omega$ contains redundant frequencies by construction but in the remainder of this work we consider that $\Omega$ denotes the set of distinct frequencies. 

 As shown in Fig.\ref{fig:quantum_Spectrum_Trees}, the choice of two \emph{paths} $(J,J')$ in the quantum spectrum tree leads to the generation of a frequency $\omega$ by computing the difference between the sum of eigenvalues over the pair of paths. One can easily notice that several pairs of paths could lead to the  same frequency. This can happen if an eigenvalue is degenerate, or if several paths of the tree end at the same leaf value (sum of eigenvalues over a path), or eventually if several pairs have the same difference value. The number of those paths evolves with the choice of the different encoding Hamiltonians,  the degeneracy of their eigenvalues, and the number $L$ of circuit layers. 

By grouping paths $(J,J')$ in Eq.\eqref{Eq:quantum_Fourier_Model_developped} leading to a certain frequency, \hela{we identify the Fourier coefficient expression, i.e.  $c_\omega = \sum_{J,J' \in R(\omega)} c_{J,J'}$}. We also formally define the \emph{Frequency Generator} $R(\omega)$ as the set of all paths leading to the generation of the frequency $\omega$ and denote the cardinality of this set by the frequency \emph{redundancy} $|R(\omega)|$.  
As we will demonstrate in Section \ref{sec:Main_Results}, the redundancy of a frequency will have a crucial role in characterizing the expressivity of QFMs.

\begin{definition}[Frequency Generator]\label{def:Redundancy}
    Consider an $L$-layer quantum Fourier model as described in Eq.(\ref{Eq:quantum_Model}-\ref{Eq:Spectrum_def}) \hela{with one-dimensional input vectors ($D=1$)}. For a given frequency $\omega$, we define its generator  $R(\omega)$ as the set of eigenvalue indices leading to the generation of $\omega$. 
    {\small
    \begin{equation}
        R(\omega)\! = \!\left\{ \!\!(J,J')\! \in\! \llbracket 1,d \rrbracket^{2L} \middle| \sum_{j_l \in J}\!\! \lambda^l_{j_l}\! -\!\! \sum_{j'_l \in J'}\! \!\lambda^{l}_{j'_{l}}\! =\! \omega  \!\right\}
    \end{equation}}
    We call the \textit{redundancy} of a frequency $\omega$ the size of its Generator, i.e. $|R(\omega)|$.
\end{definition}

Since $\sum_{\omega \in \Omega} |R(\omega)| = 2^{2nL} = d^{2L}$ by construction, the normalized redundancies $\{|\widetilde{R}(\omega)|\}_{\omega \in \Omega}:= \{\frac{|R(\omega)|}{d^{2L}}\}_{\omega \in \Omega}$ define a natural weighted probability distribution over the spectrum $\Omega$.
Therefore, by considering different encoding Hamiltonians, one can obtain different probability distributions over the spectrum that will impact the behavior of the associated quantum model.

For example, Let's consider the standard case of Pauli encoding \cite{schuld_effect_2021}, where single qubit rotation gates are used to encode the classical input $x \in \mathbb{R}$ as the rotation angle. In this case, the encoding Hamiltonian in each layer is a Pauli string. If the Pauli strings do not contain the identity, then the obtained spectrum is simply $\Omega = \llbracket -nL,nL \rrbracket$. Moreover, \hela{it has been shown} that the spectrum distribution defined by the redundancies follows a standard Gaussian distribution \cite{peters_generalization_2023}. Hence, this encoding strategy gives rise to a linear size spectrum (linear in $n$ and $L$) and concentrates the redundancies in the lower values as detailed in Appendix \ref{appendix:spec_distribution}.  

On the contrary, the exponential encoding strategy introduced in \cite{shin_exponential_2022}, which uses scaled Pauli rotations (See Appendix \ref{appendix:spec_distribution}), leads to an exponential size spectrum of consecutive integer frequencies.   Specifically, the obtained spectrum is $\Omega = \left\llbracket -\frac{3^{nL} -1}{2}, \frac{3^{nL} -1}{2} \right\rrbracket$ and some frequencies (not necessarily high frequencies) have redundancies that do not scale exponentially in $n$ and $L$. However, if one wants to obtain a fully non-degenerate spectrum (except for the null frequency), then a single circuit layer made of a non-local encoding Hamiltonian must be used as mentioned in \cite{shin_exponential_2022} and explained in Appendix \ref{appendix:spec_distribution}.
This is the case for the \emph{Golomb} encoding introduced in \cite{peters_generalization_2023} where the size of the spectrum is exponentially large ($|\Omega| = 2 \binom{d}{2} + 1$) and all non zero coefficients have a redundancy of one. A non-degenerate spectrum can also be obtained using a global Hamiltonian with distinct real eigenvalues.

\begin{figure*}[tp]
    \centering
    \input{Figures/Redundancy_Examples.tikz}
    \caption{\textbf{Spectrum distributions.} We compare two Hamiltonian encoding strategies leading to very different spectrums (x-axis) and distributions (normalized height of the bars). On the left, we present an illustration of a highly degenerate spectrum (e.g. Pauli encoding spectrum distribution)
    , and on the right an illustration of a weakly degenerate spectrum (e.g.  Golomb encoding spectrum distribution).}
    \label{fig:Pauli_Exp_encoding_distribution}
\end{figure*}
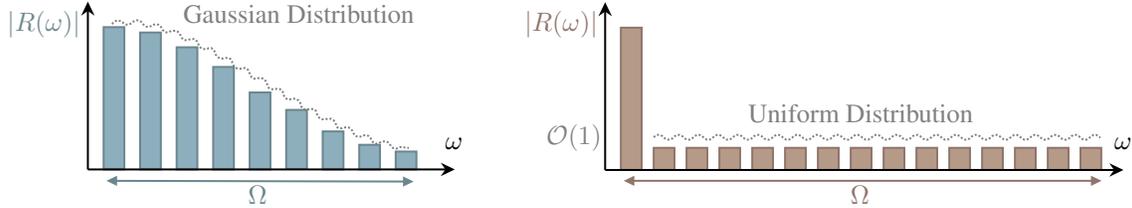

As illustrated in Fig.\ref{fig:Pauli_Exp_encoding_distribution}, one could choose a particular set of Hamiltonians to design a quantum model with a specific spectrum distribution. In this work, we will show that this choice does not only impact the spectrum of the quantum Fourier model but also the Fourier coefficients concentration. In addition, previous work \cite{landman_classically_2022} showed the possibility of classically approximating such PQCs for Machine Learning tasks. While having an exponential size spectrum may be a leeway to avoid this classical approximation, we will show in Section \ref{sec:Main_Results} that models with a large spectrum tend to have more constrained Fourier coefficients, hence limiting their expressivity and making their classical approximation potentially more efficient.

\subsection{Expressivity measure of  quantum Fourier models}\label{subsec:Mathematical_Framework}

\hela{In this section, we discuss relevant expressivity measures of quantum models of the form in Eq.\eqref{Eq:quantum_Model} using a parameterized circuit of the form in Eq.\eqref{eq:circuit_ansatz} equipped with Hamiltonian encoding. Namely, in the context of supervised learning, the model expressivity is tightly related to the class of functions (in the input data) it can generate. Consequently, it follows naturally that the model  dependence on classical input parameters and trainable parameters is conceptually different.  As shown in Eq.\eqref{Eq:quantum_Fourier_Model}, the choice of the encoding strategy dictates the general model structure expressed as a truncated Fourier series in the classical input $x \in \mathcal{X}$.
Hence, the model expressivity boils down to identifying the spectrum $\Omega$  and characterizing the attainable Fourier coefficients values as the trainable parameters $\theta \in \Theta$ vary.  This definition of model expressivity can invoke unitary expressivity measures such as the distance of the trainable unitaries distribution (induced by the trainable parameter distribution) to a 2-design on the unitary group \cite{holmes_connecting_2022,sim_expressibility_2019}. Nevertheless,  properly characterizing the expressivity of the quantum model requires further analysis of how the use of \textit{expressive} trainable unitaries $W^l(\theta)$s translates into a measure of ``expressivity'' of Fourier coefficients for a fixed spectrum $\Omega$.}
\hela{To do so, we first recall the definition of a unitary approximate 2-design and then define its plausible consequences on Fourier models expressivity. Specifically, we introduce the notion of model exponential concentration and vanishing expressivity in Definition \ref{def:Model_Concentration} and Definition \ref{def:Vanishing_Expressivity} respectively.}

By restricting our attention to the family of parametrized unitaries of the form $\{W^l(\theta)\}_{\theta \in \Theta}$ associated to each trainable unitary in the circuit expression in Eq.\eqref{eq:circuit_ansatz}, this family of unitaries is said to form an $\varepsilon$-approximate  2-design if its distribution over the unitary group induced by the uniform distribution over the parameter space $\Theta$ matches the Haar measure up to second moment for some error $\varepsilon$. 
Formally, defining the $2^{\text{nd}}$ moment superoperator of the distribution over  $\{W(\theta)\}_{\theta \in \Theta}$ as
    \begin{equation}
        M^{(2)}_{W(\Theta)} = \int_{Unif(\Theta)} dW(\theta) W(\theta)^{\otimes 2} \otimes (W(\theta)^\ast)^{\otimes 2}\;, 
    \end{equation}
the $\varepsilon$ deviation from a 2-design is given by some norm\footnote{Different norms can be used to characterize the $\varepsilon$-distance to a 2-design. They are indeed equivalent up to exponential factors in the number of qubits \cite{low_pseudo-randomness_2010}.} of the superoperator
\begin{equation}\label{Eq:Superoperator_A_2-design}
        \mathcal{A}^{(2)}_{W(\Theta)} = M^{(2)}_{W_H} - M^{(2)}_{W(\Theta)}\;,
    \end{equation}
where $M^{(2)}_{W_H}$ and  $ M^{(2)}_{W(\Theta)}$ are the second moments of the Haar distribution and the parametrization induced unitary distribution respectively. An exact 2-design is then achieved when $\|\mathcal{A}^{(2)}_{W(\Theta)}\| = 0$ and the family of parameterized unitaries is considered to be \textit{maximally} expressive.
However, such an assumption on the \hela{trainable unitaries} leads to the model \emph{exponential concentration} phenomena where the variance of the model vanishes exponentially in the number of qubits \cite{arrasmith_equivalence_2022,mcclean_barren_2018,holmes_connecting_2022}.
For completeness, we formally define the model exponential concentration as follows.

\begin{definition}[Model Exponential Concentration]\label{def:Model_Concentration}
    Consider a quantum model $f(x,\theta)$ \hela{of the form} in Eq. \eqref{Eq:quantum_Model} and assume that the trainable parameters are sampled uniformly over $\Theta$. The model is said to exhibit an exponential concentration phenomenon for the parameter hyperspace $\Theta$ when
    \begin{equation}
        \Var_{\theta \sim \Theta}[f(x,\theta)] = \mathcal{O}\left(\frac{1}{b^n}\right) \;, \forall x \in \mathcal{X}
    \end{equation}
    for some constant $b > 1$.
\end{definition}

\hela{The above definition of model concentration is defined independently of its Fourier decomposition. Nevertheless, we can obtain a similar phenomenon at the level of the model's Fourier coefficients. Specifically, one can study how the variance of each Fourier coefficient (with respect to the uniform distribution over the trainable parameter hyperspace) decays in the number of qubits. 
Notably, we say that a QFM suffers from \textit{vanishing expressivity} if some or all of its Fourier coefficients are exponentially concentrated around their mean.}

\begin{definition}[Vanishing Expressivity]\label{def:Vanishing_Expressivity}
    Consider a quantum Fourier model of the form in Eq.~\eqref{Eq:quantum_Model} with the Fourier decomposition in Eq. \eqref{Eq:quantum_Fourier_Model} and a spectrum $\Omega$. Let us also assume that the trainable parameters are sampled uniformly. The Fourier model is said to suffer from vanishing expressivity when some Fourier coefficients have an exponentially vanishing variance, i.e.
    \begin{equation}
        \exists \; \Omega_{\text{vanish}}\! \subset \Omega \ |  \ \forall \omega \!\in\! \Omega_{\text{vanish}}\ \Var_{\theta}[c_{\omega}(\theta)]\! =\! \mathcal{O}\!\left(\frac{1}{b^n}\right)
    \end{equation}
    for some constant $b > 1$.
\end{definition}

\hela{Although the above definition of vanishing expressivity resembles the model concentration phenomenon introduced in Definition \ref{def:Model_Concentration}, they do indeed differ on an operational level.}
Specifically, the model exponential concentration implies that an exponential number of shots is required to resolve exponentially small changes in the model \hela{for a fixed input $x \in \mathcal{X}$}. This makes optimization inefficient and non-scalable when considering random initialization \cite{larocca_review_2024}. However, the \textit{vanishing expressivity} phenomenon introduced in Definition \ref{def:Vanishing_Expressivity} does not directly impact the model's trainability guarantees. This is because,  in practice, we do not measure individual Fourier coefficients or use them to compute gradients. Nevertheless, one may still wonder if the trained model will be able to reach an initially exponentially vanishing Fourier coefficient. If such a correlation exists between initially  vanishing Fourier coefficients and their reachability in the final trained model, it does justify the \textit{vanishing expressivity} connotation. While we do not provide analytical evidence of this correlation, we do support our intuition  with numerical simulations, showing that vanishing coefficients are hard to train in Section \ref{subsec:train_impact}.
 
\section{Main Results}\label{sec:Main_Results}

In this section, we present our main theorems and corollaries on expressivity constraints in quantum Fourier models.
Specifically, we study the concentration of Fourier coefficients by computing their variance under different assumptions about the trainable unitaries distribution. We then show that the variance is always constrained by the frequency redundancy and that some Fourier coefficients may exhibit an exponential concentration phenomenon, leading to  \textit{vanishing expressivity}. 

To do so, we start by considering the global 2-design hypothesis on the trainable unitaries and provide an exact expression of the Fourier coefficients variance in Theorem \ref{thm:single_layer_gloabl_2design} for a single-layer model and in Theorem \ref{thm:formal_2design_reup}  for a reuploading model. 
Secondly, we relax the global 2-design assumption and provide an upper bound on  Fourier coefficients variance under the $\varepsilon$-approximate 2-design hypothesis in Theorem \ref{thm:bound_approx_2design}. Finally, we consider the brick-wise circuit architecture with local 2-design blocks and give an upper bound on the variance in this setting. This circuit architecture falls within the $\varepsilon$-approximate 2-design assumption, but with more structure that allows to take into account the locality of the observable.

We note that in the remainder of the manuscript, we focus on one-dimensional input vectors $(D=1)$. \hela{However, our results can be easily extended to the high-dimensional setting if we assume that the encoding unitaries within a single layer commute, as detailed in Appendix \ref{app:high_dim}. Beyond this assumption, the analysis becomes more intricate and we do not cover it in this work.}

    \subsection{Trainable layers as global 2-design}\label{subsec:Global_2-design}
As described in Section \ref{subsec:Mathematical_Framework},
in the unitary space, the expressivity of \hela{trainable unitaries} is often characterized by how uniformly they  explore the unitary group and \hela{a parametrized unitary} is said to be maximally expressive if its distribution approximates the Haar measure.
However, it has been shown in \cite{arrasmith_equivalence_2022,mcclean_barren_2018}, that the quantum model and its gradient exhibit an exponential concentration phenomena under the 2-design assumption, resulting in an unexpressive model in practice.

Here, we explore the implications of considering maximally expressive  \hela{trainable unitaries} (i.e. each of the  \hela{trainable} layers forms an exact 2-design) on the Fourier coefficients variance. 

First, we present an exact expression of the Fourier coefficients variance for a QFM with a single circuit layer ($L=1$) in Theorem \ref{thm:single_layer_gloabl_2design}. We then extend the result to a reuploading model with $L \geq 1$ in Theorem \ref{thm:formal_2design_reup}, Appendix \ref{app:local2design}.

    \begin{theorem}[\hela{Fourier coefficients variance with 2-design trainable unitaries, Informal}]\label{thm:single_layer_gloabl_2design}

    Consider a quantum model of the form in Eq.\eqref{Eq:quantum_Model} and a parametrized circuit of the form in Eq.\eqref{eq:circuit_ansatz} with $L=1$ layers and fixed encoding Hamiltonians resulting in a spectrum $\Omega$. We assume that each of the  trainable layers $W^l(\theta)\;,l \in \{1,2\}$ form independently a 2-design. The expectation and variance of each Fourier coefficient $c_{\omega}(\theta)$ for the frequencies $ \omega \in \Omega$ appearing in the model Fourier decomposition in Eq.\eqref{Eq:quantum_Fourier_Model} are given by
\begin{equation}\label{eq:coeff_variance_2design_single}
    \begin{aligned}
        \E_{\theta}[c_{\omega}(\theta)]& = \quad\frac{Tr(O)}{d}\delta_{\omega}^0\;,\\
        \Var_{\theta}[c_{\omega}(\theta)] &\in \Theta\left(	\alpha \frac{|\widetilde{R}(\omega)|}{d} - \frac{\alpha}{d^2} \delta_{\omega}^0\right)\;.
    \end{aligned}
\end{equation}
Here we recall that $d=2^n$ and introduce the normalized frequency redundancy $|\widetilde{R}(\omega)| := |R(\omega)|/d^2$  and the constant $\alpha:=(d||O||_2^2-Tr(O)^2)/d^2$ which depends on the observable $O$.

\end{theorem}

Theorem \ref{thm:single_layer_gloabl_2design} establishes that, \hela{under the 2-design assumption on the model's trainable unitaries}, the variance of a Fourier coefficient depends linearly on its (normalized) frequency redundancy up to some prefactor. \hela{Here we recall that the normalized frequencies sum up to one. Thus, Theorem \ref{thm:single_layer_gloabl_2design} implies that while frequencies with high redundancies exhibit a relatively large variance,  the ones with low redundancies are way more constrained. }
In other words, the distribution of the Fourier coefficients is dictated by the redundancies and hence by the encoding Hamiltonians. In Section \ref{sec:Simulations}, we illustrate this dependence through the numerical study of two \hela{models with different encoding Hamiltonians}  corresponding to  spiked and  flat frequency distributions (i.e. defined by normalized redundancies over the model's spectrum). 
\hela{ Moreover, we generalize   the result of Theorem \ref{thm:single_layer_gloabl_2design} to the setting of reuploading models with $L \geq 1$ alternating layers in Theorem \ref{thm:formal_2design_reup}. Similarly, we prove that a Fourier coefficient variance is linear in its frequency redundancy.
The formal version of Theorem \ref{thm:single_layer_gloabl_2design}, Theorem \ref{thm:formal_2design_reup} and their proof are presented in Appendix \ref{Proof:Global_2design}.}

\hela{Apart from the linear dependence of the Fourier coefficient variance on their respective redundancies,  one can also deduce the decay rate of the Fourier coefficients variance from their exact expressions given in  Theorem \ref{thm:single_layer_gloabl_2design} and Theorem \ref{thm:formal_2design_reup}. This is captured via Corollary \ref{Cor:2design_vanishing}.}

\begin{corollary}[\hela{Fourier coefficients variance decay with 2-design trainable unitaries, Informal}]\label{Cor:2design_vanishing}
 Consider a quantum model of the form in Eq.\eqref{Eq:quantum_Model} and a parametrized circuit of the form in Eq.\eqref{eq:circuit_ansatz} with $L\geq 1$ layers and fixed encoding Hamiltonians resulting in a spectrum $\Omega$. We assume that each of the trainable layers $W^l(\theta)$ form independently a 2-design. The variance of each Fourier coefficient $c_{\omega}(\theta) $ for the frequencies $ \omega \in \Omega$ appearing in the model Fourier decomposition in Eq.\eqref{Eq:quantum_Fourier_Model} is upper bounded by
\begin{align}
   \Var_{\theta}[c_{\omega}(\theta)] &\in \mathcal{O} \left(\alpha \frac{|\widetilde{R}(\omega)|}{d}\right)\;,
\end{align}
where we recall that $d=2^n$ and $\alpha$ is a constant given by $\alpha:=(d||O||_2^2-Tr(O)^2)/d^2$.
\end{corollary}

From Corollary \ref{Cor:2design_vanishing}, 
one can see that, under reasonable assumptions on the observable norm, all Fourier coefficients variance decay exponentially in the number of qubits. \hela{Specifically, the prefactor $\alpha$ can be bounded by a constant for any  observable satisfying $\norm{O}_2^2 \in \mathcal{O}(d)$. Additionally, the normalized redundancy $|\widetilde{R}(\omega)|$ is, by definition, bounded by one. This implies that irrespective of the frequency redundancy and thus of the encoding strategy, all coefficients concentrate exponentially toward their mean value. }
This result can be viewed as an exponential concentration statement of each Fourier coefficient in a reuploading model,
aligning with results in \cite{arrasmith_equivalence_2022} about the exponential concentration of the model under the 2-design assumption.

While it is useful to show the connection between the Fourier coefficients and the spectrum redundancies, considering global 2-design is a strong assumption leading to the model's exponential concentration. In practice, it is improbable that one will use trainable layers forming 2-design for learning purposes. Thus we propose in the following to relax this hypothesis by first considering trainable layers as forming approximate 2-design and then as made of local 2-design blocks and that for models with a single layer $(L=1)$.

    \subsection{Trainable layers as global \texorpdfstring{$\varepsilon$}{epsilon}-approximate 2-design}\label{subsec:approx_2-design}

    Let us now consider the broader setting \hela{where the trainable unitaries $W^l(\theta)$ form \hela{each} an $\varepsilon$-approximate 2-design.}
    By moving away from highly expressive  \hela{trainable unitaries}, one might question if it is feasible to break free from the constraining redundancy dependence of the Fourier coefficients variance established in Theorem \ref{thm:single_layer_gloabl_2design}, or if such dependency is an inductive bias of the quantum model \hela{which still hold even when the trainable unitaries are not maximally expressive (i.e. do not form exact 2-designs).}

    To do so, we further build on results from \cite{holmes_connecting_2022} about the model concentration \hela{for approximate 2-design unitaries} and explore encoding dependent concentration for single Fourier components, giving a finer interpretation of the model's expressivity through the Fourier lens.

     In the following theorem, we provide an upper bound on the Fourier coefficients variance for a single-layer circuit formed by arbitrary trainable layers.  

    \begin{theorem}[\hela{Fourier coefficients variance decay with approximate 2-design trainable unitaries, Informal}]\label{thm:bound_approx_2design_informal}
       Consider a quantum model of the form in Eq.\eqref{Eq:quantum_Model} and a parametrized circuit of the form in Eq.\eqref{eq:circuit_ansatz} with $L=1$ layers and fixed encoding Hamiltonians resulting in a spectrum $\Omega$. We assume that each of the  trainable layer $W^l(\theta)\;,l \in \{1,2\}$ form independently an $\varepsilon$-approximate 2-design. The  variance of each Fourier coefficient $c_{\omega}(\theta)$ for the frequencies $ \omega \in \Omega$ appearing in the model Fourier decomposition in Eq.\eqref{Eq:quantum_Fourier_Model} is upper bounded as
       \begin{equation}
           \Var[c_\omega] \in \mathcal{O}(Q_\varepsilon(|\widetilde{R}(\omega)|))\;,
       \end{equation}
       where $Q_\varepsilon$ is a polynomial of degree at most 2 in the normalized frequency redundancy  $|\widetilde{R}(\omega)|$ defined for different $\varepsilon$ measures as

\begin{align}
     Q_{\varepsilon_\diamond}&=  \|O\|_2^2|\widetilde{R}(\omega)|\varepsilon_{\diamond}+\|O\|_1^2 \varepsilon_{\diamond}^2  \;,\label{eq:diamond_norm_bound}\\
      Q_{\varepsilon_\infty}&= \frac{ \|O\|_2^2}{d}\sqrt{|\widetilde{R}(\omega)|}\varepsilon_{\infty} +d^2\|O\|_2^2 |\widetilde{R}(\omega)|\varepsilon_{\infty}^2\;,\label{eq:infty_norm_bound}\\
    Q_{\varepsilon_M}&= \|O\|_2^2 |\widetilde{R}(\omega)| \varepsilon_M+ d^2  \|O\|_2^2|\widetilde{R}(\omega)|^2\varepsilon_M^2\;. \label{eq:monomial_norm_bound}
\end{align}
  
\noindent Here, we use the shorthand $\varepsilon_{\diamond}:= \|\mathcal{A}^{(2)}\|_{\diamond}$ for the diamond norm defined in Eq.\eqref{eq:diamond_norm}, $\varepsilon_{\infty} :=\norm{\mathcal{A}^{(2)}}_{\infty}$ for the spectral norm and $\varepsilon_{M}:= d^2 max_{i,j} |\mathcal{A}^{(2)}|_{i,j}$. \hela{We also recall that  $\mathcal{A}^{(2)}$ is a superoperator  defined in Eq.\eqref{Eq:Superoperator_A_2-design}.} 
\end{theorem}

Theorem \ref{thm:bound_approx_2design} shows that
the variance of a Fourier coefficient in the approximate 2-design setting is constrained by the combined action of the normalized frequency redundancy $|\widetilde{R}(\omega)|$ and the $\varepsilon$-distance of the  \hela{trainable} unitaries to a 2-design. 
Specifically, for a fixed choice of the  \hela{trainable} unitaries distribution and thus for a fixed $\varepsilon$ value, the degree to which each Fourier coefficient concentrates around its mean is  \hela{constrained} by its corresponding \hela{normalized} frequency redundancy. Therefore, we prove that  \hela{the vanishing expressivity phenomenon, whereby some Fourier coefficients exhibit exponentially decaying variance,} may still hold beyond the 2-design assumption. 
We note that the bounds in Eqs.\eqref{eq:diamond_norm_bound}-\eqref{eq:monomial_norm_bound} correspond to different norms used to quantify the distance from a 2-design. \hela{These $\varepsilon$-distance definitions} are equivalent \hela{up to some prefactors \cite{low_pseudo-randomness_2010}} and we include all of them because one bound may be tighter than the other depending on the \hela{interplay between the} observable norm\footnote{We consider Schatten $p$-norms defined as $\norm{O}_p := \left(Tr\left[\left(\sqrt{O^\dagger O}\right)^p\right]\right)^{1/p}$.}, the frequency redundancy and the $\varepsilon$-distance scalings. Precisely, while $\varepsilon_{\infty}$  saturates at 1 and $\varepsilon_{\diamond}$ at 2, the monomial-based epsilon $\varepsilon_{M}$ can take up values up to $d^2$. \hela{Additionally, we recall that the normalized frequencies $|\widetilde{R}(\omega)|$ take up values within $[1/d^2,1]$. Hence, the bounds can be used to prove the vanishing expressivity phenomenon introduced in Definition \ref{def:Vanishing_Expressivity} for frequencies with relatively low redundancies. Namely, a frequency with a  normalized redundancy that counterbalance the observable norm will exhibit an exponential decay on average over trainable unitaries forming an approximate to 2-design.  } In section \ref{sec:Discussion}, we further comment on the scalings of the upper bound in Eq.\eqref{eq:monomial_norm_bound} and its dependence on the observable and encoding strategy.
We provide the proof of Theorem \ref{thm:bound_approx_2design_informal} and its formal version in Appendix \ref{Proof:approx_2design} for different norms and numerically evaluate  the Monomial bound in Eq.\eqref{eq:monomial_norm_bound} for different encoding strategies with different spectrum distributions in Section \ref{subsec:Simu_Approximate_2design}.


 \hela{Here we note that the upper bounds in Theorem \ref{thm:bound_approx_2design_informal} are looser for local observables compared to global ones, in general.} 
 This observation does not come as a surprise since the obtained bound as a function of the \textit{global} $\varepsilon$ expressivity measure of the circuit does not capture the observable-circuit interaction in finer detail. Specifically, the interaction between an $m$-local observable and the remainder of the circuit is captured by the backward light cone of the observable, i.e., the sub-circuit containing all blocks with at least one qubit causally connected to the local observable input qubits. In the next section, we explore the variance of the Fourier coefficients by taking into account the observable locality.

    \subsection{Trainable layers as local 2-design blocks}\label{subsec:local_2-design}
    
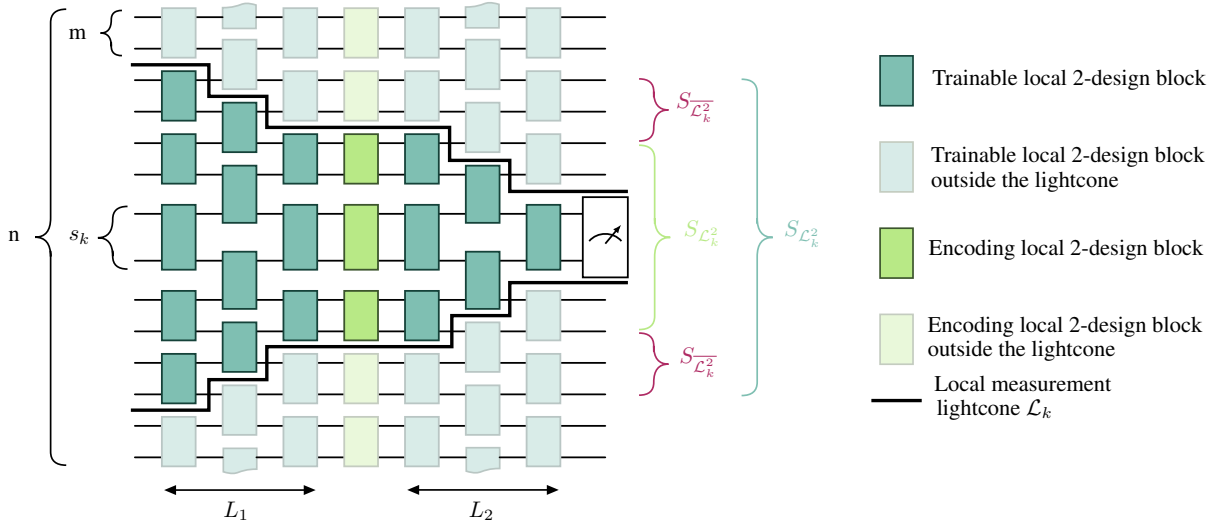
\begin{figure*}[tp]
    \centering
    \resizebox{\textwidth}{!}{\input{Figures/Framework_Local2design_Adapted.tikz}}
    \caption{\textbf{Brickwise circuit architecture made of local blocks acting on neighboring qubits.} As shown, $L_1$ is the depth of the pre-encoding trainable block and $L_2$ is the depth of the post-encoding one. 
    We consider an m-local observable acting non-trivially on subsystem $s_k$ and we denote its backward light cone by the subcircuit $\mathcal{L}_k$. We denote by $\mathcal{S}_{\mathcal{L}_k}$ the subsystem on which the backward light cone acts non-trivially and $\mathcal{S}_{E_k}$ the subspace on which the encoding layer (made of the green blocks) acts non trivially inside the light cone. we also define $\mathcal{S}_{\overline{E_k}}$ as the compliment of $\mathcal{S}_{E_k}$ in $\mathcal{S}_{\mathcal{L}_k}$.}
    \label{fig:Figure_Framework_Local2design}
\end{figure*}

 In this section, we consider a brickwise circuit architecture formed by trainable local 2-design blocks and local encoding blocks, which has been previously studied in \cite{cerezo_cost_2021}. As depicted in Fig.\ref{fig:Figure_Framework_Local2design}, the $n$-qubit circuit is made of layers of $m$-qubit unitaries (\hela{trainable} and encoding unitaries) acting on alternating groups of $m$ neighboring qubits. We consider that each of the  \hela{trainable} blocks forms an exact 2-design on the $m$-qubit subsystem on which it acts non-trivially.
 This setting is a special case of the \textit{global} $\varepsilon$-approximate 2-design \cite{harrow_approximate_2023}. However, it will give us more accurate results when one considers local observables acting non-trivially on an \hela{ $m$-qubit} subsystem $s_k$  of the form $O= \hat{O}_{s_k} \otimes \mathbb{1}_{\overline{s_k}} $. Indeed, by considering this circuit architecture, the backward \textit{causal light cone} of such local observables is well defined and one can easily notice that the effective model spectrum will be reduced. In this setting, we derive the expression of the Fourier coefficients variance in Appendix \ref{proof:local2design_coeff}. However,  since the obtained expression is quite cumbersome, we present in the following theorem an upper bound on the Fourier coefficients variance for two different assumptions on the local observable.

\begin{theorem}[\hela{Fourier coefficients variance decay with brickwise local 2-design circuit}]\label{thm:bound_local2design}
 Consider a quantum model of the form in Eq.\eqref{Eq:quantum_Model} and a parametrized circuit of the form in Eq.\eqref{eq:circuit_ansatz} using a brickwise architecture with $L=1$ layers and observable $O= \hat{O}_{s_k} \otimes \mathbb{1}_{\overline{s_k}} $ acting non trivially on the $m$-qubit subsystem $s_k$. Assume that each trainable $m$-qubit unitary forms a local 2-design. The  variance of each Fourier coefficient $c_{\omega}(\theta)$ for the frequencies $ \omega \in \Omega$ appearing in the model Fourier decomposition in Eq.\eqref{Eq:quantum_Fourier_Model} is upper bounded as
\begin{enumerate}
    \item If $||\hat{O}_{s_k}||_2^2 \leq 2^m$ , we have
\begin{equation}\label{eq:cor_bound_local_2design_Pauli}
    \Var[c_{\omega}] \leq \left(  \frac{2^{m+1}}{2^{2m}-1}\right)^{2L_2} |R_{E_k}(\omega)|^2.
\end{equation}
\item If $\hat{O}_{s_k}$ is a projector of rank $r$ , we have
\begin{equation}\label{eq:cor_bound_local_2design_proj}
    \Var[c_{\omega}] \leq \left(  \frac{2^{m+1}}{2^{2m}-1}\right)^{2L_2} \left(\frac{r}{2^m}\right)^2 |R_{E_k}(\omega)|^2.
\end{equation}
\end{enumerate}
Here $R_{E_k}(\omega)$ is the frequency generator obtained from the encoding blocks inside the observable backward light cone $\mathcal{L}_k$ (acting non trivially on $\mathcal{S}_{E_k}$) and $L_2$ is the depth of the post-encoding parameterized block.
\end{theorem}

Theorem \ref{thm:bound_local2design} gives us an upper bound on the variance of the Fourier coefficients while considering circuits made of local 2-design blocks. Once again, we observe that this quantity is constrained by the frequency redundancy. In addition, this result indicates that the vanishing Fourier coefficient phenomenon could depend on the circuit depth while considering a local observable.
\hela{Moreover, the bound in Theorem \ref{thm:bound_local2design} implies that a frequency with a relatively low frequency  (i.e. $|R_{E_k}(\omega)|= \mathcal{O}(1)$) will suffer from exponentially vanishing variance for a depth $L_2$ linear in $n$.} The proof of this Theorem is given in Appendix \ref{app:local2design}.

\section{Discussing the quantum Fourier model constraints}\label{sec:Discussion}

In this work, we established a connection between the spectrum redundancies and the statistical behaviour of Fourier coefficients for arbitrary \hela{trainable unitaries, \textit{on average}}. Namely, we showed an \textit{inductive bias} of the Fourier model where the variance of a Fourier coefficient is upper bounded by a polynome in its redundancy. we further introduced the concept of \textit{vanishing expressivity} where the variance of some Fourier coefficient is exponentially vanishing in the number of qubits. 
In this section, we further discuss these  phenomena and study their implications on \hela{model} design guidelines in \ref{subsec:Vanishing_Coeff_Cost}.
In addition, we provide a generic bound on the 2-norm of the Fourier coefficients vector and briefly discuss controllability-related constraints on the Fourier coefficients in \ref{subsec:MassConservation_Correlation_Control}. 
Finally, we discuss the limitations of the framework and the assumptions we considered and thus the limitations of the obtained results in \ref{subsec:Limitation_Framework}.

    \subsection{Vanishing Fourier coefficients and vanishing model}\label{subsec:Vanishing_Coeff_Cost}
    In this section, we discuss the \textit{vanishing expressivity} phenomenon, whereby Fourier coefficients variance decay exponentially in the system size. \hela{Specifically, we further discuss the scaling of the upper bounds established in Theorem \ref{thm:bound_approx_2design_informal} with respect to the different quantities of interest. Moreover, we relate this analysis of the Fourier coefficients decay to the analysis of the full model decay. Indeed,  we previously stressed in Section \ref{sec:Framework} that the vanishing expressivity phenomenon, whereby some Fourier coefficients exhibit exponentially decaying variance on average, is conceptually different from the model's exponential concentration introduced in Definition \ref{def:Model_Concentration}. Nevertheless, one may still wonder if these two phenomena are equivalent or if one implies the other.}

    \hela{ Under the 2-design assumption on the trainable unitaries, we showed in Theorem \ref{thm:single_layer_gloabl_2design} that the model's exponential concentration goes in hand with the exponential decay of all Fourier coefficients independently of the encoding strategy as detailed in Corollary \ref{Cor:2design_vanishing}. However, when using approximate 2-design trainable layers, the link is not trivial.}
    To better understand the relation between the model and Fourier coefficients exponential decay beyond the 2-design assumption, we provide in the following Corollary an upper bound on the model's variance when using approximate 2-design trainable layers. Ultimately, we are interested in identifying regimes where the model's variance is not exponentially vanishing whereas all or some of its Fourier coefficients suffer from exponential concentration.

\begin{corollary}\label{cor:bound_model_approx_2design_informal}
      Consider a quantum model  $f(x,\theta)$ of the form in Eq.\eqref{Eq:quantum_Model} and a parametrized circuit of the form in Eq.\eqref{eq:circuit_ansatz} with $L=1$ layers and fixed encoding Hamiltonians. We assume that each of the  trainable layer $W^l(\theta)\;,l \in \{1,2\}$ form independently an $\varepsilon_M$-approximate 2-design according to the monomial definition introduced in Definition \ref{def:monomial_norm}. 
     For a fixed $x \in \mathcal{X}$, the variance of the model $f(x,\theta)$ is upper bounded as
\begin{equation}
      \Var_{\theta}[f(x,\theta)] \in \mathcal{O}\left(\norm{O}_2^2 \varepsilon_M\right)\;.
\end{equation}
\end{corollary}

Corollary \ref{cor:bound_model_approx_2design_informal} establishes an upper bound on the full model variance similar to the one given in Theorem \ref{thm:bound_approx_2design} for each Fourier coefficient. \hela{ We note that this results have been established in previous works \cite{larocca_diagnosing_2022,larocca_review_2024,holmes_connecting_2022} but we adapt it here to the monomial distance $\varepsilon_M$ to a 2-design defined in Eq.\eqref{eq:monomialNorm}.} The proof of this Corollary is detailed in Appendix \ref{Proof:approx_2design_f}.

By combining Theorem \ref{thm:bound_approx_2design_informal} and Corollary \ref{cor:bound_model_approx_2design_informal}, we can capture scenarios where relatively low redundant frequencies are vanishing whereas there is a leeway for the global model to not be. 

\hela{Specifically, if we consider frequencies with redundancies scaling at most polynomially in system size, i.e. $|R(\omega)| \in \mathcal{O}(poly(n))$ (or equivalently $|\widetilde{R}(\omega)| \in \mathcal{O}\left(\frac{poly(n)}{d^2}\right)$), then the corresponding coefficient variance upper bound in Eq.\eqref{eq:monomial_norm_bound} scales as} 
\begin{equation}
    \Var[c_\omega] \in \mathcal{O} \left(poly(n)\frac{\ \norm{O}_2^2 \varepsilon_M}{d^2}\left(1+\varepsilon_M\right)\right)\;.
\end{equation}
\hela{This implies that as long as $\norm{O}_2^2 \varepsilon_M \in \mathcal{O}(poly(n))$, the Fourier coefficient with redundancies $|R(\omega)| \in \mathcal{O}(poly(n))$ suffer from exponentially decaying variance. On the other hand, Corollary \ref{cor:bound_model_approx_2design_informal} provides guarantees of the exponential decay of the full model only when $\norm{O}_2^2 \varepsilon_M \in \mathcal{O}\left(1/d\right)$.} Consequently, we see that there may be a leeway for the global model to be non vanishing while Fourier coefficients with polynomially big redundancies suffer from exponential concentration for a reasonably wide $\varepsilon_M$ range as depicted in Fig.\ref{fig:Figure_Vanishing_Cost_Coeffs_Epsilon}.

 \hela{A straightforward construction of an \textit{expressive} model} is to consider an encoding strategy where the size of the spectrum is exponential in the number of qubits and hence less prone to classical dequantization \cite{landman_classically_2022,sweke_potential_2023}. This implies that the spectrum is weakly degenerate with many frequencies $\omega$ such that $|R(\omega)|=\Theta(1)$. This is indeed the case for the exponential encoding and the Golomb encoding (See Appendix \ref{appendix:spec_distribution}). 
In the latter setup, we have that $|R(\omega)|=1 \quad \forall \omega \in \Omega^{\ast}$. 

\begin{figure*}[t]
    \centering
    \input{Figures/bar.tikz}
    \caption{\textbf{Illustration of the vanishing model and vanishing Fourier coefficient phenomena according to the $\varepsilon_M$ distance to a 2-design}. Considering  coefficients $c_{\omega}$ with redundancies $|R(\omega)|= \mathcal{O}(poly(n))$, one can notice that the vanishing expressivity phenomenon can happen outside of the regime with guaranteed exponential concentration of the full model (yellow part). Specifically, the dashed gray part corresponds to $\varepsilon_M$ range where the quantum model $f$ is proven to be vanishing while the black one corresponds the the vanishing Fourier coefficients regime. The yellow part indicates the regime where the Fourier coefficients are vanishing but not necessarily the case for the corresponding model. Finally, the behavior of the model and its coefficients is unknown in the blue part.}
    \label{fig:Figure_Vanishing_Cost_Coeffs_Epsilon}
\end{figure*}

Consequently, although the quantum model has theoretically access to an exponential number of frequencies, the contribution of each frequency is vanishing.

 When considering general encoding strategies, Theorem \ref{thm:bound_approx_2design} implies that  frequencies with low redundancies are more likely to suffer from exponential concentration,  limiting the expressivity of the quantum model. Specifically, for  fixed trainable unitaries and thus fixed $\varepsilon_M$, the upper bound on the Fourier coefficient variance allows high redundant frequencies to possibly escape exponential concentration while the low redundant ones will exhibit vanishing variance, leading to the \textit{vanishing expressivity} phenomenon.

\subsection{Impact of the vanishing expressivity phenomena on training \hela{and dequantization}}\label{subsec:train_impact}

In the previous section, we compared the exponential concentration of the Fourier coefficients and the exponential concentration of the full model and focused on the settings where these two phenomena can happen independently.
Indeed, we should further emphasize 
that the interpretation of these two behaviors are fundamentally different. Specifically, since the Fourier coefficients are not directly measured to evaluate the model gradient's, the statement that they initially suffer from exponential concentration cannot be directly related to a resource problem (i.e. finite number of shots) as in other exponential concentration analyses of the whole model. This means that the vanishing Fourier coefficients phenomenon cannot be directly related to trainability issues. 
This observation justifies why we opted to denote this behavior by \textit{vanishing expressivity} with the intuition that frequencies with vanishing coefficients will have a negligible contribution to the quantum model.

However, one can plausibly think of a scenario where the signal from each Fourier coefficient is exponentially small. However, it can give rise to a significant signal when merged together. Hence, the exponential concentration of Fourier coefficients does not necessarily imply a constraint on the expressivity of the quantum Fourier model in this case.

Moreover, the analysis of the vanishing expressivity phenomenon holds on average when the parameterized unitaries form approximate 2 design but it gives no guarantees on the model's effective expressivity during the training stage. Consequently, it is possible to start with exponentially small contributions of the Fourier coefficients then reach all of the theoretically accessible frequencies, given that we can efficiently train the quantum model.

\hela{\emph{Impact of vanishing expressivity on training.}} In order to have a better understanding of the consequences of vanishing Fourier coefficients, we go beyond our analytical results and perform numerical simulations to study the impact of the vanishing expressivity phenomenon on the final trained model. Specifically, we  consider the task of training a fixed quantum model (fixed trainable unitaries and encoding unitaries) to fit two different sinusoidal functions with two different target frequencies, one being highly redundant in the quantum Fourier model spectrum and the other having a relatively low redundancy as depicted in Fig.\ref{fig:training}.
Then the training results are presented, showing that the model manages to reach the high redundant frequency but not the low redundant one.
This result supports indeed the intuition that frequencies with initially vanishing Fourier coefficients are harder to reach.
We note however that this \hela{behavior} can also be due to controllability issues in which there is no configuration of parameters such that the low redundant frequency does actually have a non zero weight. In the next section, we discuss the controllability issue in more details.

\hela{\emph{Impact of vanishing expressivity on RFF\footnote{Random Fourier Features.} based dequantization.} 
For RFF-based dequantization schemes \cite{sweke_potential_2023} of QFMs, finding the optimal frequency distribution to build the classical surrogate and hence to dequantize the quantum model, requires knowledge about the spectral properties of the final model. With our results, we can propose the frequency distribution given by the redundancies as described in section \ref{sec:Framework} as a natural distribution that encodes the bias in the quantum model. However, this choice is based on the assumption that the final model's spectral properties will inherit the spectral properties of the average case model (initial model with random parameter initialization). Although, this assumption is not guaranteed to be fulfilled in general, the numerics in Fig.\ref{fig:training} show that the decaying Fourier coefficients in ``average'' models persists in the final trained model.}


\begin{figure*}[htbp]
    \centering

    \begin{minipage}[t]{0.6\linewidth} 
        \centering
        \begin{picture}(0,0)
            \put(-28,100){a)}
        \end{picture}
        \hspace{-2.5em}\resizebox{\linewidth}{!}{\input{Figures/figs_numerics/intersection.pgf}}               \label{fig:intersection}
    \end{minipage}

    \vspace{-1em} 

    \begin{minipage}[t]{0.49\linewidth}
        \centering
        \begin{picture}(0,0)
            \put(-125,-10){b)}
        \end{picture}
        \resizebox{\linewidth}{!}{\input{Figures/figs_numerics/1_training_coeffs.pgf}}
    \end{minipage}%
    \hfill
    \begin{minipage}[t]{0.49\linewidth}
        \centering
        \begin{picture}(0,0)
            \put(-125,-10){c)}
        \end{picture}
        \resizebox{\linewidth}{!}{\input{Figures/figs_numerics/122_training.pgf}}
    \end{minipage}

    \caption{\textbf{Impact of the vanishing expressivity phenomenon on a trained QFM.} We study the expressive capability of a trained QFM acting on $n=12$ qubits using the exponential encoding strategy with a) a spectrum $\Omega$ and frequency redundancies $|R(\omega)|$ . The model is trained to fit two target frequencies with different redundancies in the model's spectrum: Target 1 corresponds to a high redundant frequency (in red), and Target 2 corresponds to a low redundant frequency (in purple). b) We plot the loss and the  Fourier coefficients norm evolution fitting Target 1. In c), we plot the loss and the  Fourier coefficients norm evolution fitting Target 2. The model succeeds in fitting the high redundant frequency but not the low redundant one. }
    \label{fig:training}
\end{figure*}


    \subsection{Fourier Norm Bound and Controllability constraints}\label{subsec:MassConservation_Correlation_Control}

In this work, we focused on studying variances by considering a uniform distribution over the parameter vector $\theta$.
 \leo{In this Section, we point out that additional constraints can occur due to the lack of model \textit{controllability}, defined as the number of Fourier coefficient that one can independently control by tuning the trainable parameter vector.}

\leo{First, we establish a generic constraint on the quantum model Fourier coefficients that holds for any Hamiltonian encoding scheme and that is independent of the trainable unitaries distribution in the following theorem.}

\begin{theorem}[Fourier Norm Bound]\label{thm:Mass_Conservation}
    Consider a quantum model  $f(x,\theta)$ of the form in Eq.\eqref{Eq:quantum_Model} using an observable $O$ and a parametrized circuit of the form in Eq.\eqref{eq:circuit_ansatz} with $L\geq 1 $ layers. Also assume that the encoding Hamiltonians are fixed, giving rise to a spectrum $\Omega$.
     Then,
    \begin{equation}
        \forall x \in \mathbb{R}^d, \forall \theta \in \Theta, |f(x,\theta)|^2 \leq ||O||^2_{\infty}
    \end{equation}
    \begin{equation}\label{eq:Slimane}
        \forall \theta \in \Theta, \sum_{\omega \in \Omega}|c_\omega(\theta)|^2 \leq ||O||^2_{\infty}
    \end{equation}
\end{theorem}

The first part of Theorem \ref{thm:Mass_Conservation} is a trivial constraint that holds for any quantum model of the form in Eq.(\ref{Eq:quantum_Model}) even outside of the Fourier framework. This constraint has been mentioned in \cite{sweke_potential_2023} to highlight the fact the a quantum Fourier model can not achieve any linear function in the Fourier basis given by its spectrum.
The second part of the theorem is more subtle. \hela{While very similar to the Parseval identity, the bound in Eq.\eqref{eq:Slimane} holds for any real valued spectrum $\Omega$.}
It shows that the 2-norm of the Fourier coefficient vector is upper bounded by the observable largest eigenvalue, introducing another \hela{generic constraint on the quantum model Fourier coefficients}.
We provide the proof of Theorem \ref{thm:Mass_Conservation} in Appendix \ref{app:proof_mass_conservation}.

\leo{In addition to the previous results, it is important to stress that limitations in the controlability of the trainable unitaries can affect the controllability of the Fourier Model. By considering the expanded expression of the quantum Fourier model given in Eq.\eqref{Eq:quantum_Fourier_Model_developped}, each Fourier coefficient is defined as a sum and product of coefficients from the trainable unitary matrices. }

\leo{According to the number of parameters, and the set of gates chosen,  the number of independent matrix coefficients that one can freely control through the trainable parameters can fluctuate from one circuit to another. Previous works \cite{larocca_diagnosing_2022, larocca_theory_2023} have studied the controllability of PQCs, by analyzing the corresponding Dynamical Lie Algebra (DLA), or by considering the quantum Fisher Information matrix which characterizes the state controllability. In \cite{fontana_adjoint_2023,ragone_unified_2023}, the authors  highlight a connection between the maximal controllability of a PQC, i.e., the dimension of its DLA, and its capacity to be trained. Hence, this controllability notion will be key in characterizing the controllability of Fourier coefficients.}

Namely, by looking more closely at the Fourier coefficient expression given in Eq.\eqref{Eq:quantum_Fourier_Model_developped}, one can simply observe that a pair of paths $(J,J') \in R(\omega)$ from the frequency generator defined in Definition \ref{def:Redundancy} allocates coefficients of the trainable unitary matrices to its frequency. In addition, some unitary coefficients are shared among different Fourier coefficients as a consequence of some branches in the generating tree (see Fig.\ref{fig:quantum_Spectrum_Trees}) being shared between different frequencies. Consequently, this can potentially create correlations between the Fourier coefficients. Therefore, if the trainable layers have a low controllability, it could lead to the impossibility of controlling independently a large number of Fourier coefficients. This is particularly important \hela{due} to the fact that increasing the number of parameters seems to increase the control and decrease the distance to a 2-design (see for example the evolution of the distance to a 2-design in the case of Periodic Ansatz through the Theorem 1 in \cite{larocca_diagnosing_2022}).

In Fig.\ref{fig:training}, we offer some plots on the evolution of the Fourier coefficients when training a QFM to learn two sinusoidal functions. The first one has a \hela{target} frequency with high redundancy in the quantum Fourier model, and the second one corresponds to a low redundant frequency. One can notice that, for the second target, the PQC takes more epochs to converge and fails to minimize the loss. In addition, one can notice that, during the training, the Fourier coefficients surrounding the target frequency are changing a lot, due to the lack of controllability of the Fourier coefficient.

    \subsection{Limitations of the Framework}\label{subsec:Limitation_Framework}

In this Section, we discuss the limitations of our framework and  \hela{the used assumptions} \hela{to derive} our main results presented in Section \ref{sec:Main_Results}.

\hela{First, the statistical analysis of the Fourier coefficients established in this work holds when the trainable parameters are sampled uniformly and independently. Although we considered the case where the trainable unitaries form an approximate 2-design, the obtained constraints on the Fourier coefficients variance and their decay only hold on average. Consequently, extrapolating this average case behavior to the final trained model is not systematic. Indeed, this gap between average case and final model guarantees constrains the direct applicability of our results to rigorously study the efficiency of random Fourier Features based dequantization schemes \cite{sweke_potential_2023}.     While proving analytically the impact of the frequencies redundancies on the \textit{effective expressivity}  of the final trained model is a hard task, we provide in section \ref{subsec:train_impact} numerical evidence of this behavior. 
Moreover, the upper bounds presented in Theorem \ref{thm:bound_approx_2design_informal} are useful to establish generic theoretical guarantees on the model's expressivity. However, they only apply to quantum models with a single uploading layer. In addition, estimating the $\varepsilon$-distance of the trainable unitaries to 2-designs and its scaling is not efficient in practice. }

Finally, we would like to point out that our results could be extended to the case of subspace preserving quantum circuits. In this type of PQCs, one can restrict the computation to a particular subspace by using input states which lie in the subspace, reducing the dimension of the effective Hilbert space. These methods allow to avoid Barren Plateaus while considering subspaces of polynomial size \cite{larocca_diagnosing_2022,fontana_adjoint_2023, monbroussou_trainability_2023, ragone_unified_2023} but question the quantum advantage of such models \cite{anschuetz_efficient_2023,cerezo_does_2023}. Considering subspace preserving unitaries, our results can easily be adapted. The dependency over the frequency distribution will still hold, but the value of $d$ (the dimension of the Hilbert space) will be substituted by the dimension of the subspace. Therefore, models generated by subspace preserving circuits could exhibit similar \emph{inductive bias} arising from the redundancy constraint on the variance of its Fourier coefficients.

\section{Numerical Results}\label{sec:Simulations}

In this section, we simulate the distribution of the Fourier coefficients for several types of circuits and encoding strategies allowing us to compare the numerical simulations with our theoretical results presented in Section \ref{sec:Main_Results}. 

\begin{figure*}[tp]
\hspace{-0.4cm}\resizebox{1.05\textwidth}{!}{\input{Figures/figs_numerics/variance_redundancies.tikz}}
    \caption{For $n=5$ qubits, one circuit layer $L=1$, five repetitions of the strongly entangling ansatz per trainable layer (see Appendix \ref{ansatze}) and global observable $O_G$; relation between the variance of each Fourier coefficient $\Var\left[c_\omega\right]$ and its redundancy $|R(\omega)|$. Values shown for two different encoding strategies, a) Pauli encoding and b) exponential encoding.}
    \label{fig:variance_redundancies}
\end{figure*}
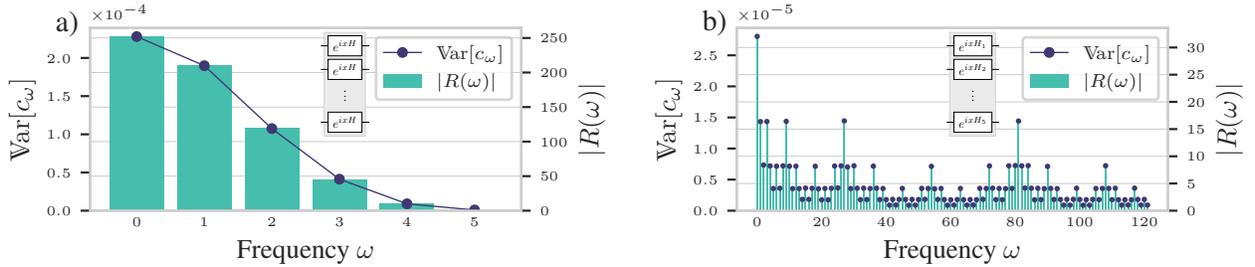
\begin{figure*}[ht]
    \resizebox{1\textwidth}{!}{\input{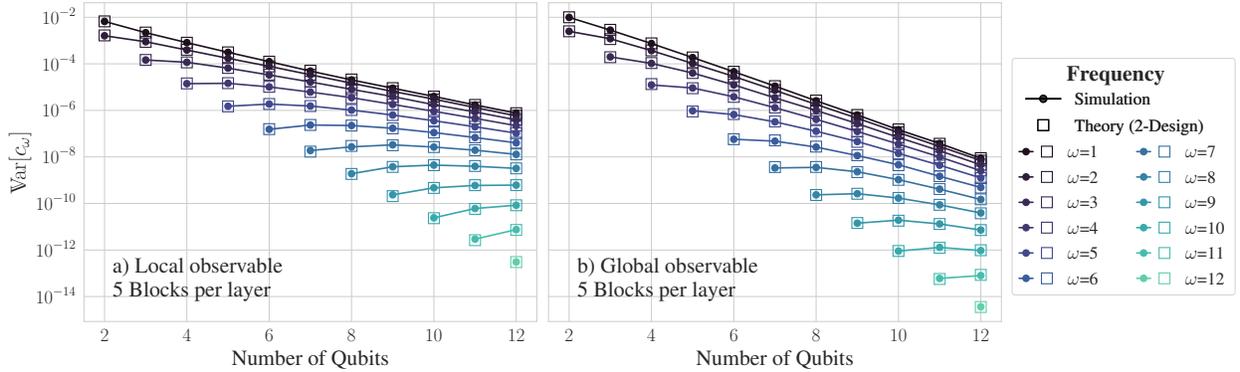}}
    \caption{Variance of the Fourier coefficients against the number of qubits for a single circuit layer ($L=1$). Five repetitions of the strongly entangling ansatz per trainable layer are used ensuring full connectivity. a) Case of local observable. b) Case of global observable. Each color corresponds to a certain frequency $\omega$, squares of the same color correspond to the variance theory for 2-design from Theorem \ref{thm:single_layer_gloabl_2design} and the dots correspond to the simulated variance.}
    \label{fig:pauli_var_n}
\end{figure*}

\subsection{Methodology}

A description of the different types of parameterized circuit architectures used can be found in Appendix \ref{ansatze}. Each circuit consists of a choice of encoding layers (encoding Hamiltonians), alternated with trainable layers, $L$ times. Each trainable layer is a repetition of the same ansatz block (See Appendix \ref{ansatze}). Finally, these blocks can act globally on all qubits,  or locally on $m$ qubits, with $m$ being constant. 

The statistical properties of the PQC are obtained with the following method: for each ansatz (parameterized circuit's architecture), we pick at random a large number of parameter vectors $\theta$s. For each vector, we evaluate the model $f(x,\theta)$ as many points $x\in \R$ as the Shannon criterion establishes. We obtain the model's Fourier coefficients $\{c_\omega(\theta)\}_{\omega \in \Omega}$ by applying the discrete Fourier transform to the measurement output. We then aggregate the results for all $\theta$s and compute the variance $\Var(c_\omega)$ for each $\omega \in \Omega$. 

As we are performing simulations, we are computing the wave function analytically and hence have access to the direct probabilities of each state.
Therefore, the shot noise from measurement is not taken into account.

We use three distinct approaches for Hamiltonian encoding. The first is \emph{Pauli encoding}, which is implemented here by applying the same single-qubit Pauli rotation gate with $H=\sigma_x$ applied on each available qubit, as seen in the inset within Fig.\figref[a]{fig:variance_redundancies}. 
Secondly, we use \emph{exponential} encoding which, as described in Appendix \ref{appendix:spec_distribution}, generates an exponentially large spectrum $\Omega$ by also using single qubit rotations but with scaling factors introduced in \cite{shin_exponential_2022} (see Fig.\figref[b]{fig:variance_redundancies}). Finally, we also consider the \emph{Golomb} encoding in the single circuit layer setting ($L=1$) which is obtained by using a global Hamiltonian encoding whose diagonal elements are those of a perfect Golomb ruler~\cite{peters_generalization_2023}. 

Furthermore, two types of observables are considered. A global observable where the measurement is acting non-trivially and simultaneously on all of the qubits: $O_G = \ket{\bzero}\!\!\bra{\bzero}$ and a local observable where an average is taken over single qubit measurements $O_L = \frac{1}{n}\sum_j \przero_j\otimes \mathds{1}_{\noj}$.



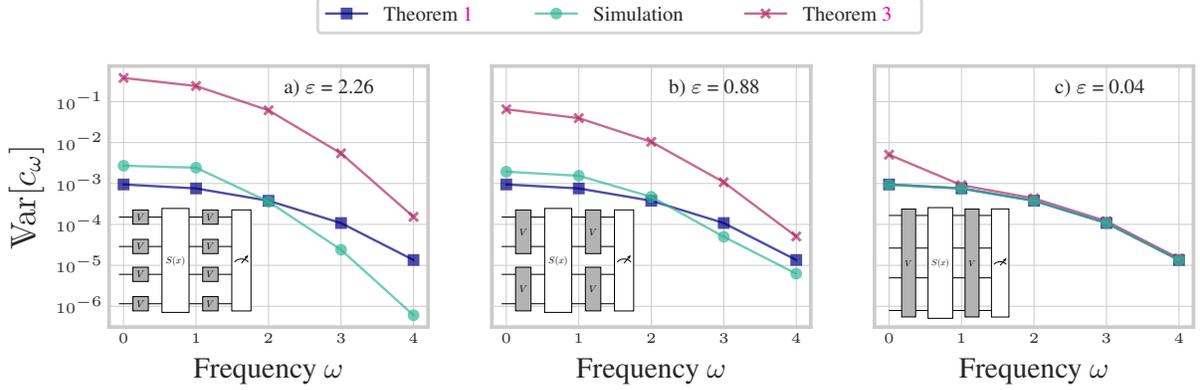
\begin{figure*}[pt]
\centering
\input{Figures/figs_numerics/figC_quad_wCircs.tikz}
\vspace{-10pt}
\caption{Comparison between simulated (in green) and two theoretical values for the variance of $c_\omega$: global 2-design setting (in blue) and $\varepsilon$-approximate 2 design setting (in pink). Parameterized layers are made of local blocks $V$ made of 5 repetitions of the \emph{Strongly Entangling} ansatz acting each on a) $m=1$ qubits b) $m=2$ qubits c) $m=4$ qubits, with a Pauli encoding layer, and the global observable $O_G$. 
}
\label{fig:lower_connectivity_quad}
\end{figure*}

\begin{figure*}[ht]
\centering
\input{Figures/figs_numerics/diff_encoding.tikz}
\caption{Same setting as in Fig.\figref[b]{fig:lower_connectivity_quad} but with a) Exponential Encoding b) Golomb encoding.} \label{fig:lower_connectivity_exp_gollomb_encoding}
\end{figure*}
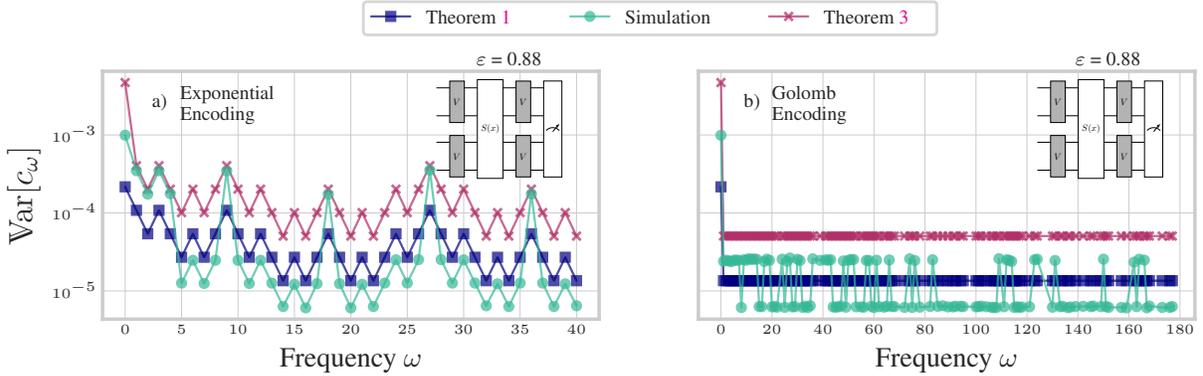

\begin{figure*}[ht]
\centering
\input{Figures/figs_numerics/diff_encoding_local.tikz}
\caption{Comparison between simulated (in green) and two theoretical values for the variance of $c_\omega$: global 2-design setting (in blue) and $\varepsilon$-approximate 2 design setting (in pink) . Parameterized layers are made of 2-local blocks $V$ made of 5 repetitions of the \emph{Strongly Entangling} ansatz with the local observable $O_L$ and a) Exponential encoding strategy b) Golomb encoding strategy.
}\label{fig:local_2design_local_meas}
\end{figure*}
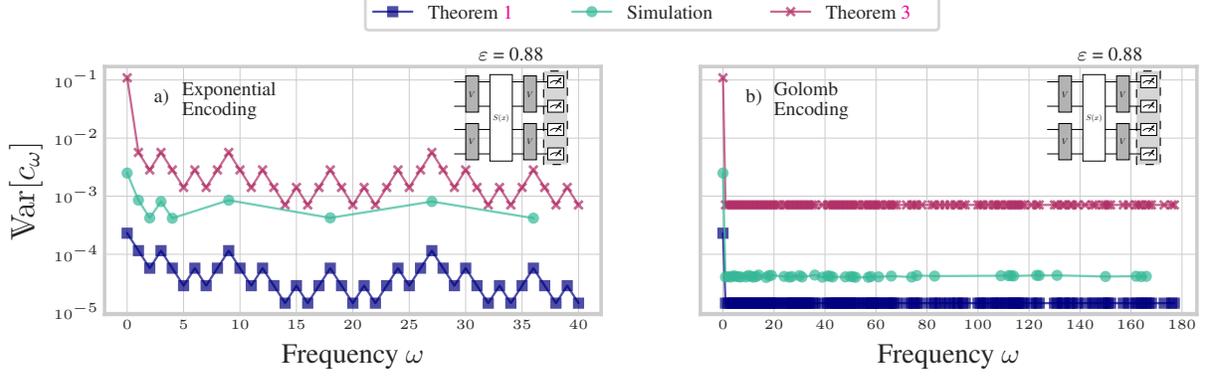
\begin{figure*}[ht]
\centering
\begin{minipage}{\textwidth}
\resizebox{\textwidth}{!}{\input{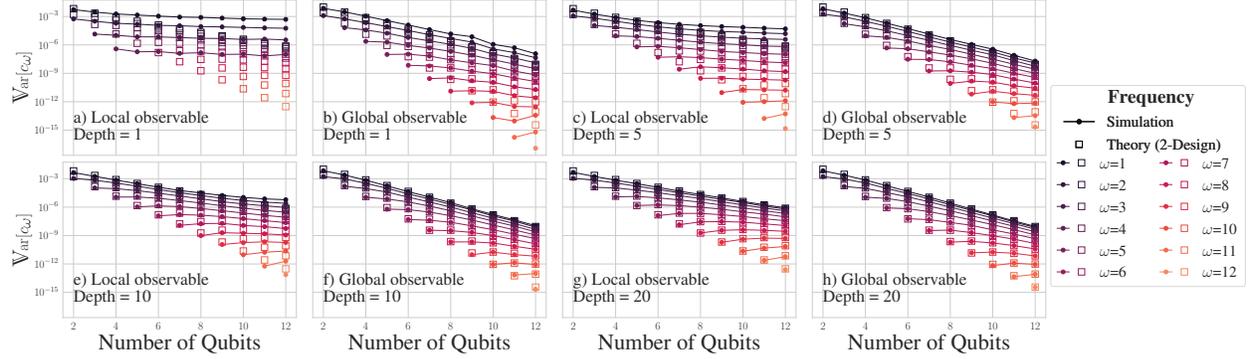}}
\end{minipage}
\caption{Variance of the Fourier coefficients against the number of qubits for a single circuit layer ($L=1$) made with the brickwise structure (Fig.\ref{fig:Figure_Framework_Local2design}) by using repetitions of the Simplified Two Design ansatz (See Fig.\ref{SimplifiedTwoDesign}), Pauli encoding and $n=2,3,\dots,12$ qubits.
We gradually increase the depth of  the trainable layers a)-f), alternating global and local observables. Each color corresponds to a certain frequency $\omega$, squares of the same color correspond to the  variance theory for 2-design of $c_{\omega}$ and the dots correspond to the simulated variance.}
\label{fig:brickwise_pauli_var_n}
\end{figure*}

\subsection{Trainable layers as global 2-design}

We consider trainable layers that form an exact 2-design, by using five repetitions of the \emph{Strongly Entangling} elementary ansatz described in Fig.\ref{circ:strongly_entangling}.

For a single-layer circuit ($L=1$),  
Theorem \ref{thm:single_layer_gloabl_2design} establishes a linear relationship between the variance of $c_\omega$ and the redundancy of the corresponding frequency. For $L>1$,  the relationship in Theorem \ref{thm:formal_2design_reup}  also involves the partial redundancies (See Definition \ref{def:Partial_Redundancy}). 

In Fig.\ref{fig:variance_redundancies}, we show that the numerics match Theorem \ref{thm:single_layer_gloabl_2design} for the Pauli and the exponential encodins respectively. 
Precisely, we see that the values of $\Var\left[c_\omega\right]$ and $|R(\omega)|$ coincide, after a linear rescaling. Fig.\figref[b]{fig:variance_redundancies} shows that the exponential encoding strategy has a broader spectrum $\Omega$ than the Pauli encoding strategy in Fig.\figref[a]{fig:variance_redundancies}. 
In addition, both encoding methods display their predicted frequency distributions defined by $|R(\omega)|$ (See appendix \ref{appendix:spec_distribution} for details).

In Fig.\ref{fig:pauli_var_n}, we simulate the same circuit, focusing on the Pauli encoding strategy, and observe the scaling of the Fourier coefficients variance with the number of qubits. In addition, we distinguish two cases: local observable in Fig.\figref[a]{fig:pauli_var_n} and  global observable in Fig.\figref[b]{fig:pauli_var_n}. 
 
In both cases, we observe that the simulated variance and the one predicted by Theorem \ref{thm:single_layer_gloabl_2design} match.
While it is less obvious why we have this behavior for the local observable in Fig.\figref[a]{fig:pauli_var_n}, it is actually due to the fact that with the strongly entangling circuit, the local observable backward light cone covers all of the circuit. However, the difference between considering a local observable and a global one can be seen through the steepness of the slopes of each frequency variance as the number of qubits increases and through the number of qubits from which we start observing a strictly decreasing variance.
Indeed, as predicted by Theorem \ref{thm:single_layer_gloabl_2design}, we observe that the Fourier coefficients variance vanish exponentially. Nonetheless, the vanishing phenomena for the high frequencies is not captured by the plots since we stop the simulations at $n=12$ qubits.

For the case of a reuploading circuit, we provide in Appendix \ref{app:additional_numerics} similar plots showing again that simulation and theory in Theorem \ref{thm:formal_2design_reup} coincide. Precisely, we reproduce in Fig.\ref{fig:thm2_sim} similar plots as in Fig.\ref{fig:variance_redundancies} for a reuploading model with $L=2$ circuit layers.

\begin{figure*}[t]
\centering
\input{Figures/figs_numerics/lightcone_Bound_global_wCircs.tikz}
\caption{Variance of the Fourier coefficients in the local observable backward lightcone with Pauli encoding generated by three different methods: simulated variance in green, variance bound from Theorem \ref{thm:bound_local2design} in pink and variance under the 2-design assumption on each of the lightcone trainable layers in blue.  We gradually decrease the depth $L_2$ of the second trainable layer , from a)-c).} \label{fig:lightcone_global_meas}
\end{figure*}
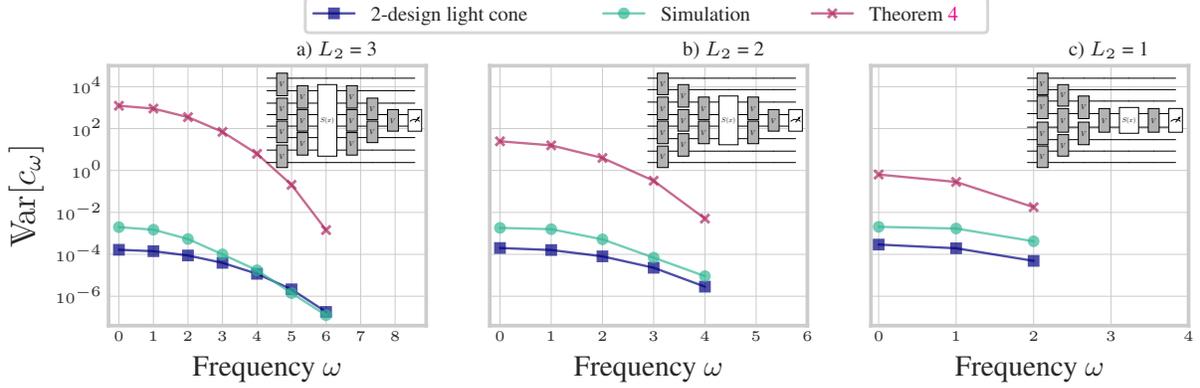

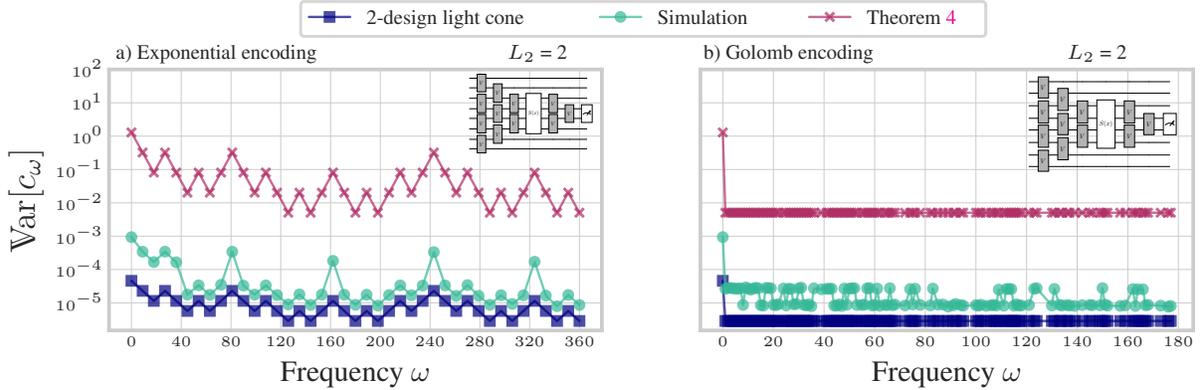
\begin{figure*}[ht]
\centering
\input{Figures/figs_numerics/lightcone_Bound_global_exp_gol.tikz}
\caption{ Variance of the Fourier coefficients in the local observable backward lightcone with $L_2 =2$ generated by three different methods: simulated variance in green, variance bound from Theorem \ref{thm:bound_local2design} in pink and variance under the 2-design assumption on each of the lightcone trainable layers in blue. a) Exponential encoding b) Golomb encoding.} \label{fig:lightcone_global_meas_exp_gol}
\end{figure*}

\subsection{Trainable layers as \texorpdfstring{$\varepsilon$}{epsilon}-approximate 2-design}\label{subsec:Simu_Approximate_2design}
In this section, we simulate parameterized circuits which gradually move away from being an exact 2-design. To do so, we decrease the number of qubits  on which the trainable blocks, within a trainable layer, are acting on (or \emph{connectivity}). 
We consider that each local trainable block (noted $V$ in the figures) forms a 2-design on the subset of qubits on which it acts non-trivially by using five repetitions of the strongly entangling ansatz applied to the corresponding $m$-qubit subsystem. We note that the smaller $m$ (low connectivity),  the entanglement between qubits decreases.
We also observe numerically that the lower the connectivity, the higher the value of $\varepsilon$, which corresponds to the monomial distance from being a 2-design (see Definition \ref{def:monomial_norm}).
The numerical computation of this distance is expensive and hence we only compute it for $n=4$ qubits. Specifically, we calculate the $2^{4n}$-dimensional matrix  $\mathcal{A}^{(2)}_{U(\Theta)}$ as given in Definition \ref{def:monomial_norm} and keep its biggest coefficient in absolute value. 
To do so, we obtain the Haar  second-moment operator with the Weingarten Calculus \cite{mele_introduction_2023} and the second-moment operator generated by the the trainable unitary by computing the empirical average over different instances of $W(\theta)$ using randomly picked parameter vectors $\theta$s.
We ensured that the sample size used to compute the empirical average is large enough to have a low variance on the computed average values.

In Fig.\ref{fig:lower_connectivity_quad}, the block $V$ acts non trivially on $m=1, m=2 \text{ and } m=4$ qubits in Fig.\figref[a]{fig:lower_connectivity_quad}, Fig.\figref[b]{fig:lower_connectivity_quad} and Fig.\figref[c]{fig:lower_connectivity_quad} respectively. For each case, we compare the simulated values of the variance of $c_\omega$ with two theoretical values: the upper bound from Theorem \ref{thm:bound_approx_2design} (approximate 2-design), and the result from Theorem \ref{thm:single_layer_gloabl_2design} (exact 2-design) using the global observable $O_G$. 
We notice that as $\varepsilon$ decreases, the simulated variance gets closer to the local 2-design result from Theorem \ref{thm:single_layer_gloabl_2design}. Moreover, the bound from Theorem \ref{thm:bound_approx_2design} is satisfied and seems to be highly correlated with the simulated variance. In addition,
it gets tighter as we approach the 2-design setting, exactly coinciding with the result from Theorem \ref{thm:single_layer_gloabl_2design} when the trainable unitaries form an exact 2-design (See Fig.\figref[c]{fig:lower_connectivity_quad}).

We also consider the exponential and Golomb encoding strategies with the same circuit architecture as the one used in Fig.\figref[b]{fig:lower_connectivity_quad} ($m=2$ and global observable $O_G$) to further assess the quality of the bound given in Theorem \ref{thm:bound_approx_2design}.
In Fig.\ref{fig:lower_connectivity_exp_gollomb_encoding}, we see again that the bound and the simulated variances are highly correlated, showing both a clear dependence on the redundancies.

In addition, if we consider the same setting but using the local observable $O_L$, we get the results plotted in Fig.\figref{fig:local_2design_local_meas}. First, we notice that the simulated spectrum does not cover all of the theoretical spectrum. This observation can be explained by the fact that the casual light cone of the local observable does not cover the encoding gates all at once, generating a sub-spectrum of the one observed in Fig.\figref{fig:lower_connectivity_exp_gollomb_encoding} where the global observable $O_G$ is used. Secondly, we see that the upper bound is still satisfied but is more loose compared to  the global observable case.

\subsection{Trainable layers made of Local 2-design blocks}

In the following, we will be using a brickwise circuit made of $m=2$ local trainable blocks  forming an exact 2-design on the corresponding 2-qubit subsystem. We also consider a single layer circuit ($L=1$) as depicted in Fig.\ref{fig:Figure_Framework_Local2design}. To do so, we take trainable layers made of repetitions of the \emph{Simplified Two Design Ansatz} described in Fig.\ref{SimplifiedTwoDesign} and we refer to the repetition number by the depth of the trainable layer.
In Fig.\ref{fig:brickwise_pauli_var_n}, we  gradually increase the depth of the trainable layers, each from $1$ to $20$, which should reach a global 2-design eventually for the number of qubits we are considering \cite{harrow_approximate_2023}. We also distinguish in this setting the case of local and global observables.

First, we observe that as the depth increases, the simulated variance gets closer to the 2 design theory with the global observable case converging more quickly, matching it exactly in Fig.\figref[f]  {fig:brickwise_pauli_var_n} and Fig.\figref[h]{fig:brickwise_pauli_var_n} (depth 10 and 20 respectively for the global observable) and in Fig.\figref[g]{fig:brickwise_pauli_var_n} (depth 20 for the local observable).

For the local observable with trainable layers of depth 1 (See Fig.\figref[a]{fig:brickwise_pauli_var_n}), we observe  the absence of some frequencies. This can be explained by the locality of the observable, the low connectivity of the trainable layers and the fact that the depth is smaller than  $log(n) \; \forall n>2$. Indeed, for larger depths (bigger than $log(n)$), we see that we do cover all of the theoretical spectrum because the light cone covers all of the encoding gates.

Secondly, the vanishing phenomenon is observed in all cases for the majority of the frequencies but with different slopes except for the case of local observable and depth 1 in Fig.\figref[a]{fig:brickwise_pauli_var_n} (consistent with \cite{cerezo_cost_2021}). For the same trainable layers depth,
The steepness of the slope is smaller when considering the local  observable and it increases with the depth for a fixed locality of the observable.

Besides, under this Brick-wise setting, we confirm the validity of the bound derived in Theorem \ref{thm:bound_local2design} illustrated in Fig.\ref{fig:lightcone_global_meas}. Here we extract the lightcone for different depths ($L_2$) of the second trainable layer $W^{(2)}$ and plot the simulated variance of each Fourier coefficient generated by the light cone, the bound from Theorem \ref{thm:bound_local2design} as well as the variance under the 2-design assumption on the light cone trainable layers from Eq.\eqref{eq:2design_lightcone}. We perform the simulations in Fig.\ref{fig:lightcone_global_meas} for the Pauli encoding strategy and replicate it for the exponential and Golomb encoding strategies in the fixed setting of $L_2=2$ in Fig.\ref{fig:lightcone_global_meas_exp_gol}.

As seen before, we observe that the bound from Theorem \ref{thm:bound_local2design} is satisfied and seems highly correlated with the simulated variance for all the encoding strategies considered.

Finally, when we assume that the restriction of each of the trainable layers to the light cone forms a 2-design (on the subsystems on which it acts non trivially), we observe that the Fourier coefficients variance in this setting is below the simulated variance. This observation supports the fact that the variance in the \emph{2-design lightcone} setting may provide a lower bound for the variance over the light cone. Thus, we leave this question for a future work.

\section{Conclusion}\label{sec:Conclusion}
In this work, we have studied the expressivity of parameterized quantum circuits (PQCs) used for learning purposes, through Fourier  lens.
Our results imply that the choice of the encoding Hamiltonians defines two important properties of the quantum model expressivity: the encoding \textit{inductive bias} and the \textit{vanishing expressivity} phenomena. The former captures the dependence of each Fourier coefficient's variance, and therefore its degree of freedom, on the corresponding frequency redundancy. The latter is a statement about the scaling of the Fourier coefficients variance as the number of qubits grows. Specifically, we showed that models containing frequencies with relatively low redundancies  suffer from a vanishing variance of those frequencies, reducing their expected expressivity . 

The learning models obtained with PQCs are therefore more constrained than expected. Even though this reduces their expressivity, it also indicates an inductive bias which might be specific to quantum models. It is left as an open question to study potential benefits of such bias as a source of quantum advantage. 
On the other hand, could one try to classically reproduce this bias, using the fact that this bias can be determined a priori by looking at the circuit?
Overall, one can use these guidelines to design more expressive PQCs, and to further study what differentiate the quantum Fourier model from its classical equivalent.

\section{Code availability}
Code based on Pennylane \cite{bergholm_pennylane_2022} to reproduce the figures and analysis is available at the following repository: \url{https://github.com/quantumsoftwarelab/DRM_fourier_expressivity}.


\section*{Acknowledgments}

This work is supported by the H2020-FETOPEN Grant PHOQUSING (GA no.: 899544), the Engineering and Physical Sciences Research Council (grants EP/T001062/1), and the Naval Group Centre of Excellence for Information Human factors and Signature Management (CEMIS).

\bibliographystyle{quantum}
\bibliography{references}

\newpage    

\appendix
\counterwithin*{equation}{section}
\renewcommand\theequation{\thesection\arabic{equation}}

\input{A0_Preliminaries.tex}

\input{A1_proof.tex}

\input{A2_additional_numerics}
\input{A3_ansatze}

\end{document}

%% file: Figures/Figure_Introduction_Merge.tikz
\tikzset{
pattern size/.store in=\mcSize, 
pattern size = 5pt,
pattern thickness/.store in=\mcThickness, 
pattern thickness = 0.3pt,
pattern radius/.store in=\mcRadius, 
pattern radius = 1pt}
\makeatletter
\pgfutil@ifundefined{pgf@pattern@name@_zwvjb3yny}{
\pgfdeclarepatternformonly[\mcThickness,\mcSize]{_zwvjb3yny}
{\pgfqpoint{0pt}{0pt}}
{\pgfpoint{\mcSize}{\mcSize}}
{\pgfpoint{\mcSize}{\mcSize}}
{
\pgfsetcolor{\tikz@pattern@color}
\pgfsetlinewidth{\mcThickness}
\pgfpathmoveto{\pgfqpoint{0pt}{\mcSize}}
\pgfpathlineto{\pgfpoint{\mcSize+\mcThickness}{-\mcThickness}}
\pgfpathmoveto{\pgfqpoint{0pt}{0pt}}
\pgfpathlineto{\pgfpoint{\mcSize+\mcThickness}{\mcSize+\mcThickness}}
\pgfusepath{stroke}
}}
\makeatother

 
\tikzset{
pattern size/.store in=\mcSize, 
pattern size = 5pt,
pattern thickness/.store in=\mcThickness, 
pattern thickness = 0.3pt,
pattern radius/.store in=\mcRadius, 
pattern radius = 1pt}
\makeatletter
\pgfutil@ifundefined{pgf@pattern@name@_bwrn4bpru}{
\pgfdeclarepatternformonly[\mcThickness,\mcSize]{_bwrn4bpru}
{\pgfqpoint{0pt}{0pt}}
{\pgfpoint{\mcSize}{\mcSize}}
{\pgfpoint{\mcSize}{\mcSize}}
{
\pgfsetcolor{\tikz@pattern@color}
\pgfsetlinewidth{\mcThickness}
\pgfpathmoveto{\pgfqpoint{0pt}{\mcSize}}
\pgfpathlineto{\pgfpoint{\mcSize+\mcThickness}{-\mcThickness}}
\pgfpathmoveto{\pgfqpoint{0pt}{0pt}}
\pgfpathlineto{\pgfpoint{\mcSize+\mcThickness}{\mcSize+\mcThickness}}
\pgfusepath{stroke}
}}
\makeatother

 
\tikzset{
pattern size/.store in=\mcSize, 
pattern size = 5pt,
pattern thickness/.store in=\mcThickness, 
pattern thickness = 0.3pt,
pattern radius/.store in=\mcRadius, 
pattern radius = 1pt}
\makeatletter
\pgfutil@ifundefined{pgf@pattern@name@_z7bg4zdlc}{
\pgfdeclarepatternformonly[\mcThickness,\mcSize]{_z7bg4zdlc}
{\pgfqpoint{0pt}{0pt}}
{\pgfpoint{\mcSize}{\mcSize}}
{\pgfpoint{\mcSize}{\mcSize}}
{
\pgfsetcolor{\tikz@pattern@color}
\pgfsetlinewidth{\mcThickness}
\pgfpathmoveto{\pgfqpoint{0pt}{\mcSize}}
\pgfpathlineto{\pgfpoint{\mcSize+\mcThickness}{-\mcThickness}}
\pgfpathmoveto{\pgfqpoint{0pt}{0pt}}
\pgfpathlineto{\pgfpoint{\mcSize+\mcThickness}{\mcSize+\mcThickness}}
\pgfusepath{stroke}
}}
\makeatother

 
\tikzset{
pattern size/.store in=\mcSize, 
pattern size = 5pt,
pattern thickness/.store in=\mcThickness, 
pattern thickness = 0.3pt,
pattern radius/.store in=\mcRadius, 
pattern radius = 1pt}
\makeatletter
\pgfutil@ifundefined{pgf@pattern@name@_0fsstxbiz}{
\pgfdeclarepatternformonly[\mcThickness,\mcSize]{_0fsstxbiz}
{\pgfqpoint{0pt}{0pt}}
{\pgfpoint{\mcSize}{\mcSize}}
{\pgfpoint{\mcSize}{\mcSize}}
{
\pgfsetcolor{\tikz@pattern@color}
\pgfsetlinewidth{\mcThickness}
\pgfpathmoveto{\pgfqpoint{0pt}{\mcSize}}
\pgfpathlineto{\pgfpoint{\mcSize+\mcThickness}{-\mcThickness}}
\pgfpathmoveto{\pgfqpoint{0pt}{0pt}}
\pgfpathlineto{\pgfpoint{\mcSize+\mcThickness}{\mcSize+\mcThickness}}
\pgfusepath{stroke}
}}
\makeatother

 
\tikzset{
pattern size/.store in=\mcSize, 
pattern size = 5pt,
pattern thickness/.store in=\mcThickness, 
pattern thickness = 0.3pt,
pattern radius/.store in=\mcRadius, 
pattern radius = 1pt}
\makeatletter
\pgfutil@ifundefined{pgf@pattern@name@_vqskh1ff5}{
\pgfdeclarepatternformonly[\mcThickness,\mcSize]{_vqskh1ff5}
{\pgfqpoint{0pt}{0pt}}
{\pgfpoint{\mcSize}{\mcSize}}
{\pgfpoint{\mcSize}{\mcSize}}
{
\pgfsetcolor{\tikz@pattern@color}
\pgfsetlinewidth{\mcThickness}
\pgfpathmoveto{\pgfqpoint{0pt}{\mcSize}}
\pgfpathlineto{\pgfpoint{\mcSize+\mcThickness}{-\mcThickness}}
\pgfpathmoveto{\pgfqpoint{0pt}{0pt}}
\pgfpathlineto{\pgfpoint{\mcSize+\mcThickness}{\mcSize+\mcThickness}}
\pgfusepath{stroke}
}}
\makeatother

 
\tikzset{
pattern size/.store in=\mcSize, 
pattern size = 5pt,
pattern thickness/.store in=\mcThickness, 
pattern thickness = 0.3pt,
pattern radius/.store in=\mcRadius, 
pattern radius = 1pt}
\makeatletter
\pgfutil@ifundefined{pgf@pattern@name@_9gfaxo5ee}{
\pgfdeclarepatternformonly[\mcThickness,\mcSize]{_9gfaxo5ee}
{\pgfqpoint{0pt}{0pt}}
{\pgfpoint{\mcSize}{\mcSize}}
{\pgfpoint{\mcSize}{\mcSize}}
{
\pgfsetcolor{\tikz@pattern@color}
\pgfsetlinewidth{\mcThickness}
\pgfpathmoveto{\pgfqpoint{0pt}{\mcSize}}
\pgfpathlineto{\pgfpoint{\mcSize+\mcThickness}{-\mcThickness}}
\pgfpathmoveto{\pgfqpoint{0pt}{0pt}}
\pgfpathlineto{\pgfpoint{\mcSize+\mcThickness}{\mcSize+\mcThickness}}
\pgfusepath{stroke}
}}
\makeatother

 
\tikzset{
pattern size/.store in=\mcSize, 
pattern size = 5pt,
pattern thickness/.store in=\mcThickness, 
pattern thickness = 0.3pt,
pattern radius/.store in=\mcRadius, 
pattern radius = 1pt}
\makeatletter
\pgfutil@ifundefined{pgf@pattern@name@_kw4dn16ew}{
\pgfdeclarepatternformonly[\mcThickness,\mcSize]{_kw4dn16ew}
{\pgfqpoint{0pt}{0pt}}
{\pgfpoint{\mcSize}{\mcSize}}
{\pgfpoint{\mcSize}{\mcSize}}
{
\pgfsetcolor{\tikz@pattern@color}
\pgfsetlinewidth{\mcThickness}
\pgfpathmoveto{\pgfqpoint{0pt}{\mcSize}}
\pgfpathlineto{\pgfpoint{\mcSize+\mcThickness}{-\mcThickness}}
\pgfpathmoveto{\pgfqpoint{0pt}{0pt}}
\pgfpathlineto{\pgfpoint{\mcSize+\mcThickness}{\mcSize+\mcThickness}}
\pgfusepath{stroke}
}}
\makeatother

 
\tikzset{
pattern size/.store in=\mcSize, 
pattern size = 5pt,
pattern thickness/.store in=\mcThickness, 
pattern thickness = 0.3pt,
pattern radius/.store in=\mcRadius, 
pattern radius = 1pt}
\makeatletter
\pgfutil@ifundefined{pgf@pattern@name@_w1fyntxf1}{
\pgfdeclarepatternformonly[\mcThickness,\mcSize]{_w1fyntxf1}
{\pgfqpoint{0pt}{0pt}}
{\pgfpoint{\mcSize}{\mcSize}}
{\pgfpoint{\mcSize}{\mcSize}}
{
\pgfsetcolor{\tikz@pattern@color}
\pgfsetlinewidth{\mcThickness}
\pgfpathmoveto{\pgfqpoint{0pt}{\mcSize}}
\pgfpathlineto{\pgfpoint{\mcSize+\mcThickness}{-\mcThickness}}
\pgfpathmoveto{\pgfqpoint{0pt}{0pt}}
\pgfpathlineto{\pgfpoint{\mcSize+\mcThickness}{\mcSize+\mcThickness}}
\pgfusepath{stroke}
}}
\makeatother

 
\tikzset{
pattern size/.store in=\mcSize, 
pattern size = 5pt,
pattern thickness/.store in=\mcThickness, 
pattern thickness = 0.3pt,
pattern radius/.store in=\mcRadius, 
pattern radius = 1pt}
\makeatletter
\pgfutil@ifundefined{pgf@pattern@name@_9nnp80jld}{
\pgfdeclarepatternformonly[\mcThickness,\mcSize]{_9nnp80jld}
{\pgfqpoint{0pt}{0pt}}
{\pgfpoint{\mcSize}{\mcSize}}
{\pgfpoint{\mcSize}{\mcSize}}
{
\pgfsetcolor{\tikz@pattern@color}
\pgfsetlinewidth{\mcThickness}
\pgfpathmoveto{\pgfqpoint{0pt}{\mcSize}}
\pgfpathlineto{\pgfpoint{\mcSize+\mcThickness}{-\mcThickness}}
\pgfpathmoveto{\pgfqpoint{0pt}{0pt}}
\pgfpathlineto{\pgfpoint{\mcSize+\mcThickness}{\mcSize+\mcThickness}}
\pgfusepath{stroke}
}}
\makeatother
\tikzset{every picture/.style={line width=0.75pt}} 
\vspace{-.2cm}
\begin{tikzpicture}[x=0.75pt,y=0.75pt,yscale=-1,xscale=1]

\draw  [color={rgb, 255:red, 65; green, 117; blue, 5 }  ,draw opacity=1 ][pattern=_zwvjb3yny,pattern size=6pt,pattern thickness=1.5pt,pattern radius=0pt, pattern color={rgb, 255:red, 177; green, 217; blue, 135}][line width=1.5]  (358.11,31.27) -- (372.36,31.27) -- (372.36,142.27) -- (358.11,142.27) -- cycle ;
\draw  [color={rgb, 255:red, 65; green, 117; blue, 5 }  ,draw opacity=1 ][pattern=_bwrn4bpru,pattern size=6pt,pattern thickness=1.5pt,pattern radius=0pt, pattern color={rgb, 255:red, 177; green, 217; blue, 135}][line width=1.5]  (380.13,44.74) -- (394.38,44.74) -- (394.38,142.27) -- (380.13,142.27) -- cycle ;
\draw  [color={rgb, 255:red, 65; green, 117; blue, 5 }  ,draw opacity=1 ][pattern=_z7bg4zdlc,pattern size=6pt,pattern thickness=1.5pt,pattern radius=0pt, pattern color={rgb, 255:red, 177; green, 217; blue, 135}][line width=1.5]  (402.15,57.34) -- (416.4,57.34) -- (416.4,142.27) -- (402.15,142.27) -- cycle ;
\draw  [color={rgb, 255:red, 65; green, 117; blue, 5 }  ,draw opacity=1 ][pattern=_0fsstxbiz,pattern size=6pt,pattern thickness=1.5pt,pattern radius=0pt, pattern color={rgb, 255:red, 177; green, 217; blue, 135}][line width=1.5]  (424.17,77) -- (438.41,77) -- (438.41,142.27) -- (424.17,142.27) -- cycle ;
\draw  [color={rgb, 255:red, 65; green, 117; blue, 5 }  ,draw opacity=1 ][pattern=_vqskh1ff5,pattern size=6pt,pattern thickness=1.5pt,pattern radius=0pt, pattern color={rgb, 255:red, 177; green, 217; blue, 135}][line width=1.5]  (446.18,92.93) -- (460.43,92.93) -- (460.43,142.27) -- (446.18,142.27) -- cycle ;
\draw  [color={rgb, 255:red, 65; green, 117; blue, 5 }  ,draw opacity=1 ][pattern=_9gfaxo5ee,pattern size=6pt,pattern thickness=1.5pt,pattern radius=0pt, pattern color={rgb, 255:red, 177; green, 217; blue, 135}][line width=1.5]  (468.2,104.93) -- (482.45,104.93) -- (482.45,142.27) -- (468.2,142.27) -- cycle ;
\draw  [color={rgb, 255:red, 65; green, 117; blue, 5 }  ,draw opacity=1 ][pattern=_kw4dn16ew,pattern size=6pt,pattern thickness=1.5pt,pattern radius=0pt, pattern color={rgb, 255:red, 177; green, 217; blue, 135}][line width=1.5]  (490.22,114.13) -- (504.47,114.13) -- (504.47,142.27) -- (490.22,142.27) -- cycle ;
\draw  [color={rgb, 255:red, 65; green, 117; blue, 5 }  ,draw opacity=1 ][pattern=_w1fyntxf1,pattern size=6pt,pattern thickness=1.5pt,pattern radius=0pt, pattern color={rgb, 255:red, 177; green, 217; blue, 135}][line width=1.5]  (512.24,123.33) -- (526.48,123.33) -- (526.48,142.27) -- (512.24,142.27) -- cycle ;
\draw  [color={rgb, 255:red, 65; green, 117; blue, 5 }  ,draw opacity=1 ][pattern=_9nnp80jld,pattern size=6pt,pattern thickness=1.5pt,pattern radius=0pt, pattern color={rgb, 255:red, 177; green, 217; blue, 135}][line width=1.5]  (534.26,128.93) -- (548.51,128.93) -- (548.51,142.27) -- (534.26,142.27) -- cycle ;

\draw [line width=1.5]    (358.11,142.27) -- (574.01,142.27) ;
\draw [shift={(578.01,142.27)}, rotate = 180] [fill={rgb, 255:red, 0; green, 0; blue, 0 }  ][line width=0.08]  [draw opacity=0] (8.13,-3.9) -- (0,0) -- (8.13,3.9) -- cycle    ;
\draw [line width=1.5]    (358.11,142.27) -- (358.11,16.27) ;
\draw [shift={(358.11,12.27)}, rotate = 90] [fill={rgb, 255:red, 0; green, 0; blue, 0 }  ][line width=0.08]  [draw opacity=0] (8.13,-3.9) -- (0,0) -- (8.13,3.9) -- cycle    ;
\draw [color={rgb, 255:red, 84; green, 142; blue, 161 }  ,draw opacity=1 ] [dash pattern={on 0.75pt off 0.75pt}]  (363.01,21.27) .. controls (366,21.35) and (367.69,22.54) .. (368.06,24.85) .. controls (368.41,27.15) and (369.73,28.09) .. (372.02,27.68) .. controls (374.3,27.26) and (375.6,28.19) .. (375.92,30.48) .. controls (376.23,32.76) and (377.51,33.68) .. (379.76,33.25) .. controls (382.01,32.82) and (383.27,33.73) .. (383.54,35.99) .. controls (383.81,38.24) and (385.36,39.37) .. (388.2,39.38) .. controls (390.41,38.93) and (391.63,39.82) .. (391.86,42.05) .. controls (392.09,44.28) and (393.29,45.16) .. (395.48,44.69) .. controls (398.25,44.66) and (399.74,45.74) .. (399.94,47.95) .. controls (400.13,50.16) and (401.6,51.23) .. (404.34,51.17) .. controls (406.49,50.67) and (407.65,51.51) .. (407.82,53.7) .. controls (408.56,56.31) and (409.99,57.35) .. (412.12,56.83) .. controls (414.23,56.3) and (415.65,57.33) .. (416.37,59.91) .. controls (416.52,62.07) and (417.64,62.88) .. (419.74,62.33) .. controls (422.38,62.18) and (423.77,63.17) .. (423.92,65.32) .. controls (424.62,67.86) and (426,68.84) .. (428.07,68.26) .. controls (430.68,68.05) and (432.05,69.02) .. (432.19,71.15) .. controls (432.89,73.66) and (434.26,74.6) .. (436.3,73.99) .. controls (438.87,73.73) and (440.24,74.66) .. (440.39,76.77) .. controls (441.1,79.26) and (442.46,80.17) .. (444.48,79.51) .. controls (447.03,79.19) and (448.39,80.08) .. (448.56,82.19) .. controls (449.29,84.65) and (450.93,85.7) .. (453.47,85.34) .. controls (455.44,84.61) and (456.81,85.47) .. (457.56,87.91) .. controls (457.79,90) and (459.16,90.84) .. (461.68,90.42) .. controls (464.19,89.97) and (465.57,90.79) .. (465.82,92.88) .. controls (466.65,95.29) and (468.04,96.1) .. (469.99,95.29) .. controls (472.48,94.78) and (474.16,95.71) .. (475.03,98.1) .. controls (475.37,100.18) and (476.79,100.94) .. (479.28,100.39) .. controls (481.19,99.51) and (482.63,100.25) .. (483.58,102.62) .. controls (483.99,104.69) and (485.44,105.42) .. (487.93,104.8) .. controls (490.41,104.15) and (491.88,104.85) .. (492.34,106.92) .. controls (493.43,109.26) and (494.92,109.95) .. (496.81,108.98) .. controls (499.3,108.27) and (500.81,108.94) .. (501.36,110.99) .. controls (502.55,113.31) and (504.09,113.96) .. (505.98,112.94) .. controls (508.47,112.15) and (510.04,112.78) .. (510.68,114.83) .. controls (511.99,117.13) and (513.59,117.74) .. (515.47,116.67) .. controls (517.34,115.58) and (518.96,116.18) .. (520.35,118.45) .. controls (521.12,120.48) and (522.78,121.06) .. (525.33,120.17) .. controls (527.2,119.03) and (528.56,119.47) .. (529.39,121.5) .. controls (530.96,123.74) and (532.69,124.28) .. (534.57,123.11) .. controls (536.44,121.94) and (537.85,122.35) .. (538.79,124.36) -- (542.01,125.27) ;
\draw  [color={rgb, 255:red, 84; green, 142; blue, 161 }  ,draw opacity=1 ][fill={rgb, 255:red, 84; green, 142; blue, 161 }  ,fill opacity=0.85 ] (367.24,20.27) -- (368.17,21.21) -- (366.06,23.32) -- (368.17,25.43) -- (367.24,26.37) -- (365.13,24.25) -- (363.01,26.37) -- (362.08,25.43) -- (364.19,23.32) -- (362.08,21.21) -- (363.01,20.27) -- (365.13,22.39) -- cycle ;
\draw  [color={rgb, 255:red, 84; green, 142; blue, 161 }  ,draw opacity=1 ][fill={rgb, 255:red, 84; green, 142; blue, 161 }  ,fill opacity=0.85 ] (388.84,35.47) -- (389.77,36.41) -- (387.66,38.52) -- (389.77,40.63) -- (388.84,41.57) -- (386.73,39.45) -- (384.61,41.57) -- (383.68,40.63) -- (385.79,38.52) -- (383.68,36.41) -- (384.61,35.47) -- (386.73,37.59) -- cycle ;
\draw  [color={rgb, 255:red, 84; green, 142; blue, 161 }  ,draw opacity=1 ][fill={rgb, 255:red, 84; green, 142; blue, 161 }  ,fill opacity=0.85 ] (411.64,49.47) -- (412.57,50.41) -- (410.46,52.52) -- (412.57,54.63) -- (411.64,55.57) -- (409.53,53.45) -- (407.41,55.57) -- (406.48,54.63) -- (408.59,52.52) -- (406.48,50.41) -- (407.41,49.47) -- (409.53,51.59) -- cycle ;
\draw  [color={rgb, 255:red, 84; green, 142; blue, 161 }  ,draw opacity=1 ][fill={rgb, 255:red, 84; green, 142; blue, 161 }  ,fill opacity=0.85 ] (433.24,67.87) -- (434.17,68.81) -- (432.06,70.92) -- (434.17,73.03) -- (433.24,73.97) -- (431.13,71.85) -- (429.01,73.97) -- (428.08,73.03) -- (430.19,70.92) -- (428.08,68.81) -- (429.01,67.87) -- (431.13,69.99) -- cycle ;
\draw  [color={rgb, 255:red, 84; green, 142; blue, 161 }  ,draw opacity=1 ][fill={rgb, 255:red, 84; green, 142; blue, 161 }  ,fill opacity=0.85 ] (455.24,82.67) -- (456.17,83.61) -- (454.06,85.72) -- (456.17,87.83) -- (455.24,88.77) -- (453.13,86.65) -- (451.01,88.77) -- (450.08,87.83) -- (452.19,85.72) -- (450.08,83.61) -- (451.01,82.67) -- (453.13,84.79) -- cycle ;
\draw  [color={rgb, 255:red, 84; green, 142; blue, 161 }  ,draw opacity=1 ][fill={rgb, 255:red, 84; green, 142; blue, 161 }  ,fill opacity=0.85 ] (477.64,95.87) -- (478.57,96.81) -- (476.46,98.92) -- (478.57,101.03) -- (477.64,101.97) -- (475.53,99.85) -- (473.41,101.97) -- (472.48,101.03) -- (474.59,98.92) -- (472.48,96.81) -- (473.41,95.87) -- (475.53,97.99) -- cycle ;
\draw  [color={rgb, 255:red, 84; green, 142; blue, 161 }  ,draw opacity=1 ][fill={rgb, 255:red, 84; green, 142; blue, 161 }  ,fill opacity=0.85 ] (498.84,105.87) -- (499.77,106.81) -- (497.66,108.92) -- (499.77,111.03) -- (498.84,111.97) -- (496.73,109.85) -- (494.61,111.97) -- (493.68,111.03) -- (495.79,108.92) -- (493.68,106.81) -- (494.61,105.87) -- (496.73,107.99) -- cycle ;
\draw  [color={rgb, 255:red, 84; green, 142; blue, 161 }  ,draw opacity=1 ][fill={rgb, 255:red, 84; green, 142; blue, 161 }  ,fill opacity=0.85 ] (521.64,114.27) -- (522.57,115.21) -- (520.46,117.32) -- (522.57,119.43) -- (521.64,120.37) -- (519.53,118.25) -- (517.41,120.37) -- (516.48,119.43) -- (518.59,117.32) -- (516.48,115.21) -- (517.41,114.27) -- (519.53,116.39) -- cycle ;
\draw  [color={rgb, 255:red, 84; green, 142; blue, 161 }  ,draw opacity=1 ][fill={rgb, 255:red, 84; green, 142; blue, 161 }  ,fill opacity=0.85 ] (543.19,121.23) -- (544.13,122.16) -- (542.01,124.27) -- (544.13,126.39) -- (543.19,127.32) -- (541.08,125.21) -- (538.97,127.32) -- (538.03,126.39) -- (540.15,124.27) -- (538.03,122.16) -- (538.97,121.23) -- (541.08,123.34) -- cycle ;
\draw [color={rgb, 255:red, 177; green, 217; blue, 135 }  ,draw opacity=1 ]   (361.99,149.96) -- (546.79,149.96) ;
\draw [shift={(549.79,149.96)}, rotate = 180] [fill={rgb, 255:red, 177; green, 217; blue, 135 }  ,fill opacity=1 ][line width=0.08]  [draw opacity=0] (7.14,-3.43) -- (0,0) -- (7.14,3.43) -- (4.74,0) -- cycle    ;
\draw [shift={(358.99,149.96)}, rotate = 0] [fill={rgb, 255:red, 177; green, 217; blue, 135 }  ,fill opacity=1 ][line width=0.08]  [draw opacity=0] (7.14,-3.43) -- (0,0) -- (7.14,3.43) -- (4.74,0) -- cycle    ;

\draw (55.7,0) node [anchor=north west][inner sep=0.75pt]    {$f( x,\theta ) =\displaystyle\sum _{\textcolor[rgb]{0.69,0.85,0.53}{\omega \in \Omega }}\textcolor[rgb]{0.33,0.56,0.63}{c_{\omega }(\theta)} e^{i\textcolor[rgb]{0.69,0.85,0.53}{\omega^T } x}$};
\draw (573.86,124.73) node [anchor=north west][inner sep=0.75pt]  [font=\footnotesize] [align=left] {$\displaystyle \omega $};
\draw (304.8,13.41) node [anchor=north west][inner sep=0.75pt]  [color={rgb, 255:red, 177; green, 217; blue, 135 }  ,opacity=1 ]  {$| R( \omega ) |$};
\draw (371.49,12.41) node [anchor=north west][inner sep=0.75pt]    {$\textcolor[rgb]{0.33,0.56,0.63}{\Var[ c_{\omega }(\theta)]}$};
\draw (446.39,154.2) node [anchor=north west][inner sep=0.75pt]  [color={rgb, 255:red, 177; green, 217; blue, 135 }  ,opacity=1 ]  {$\Omega $};

\node at (290,10) {b)};
\node at (0,10) {a)};
\end{tikzpicture}
\begin{tikzpicture}
\begin{adjustbox}{scale=0.75}
\node at (16.5,-16) {
\begin{quantikz}[row sep = 6pt, column sep = 12pt]
    &   \gate[4, style={fill=blue_ansatz ,draw = border_blue_ansatz}]{W^1(\theta)} \gategroup[4,steps=2,style={inner sep=1pt,fill=lightgray!20},background]{Circuit Layer 1} & \gate[4, style={fill=green_encoding ,draw = border_green_encoding}]{S^1(x)} & & \gate[4, style={fill=blue_ansatz ,draw = border_blue_ansatz}]{W^L(\theta)}\gategroup[4,steps=2,style={inner sep=1pt,fill=lightgray!20},background]{Circuit Layer L} & \gate[4, style={fill=green_encoding ,draw = border_green_encoding}]{S^L(x)} & \gate[4, style={fill=blue_ansatz ,draw = border_blue_ansatz}]{W^{L+1}(\theta)} &  \meter{} \gategroup[4,steps=1,style={inner sep=1pt,fill=lightgray!20},background]{$O$}\\
    &   &           &             &  & &  &\meter{}\\
    \wireoverride{n}\qquad\vdots&\wireoverride{n}&\wireoverride{n}&\wireoverride{n}\ldots& \wireoverride{n} & \wireoverride{n} \vdots & \wireoverride{n}& \wireoverride{n} \vdots\\
    &   &   & & & & &  \meter{}
\end{quantikz} 

};
\end{adjustbox}
\end{tikzpicture}
\vspace{-4.1cm}

%% file: Figures/Quantum_Spectrum_Trees.tikz
\tikzset{every picture/.style={line width=0.75pt}} 

\begin{tikzpicture}[x=0.75pt,y=0.75pt,yscale=-1,xscale=1]

\draw  [color={rgb, 255:red, 128; green, 128; blue, 128 }  ,draw opacity=1 ][fill={rgb, 255:red, 242; green, 242; blue, 242 }  ,fill opacity=1 ] (2.4,2) -- (281.63,2) -- (281.63,163.97) -- (2.4,163.97) -- cycle ;
\draw    (157.28,8.91) -- (101.66,55.84) ;
\draw    (213.36,56.34) -- (157.28,8.91) ;
\draw    (213.36,56.34) -- (157.73,102.35) ;
\draw    (46.04,102.78) -- (101.66,55.84) ;
\draw    (46.04,102.78) -- (28.19,148.24) ;
\draw    (101.66,55.84) .. controls (114.9,37.23) and (141.49,14.96) .. (157.28,8.91) ;
\draw [color={rgb, 255:red, 208; green, 2; blue, 27 }  ,draw opacity=1 ][line width=1.5]    (157.28,8.91) .. controls (171.79,14.34) and (205.81,43.41) .. (213.36,56.34) ;
\draw    (101.66,55.84) -- (157.73,102.35) ;
\draw    (213.36,56.34) -- (268.98,101.98) ;
\draw    (101.66,101.86) .. controls (100.67,95.98) and (98.2,62.58) .. (101.66,55.84) ;
\draw    (101.66,101.86) .. controls (105,93.51) and (103.15,62.58) .. (101.66,55.84) ;
\draw    (213.36,102.35) .. controls (211.37,95.36) and (210.14,63.82) .. (213.36,56.34) ;
\draw [color={rgb, 255:red, 208; green, 2; blue, 27 }  ,draw opacity=1 ][line width=1.5]    (213.76,102.54) .. controls (216.32,94.74) and (215.08,64.44) .. (213.36,56.34) ;
\draw    (46.04,102.78) -- (46.04,148.24) ;
\draw    (46.04,148.79) .. controls (51.61,139.89) and (49.75,112.06) .. (46.04,102.78) ;
\draw    (46.04,102.78) -- (9.26,148.24) ;
\draw    (9.26,148.24) -- (9.26,157.89) ;
\draw    (28.19,148.24) -- (28.19,157.89) ;
\draw    (46.04,147.86) -- (46.04,157.51) ;
\draw    (101.7,101.72) -- (83.48,148.24) ;
\draw    (101.7,101.98) -- (101.7,147.44) ;
\draw    (101.7,147.99) .. controls (107.27,139.09) and (105.41,111.26) .. (101.7,101.98) ;
\draw    (101.7,101.98) -- (65.67,148.55) ;
\draw    (65.67,148.55) -- (65.67,158.2) ;
\draw    (83.48,148.24) -- (83.48,157.89) ;
\draw    (101.7,147.07) -- (101.7,156.71) ;
\draw    (157.73,102.09) -- (139.51,148.18) ;
\draw    (157.73,102.35) -- (157.73,147.81) ;
\draw    (157.73,148.37) .. controls (163.3,139.46) and (161.44,111.63) .. (157.73,102.35) ;
\draw    (157.73,102.35) -- (120.96,148.55) ;
\draw    (120.96,148.55) -- (120.96,158.2) ;
\draw    (139.51,148.18) -- (139.51,157.83) ;
\draw    (157.73,147.44) -- (157.73,157.09) ;
\draw    (213.76,102.28) -- (195.17,148.37) ;
\draw    (213.76,148.55) .. controls (219.33,139.65) and (217.48,111.82) .. (213.76,102.54) ;
\draw    (213.76,102.54) -- (176.25,148.74) ;
\draw    (176.25,148.74) -- (176.25,158.39) ;
\draw    (195.17,148.37) -- (195.17,158.02) ;
\draw [color={rgb, 255:red, 208; green, 2; blue, 27 }  ,draw opacity=1 ][line width=1.5]    (213.76,147.62) -- (213.76,157.27) ;
\draw    (269.43,101.9) -- (250.83,148) ;
\draw    (269.43,102.17) -- (269.43,147.63) ;
\draw    (269.43,148.18) .. controls (274.99,139.27) and (273.14,111.44) .. (269.43,102.17) ;
\draw    (269.43,102.17) -- (232.28,148) ;
\draw    (232.28,148) -- (232.28,157.65) ;
\draw    (250.83,148) -- (250.83,157.65) ;
\draw    (269.43,147.25) -- (269.43,156.9) ;
\draw [color={rgb, 255:red, 208; green, 2; blue, 27 }  ,draw opacity=1 ][line width=1.5]    (213.76,102.54) -- (213.76,148) ;
\draw  [color={rgb, 255:red, 128; green, 128; blue, 128 }  ,draw opacity=1 ][fill={rgb, 255:red, 242; green, 242; blue, 242 }  ,fill opacity=1 ] (320.28,1.67) -- (599.51,1.67) -- (599.51,163.64) -- (320.28,163.64) -- cycle ;
\draw    (475.16,8.91) -- (419.54,55.84) ;
\draw    (531.24,56.34) -- (475.16,8.91) ;
\draw [color={rgb, 255:red, 208; green, 2; blue, 27 }  ,draw opacity=1 ][line width=1.5]    (531.24,56.34) -- (475.62,102.35) ;
\draw    (363.92,102.78) -- (419.54,55.84) ;
\draw    (363.92,102.78) -- (346.07,148.24) ;
\draw    (419.54,55.84) .. controls (432.78,37.23) and (459.37,14.96) .. (475.16,8.91) ;
\draw [color={rgb, 255:red, 208; green, 2; blue, 27 }  ,draw opacity=1 ][line width=1.5]    (475.16,8.91) .. controls (489.68,14.34) and (523.69,43.41) .. (531.24,56.34) ;
\draw    (419.54,55.84) -- (475.62,102.35) ;
\draw    (531.24,56.34) -- (586.86,101.98) ;
\draw    (419.54,101.86) .. controls (418.55,95.98) and (416.08,62.58) .. (419.54,55.84) ;
\draw    (419.54,101.86) .. controls (422.88,93.51) and (421.03,62.58) .. (419.54,55.84) ;
\draw    (531.24,102.35) .. controls (529.26,95.36) and (528.02,63.82) .. (531.24,56.34) ;
\draw [color={rgb, 255:red, 0; green, 0; blue, 0 }  ,draw opacity=1 ][line width=0.75]    (531.65,102.54) .. controls (534.2,94.74) and (532.97,64.44) .. (531.24,56.34) ;
\draw    (363.92,102.78) -- (363.92,148.24) ;
\draw    (363.92,148.79) .. controls (369.49,139.89) and (367.63,112.06) .. (363.92,102.78) ;
\draw    (363.92,102.78) -- (327.15,148.24) ;
\draw    (327.15,148.24) -- (327.15,157.89) ;
\draw    (346.07,148.24) -- (346.07,157.89) ;
\draw    (363.92,147.86) -- (363.92,157.51) ;
\draw    (419.58,101.72) -- (401.36,148.24) ;
\draw    (419.58,101.98) -- (419.58,147.44) ;
\draw    (419.58,147.99) .. controls (425.15,139.09) and (423.3,111.26) .. (419.58,101.98) ;
\draw    (419.58,101.98) -- (383.55,148.55) ;
\draw    (383.55,148.55) -- (383.55,158.2) ;
\draw    (401.36,148.24) -- (401.36,157.89) ;
\draw    (419.58,147.07) -- (419.58,156.71) ;
\draw    (475.62,102.09) -- (457.39,148.18) ;
\draw    (475.62,102.35) -- (475.62,147.81) ;
\draw    (475.62,148.37) .. controls (481.18,139.46) and (479.33,111.63) .. (475.62,102.35) ;
\draw [color={rgb, 255:red, 208; green, 2; blue, 27 }  ,draw opacity=1 ][line width=1.5]    (475.62,102.35) -- (438.84,148.55) ;
\draw [color={rgb, 255:red, 208; green, 2; blue, 27 }  ,draw opacity=1 ][line width=1.5]    (438.84,147.63) -- (438.84,157.28) ;
\draw    (457.39,148.18) -- (457.39,157.83) ;
\draw    (475.62,147.44) -- (475.62,157.09) ;
\draw    (531.65,102.28) -- (513.05,148.37) ;
\draw    (531.65,148.55) .. controls (537.21,139.65) and (535.36,111.82) .. (531.65,102.54) ;
\draw    (531.65,102.54) -- (494.13,148.74) ;
\draw    (494.13,148.74) -- (494.13,158.39) ;
\draw    (513.05,148.37) -- (513.05,158.02) ;
\draw [color={rgb, 255:red, 0; green, 0; blue, 0 }  ,draw opacity=1 ][line width=0.75]    (531.65,147.62) -- (531.65,157.27) ;
\draw    (587.31,101.9) -- (568.71,148) ;
\draw    (587.31,102.17) -- (587.31,147.63) ;
\draw    (587.31,148.18) .. controls (592.87,139.27) and (591.02,111.44) .. (587.31,102.17) ;
\draw    (587.31,102.17) -- (550.16,148) ;
\draw    (550.16,148) -- (550.16,157.65) ;
\draw    (568.71,148) -- (568.71,157.65) ;
\draw    (587.31,147.25) -- (587.31,156.9) ;
\draw [color={rgb, 255:red, 0; green, 0; blue, 0 }  ,draw opacity=1 ][line width=0.75]    (531.65,102.54) -- (531.65,148) ;
\draw  [dash pattern={on 0.84pt off 2.51pt}]  (46.04,102.78) -- (587.31,102.17) ;
\draw  [dash pattern={on 0.84pt off 2.51pt}]  (101.66,55.84) -- (531.24,56.34) ;
\draw  [dash pattern={on 0.84pt off 2.51pt}]  (157.28,8.91) -- (475.16,8.91) ;
\draw  [dash pattern={on 0.84pt off 2.51pt}]  (9.26,148.24) -- (587.31,147.63) ;
\draw [color={rgb, 255:red, 208; green, 2; blue, 27 }  ,draw opacity=1 ]   (213.76,157.27) .. controls (211,179.75) and (250,157.75) .. (250,175.25) ;
\draw [color={rgb, 255:red, 208; green, 2; blue, 27 }  ,draw opacity=1 ]   (404,175.25) .. controls (404,159.25) and (439.5,177.75) .. (438.84,157.28) ;
\draw  [color={rgb, 255:red, 208; green, 2; blue, 27 }  ,draw opacity=1 ] (231.5,175.57) -- (318.5,175.57) -- (318.5,197.07) -- (231.5,197.07) -- cycle ;
\draw  [color={rgb, 255:red, 208; green, 2; blue, 27 }  ,draw opacity=1 ] (334.5,175.57) -- (422,175.57) -- (422,197.07) -- (334.5,197.07) -- cycle ;

\draw (197.84,177.17) node [anchor=north west][inner sep=0.75pt]  [font=\fontsize{9.8pt}{9pt}\selectfont,color={rgb, 255:red, 0; green, 0; blue, 0 }  ,opacity=1 ]  {$\omega \ =\ \left(\textcolor[rgb]{0.61,0.41,0.11}{\lambda }\textcolor[rgb]{0.61,0.41,0.11}{_{4}^{1}} +\textcolor[rgb]{0.25,0.46,0.02}{\lambda }\textcolor[rgb]{0.25,0.46,0.02}{_{3}^{2}} +\textcolor[rgb]{0.62,0.36,0.68}{\lambda }\textcolor[rgb]{0.62,0.36,0.68}{_{3}^{3}}\right) -\left(\textcolor[rgb]{0.61,0.41,0.11}{\lambda }\textcolor[rgb]{0.61,0.41,0.11}{_{4}^{1}} +\textcolor[rgb]{0.25,0.46,0.02}{\lambda }\textcolor[rgb]{0.25,0.46,0.02}{_{1}^{2}} +\textcolor[rgb]{0.62,0.36,0.68}{\lambda }\textcolor[rgb]{0.62,0.36,0.68}{_{1}^{3}}\right)$};
\draw (290.95,25.99) node [anchor=north west][inner sep=0.75pt]  [font=\large,color={rgb, 255:red, 156; green, 105; blue, 29 }  ,opacity=1 ]  {$H_{1}$};
\draw (290.95,119.68) node [anchor=north west][inner sep=0.75pt]  [font=\large,color={rgb, 255:red, 158; green, 91; blue, 173 }  ,opacity=1 ]  {$H_{3}$};
\draw (290.95,72.99) node [anchor=north west][inner sep=0.75pt]  [font=\large,color={rgb, 255:red, 65; green, 117; blue, 5 }  ,opacity=1 ]  {$H_{2}$};
\draw  [color={rgb, 255:red, 0; green, 0; blue, 0 }  ,draw opacity=1 ][fill={rgb, 255:red, 255; green, 255; blue, 255 }  ,fill opacity=1 ]  (118.59,19.57) -- (142.59,19.57) -- (142.59,36.57) -- (118.59,36.57) -- cycle  ;
\draw (130.59,28.07) node  [font=\fontsize{9pt}{9pt}\selectfont,color={rgb, 255:red, 156; green, 105; blue, 29 }  ,opacity=1 ]  {$\lambda _{1,2}^{1}$};
\draw  [color={rgb, 255:red, 0; green, 0; blue, 0 }  ,draw opacity=1 ][fill={rgb, 255:red, 255; green, 255; blue, 255 }  ,fill opacity=1 ]  (175.32,19.57) -- (199.32,19.57) -- (199.32,36.57) -- (175.32,36.57) -- cycle  ;
\draw (187.32,28.07) node  [font=\fontsize{9pt}{9pt}\selectfont,color={rgb, 255:red, 156; green, 105; blue, 29 }  ,opacity=1 ]  {$\lambda _{3,4}^{1}$};
\draw  [color={rgb, 255:red, 0; green, 0; blue, 0 }  ,draw opacity=1 ][fill={rgb, 255:red, 255; green, 255; blue, 255 }  ,fill opacity=1 ]  (89.89,70.81) -- (113.89,70.81) -- (113.89,87.81) -- (89.89,87.81) -- cycle  ;
\draw (101.89,79.31) node  [font=\fontsize{9pt}{9pt}\selectfont,color={rgb, 255:red, 65; green, 117; blue, 5 }  ,opacity=1 ]  {$\lambda _{2,3}^{2}$};
\draw  [color={rgb, 255:red, 0; green, 0; blue, 0 }  ,draw opacity=1 ][fill={rgb, 255:red, 255; green, 255; blue, 255 }  ,fill opacity=1 ]  (64.85,70.81) -- (82.85,70.81) -- (82.85,87.81) -- (64.85,87.81) -- cycle  ;
\draw (73.85,79.31) node  [font=\fontsize{9pt}{9pt}\selectfont,color={rgb, 255:red, 65; green, 117; blue, 5 }  ,opacity=1 ]  {$\lambda _{1}^{2}$};
\draw  [color={rgb, 255:red, 0; green, 0; blue, 0 }  ,draw opacity=1 ][fill={rgb, 255:red, 255; green, 255; blue, 255 }  ,fill opacity=1 ]  (120.7,70.6) -- (138.7,70.6) -- (138.7,87.6) -- (120.7,87.6) -- cycle  ;
\draw (129.7,79.1) node  [font=\fontsize{9pt}{9pt}\selectfont,color={rgb, 255:red, 65; green, 117; blue, 5 }  ,opacity=1 ]  {$\lambda _{4}^{2}$};
\draw  [color={rgb, 255:red, 0; green, 0; blue, 0 }  ,draw opacity=1 ][fill={rgb, 255:red, 255; green, 255; blue, 255 }  ,fill opacity=1 ]  (75.8,128.93) -- (93.8,128.93) -- (93.8,145.93) -- (75.8,145.93) -- cycle  ;
\draw (84.8,137.43) node  [font=\fontsize{9pt}{9pt}\selectfont,color={rgb, 255:red, 158; green, 91; blue, 173 }  ,opacity=1 ]  {$\lambda _{2}^{3}$};
\draw  [color={rgb, 255:red, 0; green, 0; blue, 0 }  ,draw opacity=1 ][fill={rgb, 255:red, 255; green, 255; blue, 255 }  ,fill opacity=1 ]  (55.94,128.93) -- (73.94,128.93) -- (73.94,145.93) -- (55.94,145.93) -- cycle  ;
\draw (64.94,137.43) node  [font=\fontsize{9pt}{9pt}\selectfont,color={rgb, 255:red, 158; green, 91; blue, 173 }  ,opacity=1 ]  {$\lambda _{1}^{3}$};
\draw  [color={rgb, 255:red, 0; green, 0; blue, 0 }  ,draw opacity=1 ][fill={rgb, 255:red, 255; green, 255; blue, 255 }  ,fill opacity=1 ]  (95.63,128.93) -- (119.63,128.93) -- (119.63,145.93) -- (95.63,145.93) -- cycle  ;
\draw (107.63,137.43) node  [font=\fontsize{9pt}{9pt}\selectfont,color={rgb, 255:red, 158; green, 91; blue, 173 }  ,opacity=1 ]  {$\lambda _{3,4}^{3}$};
\draw  [color={rgb, 255:red, 0; green, 0; blue, 0 }  ,draw opacity=1 ][fill={rgb, 255:red, 255; green, 255; blue, 255 }  ,fill opacity=1 ]  (436.92,19.24) -- (460.92,19.24) -- (460.92,36.24) -- (436.92,36.24) -- cycle  ;
\draw (448.92,27.74) node  [font=\fontsize{9pt}{9pt}\selectfont,color={rgb, 255:red, 156; green, 105; blue, 29 }  ,opacity=1 ]  {$\lambda _{1,2}^{1}$};
\draw  [color={rgb, 255:red, 0; green, 0; blue, 0 }  ,draw opacity=1 ][fill={rgb, 255:red, 255; green, 255; blue, 255 }  ,fill opacity=1 ]  (493.66,19.24) -- (517.66,19.24) -- (517.66,36.24) -- (493.66,36.24) -- cycle  ;
\draw (505.66,27.74) node  [font=\fontsize{9pt}{9pt}\selectfont,color={rgb, 255:red, 156; green, 105; blue, 29 }  ,opacity=1 ]  {$\lambda _{3,4}^{1}$};
\draw  [color={rgb, 255:red, 0; green, 0; blue, 0 }  ,draw opacity=1 ][fill={rgb, 255:red, 255; green, 255; blue, 255 }  ,fill opacity=1 ]  (407.7,70.81) -- (431.7,70.81) -- (431.7,87.81) -- (407.7,87.81) -- cycle  ;
\draw (419.7,79.31) node  [font=\fontsize{9pt}{9pt}\selectfont,color={rgb, 255:red, 65; green, 117; blue, 5 }  ,opacity=1 ]  {$\lambda _{2,3}^{2}$};
\draw  [color={rgb, 255:red, 0; green, 0; blue, 0 }  ,draw opacity=1 ][fill={rgb, 255:red, 255; green, 255; blue, 255 }  ,fill opacity=1 ]  (382.67,70.81) -- (400.67,70.81) -- (400.67,87.81) -- (382.67,87.81) -- cycle  ;
\draw (391.67,79.31) node  [font=\fontsize{9pt}{9pt}\selectfont,color={rgb, 255:red, 65; green, 117; blue, 5 }  ,opacity=1 ]  {$\lambda _{1}^{2}$};
\draw  [color={rgb, 255:red, 0; green, 0; blue, 0 }  ,draw opacity=1 ][fill={rgb, 255:red, 255; green, 255; blue, 255 }  ,fill opacity=1 ]  (438.51,70.81) -- (456.51,70.81) -- (456.51,87.81) -- (438.51,87.81) -- cycle  ;
\draw (447.51,79.31) node  [font=\fontsize{9pt}{9pt}\selectfont,color={rgb, 255:red, 65; green, 117; blue, 5 }  ,opacity=1 ]  {$\lambda _{4}^{2}$};
\draw  [color={rgb, 255:red, 0; green, 0; blue, 0 }  ,draw opacity=1 ][fill={rgb, 255:red, 255; green, 255; blue, 255 }  ,fill opacity=1 ]  (395.3,128.93) -- (413.3,128.93) -- (413.3,145.93) -- (395.3,145.93) -- cycle  ;
\draw (404.3,137.43) node  [font=\fontsize{9pt}{9pt}\selectfont,color={rgb, 255:red, 158; green, 91; blue, 173 }  ,opacity=1 ]  {$\lambda _{2}^{3}$};
\draw  [color={rgb, 255:red, 0; green, 0; blue, 0 }  ,draw opacity=1 ][fill={rgb, 255:red, 255; green, 255; blue, 255 }  ,fill opacity=1 ]  (375.44,128.93) -- (393.44,128.93) -- (393.44,145.93) -- (375.44,145.93) -- cycle  ;
\draw (384.44,137.43) node  [font=\fontsize{9pt}{9pt}\selectfont,color={rgb, 255:red, 158; green, 91; blue, 173 }  ,opacity=1 ]  {$\lambda _{1}^{3}$};
\draw  [color={rgb, 255:red, 0; green, 0; blue, 0 }  ,draw opacity=1 ][fill={rgb, 255:red, 255; green, 255; blue, 255 }  ,fill opacity=1 ]  (415.13,128.93) -- (439.13,128.93) -- (439.13,145.93) -- (415.13,145.93) -- cycle  ;
\draw (427.13,137.43) node  [font=\fontsize{9pt}{9pt}\selectfont,color={rgb, 255:red, 158; green, 91; blue, 173 }  ,opacity=1 ]  {$\lambda _{3,4}^{3}$};
\draw (6.5,167.75) node [anchor=north west][inner sep=0.75pt]  [font=\large,color={rgb, 255:red, 208; green, 2; blue, 27 }  ,opacity=1 ] [align=left] {$\displaystyle J=( 4,3,3)$};
\draw (512,168.25) node [anchor=north west][inner sep=0.75pt]  [font=\large,color={rgb, 255:red, 208; green, 2; blue, 27 }  ,opacity=1 ] [align=left] {$\displaystyle J'=( 4,1,1)$};

\end{tikzpicture}

%% file: Figures/Redundancy_Examples.tikz
\tikzset{every picture/.style={line width=0.75pt}} 

\begin{tikzpicture}[x=0.75pt,y=0.75pt,yscale=-1,xscale=1,font=\fontsize{10pt}{10pt}\selectfont]

\draw  [color={rgb, 255:red, 143; green, 117; blue, 108 }  ,draw opacity=1 ][fill={rgb, 255:red, 181; green, 154; blue, 135 }  ,fill opacity=1 ] (330.92,79.66) -- (341.38,79.66) -- (341.38,90.83) -- (330.92,90.83) -- cycle ;
\draw  [color={rgb, 255:red, 143; green, 117; blue, 108 }  ,draw opacity=1 ][fill={rgb, 255:red, 181; green, 154; blue, 135 }  ,fill opacity=1 ] (347.48,79.66) -- (357.94,79.66) -- (357.94,90.83) -- (347.48,90.83) -- cycle ;
\draw  [color={rgb, 255:red, 143; green, 117; blue, 108 }  ,draw opacity=1 ][fill={rgb, 255:red, 181; green, 154; blue, 135 }  ,fill opacity=1 ] (546.17,79.66) -- (556.63,79.66) -- (556.63,90.83) -- (546.17,90.83) -- cycle ;
\draw  [color={rgb, 255:red, 143; green, 117; blue, 108 }  ,draw opacity=1 ][fill={rgb, 255:red, 181; green, 154; blue, 135 }  ,fill opacity=1 ] (380.61,79.66) -- (391.07,79.66) -- (391.07,90.83) -- (380.61,90.83) -- cycle ;
\draw  [color={rgb, 255:red, 143; green, 117; blue, 108 }  ,draw opacity=1 ][fill={rgb, 255:red, 181; green, 154; blue, 135 }  ,fill opacity=1 ] (529.67,79.66) -- (540.13,79.66) -- (540.13,90.83) -- (529.67,90.83) -- cycle ;
\draw  [color={rgb, 255:red, 143; green, 117; blue, 108 }  ,draw opacity=1 ][fill={rgb, 255:red, 181; green, 154; blue, 135 }  ,fill opacity=1 ] (513.11,79.66) -- (523.57,79.66) -- (523.57,90.83) -- (513.11,90.83) -- cycle ;
\draw  [color={rgb, 255:red, 143; green, 117; blue, 108 }  ,draw opacity=1 ][fill={rgb, 255:red, 181; green, 154; blue, 135 }  ,fill opacity=1 ] (496.54,79.66) -- (507,79.66) -- (507,90.83) -- (496.54,90.83) -- cycle ;
\draw  [color={rgb, 255:red, 143; green, 117; blue, 108 }  ,draw opacity=1 ][fill={rgb, 255:red, 181; green, 154; blue, 135 }  ,fill opacity=1 ] (479.98,79.66) -- (490.44,79.66) -- (490.44,90.83) -- (479.98,90.83) -- cycle ;
\draw  [color={rgb, 255:red, 143; green, 117; blue, 108 }  ,draw opacity=1 ][fill={rgb, 255:red, 181; green, 154; blue, 135 }  ,fill opacity=1 ] (413.73,79.66) -- (424.19,79.66) -- (424.19,90.83) -- (413.73,90.83) -- cycle ;
\draw  [color={rgb, 255:red, 143; green, 117; blue, 108 }  ,draw opacity=1 ][fill={rgb, 255:red, 181; green, 154; blue, 135 }  ,fill opacity=1 ] (446.86,79.66) -- (457.32,79.66) -- (457.32,90.83) -- (446.86,90.83) -- cycle ;
\draw  [color={rgb, 255:red, 143; green, 117; blue, 108 }  ,draw opacity=1 ][fill={rgb, 255:red, 181; green, 154; blue, 135 }  ,fill opacity=1 ] (463.42,79.66) -- (473.88,79.66) -- (473.88,90.83) -- (463.42,90.83) -- cycle ;
\draw  [color={rgb, 255:red, 143; green, 117; blue, 108 }  ,draw opacity=1 ][fill={rgb, 255:red, 181; green, 154; blue, 135 }  ,fill opacity=1 ] (430.29,79.66) -- (440.75,79.66) -- (440.75,90.83) -- (430.29,90.83) -- cycle ;
\draw  [color={rgb, 255:red, 143; green, 117; blue, 108 }  ,draw opacity=1 ][fill={rgb, 255:red, 181; green, 154; blue, 135 }  ,fill opacity=1 ] (397.17,79.66) -- (407.63,79.66) -- (407.63,90.83) -- (397.17,90.83) -- cycle ;
\draw  [color={rgb, 255:red, 143; green, 117; blue, 108 }  ,draw opacity=1 ][fill={rgb, 255:red, 181; green, 154; blue, 135 }  ,fill opacity=1 ] (364.04,79.66) -- (374.5,79.66) -- (374.5,90.83) -- (364.04,90.83) -- cycle ;
\draw  [color={rgb, 255:red, 143; green, 117; blue, 108 }  ,draw opacity=1 ][fill={rgb, 255:red, 181; green, 154; blue, 135 }  ,fill opacity=1 ] (314.36,19.01) -- (324.82,19.01) -- (324.82,90.83) -- (314.36,90.83) -- cycle ;
\draw    (306.43,90.83) -- (306.43,9.14) ;
\draw [shift={(306.43,6.14)}, rotate = 90] [fill={rgb, 255:red, 0; green, 0; blue, 0 }  ][line width=0.08]  [draw opacity=0] (7.14,-3.43) -- (0,0) -- (7.14,3.43) -- (4.74,0) -- cycle    ;
\draw    (306.43,90.83) -- (564.84,90.83) ;
\draw [shift={(567.84,90.83)}, rotate = 180] [fill={rgb, 255:red, 0; green, 0; blue, 0 }  ][line width=0.08]  [draw opacity=0] (7.14,-3.43) -- (0,0) -- (7.14,3.43) -- (4.74,0) -- cycle    ;
\draw [color={rgb, 255:red, 143; green, 117; blue, 108 }  ,draw opacity=1 ]   (318.78,96.3) -- (553.8,96.3) ;
\draw [shift={(556.8,96.3)}, rotate = 180] [fill={rgb, 255:red, 143; green, 117; blue, 108 }  ,fill opacity=1 ][line width=0.08]  [draw opacity=0] (5.36,-2.57) -- (0,0) -- (5.36,2.57) -- cycle    ;
\draw [shift={(315.78,96.3)}, rotate = 0] [fill={rgb, 255:red, 143; green, 117; blue, 108 }  ,fill opacity=1 ][line width=0.08]  [draw opacity=0] (5.36,-2.57) -- (0,0) -- (5.36,2.57) -- cycle    ;
\draw [color={rgb, 255:red, 128; green, 128; blue, 128 }  ,draw opacity=1 ] [dash pattern={on 0.75pt off 0.75pt}]  (330.94,74.33) .. controls (332.61,72.66) and (334.27,72.66) .. (335.94,74.33) .. controls (337.61,76) and (339.27,76) .. (340.94,74.33) .. controls (342.61,72.66) and (344.27,72.66) .. (345.94,74.33) .. controls (347.61,76) and (349.27,76) .. (350.94,74.33) .. controls (352.61,72.66) and (354.27,72.66) .. (355.94,74.33) .. controls (357.61,76) and (359.27,76) .. (360.94,74.33) .. controls (362.61,72.66) and (364.27,72.66) .. (365.94,74.33) .. controls (367.61,76) and (369.27,76) .. (370.94,74.33) .. controls (372.61,72.66) and (374.27,72.66) .. (375.94,74.33) .. controls (377.61,76) and (379.27,76) .. (380.94,74.33) .. controls (382.61,72.66) and (384.27,72.66) .. (385.94,74.33) .. controls (387.61,76) and (389.27,76) .. (390.94,74.33) .. controls (392.61,72.66) and (394.27,72.66) .. (395.94,74.33) .. controls (397.61,76) and (399.27,76) .. (400.94,74.33) .. controls (402.61,72.66) and (404.27,72.66) .. (405.94,74.33) .. controls (407.61,76) and (409.27,76) .. (410.94,74.33) .. controls (412.61,72.66) and (414.27,72.66) .. (415.94,74.33) .. controls (417.61,76) and (419.27,76) .. (420.94,74.33) .. controls (422.61,72.66) and (424.27,72.66) .. (425.94,74.33) .. controls (427.61,76) and (429.27,76) .. (430.94,74.33) .. controls (432.61,72.66) and (434.27,72.66) .. (435.94,74.33) .. controls (437.61,76) and (439.27,76) .. (440.94,74.33) .. controls (442.61,72.66) and (444.27,72.66) .. (445.94,74.33) .. controls (447.61,76) and (449.27,76) .. (450.94,74.33) .. controls (452.61,72.66) and (454.27,72.66) .. (455.94,74.33) .. controls (457.61,76) and (459.27,76) .. (460.94,74.33) .. controls (462.61,72.66) and (464.27,72.66) .. (465.94,74.33) .. controls (467.61,76) and (469.27,76) .. (470.94,74.33) .. controls (472.61,72.66) and (474.27,72.66) .. (475.94,74.33) .. controls (477.61,76) and (479.27,76) .. (480.94,74.33) .. controls (482.61,72.66) and (484.27,72.66) .. (485.94,74.33) .. controls (487.61,76) and (489.27,76) .. (490.94,74.33) .. controls (492.61,72.66) and (494.27,72.66) .. (495.94,74.33) .. controls (497.61,76) and (499.27,76) .. (500.94,74.33) .. controls (502.61,72.66) and (504.27,72.66) .. (505.94,74.33) .. controls (507.61,76) and (509.27,76) .. (510.94,74.33) .. controls (512.61,72.66) and (514.27,72.66) .. (515.94,74.33) .. controls (517.61,76) and (519.27,76) .. (520.94,74.33) .. controls (522.61,72.66) and (524.27,72.66) .. (525.94,74.33) .. controls (527.61,76) and (529.27,76) .. (530.94,74.33) .. controls (532.61,72.66) and (534.27,72.66) .. (535.94,74.33) .. controls (537.61,76) and (539.27,76) .. (540.94,74.33) .. controls (542.61,72.66) and (544.27,72.66) .. (545.94,74.33) .. controls (547.61,76) and (549.27,76) .. (550.94,74.33) .. controls (552.61,72.66) and (554.27,72.66) .. (555.94,74.33) -- (556.21,74.33) -- (556.21,74.33) ;
\draw  [color={rgb, 255:red, 108; green, 134; blue, 143 }  ,draw opacity=1 ][fill={rgb, 255:red, 141; green, 174; blue, 188 }  ,fill opacity=1 ] (53.47,18.74) -- (63.93,18.74) -- (63.93,90.56) -- (53.47,90.56) -- cycle ;
\draw  [color={rgb, 255:red, 108; green, 134; blue, 143 }  ,draw opacity=1 ][fill={rgb, 255:red, 141; green, 174; blue, 188 }  ,fill opacity=1 ] (71.89,21.45) -- (82.35,21.45) -- (82.35,90.56) -- (71.89,90.56) -- cycle ;
\draw  [color={rgb, 255:red, 108; green, 134; blue, 143 }  ,draw opacity=1 ][fill={rgb, 255:red, 141; green, 174; blue, 188 }  ,fill opacity=1 ] (90.31,28.9) -- (100.77,28.9) -- (100.77,90.56) -- (90.31,90.56) -- cycle ;
\draw  [color={rgb, 255:red, 108; green, 134; blue, 143 }  ,draw opacity=1 ][fill={rgb, 255:red, 141; green, 174; blue, 188 }  ,fill opacity=1 ] (163.98,71.36) -- (174.44,71.36) -- (174.44,90.56) -- (163.98,90.56) -- cycle ;
\draw  [color={rgb, 255:red, 108; green, 134; blue, 143 }  ,draw opacity=1 ][fill={rgb, 255:red, 141; green, 174; blue, 188 }  ,fill opacity=1 ] (182.4,78.13) -- (192.86,78.13) -- (192.86,90.56) -- (182.4,90.56) -- cycle ;
\draw  [color={rgb, 255:red, 108; green, 134; blue, 143 }  ,draw opacity=1 ][fill={rgb, 255:red, 141; green, 174; blue, 188 }  ,fill opacity=1 ] (200.79,81.52) -- (211.25,81.52) -- (211.25,90.56) -- (200.79,90.56) -- cycle ;
\draw  [color={rgb, 255:red, 108; green, 134; blue, 143 }  ,draw opacity=1 ][fill={rgb, 255:red, 141; green, 174; blue, 188 }  ,fill opacity=1 ] (108.73,38.84) -- (119.19,38.84) -- (119.19,90.56) -- (108.73,90.56) -- cycle ;
\draw  [color={rgb, 255:red, 108; green, 134; blue, 143 }  ,draw opacity=1 ][fill={rgb, 255:red, 141; green, 174; blue, 188 }  ,fill opacity=1 ] (127.15,51.71) -- (137.61,51.71) -- (137.61,90.56) -- (127.15,90.56) -- cycle ;
\draw  [color={rgb, 255:red, 108; green, 134; blue, 143 }  ,draw opacity=1 ][fill={rgb, 255:red, 141; green, 174; blue, 188 }  ,fill opacity=1 ] (145.56,60.52) -- (156.03,60.52) -- (156.03,90.33) -- (145.56,90.33) -- cycle ;
\draw    (45.54,90.56) -- (45.54,8.86) ;
\draw [shift={(45.54,5.86)}, rotate = 90] [fill={rgb, 255:red, 0; green, 0; blue, 0 }  ][line width=0.08]  [draw opacity=0] (7.14,-3.43) -- (0,0) -- (7.14,3.43) -- (4.74,0) -- cycle    ;
\draw    (45.54,90.56) -- (227.61,90.56) ;
\draw [shift={(230.61,90.56)}, rotate = 180] [fill={rgb, 255:red, 0; green, 0; blue, 0 }  ][line width=0.08]  [draw opacity=0] (7.14,-3.43) -- (0,0) -- (7.14,3.43) -- (4.74,0) -- cycle    ;
\draw [color={rgb, 255:red, 128; green, 128; blue, 128 }  ,draw opacity=1 ][line width=0.75]  [dash pattern={on 0.75pt off 0.75pt}]  (57.75,17.19) .. controls (59.44,15.14) and (61.22,14.85) .. (63.08,16.32) .. controls (64.74,17.91) and (66.41,17.9) .. (68.08,16.28) .. controls (69.99,14.77) and (71.73,14.98) .. (73.29,16.89) .. controls (74.29,18.78) and (75.87,19.11) .. (78.02,17.9) .. controls (79.86,16.67) and (81.48,17.15) .. (82.88,19.33) .. controls (83.74,21.38) and (85.17,21.89) .. (87.16,20.85) .. controls (89.73,20.08) and (91.43,20.77) .. (92.26,22.94) .. controls (93.04,25.12) and (94.52,25.8) .. (96.71,24.98) .. controls (98.96,24.21) and (100.47,24.96) .. (101.24,27.21) .. controls (101.99,29.48) and (103.53,30.28) .. (105.84,29.63) .. controls (107.67,28.73) and (108.96,29.44) .. (109.71,31.76) .. controls (110.44,34.09) and (112.01,34.98) .. (114.41,34.44) .. controls (116.83,33.92) and (118.15,34.69) .. (118.36,36.75) .. controls (119.09,39.13) and (120.68,40.08) .. (123.13,39.61) .. controls (125.06,38.84) and (126.4,39.65) .. (127.13,42.05) .. controls (127.86,44.46) and (129.2,45.28) .. (131.15,44.52) .. controls (133.64,44.09) and (135.25,45.09) .. (135.99,47.51) .. controls (136.19,49.59) and (137.54,50.42) .. (140.03,50) .. controls (142.52,49.57) and (143.87,50.4) .. (144.08,52.48) .. controls (144.84,54.89) and (146.19,55.71) .. (148.13,54.94) .. controls (150.6,54.48) and (152.22,55.45) .. (152.98,57.84) .. controls (153.21,59.91) and (154.56,60.7) .. (157.02,60.19) .. controls (159.45,59.66) and (160.79,60.43) .. (161.04,62.48) .. controls (161.84,64.83) and (163.44,65.71) .. (165.84,65.12) .. controls (168.21,64.49) and (169.8,65.32) .. (170.61,67.62) .. controls (170.92,69.64) and (172.23,70.29) .. (174.55,69.57) .. controls (176.82,68.8) and (178.38,69.52) .. (179.24,71.74) .. controls (180.14,73.95) and (181.69,74.61) .. (183.88,73.71) .. controls (186,72.75) and (187.52,73.33) .. (188.45,75.45) .. controls (189.44,77.55) and (190.95,78.04) .. (192.96,76.93) .. controls (195.36,75.87) and (197.08,76.33) .. (198.13,78.3) .. controls (199.26,80.25) and (200.95,80.57) .. (203.18,79.26) -- (207.42,79.72) ;
\draw [color={rgb, 255:red, 108; green, 134; blue, 143 }  ,draw opacity=1 ]   (57.89,96.31) -- (206.85,96.67) ;
\draw [shift={(209.85,96.67)}, rotate = 180.14] [fill={rgb, 255:red, 108; green, 134; blue, 143 }  ,fill opacity=1 ][line width=0.08]  [draw opacity=0] (5.36,-2.57) -- (0,0) -- (5.36,2.57) -- cycle    ;
\draw [shift={(54.89,96.3)}, rotate = 0.14] [fill={rgb, 255:red, 108; green, 134; blue, 143 }  ,fill opacity=1 ][line width=0.08]  [draw opacity=0] (5.36,-2.57) -- (0,0) -- (5.36,2.57) -- cycle    ;

\draw (224.44,73.12) node [anchor=north west][inner sep=0.75pt]    {$\omega $};
\draw (275.22,66.05) node [anchor=north west][inner sep=0.75pt]  [color={rgb, 255:red, 128; green, 128; blue, 128 }  ,opacity=1 ]  {$\mathcal{O}( 1)$};
\draw (124.81,97.5) node [anchor=north west][inner sep=0.75pt]  [color={rgb, 255:red, 108; green, 134; blue, 143 }  ,opacity=1 ]  {$\Omega $};
\draw (92.22,6.1) node [anchor=north west][inner sep=0.75pt]  [color={rgb, 255:red, 128; green, 128; blue, 128 }  ,opacity=1 ] [align=left] {Gaussian Distribution};
\draw (3.46,8.92) node [anchor=north west][inner sep=0.75pt]  [color={rgb, 255:red, 108; green, 134; blue, 143 }  ,opacity=1 ]  {$| R( \omega ) |$};
\draw (428.72,97.5) node [anchor=north west][inner sep=0.75pt]  [color={rgb, 255:red, 143; green, 117; blue, 108 }  ,opacity=1 ]  {$\Omega $};
\draw (377.13,56.23) node [anchor=north west][inner sep=0.75pt]  [color={rgb, 255:red, 128; green, 128; blue, 128 }  ,opacity=1 ] [align=left] {Uniform Distribution};
\draw (561.85,73.19) node [anchor=north west][inner sep=0.75pt]    {$\omega $};
\draw (264.35,8.92) node [anchor=north west][inner sep=0.75pt]  [color={rgb, 255:red, 143; green, 117; blue, 108 }  ,opacity=1 ]  {$| R( \omega ) |$};

\end{tikzpicture}

%% file: Figures/Framework_Local2design_Adapted.tikz
\newlength{\wgate}

\newlength{\hgate}
\setlength{\wgate}{15pt}
\setlength{\hgate}{.5pt}

\tikzset{every picture/.style={line width=0.75pt}} 

\begin{tikzpicture}[x=.9pt,y=0.75pt,yscale=-1.15,xscale=1.1][remember picture]

\draw  [draw opacity=0][fill={rgb, 255:red, 255; green, 255; blue, 255 }  ,fill opacity=1 ] (111.48,255.69) -- (126.35,255.69) -- (126.35,264.68) -- (111.48,264.68) -- cycle ;
\draw  [color={rgb, 255:red, 21; green, 66; blue, 57 }  ,draw opacity=0.3 ][fill={rgb, 255:red, 127; green, 191; blue, 178 }  ,fill opacity=0.3 ] (111.48,255.69) -- (126.48,255.69) -- (126.48,266.41) .. controls (117.11,266.41) and (118.98,270.27) .. (111.48,267.77) -- cycle ;

\draw   (41,29.36) .. controls (36.33,29.36) and (34,31.69) .. (34,36.36) -- (34,136.86) .. controls (34,143.53) and (31.67,146.86) .. (27,146.86) .. controls (31.67,146.86) and (34,150.19) .. (34,156.86)(34,153.86) -- (34,257.36) .. controls (34,262.03) and (36.33,264.36) .. (41,264.36) ;
\draw   (66,30) .. controls (62.81,30) and (61.21,31.6) .. (61.21,34.79) -- (61.21,34.79) .. controls (61.21,39.36) and (59.61,41.64) .. (56.42,41.64) .. controls (59.61,41.64) and (61.21,43.92) .. (61.21,48.48)(61.21,46.43) -- (61.21,48.48) .. controls (61.21,51.67) and (62.81,53.27) .. (66,53.27) ;
\draw  [color={rgb, 255:red, 184; green, 233; blue, 134 }  ,draw opacity=1 ] (298,194.46) .. controls (302.67,194.46) and (305,192.13) .. (305,187.46) -- (305,156.96) .. controls (305,150.29) and (307.33,146.96) .. (312,146.96) .. controls (307.33,146.96) and (305,143.63) .. (305,136.96)(305,139.96) -- (305,106.46) .. controls (305,101.79) and (302.67,99.46) .. (298,99.46) ;
\draw  [color={rgb, 255:red, 175; green, 42; blue, 99 }  ,draw opacity=1 ] (298,228.27) .. controls (302.39,228.27) and (304.59,226.07) .. (304.59,221.68) -- (304.59,221.68) .. controls (304.59,215.41) and (306.79,212.27) .. (311.18,212.27) .. controls (306.79,212.27) and (304.59,209.13) .. (304.59,202.86)(304.59,205.68) -- (304.59,202.86) .. controls (304.59,198.47) and (302.39,196.27) .. (298,196.27) ;
\draw  [color={rgb, 255:red, 175; green, 42; blue, 99 }  ,draw opacity=1 ] (298,97.27) .. controls (302.39,97.27) and (304.59,95.07) .. (304.59,90.68) -- (304.59,90.68) .. controls (304.59,84.41) and (306.79,81.27) .. (311.18,81.27) .. controls (306.79,81.27) and (304.59,78.13) .. (304.59,71.86)(304.59,74.68) -- (304.59,71.86) .. controls (304.59,67.47) and (302.39,65.27) .. (298,65.27) ;
\draw  [color={rgb, 255:red, 127; green, 191; blue, 178 }  ,draw opacity=1 ] (344,228.46) .. controls (348.67,228.46) and (351,226.13) .. (351,221.46) -- (351,156.96) .. controls (351,150.29) and (353.33,146.96) .. (358,146.96) .. controls (353.33,146.96) and (351,143.63) .. (351,136.96)(351,139.96) -- (351,72.46) .. controls (351,67.79) and (348.67,65.46) .. (344,65.46) ;
\draw [color={rgb, 255:red, 0; green, 0; blue, 0 }  ,draw opacity=1 ][line width=1.5]    (70,58) -- (105,58) -- (105,74.27) -- (131,74.27) -- (131,90.27) -- (213,90.27) -- (213,107.27) -- (240,107.27) -- (240,123.27) -- (293,123.27) ;
\draw [color={rgb, 255:red, 0; green, 0; blue, 0 }  ,draw opacity=1 ][line width=1.5]    (70,236) -- (105,236) -- (105,220.27) -- (131,220.27) -- (131,203.27) -- (214,203.27) -- (214,187.27) -- (240,187.27) -- (240,170.27) -- (293,170.27) ;
\draw  [color={rgb, 255:red, 21; green, 66; blue, 57 }  ,draw opacity=1 ][fill={rgb, 255:red, 127; green, 191; blue, 178 }  ,fill opacity=1 ] (406.14,53.41) -- (420.91,53.41) -- (420.91,79.09) -- (406.14,79.09) -- cycle ;
\draw  [color={rgb, 255:red, 60; green, 85; blue, 45 }  ,draw opacity=0.3 ][fill={rgb, 255:red, 127; green, 191; blue, 178 }  ,fill opacity=0.3 ] (406.14,97.82) -- (420.91,97.82) -- (420.91,123.5) -- (406.14,123.5) -- cycle ;
\draw  [color={rgb, 255:red, 60; green, 85; blue, 45 }  ,draw opacity=1 ][fill={rgb, 255:red, 184; green, 233; blue, 134 }  ,fill opacity=1 ] (406.14,142.23) -- (420.91,142.23) -- (420.91,167.91) -- (406.14,167.91) -- cycle ;
\draw  [color={rgb, 255:red, 60; green, 85; blue, 45 }  ,draw opacity=0.3 ][fill={rgb, 255:red, 184; green, 233; blue, 134 }  ,fill opacity=0.3 ] (406.14,186.64) -- (420.91,186.64) -- (420.91,212.33) -- (406.14,212.33) -- cycle ;
\draw [line width=1.5]    (402.23,231.05) -- (424.82,231.05) ;
\draw  [draw opacity=0][fill={rgb, 255:red, 255; green, 255; blue, 255 }  ,fill opacity=1 ] (126.15,39.28) -- (111.28,39.28) -- (111.28,30.29) -- (126.15,30.29) -- cycle ;
\draw  [color={rgb, 255:red, 21; green, 66; blue, 57 }  ,draw opacity=0.3 ][fill={rgb, 255:red, 127; green, 191; blue, 178 }  ,fill opacity=0.3 ] (126.15,39.28) -- (111.15,39.28) -- (111.15,28.56) .. controls (120.52,28.56) and (118.65,24.7) .. (126.15,27.2) -- cycle ;

\draw  [draw opacity=0][fill={rgb, 255:red, 255; green, 255; blue, 255 }  ,fill opacity=1 ] (220.48,255.35) -- (235.35,255.35) -- (235.35,264.35) -- (220.48,264.35) -- cycle ;
\draw  [color={rgb, 255:red, 21; green, 66; blue, 57 }  ,draw opacity=0.3 ][fill={rgb, 255:red, 127; green, 191; blue, 178 }  ,fill opacity=0.3 ] (220.48,255.35) -- (235.48,255.35) -- (235.48,266.07) .. controls (226.11,266.07) and (227.98,269.94) .. (220.48,267.44) -- cycle ;

\draw  [draw opacity=0][fill={rgb, 255:red, 255; green, 255; blue, 255 }  ,fill opacity=1 ] (235.15,38.95) -- (220.28,38.95) -- (220.28,29.95) -- (235.15,29.95) -- cycle ;
\draw  [color={rgb, 255:red, 21; green, 66; blue, 57 }  ,draw opacity=0.3 ][fill={rgb, 255:red, 127; green, 191; blue, 178 }  ,fill opacity=0.3 ] (235.15,38.95) -- (220.15,38.95) -- (220.15,28.23) .. controls (229.52,28.23) and (227.65,24.36) .. (235.15,26.86) -- cycle ;

\draw    (196.98,277.1) -- (258.98,277.1) ;
\draw [shift={(261.98,277.1)}, rotate = 180] [fill={rgb, 255:red, 0; green, 0; blue, 0 }  ][line width=0.08]  [draw opacity=0] (5.36,-2.57) -- (0,0) -- (5.36,2.57) -- cycle    ;
\draw [shift={(193.98,277.1)}, rotate = 0] [fill={rgb, 255:red, 0; green, 0; blue, 0 }  ][line width=0.08]  [draw opacity=0] (5.36,-2.57) -- (0,0) -- (5.36,2.57) -- cycle    ;
\draw    (87.92,277.18) -- (149.92,277.18) ;
\draw [shift={(152.92,277.18)}, rotate = 180] [fill={rgb, 255:red, 0; green, 0; blue, 0 }  ][line width=0.08]  [draw opacity=0] (5.36,-2.57) -- (0,0) -- (5.36,2.57) -- cycle    ;
\draw [shift={(84.92,277.18)}, rotate = 0] [fill={rgb, 255:red, 0; green, 0; blue, 0 }  ][line width=0.08]  [draw opacity=0] (5.36,-2.57) -- (0,0) -- (5.36,2.57) -- cycle    ;
\draw   (68.5,131) .. controls (64.07,131) and (61.86,133.21) .. (61.86,137.64) -- (61.86,137.64) .. controls (61.86,143.97) and (59.64,147.14) .. (55.21,147.14) .. controls (59.64,147.14) and (61.86,150.3) .. (61.86,156.63)(61.86,153.78) -- (61.86,156.63) .. controls (61.86,161.06) and (64.07,163.27) .. (68.5,163.27) ;

\draw (41,38.2) node [anchor=north west][inner sep=0.75pt]   [align=left] {m};
\draw (14,142.16) node [anchor=north west][inner sep=0.75pt]   [align=left] {n};
\draw (317,136.86) node [anchor=north west][inner sep=0.75pt]  [color={rgb, 255:red, 184; green, 233; blue, 134 }  ,opacity=1 ]  {$S_{E_{k}}$};
\draw (362,137.13) node [anchor=north west][inner sep=0.75pt]  [color={rgb, 255:red, 127; green, 191; blue, 178 }  ,opacity=1 ]  {$S_{E_k}$};
\draw (315,202.4) node [anchor=north west][inner sep=0.75pt]  [color={rgb, 255:red, 175; green, 42; blue, 99 }  ,opacity=1 ]  {$S_{\overline{E_k}}$};
\draw (313.5,70.9) node [anchor=north west][inner sep=0.75pt]  [color={rgb, 255:red, 175; green, 42; blue, 99 }  ,opacity=1 ]  {$S_{\overline{E_k}}$};
\draw (428.45,59.38) node [anchor=north west][inner sep=0.75pt]   [align=left] {Trainable local 2-design block};
\draw (427.91,99.02) node [anchor=north west][inner sep=0.75pt]   [align=left] {Trainable local 2-design block};
\draw (427.91,111.52) node [anchor=north west][inner sep=0.75pt]   [align=left] {outside the lightcone};
\draw (426.95,146.88) node [anchor=north west][inner sep=0.75pt]   [align=left] {Local encoding block};
\draw (426.41,186.52) node [anchor=north west][inner sep=0.75pt]   [align=left] {Local encoding block};
\draw (426.41,199.02) node [anchor=north west][inner sep=0.75pt]   [align=left] {outside the lightcone};
\draw (429.41,217.02) node [anchor=north west][inner sep=0.75pt]   [align=left] {Local measurement};
\draw (430.41,229.02) node [anchor=north west][inner sep=0.75pt]   [align=left] {lightcone $\displaystyle \mathcal{L}_{k}$};
\draw (110.44,281.76) node [anchor=north west][inner sep=0.75pt]    {$L_{1}$};
\draw (220.44,281.76) node [anchor=north west][inner sep=0.75pt]    {$L_{2}$};
\draw (41,142.13) node [anchor=north west][inner sep=0.75pt]  [color={rgb, 255:red, 0; green, 0; blue, 0 }  ,opacity=1 ]  {$s_k$};

\end{tikzpicture}

\begin{tikzpicture}[overlay, remember picture]
\node at (-13.82,4.31){
\begin{quantikz}[row sep = 6pt, column sep = 12pt]
& \gate[2, style={fill=blue_ansatz!30, draw=border_blue_ansatz!30}][\wgate][\hgate]{} &  \gate[1, style={fill opacity=0,draw opacity=0}][\wgate][\hgate]{}          & \gate[2, style={fill=blue_ansatz!30, draw=border_blue_ansatz!30}][\wgate][\hgate]{} &\gate[2, style={fill=green_encoding!30, draw=border_green_encoding!30}][\wgate][\hgate]{} & \gate[2, style={fill=blue_ansatz!30, draw=border_blue_ansatz!30}][\wgate][\hgate]{} &\gate[1, style={fill opacity=0,draw opacity=0}][\wgate][\hgate]{}& \gate[2, style={fill=blue_ansatz!30, draw=border_blue_ansatz!30}][\wgate][\hgate]{}& \\
&                   & \gate[2, style={fill=blue_ansatz!30, draw=border_blue_ansatz!30}][\wgate][\hgate]{} &            & &            & \gate[2, style={fill=blue_ansatz!30, draw=border_blue_ansatz!30}][\wgate][\hgate]{} &           & \\
& \gate[2, style={fill=blue_ansatz, draw=border_blue_ansatz}][\wgate][\hgate]{} &            & \gate[2, style={fill=blue_ansatz!30, draw=border_blue_ansatz!30}][\wgate][\hgate]{} & \gate[2, style={fill=green_encoding!30, draw=border_green_encoding!30}][\wgate][\hgate]{} & \gate[2, style={fill=blue_ansatz!30, draw=border_blue_ansatz!30}][\wgate][\hgate]{} &            & \gate[2, style={fill=blue_ansatz!30, draw=border_blue_ansatz!30}][\wgate][\hgate]{}& \\
&            & \gate[2, style={fill=blue_ansatz, draw=border_blue_ansatz}][\wgate][\hgate]{} &            & &            & \gate[2, style={fill=blue_ansatz!30, draw=border_blue_ansatz!30}][\wgate][\hgate]{} &           & \\
& \gate[2, style={fill=blue_ansatz, draw=border_blue_ansatz}][\wgate][\hgate]{} &            & \gate[2, style={fill=blue_ansatz, draw=border_blue_ansatz}][\wgate][\hgate]{} &\gate[2, style={fill=green_encoding, draw=border_green_encoding}][\wgate][\hgate]{} & \gate[2, style={fill=blue_ansatz, draw=border_blue_ansatz}][\wgate][\hgate]{} &            & \gate[2, style={fill=blue_ansatz!30, draw=border_blue_ansatz!30}][\wgate][\hgate]{}& \\
&            & \gate[2, style={fill=blue_ansatz, draw=border_blue_ansatz}][\wgate][\hgate]{} &            & &            & \gate[2, style={fill=blue_ansatz, draw=border_blue_ansatz}][\wgate][\hgate]{} &           & \\
& \gate[2, style={fill=blue_ansatz, draw=border_blue_ansatz}][\wgate][\hgate]{} &            & \gate[2, style={fill=blue_ansatz, draw=border_blue_ansatz}][\wgate][\hgate]{} & \gate[2, style={fill=green_encoding, draw=border_green_encoding}][\wgate][\hgate]{}& \gate[2, style={fill=blue_ansatz, draw=border_blue_ansatz}][\wgate][\hgate]{} &            & \gate[2, style={fill=blue_ansatz, draw=border_blue_ansatz}][\wgate][\hgate]{}&\meter[2][1pt][1pt]{} \\
&            & \gate[2, style={fill=blue_ansatz, draw=border_blue_ansatz}][\wgate][\hgate]{} &            & &            & \gate[2, style={fill=blue_ansatz, draw=border_blue_ansatz}][\wgate][\hgate]{} &           & \\
& \gate[2, style={fill=blue_ansatz, draw=border_blue_ansatz}][\wgate][\hgate]{} &            & \gate[2, style={fill=blue_ansatz, draw=border_blue_ansatz}][\wgate][\hgate]{} &\gate[2, style={fill=green_encoding, draw=border_green_encoding}][\wgate][\hgate]{} & \gate[2, style={fill=blue_ansatz, draw=border_blue_ansatz}][\wgate][\hgate]{} &            & \gate[2, style={fill=blue_ansatz!30, draw=border_blue_ansatz!30}][\wgate][\hgate]{}& \\
&            & \gate[2, style={fill=blue_ansatz, draw=border_blue_ansatz}][\wgate][\hgate]{} &            & &            & \gate[2, style={fill=blue_ansatz!30, draw=border_blue_ansatz!30}][\wgate][\hgate]{} &           & \\
& \gate[2, style={fill=blue_ansatz, draw=border_blue_ansatz}][\wgate][\hgate]{} &            & \gate[2, style={fill=blue_ansatz!30, draw=border_blue_ansatz!30}][\wgate][\hgate]{} & \gate[2, style={fill=green_encoding!30, draw=border_green_encoding!30}][\wgate][\hgate]{}  & \gate[2, style={fill=blue_ansatz!30, draw=border_blue_ansatz!30}][\wgate][\hgate]{} &            & \gate[2, style={fill=blue_ansatz!30, draw=border_blue_ansatz!30}][\wgate][\hgate]{}& \\
&            & \gate[2, style={fill=blue_ansatz!30, draw=border_blue_ansatz!30}][\wgate][\hgate]{} &            & &            & \gate[2, style={fill=blue_ansatz!30, draw=border_blue_ansatz!30}][\wgate][\hgate]{} &           & \\
& \gate[2, style={fill=blue_ansatz!30, draw=border_blue_ansatz!30}][\wgate][\hgate]{} &            & \gate[2, style={fill=blue_ansatz!30, draw=border_blue_ansatz!30}][\wgate][\hgate]{} & \gate[2, style={fill=green_encoding!30, draw=border_green_encoding!30}][\wgate][\hgate]{} & \gate[2, style={fill=blue_ansatz!30, draw=border_blue_ansatz!30}][\wgate][\hgate]{} &            & \gate[2, style={fill=blue_ansatz!30, draw=border_blue_ansatz!30}][\wgate][\hgate]{}& \\
&            &\gate[1, style={fill opacity=0,draw opacity=0}][\wgate][\hgate]{}&            & &            & \gate[1, style={fill opacity=0,draw opacity=0}][\wgate][\hgate]{} &           & \\
\end{quantikz}};
\end{tikzpicture}

%% file: Figures/bar.tikz
 
\tikzset{
pattern size/.store in=\mcSize, 
pattern size = 5pt,
pattern thickness/.store in=\mcThickness, 
pattern thickness = 0.3pt,
pattern radius/.store in=\mcRadius, 
pattern radius = 1pt}
\makeatletter
\pgfutil@ifundefined{pgf@pattern@name@_xl115j1h8}{
\pgfdeclarepatternformonly[\mcThickness,\mcSize]{_xl115j1h8}
{\pgfqpoint{0pt}{0pt}}
{\pgfpoint{\mcSize}{\mcSize}}
{\pgfpoint{\mcSize}{\mcSize}}
{
\pgfsetcolor{\tikz@pattern@color}
\pgfsetlinewidth{\mcThickness}
\pgfpathmoveto{\pgfqpoint{0pt}{\mcSize}}
\pgfpathlineto{\pgfpoint{\mcSize+\mcThickness}{-\mcThickness}}
\pgfpathmoveto{\pgfqpoint{0pt}{0pt}}
\pgfpathlineto{\pgfpoint{\mcSize+\mcThickness}{\mcSize+\mcThickness}}
\pgfusepath{stroke}
}}
\makeatother

 
\tikzset{
pattern size/.store in=\mcSize, 
pattern size = 5pt,
pattern thickness/.store in=\mcThickness, 
pattern thickness = 0.3pt,
pattern radius/.store in=\mcRadius, 
pattern radius = 1pt}
\makeatletter
\pgfutil@ifundefined{pgf@pattern@name@_otr8ldqp9}{
\pgfdeclarepatternformonly[\mcThickness,\mcSize]{_otr8ldqp9}
{\pgfqpoint{0pt}{0pt}}
{\pgfpoint{\mcSize}{\mcSize}}
{\pgfpoint{\mcSize}{\mcSize}}
{
\pgfsetcolor{\tikz@pattern@color}
\pgfsetlinewidth{\mcThickness}
\pgfpathmoveto{\pgfqpoint{0pt}{\mcSize}}
\pgfpathlineto{\pgfpoint{\mcSize+\mcThickness}{-\mcThickness}}
\pgfpathmoveto{\pgfqpoint{0pt}{0pt}}
\pgfpathlineto{\pgfpoint{\mcSize+\mcThickness}{\mcSize+\mcThickness}}
\pgfusepath{stroke}
}}
\makeatother

 
\tikzset{
pattern size/.store in=\mcSize, 
pattern size = 5pt,
pattern thickness/.store in=\mcThickness, 
pattern thickness = 0.3pt,
pattern radius/.store in=\mcRadius, 
pattern radius = 1pt}
\makeatletter
\pgfutil@ifundefined{pgf@pattern@name@_0hypbxzm3}{
\pgfdeclarepatternformonly[\mcThickness,\mcSize]{_0hypbxzm3}
{\pgfqpoint{0pt}{0pt}}
{\pgfpoint{\mcSize}{\mcSize}}
{\pgfpoint{\mcSize}{\mcSize}}
{
\pgfsetcolor{\tikz@pattern@color}
\pgfsetlinewidth{\mcThickness}
\pgfpathmoveto{\pgfqpoint{0pt}{\mcSize}}
\pgfpathlineto{\pgfpoint{\mcSize+\mcThickness}{-\mcThickness}}
\pgfpathmoveto{\pgfqpoint{0pt}{0pt}}
\pgfpathlineto{\pgfpoint{\mcSize+\mcThickness}{\mcSize+\mcThickness}}
\pgfusepath{stroke}
}}
\makeatother

 
\tikzset{
pattern size/.store in=\mcSize, 
pattern size = 5pt,
pattern thickness/.store in=\mcThickness, 
pattern thickness = 0.3pt,
pattern radius/.store in=\mcRadius, 
pattern radius = 1pt}
\makeatletter
\pgfutil@ifundefined{pgf@pattern@name@_83q7xoyq6}{
\pgfdeclarepatternformonly[\mcThickness,\mcSize]{_83q7xoyq6}
{\pgfqpoint{0pt}{0pt}}
{\pgfpoint{\mcSize}{\mcSize}}
{\pgfpoint{\mcSize}{\mcSize}}
{
\pgfsetcolor{\tikz@pattern@color}
\pgfsetlinewidth{\mcThickness}
\pgfpathmoveto{\pgfqpoint{0pt}{\mcSize}}
\pgfpathlineto{\pgfpoint{\mcSize+\mcThickness}{-\mcThickness}}
\pgfpathmoveto{\pgfqpoint{0pt}{0pt}}
\pgfpathlineto{\pgfpoint{\mcSize+\mcThickness}{\mcSize+\mcThickness}}
\pgfusepath{stroke}
}}
\makeatother
\tikzset{every picture/.style={line width=0.75pt}} 

\begin{tikzpicture}[x=0.75pt,y=0.75pt,yscale=-1,xscale=1]

\draw  [color={rgb, 255:red, 21; green, 55; blue, 66 }  ,draw opacity=1 ][fill={rgb, 255:red, 133; green, 153; blue, 182 }  ,fill opacity=1 ] (315.29,10.32) .. controls (315.29,7.17) and (317.85,4.61) .. (321.01,4.61) -- (406.29,4.61) .. controls (409.44,4.61) and (412,7.17) .. (412,10.32) -- (412,45.89) .. controls (412,49.05) and (409.44,51.61) .. (406.29,51.61) -- (321.01,51.61) .. controls (317.85,51.61) and (315.29,49.05) .. (315.29,45.89) -- cycle ;
\draw  [color={rgb, 255:red, 185; green, 157; blue, 61 }  ,draw opacity=1 ][fill={rgb, 255:red, 241; green, 207; blue, 140 }  ,fill opacity=1 ] (176,9.32) .. controls (176,6.17) and (178.56,3.61) .. (181.71,3.61) -- (304.29,3.61) .. controls (307.45,3.61) and (310.01,6.17) .. (310.01,9.32) -- (310.01,44.89) .. controls (310.01,48.05) and (307.45,50.61) .. (304.29,50.61) -- (181.71,50.61) .. controls (178.56,50.61) and (176,48.05) .. (176,44.89) -- cycle ;
\draw    (95.48,26.75) -- (451.42,27.1) ;
\draw [shift={(454.42,27.1)}, rotate = 180.06] [fill={rgb, 255:red, 0; green, 0; blue, 0 }  ][line width=0.08]  [draw opacity=0] (8.93,-4.29) -- (0,0) -- (8.93,4.29) -- cycle    ;
\draw    (173.07,22.3) -- (173.07,52.69) ;
\draw    (312.67,22.3) -- (312.67,52.69) ;
\draw  [color={rgb, 255:red, 155; green, 155; blue, 155 }  ,draw opacity=1 ][pattern=_xl115j1h8,pattern size=6pt,pattern thickness=1.5pt,pattern radius=0pt, pattern color={rgb, 255:red, 155; green, 155; blue, 155}] (486.51,12.68) .. controls (486.51,11.58) and (487.4,10.68) .. (488.51,10.68) -- (523.15,10.68) .. controls (524.25,10.68) and (525.15,11.58) .. (525.15,12.68) -- (525.15,18.68) .. controls (525.15,19.79) and (524.25,20.68) .. (523.15,20.68) -- (488.51,20.68) .. controls (487.4,20.68) and (486.51,19.79) .. (486.51,18.68) -- cycle ;
\draw  [color={rgb, 255:red, 56; green, 53; blue, 53 }  ,draw opacity=1 ][pattern=_otr8ldqp9,pattern size=6pt,pattern thickness=1.5pt,pattern radius=0pt, pattern color={rgb, 255:red, 56; green, 53; blue, 53}] (101,33.64) .. controls (101,31.7) and (102.57,30.13) .. (104.51,30.13) -- (302.95,30.13) .. controls (304.89,30.13) and (306.46,31.7) .. (306.46,33.64) -- (306.46,44.17) .. controls (306.46,46.11) and (304.89,47.68) .. (302.95,47.68) -- (104.51,47.68) .. controls (102.57,47.68) and (101,46.11) .. (101,44.17) -- cycle ;
\draw  [color={rgb, 255:red, 56; green, 53; blue, 53 }  ,draw opacity=1 ][pattern=_0hypbxzm3,pattern size=6pt,pattern thickness=1.5pt,pattern radius=0pt, pattern color={rgb, 255:red, 56; green, 53; blue, 53}] (487.22,46.88) .. controls (487.22,45.83) and (488.07,44.98) .. (489.11,44.98) -- (523.25,44.98) .. controls (524.3,44.98) and (525.15,45.83) .. (525.15,46.88) -- (525.15,52.57) .. controls (525.15,53.62) and (524.3,54.47) .. (523.25,54.47) -- (489.11,54.47) .. controls (488.07,54.47) and (487.22,53.62) .. (487.22,52.57) -- cycle ;
\draw  [color={rgb, 255:red, 155; green, 155; blue, 155 }  ,draw opacity=1 ][pattern=_83q7xoyq6,pattern size=6pt,pattern thickness=1.5pt,pattern radius=0pt, pattern color={rgb, 255:red, 155; green, 155; blue, 155}] (100.07,9.73) .. controls (100.07,7.82) and (101.62,6.26) .. (103.53,6.26) -- (167.53,6.26) .. controls (169.45,6.26) and (171,7.82) .. (171,9.73) -- (171,20.13) .. controls (171,22.05) and (169.45,23.6) .. (167.53,23.6) -- (103.53,23.6) .. controls (101.62,23.6) and (100.07,22.05) .. (100.07,20.13) -- cycle ;

\draw (20.19,7.52) node [anchor=north west][inner sep=0.75pt]   [align=left] {Model $\displaystyle f$};
\draw (2.19,30.89) node [anchor=north west][inner sep=0.75pt]   [align=left] {Coefficient $\displaystyle c_{\omega }$};
\draw (135.75,56.25) node [anchor=north west][inner sep=0.75pt]    {$O\left( 1/2^{n}\right)$};
\draw (228.64,57.75) node [anchor=north west][inner sep=0.75pt]    {$.\ .\ .\ \ \ \ \ O( poly( n))$};
\draw (418.34,33.92) node [anchor=north west][inner sep=0.75pt]    {$\varepsilon _{M} \ \| O\| _{2}^{2}$};
\draw (527.58,8.45) node [anchor=north west][inner sep=0.75pt]   [align=left] {Proven vanishing\\};
\draw (556.35,39.16) node [anchor=north west][inner sep=0.75pt]  [font=\normalsize] [align=left] {};
\draw (529.47,35.96) node [anchor=north west][inner sep=0.75pt]   [align=left] {Proven vanishing };
\draw (529.46,50.63) node [anchor=north west][inner sep=0.75pt]   [align=left] {Fourier coefficients};
\draw (529.46,20.82) node [anchor=north west][inner sep=0.75pt]   [align=left] {model};

\end{tikzpicture}

%% file: Figures/figs_numerics/intersection.pgf
\begingroup%
\makeatletter%
\begin{pgfpicture}%
\pgfpathrectangle{\pgfpointorigin}{\pgfqpoint{9.843790in}{3.259738in}}%
\pgfusepath{use as bounding box, clip}%
\begin{pgfscope}%
\pgfsetbuttcap%
\pgfsetmiterjoin%
\pgfsetlinewidth{0.000000pt}%
\definecolor{currentstroke}{rgb}{1.000000,1.000000,1.000000}%
\pgfsetstrokecolor{currentstroke}%
\pgfsetdash{}{0pt}%
\pgfpathmoveto{\pgfqpoint{0.000000in}{0.000000in}}%
\pgfpathlineto{\pgfqpoint{9.843790in}{0.000000in}}%
\pgfpathlineto{\pgfqpoint{9.843790in}{3.259738in}}%
\pgfpathlineto{\pgfqpoint{0.000000in}{3.259738in}}%
\pgfpathlineto{\pgfqpoint{0.000000in}{0.000000in}}%
\pgfpathclose%
\pgfusepath{}%
\end{pgfscope}%
\begin{pgfscope}%
\pgfsetbuttcap%
\pgfsetmiterjoin%
\pgfsetlinewidth{0.000000pt}%
\definecolor{currentstroke}{rgb}{0.000000,0.000000,0.000000}%
\pgfsetstrokecolor{currentstroke}%
\pgfsetstrokeopacity{0.000000}%
\pgfsetdash{}{0pt}%
\pgfpathmoveto{\pgfqpoint{0.752497in}{0.631404in}}%
\pgfpathlineto{\pgfqpoint{5.084303in}{0.631404in}}%
\pgfpathlineto{\pgfqpoint{5.084303in}{2.702875in}}%
\pgfpathlineto{\pgfqpoint{0.752497in}{2.702875in}}%
\pgfpathlineto{\pgfqpoint{0.752497in}{0.631404in}}%
\pgfpathclose%
\pgfusepath{}%
\end{pgfscope}%
\begin{pgfscope}%
\pgfpathrectangle{\pgfqpoint{0.752497in}{0.631404in}}{\pgfqpoint{4.331806in}{2.071471in}}%
\pgfusepath{clip}%
\pgfsetroundcap%
\pgfsetroundjoin%
\pgfsetlinewidth{1.003750pt}%
\definecolor{currentstroke}{rgb}{0.800000,0.800000,0.800000}%
\pgfsetstrokecolor{currentstroke}%
\pgfsetdash{}{0pt}%
\pgfpathmoveto{\pgfqpoint{0.958774in}{0.631404in}}%
\pgfpathlineto{\pgfqpoint{0.958774in}{2.702875in}}%
\pgfusepath{stroke}%
\end{pgfscope}%
\begin{pgfscope}%
\definecolor{textcolor}{rgb}{0.150000,0.150000,0.150000}%
\pgfsetstrokecolor{textcolor}%
\pgfsetfillcolor{textcolor}%
\pgftext[x=0.958774in,y=0.546682in,,top]{\color{textcolor}\rmfamily\fontsize{14.000000}{16.800000}\selectfont \(\displaystyle {0}\)}%
\end{pgfscope}%
\begin{pgfscope}%
\pgfpathrectangle{\pgfqpoint{0.752497in}{0.631404in}}{\pgfqpoint{4.331806in}{2.071471in}}%
\pgfusepath{clip}%
\pgfsetroundcap%
\pgfsetroundjoin%
\pgfsetlinewidth{1.003750pt}%
\definecolor{currentstroke}{rgb}{0.800000,0.800000,0.800000}%
\pgfsetstrokecolor{currentstroke}%
\pgfsetdash{}{0pt}%
\pgfpathmoveto{\pgfqpoint{1.783880in}{0.631404in}}%
\pgfpathlineto{\pgfqpoint{1.783880in}{2.702875in}}%
\pgfusepath{stroke}%
\end{pgfscope}%
\begin{pgfscope}%
\definecolor{textcolor}{rgb}{0.150000,0.150000,0.150000}%
\pgfsetstrokecolor{textcolor}%
\pgfsetfillcolor{textcolor}%
\pgftext[x=1.783880in,y=0.546682in,,top]{\color{textcolor}\rmfamily\fontsize{14.000000}{16.800000}\selectfont \(\displaystyle {2}\)}%
\end{pgfscope}%
\begin{pgfscope}%
\pgfpathrectangle{\pgfqpoint{0.752497in}{0.631404in}}{\pgfqpoint{4.331806in}{2.071471in}}%
\pgfusepath{clip}%
\pgfsetroundcap%
\pgfsetroundjoin%
\pgfsetlinewidth{1.003750pt}%
\definecolor{currentstroke}{rgb}{0.800000,0.800000,0.800000}%
\pgfsetstrokecolor{currentstroke}%
\pgfsetdash{}{0pt}%
\pgfpathmoveto{\pgfqpoint{2.608986in}{0.631404in}}%
\pgfpathlineto{\pgfqpoint{2.608986in}{2.702875in}}%
\pgfusepath{stroke}%
\end{pgfscope}%
\begin{pgfscope}%
\definecolor{textcolor}{rgb}{0.150000,0.150000,0.150000}%
\pgfsetstrokecolor{textcolor}%
\pgfsetfillcolor{textcolor}%
\pgftext[x=2.608986in,y=0.546682in,,top]{\color{textcolor}\rmfamily\fontsize{14.000000}{16.800000}\selectfont \(\displaystyle {4}\)}%
\end{pgfscope}%
\begin{pgfscope}%
\pgfpathrectangle{\pgfqpoint{0.752497in}{0.631404in}}{\pgfqpoint{4.331806in}{2.071471in}}%
\pgfusepath{clip}%
\pgfsetroundcap%
\pgfsetroundjoin%
\pgfsetlinewidth{1.003750pt}%
\definecolor{currentstroke}{rgb}{0.800000,0.800000,0.800000}%
\pgfsetstrokecolor{currentstroke}%
\pgfsetdash{}{0pt}%
\pgfpathmoveto{\pgfqpoint{3.434091in}{0.631404in}}%
\pgfpathlineto{\pgfqpoint{3.434091in}{2.702875in}}%
\pgfusepath{stroke}%
\end{pgfscope}%
\begin{pgfscope}%
\definecolor{textcolor}{rgb}{0.150000,0.150000,0.150000}%
\pgfsetstrokecolor{textcolor}%
\pgfsetfillcolor{textcolor}%
\pgftext[x=3.434091in,y=0.546682in,,top]{\color{textcolor}\rmfamily\fontsize{14.000000}{16.800000}\selectfont \(\displaystyle {6}\)}%
\end{pgfscope}%
\begin{pgfscope}%
\pgfpathrectangle{\pgfqpoint{0.752497in}{0.631404in}}{\pgfqpoint{4.331806in}{2.071471in}}%
\pgfusepath{clip}%
\pgfsetroundcap%
\pgfsetroundjoin%
\pgfsetlinewidth{1.003750pt}%
\definecolor{currentstroke}{rgb}{0.800000,0.800000,0.800000}%
\pgfsetstrokecolor{currentstroke}%
\pgfsetdash{}{0pt}%
\pgfpathmoveto{\pgfqpoint{4.259197in}{0.631404in}}%
\pgfpathlineto{\pgfqpoint{4.259197in}{2.702875in}}%
\pgfusepath{stroke}%
\end{pgfscope}%
\begin{pgfscope}%
\definecolor{textcolor}{rgb}{0.150000,0.150000,0.150000}%
\pgfsetstrokecolor{textcolor}%
\pgfsetfillcolor{textcolor}%
\pgftext[x=4.259197in,y=0.546682in,,top]{\color{textcolor}\rmfamily\fontsize{14.000000}{16.800000}\selectfont \(\displaystyle {8}\)}%
\end{pgfscope}%
\begin{pgfscope}%
\pgfpathrectangle{\pgfqpoint{0.752497in}{0.631404in}}{\pgfqpoint{4.331806in}{2.071471in}}%
\pgfusepath{clip}%
\pgfsetroundcap%
\pgfsetroundjoin%
\pgfsetlinewidth{1.003750pt}%
\definecolor{currentstroke}{rgb}{0.800000,0.800000,0.800000}%
\pgfsetstrokecolor{currentstroke}%
\pgfsetdash{}{0pt}%
\pgfpathmoveto{\pgfqpoint{5.084303in}{0.631404in}}%
\pgfpathlineto{\pgfqpoint{5.084303in}{2.702875in}}%
\pgfusepath{stroke}%
\end{pgfscope}%
\begin{pgfscope}%
\definecolor{textcolor}{rgb}{0.150000,0.150000,0.150000}%
\pgfsetstrokecolor{textcolor}%
\pgfsetfillcolor{textcolor}%
\pgftext[x=5.084303in,y=0.546682in,,top]{\color{textcolor}\rmfamily\fontsize{14.000000}{16.800000}\selectfont \(\displaystyle {10}\)}%
\end{pgfscope}%
\begin{pgfscope}%
\definecolor{textcolor}{rgb}{0.150000,0.150000,0.150000}%
\pgfsetstrokecolor{textcolor}%
\pgfsetfillcolor{textcolor}%
\pgftext[x=2.918400in,y=0.313349in,,top]{\color{textcolor}\rmfamily\fontsize{18.000000}{21.600000}\selectfont \(\displaystyle \omega\)}%
\end{pgfscope}%
\begin{pgfscope}%
\pgfpathrectangle{\pgfqpoint{0.752497in}{0.631404in}}{\pgfqpoint{4.331806in}{2.071471in}}%
\pgfusepath{clip}%
\pgfsetroundcap%
\pgfsetroundjoin%
\pgfsetlinewidth{1.003750pt}%
\definecolor{currentstroke}{rgb}{0.800000,0.800000,0.800000}%
\pgfsetstrokecolor{currentstroke}%
\pgfsetdash{}{0pt}%
\pgfpathmoveto{\pgfqpoint{0.752497in}{0.927642in}}%
\pgfpathlineto{\pgfqpoint{5.084303in}{0.927642in}}%
\pgfusepath{stroke}%
\end{pgfscope}%
\begin{pgfscope}%
\definecolor{textcolor}{rgb}{0.150000,0.150000,0.150000}%
\pgfsetstrokecolor{textcolor}%
\pgfsetfillcolor{textcolor}%
\pgftext[x=0.395555in, y=0.858197in, left, base]{\color{textcolor}\rmfamily\fontsize{14.000000}{16.800000}\selectfont \(\displaystyle {10^{2}}\)}%
\end{pgfscope}%
\begin{pgfscope}%
\pgfpathrectangle{\pgfqpoint{0.752497in}{0.631404in}}{\pgfqpoint{4.331806in}{2.071471in}}%
\pgfusepath{clip}%
\pgfsetroundcap%
\pgfsetroundjoin%
\pgfsetlinewidth{1.003750pt}%
\definecolor{currentstroke}{rgb}{0.800000,0.800000,0.800000}%
\pgfsetstrokecolor{currentstroke}%
\pgfsetdash{}{0pt}%
\pgfpathmoveto{\pgfqpoint{0.752497in}{1.970260in}}%
\pgfpathlineto{\pgfqpoint{5.084303in}{1.970260in}}%
\pgfusepath{stroke}%
\end{pgfscope}%
\begin{pgfscope}%
\definecolor{textcolor}{rgb}{0.150000,0.150000,0.150000}%
\pgfsetstrokecolor{textcolor}%
\pgfsetfillcolor{textcolor}%
\pgftext[x=0.395555in, y=1.900815in, left, base]{\color{textcolor}\rmfamily\fontsize{14.000000}{16.800000}\selectfont \(\displaystyle {10^{3}}\)}%
\end{pgfscope}%
\begin{pgfscope}%
\definecolor{textcolor}{rgb}{0.150000,0.150000,0.150000}%
\pgfsetstrokecolor{textcolor}%
\pgfsetfillcolor{textcolor}%
\pgftext[x=0.340000in,y=1.667139in,,bottom,rotate=90.000000]{\color{textcolor}\rmfamily\fontsize{18.000000}{21.600000}\selectfont \(\displaystyle |R(\omega)|\)}%
\end{pgfscope}%
\begin{pgfscope}%
\pgfpathrectangle{\pgfqpoint{0.752497in}{0.631404in}}{\pgfqpoint{4.331806in}{2.071471in}}%
\pgfusepath{clip}%
\pgfsetbuttcap%
\pgfsetmiterjoin%
\definecolor{currentfill}{rgb}{0.333333,0.556863,0.631373}%
\pgfsetfillcolor{currentfill}%
\pgfsetlinewidth{0.000000pt}%
\definecolor{currentstroke}{rgb}{0.000000,0.000000,0.000000}%
\pgfsetstrokecolor{currentstroke}%
\pgfsetstrokeopacity{0.000000}%
\pgfsetdash{}{0pt}%
\pgfpathmoveto{\pgfqpoint{1.123795in}{-226.701573in}}%
\pgfpathlineto{\pgfqpoint{1.123795in}{2.608717in}}%
\pgfpathlineto{\pgfqpoint{0.793753in}{2.608717in}}%
\pgfpathlineto{\pgfqpoint{0.793753in}{-226.701573in}}%
\pgfusepath{fill}%
\end{pgfscope}%
\begin{pgfscope}%
\pgfpathrectangle{\pgfqpoint{0.752497in}{0.631404in}}{\pgfqpoint{4.331806in}{2.071471in}}%
\pgfusepath{clip}%
\pgfsetbuttcap%
\pgfsetmiterjoin%
\definecolor{currentfill}{rgb}{0.333333,0.556863,0.631373}%
\pgfsetfillcolor{currentfill}%
\pgfsetlinewidth{0.000000pt}%
\definecolor{currentstroke}{rgb}{0.000000,0.000000,0.000000}%
\pgfsetstrokecolor{currentstroke}%
\pgfsetstrokeopacity{0.000000}%
\pgfsetdash{}{0pt}%
\pgfpathmoveto{\pgfqpoint{1.536348in}{-226.701573in}}%
\pgfpathlineto{\pgfqpoint{1.536348in}{2.294858in}}%
\pgfpathlineto{\pgfqpoint{1.206306in}{2.294858in}}%
\pgfpathlineto{\pgfqpoint{1.206306in}{-226.701573in}}%
\pgfusepath{fill}%
\end{pgfscope}%
\begin{pgfscope}%
\pgfpathrectangle{\pgfqpoint{0.752497in}{0.631404in}}{\pgfqpoint{4.331806in}{2.071471in}}%
\pgfusepath{clip}%
\pgfsetbuttcap%
\pgfsetmiterjoin%
\definecolor{currentfill}{rgb}{0.333333,0.556863,0.631373}%
\pgfsetfillcolor{currentfill}%
\pgfsetlinewidth{0.000000pt}%
\definecolor{currentstroke}{rgb}{0.000000,0.000000,0.000000}%
\pgfsetstrokecolor{currentstroke}%
\pgfsetstrokeopacity{0.000000}%
\pgfsetdash{}{0pt}%
\pgfpathmoveto{\pgfqpoint{1.948901in}{-226.701573in}}%
\pgfpathlineto{\pgfqpoint{1.948901in}{1.980999in}}%
\pgfpathlineto{\pgfqpoint{1.618859in}{1.980999in}}%
\pgfpathlineto{\pgfqpoint{1.618859in}{-226.701573in}}%
\pgfusepath{fill}%
\end{pgfscope}%
\begin{pgfscope}%
\pgfpathrectangle{\pgfqpoint{0.752497in}{0.631404in}}{\pgfqpoint{4.331806in}{2.071471in}}%
\pgfusepath{clip}%
\pgfsetbuttcap%
\pgfsetmiterjoin%
\definecolor{currentfill}{rgb}{0.333333,0.556863,0.631373}%
\pgfsetfillcolor{currentfill}%
\pgfsetlinewidth{0.000000pt}%
\definecolor{currentstroke}{rgb}{0.000000,0.000000,0.000000}%
\pgfsetstrokecolor{currentstroke}%
\pgfsetstrokeopacity{0.000000}%
\pgfsetdash{}{0pt}%
\pgfpathmoveto{\pgfqpoint{2.361454in}{-226.701573in}}%
\pgfpathlineto{\pgfqpoint{2.361454in}{2.294858in}}%
\pgfpathlineto{\pgfqpoint{2.031412in}{2.294858in}}%
\pgfpathlineto{\pgfqpoint{2.031412in}{-226.701573in}}%
\pgfusepath{fill}%
\end{pgfscope}%
\begin{pgfscope}%
\pgfpathrectangle{\pgfqpoint{0.752497in}{0.631404in}}{\pgfqpoint{4.331806in}{2.071471in}}%
\pgfusepath{clip}%
\pgfsetbuttcap%
\pgfsetmiterjoin%
\definecolor{currentfill}{rgb}{0.333333,0.556863,0.631373}%
\pgfsetfillcolor{currentfill}%
\pgfsetlinewidth{0.000000pt}%
\definecolor{currentstroke}{rgb}{0.000000,0.000000,0.000000}%
\pgfsetstrokecolor{currentstroke}%
\pgfsetstrokeopacity{0.000000}%
\pgfsetdash{}{0pt}%
\pgfpathmoveto{\pgfqpoint{2.774007in}{-226.701573in}}%
\pgfpathlineto{\pgfqpoint{2.774007in}{1.980999in}}%
\pgfpathlineto{\pgfqpoint{2.443964in}{1.980999in}}%
\pgfpathlineto{\pgfqpoint{2.443964in}{-226.701573in}}%
\pgfusepath{fill}%
\end{pgfscope}%
\begin{pgfscope}%
\pgfpathrectangle{\pgfqpoint{0.752497in}{0.631404in}}{\pgfqpoint{4.331806in}{2.071471in}}%
\pgfusepath{clip}%
\pgfsetbuttcap%
\pgfsetmiterjoin%
\definecolor{currentfill}{rgb}{0.333333,0.556863,0.631373}%
\pgfsetfillcolor{currentfill}%
\pgfsetlinewidth{0.000000pt}%
\definecolor{currentstroke}{rgb}{0.000000,0.000000,0.000000}%
\pgfsetstrokecolor{currentstroke}%
\pgfsetstrokeopacity{0.000000}%
\pgfsetdash{}{0pt}%
\pgfpathmoveto{\pgfqpoint{3.186560in}{-226.701573in}}%
\pgfpathlineto{\pgfqpoint{3.186560in}{1.667139in}}%
\pgfpathlineto{\pgfqpoint{2.856517in}{1.667139in}}%
\pgfpathlineto{\pgfqpoint{2.856517in}{-226.701573in}}%
\pgfusepath{fill}%
\end{pgfscope}%
\begin{pgfscope}%
\pgfpathrectangle{\pgfqpoint{0.752497in}{0.631404in}}{\pgfqpoint{4.331806in}{2.071471in}}%
\pgfusepath{clip}%
\pgfsetbuttcap%
\pgfsetmiterjoin%
\definecolor{currentfill}{rgb}{0.333333,0.556863,0.631373}%
\pgfsetfillcolor{currentfill}%
\pgfsetlinewidth{0.000000pt}%
\definecolor{currentstroke}{rgb}{0.000000,0.000000,0.000000}%
\pgfsetstrokecolor{currentstroke}%
\pgfsetstrokeopacity{0.000000}%
\pgfsetdash{}{0pt}%
\pgfpathmoveto{\pgfqpoint{3.599113in}{-226.701573in}}%
\pgfpathlineto{\pgfqpoint{3.599113in}{1.980999in}}%
\pgfpathlineto{\pgfqpoint{3.269070in}{1.980999in}}%
\pgfpathlineto{\pgfqpoint{3.269070in}{-226.701573in}}%
\pgfusepath{fill}%
\end{pgfscope}%
\begin{pgfscope}%
\pgfpathrectangle{\pgfqpoint{0.752497in}{0.631404in}}{\pgfqpoint{4.331806in}{2.071471in}}%
\pgfusepath{clip}%
\pgfsetbuttcap%
\pgfsetmiterjoin%
\definecolor{currentfill}{rgb}{0.333333,0.556863,0.631373}%
\pgfsetfillcolor{currentfill}%
\pgfsetlinewidth{0.000000pt}%
\definecolor{currentstroke}{rgb}{0.000000,0.000000,0.000000}%
\pgfsetstrokecolor{currentstroke}%
\pgfsetstrokeopacity{0.000000}%
\pgfsetdash{}{0pt}%
\pgfpathmoveto{\pgfqpoint{4.011666in}{-226.701573in}}%
\pgfpathlineto{\pgfqpoint{4.011666in}{1.667139in}}%
\pgfpathlineto{\pgfqpoint{3.681623in}{1.667139in}}%
\pgfpathlineto{\pgfqpoint{3.681623in}{-226.701573in}}%
\pgfusepath{fill}%
\end{pgfscope}%
\begin{pgfscope}%
\pgfpathrectangle{\pgfqpoint{0.752497in}{0.631404in}}{\pgfqpoint{4.331806in}{2.071471in}}%
\pgfusepath{clip}%
\pgfsetbuttcap%
\pgfsetmiterjoin%
\definecolor{currentfill}{rgb}{0.333333,0.556863,0.631373}%
\pgfsetfillcolor{currentfill}%
\pgfsetlinewidth{0.000000pt}%
\definecolor{currentstroke}{rgb}{0.000000,0.000000,0.000000}%
\pgfsetstrokecolor{currentstroke}%
\pgfsetstrokeopacity{0.000000}%
\pgfsetdash{}{0pt}%
\pgfpathmoveto{\pgfqpoint{4.424218in}{-226.701573in}}%
\pgfpathlineto{\pgfqpoint{4.424218in}{1.980999in}}%
\pgfpathlineto{\pgfqpoint{4.094176in}{1.980999in}}%
\pgfpathlineto{\pgfqpoint{4.094176in}{-226.701573in}}%
\pgfusepath{fill}%
\end{pgfscope}%
\begin{pgfscope}%
\pgfpathrectangle{\pgfqpoint{0.752497in}{0.631404in}}{\pgfqpoint{4.331806in}{2.071471in}}%
\pgfusepath{clip}%
\pgfsetbuttcap%
\pgfsetmiterjoin%
\definecolor{currentfill}{rgb}{0.333333,0.556863,0.631373}%
\pgfsetfillcolor{currentfill}%
\pgfsetlinewidth{0.000000pt}%
\definecolor{currentstroke}{rgb}{0.000000,0.000000,0.000000}%
\pgfsetstrokecolor{currentstroke}%
\pgfsetstrokeopacity{0.000000}%
\pgfsetdash{}{0pt}%
\pgfpathmoveto{\pgfqpoint{4.836771in}{-226.701573in}}%
\pgfpathlineto{\pgfqpoint{4.836771in}{2.294858in}}%
\pgfpathlineto{\pgfqpoint{4.506729in}{2.294858in}}%
\pgfpathlineto{\pgfqpoint{4.506729in}{-226.701573in}}%
\pgfusepath{fill}%
\end{pgfscope}%
\begin{pgfscope}%
\pgfpathrectangle{\pgfqpoint{0.752497in}{0.631404in}}{\pgfqpoint{4.331806in}{2.071471in}}%
\pgfusepath{clip}%
\pgfsetbuttcap%
\pgfsetmiterjoin%
\definecolor{currentfill}{rgb}{0.333333,0.556863,0.631373}%
\pgfsetfillcolor{currentfill}%
\pgfsetlinewidth{0.000000pt}%
\definecolor{currentstroke}{rgb}{0.000000,0.000000,0.000000}%
\pgfsetstrokecolor{currentstroke}%
\pgfsetstrokeopacity{0.000000}%
\pgfsetdash{}{0pt}%
\pgfpathmoveto{\pgfqpoint{5.249324in}{-226.701573in}}%
\pgfpathlineto{\pgfqpoint{5.249324in}{1.980999in}}%
\pgfpathlineto{\pgfqpoint{4.919282in}{1.980999in}}%
\pgfpathlineto{\pgfqpoint{4.919282in}{-226.701573in}}%
\pgfusepath{fill}%
\end{pgfscope}%
\begin{pgfscope}%
\pgfpathrectangle{\pgfqpoint{0.752497in}{0.631404in}}{\pgfqpoint{4.331806in}{2.071471in}}%
\pgfusepath{clip}%
\pgfsetbuttcap%
\pgfsetmiterjoin%
\definecolor{currentfill}{rgb}{0.992157,0.384314,0.384314}%
\pgfsetfillcolor{currentfill}%
\pgfsetlinewidth{0.000000pt}%
\definecolor{currentstroke}{rgb}{0.000000,0.000000,0.000000}%
\pgfsetstrokecolor{currentstroke}%
\pgfsetstrokeopacity{0.000000}%
\pgfsetdash{}{0pt}%
\pgfpathmoveto{\pgfqpoint{1.536348in}{-226.701573in}}%
\pgfpathlineto{\pgfqpoint{1.536348in}{2.294858in}}%
\pgfpathlineto{\pgfqpoint{1.206306in}{2.294858in}}%
\pgfpathlineto{\pgfqpoint{1.206306in}{-226.701573in}}%
\pgfusepath{fill}%
\end{pgfscope}%
\begin{pgfscope}%
\pgfpathrectangle{\pgfqpoint{0.752497in}{0.631404in}}{\pgfqpoint{4.331806in}{2.071471in}}%
\pgfusepath{clip}%
\pgfsetbuttcap%
\pgfsetmiterjoin%
\definecolor{currentfill}{rgb}{1.000000,0.647059,0.000000}%
\pgfsetfillcolor{currentfill}%
\pgfsetlinewidth{0.000000pt}%
\definecolor{currentstroke}{rgb}{0.000000,0.000000,0.000000}%
\pgfsetstrokecolor{currentstroke}%
\pgfsetstrokeopacity{0.000000}%
\pgfsetdash{}{0pt}%
\pgfpathmoveto{\pgfqpoint{51.455251in}{-226.701573in}}%
\pgfpathlineto{\pgfqpoint{51.455251in}{0.725562in}}%
\pgfpathlineto{\pgfqpoint{51.125209in}{0.725562in}}%
\pgfpathlineto{\pgfqpoint{51.125209in}{-226.701573in}}%
\pgfusepath{fill}%
\end{pgfscope}%
\begin{pgfscope}%
\pgfsetroundcap%
\pgfsetroundjoin%
\pgfsetlinewidth{1.505625pt}%
\definecolor{currentstroke}{rgb}{0.100000,0.100000,0.100000}%
\pgfsetstrokecolor{currentstroke}%
\pgfsetdash{}{0pt}%
\pgfpathmoveto{\pgfqpoint{5.019326in}{0.600332in}}%
\pgfpathlineto{\pgfqpoint{5.149280in}{0.662476in}}%
\pgfusepath{stroke}%
\end{pgfscope}%
\begin{pgfscope}%
\pgfsetroundcap%
\pgfsetroundjoin%
\pgfsetlinewidth{1.505625pt}%
\definecolor{currentstroke}{rgb}{0.100000,0.100000,0.100000}%
\pgfsetstrokecolor{currentstroke}%
\pgfsetdash{}{0pt}%
\pgfpathmoveto{\pgfqpoint{5.019326in}{2.671803in}}%
\pgfpathlineto{\pgfqpoint{5.149280in}{2.733947in}}%
\pgfusepath{stroke}%
\end{pgfscope}%
\begin{pgfscope}%
\pgfsetrectcap%
\pgfsetmiterjoin%
\pgfsetlinewidth{1.254687pt}%
\definecolor{currentstroke}{rgb}{0.800000,0.800000,0.800000}%
\pgfsetstrokecolor{currentstroke}%
\pgfsetdash{}{0pt}%
\pgfpathmoveto{\pgfqpoint{0.752497in}{0.631404in}}%
\pgfpathlineto{\pgfqpoint{0.752497in}{2.702875in}}%
\pgfusepath{stroke}%
\end{pgfscope}%
\begin{pgfscope}%
\pgfsetrectcap%
\pgfsetmiterjoin%
\pgfsetlinewidth{1.254687pt}%
\definecolor{currentstroke}{rgb}{0.800000,0.800000,0.800000}%
\pgfsetstrokecolor{currentstroke}%
\pgfsetdash{}{0pt}%
\pgfpathmoveto{\pgfqpoint{5.084303in}{0.631404in}}%
\pgfpathlineto{\pgfqpoint{5.084303in}{2.702875in}}%
\pgfusepath{stroke}%
\end{pgfscope}%
\begin{pgfscope}%
\pgfsetrectcap%
\pgfsetmiterjoin%
\pgfsetlinewidth{1.254687pt}%
\definecolor{currentstroke}{rgb}{0.800000,0.800000,0.800000}%
\pgfsetstrokecolor{currentstroke}%
\pgfsetdash{}{0pt}%
\pgfpathmoveto{\pgfqpoint{0.752497in}{0.631404in}}%
\pgfpathlineto{\pgfqpoint{5.084303in}{0.631404in}}%
\pgfusepath{stroke}%
\end{pgfscope}%
\begin{pgfscope}%
\pgfsetrectcap%
\pgfsetmiterjoin%
\pgfsetlinewidth{1.254687pt}%
\definecolor{currentstroke}{rgb}{0.800000,0.800000,0.800000}%
\pgfsetstrokecolor{currentstroke}%
\pgfsetdash{}{0pt}%
\pgfpathmoveto{\pgfqpoint{0.752497in}{2.702875in}}%
\pgfpathlineto{\pgfqpoint{5.084303in}{2.702875in}}%
\pgfusepath{stroke}%
\end{pgfscope}%
\begin{pgfscope}%
\pgfsetbuttcap%
\pgfsetmiterjoin%
\pgfsetlinewidth{0.000000pt}%
\definecolor{currentstroke}{rgb}{0.000000,0.000000,0.000000}%
\pgfsetstrokecolor{currentstroke}%
\pgfsetstrokeopacity{0.000000}%
\pgfsetdash{}{0pt}%
\pgfpathmoveto{\pgfqpoint{5.411984in}{0.631404in}}%
\pgfpathlineto{\pgfqpoint{9.743790in}{0.631404in}}%
\pgfpathlineto{\pgfqpoint{9.743790in}{2.702875in}}%
\pgfpathlineto{\pgfqpoint{5.411984in}{2.702875in}}%
\pgfpathlineto{\pgfqpoint{5.411984in}{0.631404in}}%
\pgfpathclose%
\pgfusepath{}%
\end{pgfscope}%
\begin{pgfscope}%
\pgfpathrectangle{\pgfqpoint{5.411984in}{0.631404in}}{\pgfqpoint{4.331806in}{2.071471in}}%
\pgfusepath{clip}%
\pgfsetroundcap%
\pgfsetroundjoin%
\pgfsetlinewidth{1.003750pt}%
\definecolor{currentstroke}{rgb}{0.800000,0.800000,0.800000}%
\pgfsetstrokecolor{currentstroke}%
\pgfsetdash{}{0pt}%
\pgfpathmoveto{\pgfqpoint{5.845165in}{0.631404in}}%
\pgfpathlineto{\pgfqpoint{5.845165in}{2.702875in}}%
\pgfusepath{stroke}%
\end{pgfscope}%
\begin{pgfscope}%
\definecolor{textcolor}{rgb}{0.150000,0.150000,0.150000}%
\pgfsetstrokecolor{textcolor}%
\pgfsetfillcolor{textcolor}%
\pgftext[x=5.845165in,y=0.546682in,,top]{\color{textcolor}\rmfamily\fontsize{14.000000}{16.800000}\selectfont \(\displaystyle {118}\)}%
\end{pgfscope}%
\begin{pgfscope}%
\pgfpathrectangle{\pgfqpoint{5.411984in}{0.631404in}}{\pgfqpoint{4.331806in}{2.071471in}}%
\pgfusepath{clip}%
\pgfsetroundcap%
\pgfsetroundjoin%
\pgfsetlinewidth{1.003750pt}%
\definecolor{currentstroke}{rgb}{0.800000,0.800000,0.800000}%
\pgfsetstrokecolor{currentstroke}%
\pgfsetdash{}{0pt}%
\pgfpathmoveto{\pgfqpoint{6.711526in}{0.631404in}}%
\pgfpathlineto{\pgfqpoint{6.711526in}{2.702875in}}%
\pgfusepath{stroke}%
\end{pgfscope}%
\begin{pgfscope}%
\definecolor{textcolor}{rgb}{0.150000,0.150000,0.150000}%
\pgfsetstrokecolor{textcolor}%
\pgfsetfillcolor{textcolor}%
\pgftext[x=6.711526in,y=0.546682in,,top]{\color{textcolor}\rmfamily\fontsize{14.000000}{16.800000}\selectfont \(\displaystyle {120}\)}%
\end{pgfscope}%
\begin{pgfscope}%
\pgfpathrectangle{\pgfqpoint{5.411984in}{0.631404in}}{\pgfqpoint{4.331806in}{2.071471in}}%
\pgfusepath{clip}%
\pgfsetroundcap%
\pgfsetroundjoin%
\pgfsetlinewidth{1.003750pt}%
\definecolor{currentstroke}{rgb}{0.800000,0.800000,0.800000}%
\pgfsetstrokecolor{currentstroke}%
\pgfsetdash{}{0pt}%
\pgfpathmoveto{\pgfqpoint{7.577887in}{0.631404in}}%
\pgfpathlineto{\pgfqpoint{7.577887in}{2.702875in}}%
\pgfusepath{stroke}%
\end{pgfscope}%
\begin{pgfscope}%
\definecolor{textcolor}{rgb}{0.150000,0.150000,0.150000}%
\pgfsetstrokecolor{textcolor}%
\pgfsetfillcolor{textcolor}%
\pgftext[x=7.577887in,y=0.546682in,,top]{\color{textcolor}\rmfamily\fontsize{14.000000}{16.800000}\selectfont \(\displaystyle {122}\)}%
\end{pgfscope}%
\begin{pgfscope}%
\pgfpathrectangle{\pgfqpoint{5.411984in}{0.631404in}}{\pgfqpoint{4.331806in}{2.071471in}}%
\pgfusepath{clip}%
\pgfsetroundcap%
\pgfsetroundjoin%
\pgfsetlinewidth{1.003750pt}%
\definecolor{currentstroke}{rgb}{0.800000,0.800000,0.800000}%
\pgfsetstrokecolor{currentstroke}%
\pgfsetdash{}{0pt}%
\pgfpathmoveto{\pgfqpoint{8.444248in}{0.631404in}}%
\pgfpathlineto{\pgfqpoint{8.444248in}{2.702875in}}%
\pgfusepath{stroke}%
\end{pgfscope}%
\begin{pgfscope}%
\definecolor{textcolor}{rgb}{0.150000,0.150000,0.150000}%
\pgfsetstrokecolor{textcolor}%
\pgfsetfillcolor{textcolor}%
\pgftext[x=8.444248in,y=0.546682in,,top]{\color{textcolor}\rmfamily\fontsize{14.000000}{16.800000}\selectfont \(\displaystyle {124}\)}%
\end{pgfscope}%
\begin{pgfscope}%
\pgfpathrectangle{\pgfqpoint{5.411984in}{0.631404in}}{\pgfqpoint{4.331806in}{2.071471in}}%
\pgfusepath{clip}%
\pgfsetroundcap%
\pgfsetroundjoin%
\pgfsetlinewidth{1.003750pt}%
\definecolor{currentstroke}{rgb}{0.800000,0.800000,0.800000}%
\pgfsetstrokecolor{currentstroke}%
\pgfsetdash{}{0pt}%
\pgfpathmoveto{\pgfqpoint{9.310609in}{0.631404in}}%
\pgfpathlineto{\pgfqpoint{9.310609in}{2.702875in}}%
\pgfusepath{stroke}%
\end{pgfscope}%
\begin{pgfscope}%
\definecolor{textcolor}{rgb}{0.150000,0.150000,0.150000}%
\pgfsetstrokecolor{textcolor}%
\pgfsetfillcolor{textcolor}%
\pgftext[x=9.310609in,y=0.546682in,,top]{\color{textcolor}\rmfamily\fontsize{14.000000}{16.800000}\selectfont \(\displaystyle {126}\)}%
\end{pgfscope}%
\begin{pgfscope}%
\definecolor{textcolor}{rgb}{0.150000,0.150000,0.150000}%
\pgfsetstrokecolor{textcolor}%
\pgfsetfillcolor{textcolor}%
\pgftext[x=7.577887in,y=0.313349in,,top]{\color{textcolor}\rmfamily\fontsize{18.000000}{21.600000}\selectfont \(\displaystyle \omega\)}%
\end{pgfscope}%
\begin{pgfscope}%
\pgfpathrectangle{\pgfqpoint{5.411984in}{0.631404in}}{\pgfqpoint{4.331806in}{2.071471in}}%
\pgfusepath{clip}%
\pgfsetroundcap%
\pgfsetroundjoin%
\pgfsetlinewidth{1.003750pt}%
\definecolor{currentstroke}{rgb}{0.800000,0.800000,0.800000}%
\pgfsetstrokecolor{currentstroke}%
\pgfsetdash{}{0pt}%
\pgfpathmoveto{\pgfqpoint{5.411984in}{0.927642in}}%
\pgfpathlineto{\pgfqpoint{9.743790in}{0.927642in}}%
\pgfusepath{stroke}%
\end{pgfscope}%
\begin{pgfscope}%
\pgfpathrectangle{\pgfqpoint{5.411984in}{0.631404in}}{\pgfqpoint{4.331806in}{2.071471in}}%
\pgfusepath{clip}%
\pgfsetroundcap%
\pgfsetroundjoin%
\pgfsetlinewidth{1.003750pt}%
\definecolor{currentstroke}{rgb}{0.800000,0.800000,0.800000}%
\pgfsetstrokecolor{currentstroke}%
\pgfsetdash{}{0pt}%
\pgfpathmoveto{\pgfqpoint{5.411984in}{1.970260in}}%
\pgfpathlineto{\pgfqpoint{9.743790in}{1.970260in}}%
\pgfusepath{stroke}%
\end{pgfscope}%
\begin{pgfscope}%
\pgfpathrectangle{\pgfqpoint{5.411984in}{0.631404in}}{\pgfqpoint{4.331806in}{2.071471in}}%
\pgfusepath{clip}%
\pgfsetbuttcap%
\pgfsetmiterjoin%
\definecolor{currentfill}{rgb}{0.333333,0.556863,0.631373}%
\pgfsetfillcolor{currentfill}%
\pgfsetlinewidth{0.000000pt}%
\definecolor{currentstroke}{rgb}{0.000000,0.000000,0.000000}%
\pgfsetstrokecolor{currentstroke}%
\pgfsetstrokeopacity{0.000000}%
\pgfsetdash{}{0pt}%
\pgfpathmoveto{\pgfqpoint{5.585256in}{-226.701573in}}%
\pgfpathlineto{\pgfqpoint{5.585256in}{1.667139in}}%
\pgfpathlineto{\pgfqpoint{5.238712in}{1.667139in}}%
\pgfpathlineto{\pgfqpoint{5.238712in}{-226.701573in}}%
\pgfusepath{fill}%
\end{pgfscope}%
\begin{pgfscope}%
\pgfpathrectangle{\pgfqpoint{5.411984in}{0.631404in}}{\pgfqpoint{4.331806in}{2.071471in}}%
\pgfusepath{clip}%
\pgfsetbuttcap%
\pgfsetmiterjoin%
\definecolor{currentfill}{rgb}{0.333333,0.556863,0.631373}%
\pgfsetfillcolor{currentfill}%
\pgfsetlinewidth{0.000000pt}%
\definecolor{currentstroke}{rgb}{0.000000,0.000000,0.000000}%
\pgfsetstrokecolor{currentstroke}%
\pgfsetstrokeopacity{0.000000}%
\pgfsetdash{}{0pt}%
\pgfpathmoveto{\pgfqpoint{6.018437in}{-226.701573in}}%
\pgfpathlineto{\pgfqpoint{6.018437in}{1.353280in}}%
\pgfpathlineto{\pgfqpoint{5.671892in}{1.353280in}}%
\pgfpathlineto{\pgfqpoint{5.671892in}{-226.701573in}}%
\pgfusepath{fill}%
\end{pgfscope}%
\begin{pgfscope}%
\pgfpathrectangle{\pgfqpoint{5.411984in}{0.631404in}}{\pgfqpoint{4.331806in}{2.071471in}}%
\pgfusepath{clip}%
\pgfsetbuttcap%
\pgfsetmiterjoin%
\definecolor{currentfill}{rgb}{0.333333,0.556863,0.631373}%
\pgfsetfillcolor{currentfill}%
\pgfsetlinewidth{0.000000pt}%
\definecolor{currentstroke}{rgb}{0.000000,0.000000,0.000000}%
\pgfsetstrokecolor{currentstroke}%
\pgfsetstrokeopacity{0.000000}%
\pgfsetdash{}{0pt}%
\pgfpathmoveto{\pgfqpoint{6.451618in}{-226.701573in}}%
\pgfpathlineto{\pgfqpoint{6.451618in}{1.039421in}}%
\pgfpathlineto{\pgfqpoint{6.105073in}{1.039421in}}%
\pgfpathlineto{\pgfqpoint{6.105073in}{-226.701573in}}%
\pgfusepath{fill}%
\end{pgfscope}%
\begin{pgfscope}%
\pgfpathrectangle{\pgfqpoint{5.411984in}{0.631404in}}{\pgfqpoint{4.331806in}{2.071471in}}%
\pgfusepath{clip}%
\pgfsetbuttcap%
\pgfsetmiterjoin%
\definecolor{currentfill}{rgb}{0.333333,0.556863,0.631373}%
\pgfsetfillcolor{currentfill}%
\pgfsetlinewidth{0.000000pt}%
\definecolor{currentstroke}{rgb}{0.000000,0.000000,0.000000}%
\pgfsetstrokecolor{currentstroke}%
\pgfsetstrokeopacity{0.000000}%
\pgfsetdash{}{0pt}%
\pgfpathmoveto{\pgfqpoint{6.884798in}{-226.701573in}}%
\pgfpathlineto{\pgfqpoint{6.884798in}{1.353280in}}%
\pgfpathlineto{\pgfqpoint{6.538254in}{1.353280in}}%
\pgfpathlineto{\pgfqpoint{6.538254in}{-226.701573in}}%
\pgfusepath{fill}%
\end{pgfscope}%
\begin{pgfscope}%
\pgfpathrectangle{\pgfqpoint{5.411984in}{0.631404in}}{\pgfqpoint{4.331806in}{2.071471in}}%
\pgfusepath{clip}%
\pgfsetbuttcap%
\pgfsetmiterjoin%
\definecolor{currentfill}{rgb}{0.333333,0.556863,0.631373}%
\pgfsetfillcolor{currentfill}%
\pgfsetlinewidth{0.000000pt}%
\definecolor{currentstroke}{rgb}{0.000000,0.000000,0.000000}%
\pgfsetstrokecolor{currentstroke}%
\pgfsetstrokeopacity{0.000000}%
\pgfsetdash{}{0pt}%
\pgfpathmoveto{\pgfqpoint{7.317979in}{-226.701573in}}%
\pgfpathlineto{\pgfqpoint{7.317979in}{1.039421in}}%
\pgfpathlineto{\pgfqpoint{6.971434in}{1.039421in}}%
\pgfpathlineto{\pgfqpoint{6.971434in}{-226.701573in}}%
\pgfusepath{fill}%
\end{pgfscope}%
\begin{pgfscope}%
\pgfpathrectangle{\pgfqpoint{5.411984in}{0.631404in}}{\pgfqpoint{4.331806in}{2.071471in}}%
\pgfusepath{clip}%
\pgfsetbuttcap%
\pgfsetmiterjoin%
\definecolor{currentfill}{rgb}{0.333333,0.556863,0.631373}%
\pgfsetfillcolor{currentfill}%
\pgfsetlinewidth{0.000000pt}%
\definecolor{currentstroke}{rgb}{0.000000,0.000000,0.000000}%
\pgfsetstrokecolor{currentstroke}%
\pgfsetstrokeopacity{0.000000}%
\pgfsetdash{}{0pt}%
\pgfpathmoveto{\pgfqpoint{7.751159in}{-226.701573in}}%
\pgfpathlineto{\pgfqpoint{7.751159in}{0.725562in}}%
\pgfpathlineto{\pgfqpoint{7.404615in}{0.725562in}}%
\pgfpathlineto{\pgfqpoint{7.404615in}{-226.701573in}}%
\pgfusepath{fill}%
\end{pgfscope}%
\begin{pgfscope}%
\pgfpathrectangle{\pgfqpoint{5.411984in}{0.631404in}}{\pgfqpoint{4.331806in}{2.071471in}}%
\pgfusepath{clip}%
\pgfsetbuttcap%
\pgfsetmiterjoin%
\definecolor{currentfill}{rgb}{0.333333,0.556863,0.631373}%
\pgfsetfillcolor{currentfill}%
\pgfsetlinewidth{0.000000pt}%
\definecolor{currentstroke}{rgb}{0.000000,0.000000,0.000000}%
\pgfsetstrokecolor{currentstroke}%
\pgfsetstrokeopacity{0.000000}%
\pgfsetdash{}{0pt}%
\pgfpathmoveto{\pgfqpoint{8.184340in}{-226.701573in}}%
\pgfpathlineto{\pgfqpoint{8.184340in}{1.039421in}}%
\pgfpathlineto{\pgfqpoint{7.837795in}{1.039421in}}%
\pgfpathlineto{\pgfqpoint{7.837795in}{-226.701573in}}%
\pgfusepath{fill}%
\end{pgfscope}%
\begin{pgfscope}%
\pgfpathrectangle{\pgfqpoint{5.411984in}{0.631404in}}{\pgfqpoint{4.331806in}{2.071471in}}%
\pgfusepath{clip}%
\pgfsetbuttcap%
\pgfsetmiterjoin%
\definecolor{currentfill}{rgb}{0.333333,0.556863,0.631373}%
\pgfsetfillcolor{currentfill}%
\pgfsetlinewidth{0.000000pt}%
\definecolor{currentstroke}{rgb}{0.000000,0.000000,0.000000}%
\pgfsetstrokecolor{currentstroke}%
\pgfsetstrokeopacity{0.000000}%
\pgfsetdash{}{0pt}%
\pgfpathmoveto{\pgfqpoint{8.617520in}{-226.701573in}}%
\pgfpathlineto{\pgfqpoint{8.617520in}{0.725562in}}%
\pgfpathlineto{\pgfqpoint{8.270976in}{0.725562in}}%
\pgfpathlineto{\pgfqpoint{8.270976in}{-226.701573in}}%
\pgfusepath{fill}%
\end{pgfscope}%
\begin{pgfscope}%
\pgfpathrectangle{\pgfqpoint{5.411984in}{0.631404in}}{\pgfqpoint{4.331806in}{2.071471in}}%
\pgfusepath{clip}%
\pgfsetbuttcap%
\pgfsetmiterjoin%
\definecolor{currentfill}{rgb}{0.333333,0.556863,0.631373}%
\pgfsetfillcolor{currentfill}%
\pgfsetlinewidth{0.000000pt}%
\definecolor{currentstroke}{rgb}{0.000000,0.000000,0.000000}%
\pgfsetstrokecolor{currentstroke}%
\pgfsetstrokeopacity{0.000000}%
\pgfsetdash{}{0pt}%
\pgfpathmoveto{\pgfqpoint{9.050701in}{-226.701573in}}%
\pgfpathlineto{\pgfqpoint{9.050701in}{1.039421in}}%
\pgfpathlineto{\pgfqpoint{8.704156in}{1.039421in}}%
\pgfpathlineto{\pgfqpoint{8.704156in}{-226.701573in}}%
\pgfusepath{fill}%
\end{pgfscope}%
\begin{pgfscope}%
\pgfpathrectangle{\pgfqpoint{5.411984in}{0.631404in}}{\pgfqpoint{4.331806in}{2.071471in}}%
\pgfusepath{clip}%
\pgfsetbuttcap%
\pgfsetmiterjoin%
\definecolor{currentfill}{rgb}{0.333333,0.556863,0.631373}%
\pgfsetfillcolor{currentfill}%
\pgfsetlinewidth{0.000000pt}%
\definecolor{currentstroke}{rgb}{0.000000,0.000000,0.000000}%
\pgfsetstrokecolor{currentstroke}%
\pgfsetstrokeopacity{0.000000}%
\pgfsetdash{}{0pt}%
\pgfpathmoveto{\pgfqpoint{9.483881in}{-226.701573in}}%
\pgfpathlineto{\pgfqpoint{9.483881in}{1.353280in}}%
\pgfpathlineto{\pgfqpoint{9.137337in}{1.353280in}}%
\pgfpathlineto{\pgfqpoint{9.137337in}{-226.701573in}}%
\pgfusepath{fill}%
\end{pgfscope}%
\begin{pgfscope}%
\pgfpathrectangle{\pgfqpoint{5.411984in}{0.631404in}}{\pgfqpoint{4.331806in}{2.071471in}}%
\pgfusepath{clip}%
\pgfsetbuttcap%
\pgfsetmiterjoin%
\definecolor{currentfill}{rgb}{0.333333,0.556863,0.631373}%
\pgfsetfillcolor{currentfill}%
\pgfsetlinewidth{0.000000pt}%
\definecolor{currentstroke}{rgb}{0.000000,0.000000,0.000000}%
\pgfsetstrokecolor{currentstroke}%
\pgfsetstrokeopacity{0.000000}%
\pgfsetdash{}{0pt}%
\pgfpathmoveto{\pgfqpoint{9.917062in}{-226.701573in}}%
\pgfpathlineto{\pgfqpoint{9.917062in}{1.039421in}}%
\pgfpathlineto{\pgfqpoint{9.570518in}{1.039421in}}%
\pgfpathlineto{\pgfqpoint{9.570518in}{-226.701573in}}%
\pgfusepath{fill}%
\end{pgfscope}%
\begin{pgfscope}%
\pgfpathrectangle{\pgfqpoint{5.411984in}{0.631404in}}{\pgfqpoint{4.331806in}{2.071471in}}%
\pgfusepath{clip}%
\pgfsetbuttcap%
\pgfsetmiterjoin%
\definecolor{currentfill}{rgb}{1.000000,0.647059,0.000000}%
\pgfsetfillcolor{currentfill}%
\pgfsetlinewidth{0.000000pt}%
\definecolor{currentstroke}{rgb}{0.000000,0.000000,0.000000}%
\pgfsetstrokecolor{currentstroke}%
\pgfsetstrokeopacity{0.000000}%
\pgfsetdash{}{0pt}%
\pgfpathmoveto{\pgfqpoint{7.751159in}{-226.701573in}}%
\pgfpathlineto{\pgfqpoint{7.751159in}{0.725562in}}%
\pgfpathlineto{\pgfqpoint{7.404615in}{0.725562in}}%
\pgfpathlineto{\pgfqpoint{7.404615in}{-226.701573in}}%
\pgfusepath{fill}%
\end{pgfscope}%
\begin{pgfscope}%
\pgfsetroundcap%
\pgfsetroundjoin%
\pgfsetlinewidth{1.505625pt}%
\definecolor{currentstroke}{rgb}{0.100000,0.100000,0.100000}%
\pgfsetstrokecolor{currentstroke}%
\pgfsetdash{}{0pt}%
\pgfpathmoveto{\pgfqpoint{5.347007in}{0.600332in}}%
\pgfpathlineto{\pgfqpoint{5.476961in}{0.662476in}}%
\pgfusepath{stroke}%
\end{pgfscope}%
\begin{pgfscope}%
\pgfsetroundcap%
\pgfsetroundjoin%
\pgfsetlinewidth{1.505625pt}%
\definecolor{currentstroke}{rgb}{0.100000,0.100000,0.100000}%
\pgfsetstrokecolor{currentstroke}%
\pgfsetdash{}{0pt}%
\pgfpathmoveto{\pgfqpoint{5.347007in}{2.671803in}}%
\pgfpathlineto{\pgfqpoint{5.476961in}{2.733947in}}%
\pgfusepath{stroke}%
\end{pgfscope}%
\begin{pgfscope}%
\pgfsetrectcap%
\pgfsetmiterjoin%
\pgfsetlinewidth{1.254687pt}%
\definecolor{currentstroke}{rgb}{0.800000,0.800000,0.800000}%
\pgfsetstrokecolor{currentstroke}%
\pgfsetdash{}{0pt}%
\pgfpathmoveto{\pgfqpoint{5.411984in}{0.631404in}}%
\pgfpathlineto{\pgfqpoint{5.411984in}{2.702875in}}%
\pgfusepath{stroke}%
\end{pgfscope}%
\begin{pgfscope}%
\pgfsetrectcap%
\pgfsetmiterjoin%
\pgfsetlinewidth{1.254687pt}%
\definecolor{currentstroke}{rgb}{0.800000,0.800000,0.800000}%
\pgfsetstrokecolor{currentstroke}%
\pgfsetdash{}{0pt}%
\pgfpathmoveto{\pgfqpoint{9.743790in}{0.631404in}}%
\pgfpathlineto{\pgfqpoint{9.743790in}{2.702875in}}%
\pgfusepath{stroke}%
\end{pgfscope}%
\begin{pgfscope}%
\pgfsetrectcap%
\pgfsetmiterjoin%
\pgfsetlinewidth{1.254687pt}%
\definecolor{currentstroke}{rgb}{0.800000,0.800000,0.800000}%
\pgfsetstrokecolor{currentstroke}%
\pgfsetdash{}{0pt}%
\pgfpathmoveto{\pgfqpoint{5.411984in}{0.631404in}}%
\pgfpathlineto{\pgfqpoint{9.743790in}{0.631404in}}%
\pgfusepath{stroke}%
\end{pgfscope}%
\begin{pgfscope}%
\pgfsetrectcap%
\pgfsetmiterjoin%
\pgfsetlinewidth{1.254687pt}%
\definecolor{currentstroke}{rgb}{0.800000,0.800000,0.800000}%
\pgfsetstrokecolor{currentstroke}%
\pgfsetdash{}{0pt}%
\pgfpathmoveto{\pgfqpoint{5.411984in}{2.702875in}}%
\pgfpathlineto{\pgfqpoint{9.743790in}{2.702875in}}%
\pgfusepath{stroke}%
\end{pgfscope}%
\begin{pgfscope}%
\definecolor{textcolor}{rgb}{0.150000,0.150000,0.150000}%
\pgfsetstrokecolor{textcolor}%
\pgfsetfillcolor{textcolor}%
\pgftext[x=5.253790in,y=0.967525in,,]{\color{textcolor}\rmfamily\fontsize{26.000000}{31.200000}\selectfont ...}%
\end{pgfscope}%
\begin{pgfscope}%
\definecolor{textcolor}{rgb}{0.150000,0.150000,0.150000}%
\pgfsetstrokecolor{textcolor}%
\pgfsetfillcolor{textcolor}%
\pgftext[x=1.623790in,y=2.467525in,,]{\color{textcolor}\rmfamily\fontsize{16.000000}{19.200000}\selectfont Target 1}%
\end{pgfscope}%
\begin{pgfscope}%
\definecolor{textcolor}{rgb}{0.150000,0.150000,0.150000}%
\pgfsetstrokecolor{textcolor}%
\pgfsetfillcolor{textcolor}%
\pgftext[x=7.623790in,y=1.237525in,,]{\color{textcolor}\rmfamily\fontsize{16.000000}{19.200000}\selectfont Target 2}%
\end{pgfscope}%
\begin{pgfscope}%
\pgfsetbuttcap%
\pgfsetmiterjoin%
\pgfsetlinewidth{1.003750pt}%
\definecolor{currentstroke}{rgb}{0.800000,0.800000,0.800000}%
\pgfsetstrokecolor{currentstroke}%
\pgfsetstrokeopacity{0.800000}%
\pgfsetdash{}{0pt}%
\pgfpathmoveto{\pgfqpoint{2.052769in}{2.843914in}}%
\pgfpathlineto{\pgfqpoint{7.794810in}{2.843914in}}%
\pgfpathquadraticcurveto{\pgfqpoint{7.825366in}{2.843914in}}{\pgfqpoint{7.825366in}{2.874469in}}%
\pgfpathlineto{\pgfqpoint{7.825366in}{3.129182in}}%
\pgfpathquadraticcurveto{\pgfqpoint{7.825366in}{3.159738in}}{\pgfqpoint{7.794810in}{3.159738in}}%
\pgfpathlineto{\pgfqpoint{2.052769in}{3.159738in}}%
\pgfpathquadraticcurveto{\pgfqpoint{2.022214in}{3.159738in}}{\pgfqpoint{2.022214in}{3.129182in}}%
\pgfpathlineto{\pgfqpoint{2.022214in}{2.874469in}}%
\pgfpathquadraticcurveto{\pgfqpoint{2.022214in}{2.843914in}}{\pgfqpoint{2.052769in}{2.843914in}}%
\pgfpathlineto{\pgfqpoint{2.052769in}{2.843914in}}%
\pgfpathclose%
\pgfusepath{stroke}%
\end{pgfscope}%
\begin{pgfscope}%
\pgfsetbuttcap%
\pgfsetmiterjoin%
\definecolor{currentfill}{rgb}{0.333333,0.556863,0.631373}%
\pgfsetfillcolor{currentfill}%
\pgfsetlinewidth{0.000000pt}%
\definecolor{currentstroke}{rgb}{0.000000,0.000000,0.000000}%
\pgfsetstrokecolor{currentstroke}%
\pgfsetstrokeopacity{0.000000}%
\pgfsetdash{}{0pt}%
\pgfpathmoveto{\pgfqpoint{2.083325in}{2.964227in}}%
\pgfpathlineto{\pgfqpoint{2.388880in}{2.964227in}}%
\pgfpathlineto{\pgfqpoint{2.388880in}{3.071172in}}%
\pgfpathlineto{\pgfqpoint{2.083325in}{3.071172in}}%
\pgfpathlineto{\pgfqpoint{2.083325in}{2.964227in}}%
\pgfpathclose%
\pgfusepath{fill}%
\end{pgfscope}%
\begin{pgfscope}%
\definecolor{textcolor}{rgb}{0.150000,0.150000,0.150000}%
\pgfsetstrokecolor{textcolor}%
\pgfsetfillcolor{textcolor}%
\pgftext[x=2.511102in,y=2.964227in,left,base]{\color{textcolor}\rmfamily\fontsize{11.000000}{13.200000}\selectfont \(\displaystyle |R(\omega)|\ \omega \in \Omega\)}%
\end{pgfscope}%
\begin{pgfscope}%
\pgfsetbuttcap%
\pgfsetmiterjoin%
\definecolor{currentfill}{rgb}{0.992157,0.384314,0.384314}%
\pgfsetfillcolor{currentfill}%
\pgfsetlinewidth{0.000000pt}%
\definecolor{currentstroke}{rgb}{0.000000,0.000000,0.000000}%
\pgfsetstrokecolor{currentstroke}%
\pgfsetstrokeopacity{0.000000}%
\pgfsetdash{}{0pt}%
\pgfpathmoveto{\pgfqpoint{3.682740in}{2.964227in}}%
\pgfpathlineto{\pgfqpoint{3.988295in}{2.964227in}}%
\pgfpathlineto{\pgfqpoint{3.988295in}{3.071172in}}%
\pgfpathlineto{\pgfqpoint{3.682740in}{3.071172in}}%
\pgfpathlineto{\pgfqpoint{3.682740in}{2.964227in}}%
\pgfpathclose%
\pgfusepath{fill}%
\end{pgfscope}%
\begin{pgfscope}%
\definecolor{textcolor}{rgb}{0.150000,0.150000,0.150000}%
\pgfsetstrokecolor{textcolor}%
\pgfsetfillcolor{textcolor}%
\pgftext[x=4.110518in,y=2.964227in,left,base]{\color{textcolor}\rmfamily\fontsize{11.000000}{13.200000}\selectfont \(\displaystyle |R(\omega)|\ \omega \in \Omega\cap \Omega^1_{\text{target}}\)}%
\end{pgfscope}%
\begin{pgfscope}%
\pgfsetbuttcap%
\pgfsetmiterjoin%
\definecolor{currentfill}{rgb}{1.000000,0.647059,0.000000}%
\pgfsetfillcolor{currentfill}%
\pgfsetlinewidth{0.000000pt}%
\definecolor{currentstroke}{rgb}{0.000000,0.000000,0.000000}%
\pgfsetstrokecolor{currentstroke}%
\pgfsetstrokeopacity{0.000000}%
\pgfsetdash{}{0pt}%
\pgfpathmoveto{\pgfqpoint{5.876275in}{2.964227in}}%
\pgfpathlineto{\pgfqpoint{6.181831in}{2.964227in}}%
\pgfpathlineto{\pgfqpoint{6.181831in}{3.071172in}}%
\pgfpathlineto{\pgfqpoint{5.876275in}{3.071172in}}%
\pgfpathlineto{\pgfqpoint{5.876275in}{2.964227in}}%
\pgfpathclose%
\pgfusepath{fill}%
\end{pgfscope}%
\begin{pgfscope}%
\definecolor{textcolor}{rgb}{0.150000,0.150000,0.150000}%
\pgfsetstrokecolor{textcolor}%
\pgfsetfillcolor{textcolor}%
\pgftext[x=6.304053in,y=2.964227in,left,base]{\color{textcolor}\rmfamily\fontsize{11.000000}{13.200000}\selectfont \(\displaystyle |R(\omega)|\ \omega \in \Omega\cap \Omega^2_{\text{target}}\)}%
\end{pgfscope}%
\end{pgfpicture}%
\makeatother%
\endgroup%

%% file: Figures/figs_numerics/1_training_coeffs.pgf
\begingroup%
\makeatletter%
\begin{pgfpicture}%
\pgfpathrectangle{\pgfpointorigin}{\pgfqpoint{9.742776in}{3.335208in}}%
\pgfusepath{use as bounding box, clip}%
\begin{pgfscope}%
\pgfsetbuttcap%
\pgfsetmiterjoin%
\pgfsetlinewidth{0.000000pt}%
\definecolor{currentstroke}{rgb}{1.000000,1.000000,1.000000}%
\pgfsetstrokecolor{currentstroke}%
\pgfsetdash{}{0pt}%
\pgfpathmoveto{\pgfqpoint{0.000000in}{0.000000in}}%
\pgfpathlineto{\pgfqpoint{9.742776in}{0.000000in}}%
\pgfpathlineto{\pgfqpoint{9.742776in}{3.335208in}}%
\pgfpathlineto{\pgfqpoint{0.000000in}{3.335208in}}%
\pgfpathlineto{\pgfqpoint{0.000000in}{0.000000in}}%
\pgfpathclose%
\pgfusepath{}%
\end{pgfscope}%
\begin{pgfscope}%
\pgfsetbuttcap%
\pgfsetmiterjoin%
\pgfsetlinewidth{0.000000pt}%
\definecolor{currentstroke}{rgb}{0.000000,0.000000,0.000000}%
\pgfsetstrokecolor{currentstroke}%
\pgfsetstrokeopacity{0.000000}%
\pgfsetdash{}{0pt}%
\pgfpathmoveto{\pgfqpoint{0.703855in}{0.631404in}}%
\pgfpathlineto{\pgfqpoint{4.738036in}{0.631404in}}%
\pgfpathlineto{\pgfqpoint{4.738036in}{3.040116in}}%
\pgfpathlineto{\pgfqpoint{0.703855in}{3.040116in}}%
\pgfpathlineto{\pgfqpoint{0.703855in}{0.631404in}}%
\pgfpathclose%
\pgfusepath{}%
\end{pgfscope}%
\begin{pgfscope}%
\pgfpathrectangle{\pgfqpoint{0.703855in}{0.631404in}}{\pgfqpoint{4.034181in}{2.408712in}}%
\pgfusepath{clip}%
\pgfsetroundcap%
\pgfsetroundjoin%
\pgfsetlinewidth{1.003750pt}%
\definecolor{currentstroke}{rgb}{0.800000,0.800000,0.800000}%
\pgfsetstrokecolor{currentstroke}%
\pgfsetdash{}{0pt}%
\pgfpathmoveto{\pgfqpoint{0.887226in}{0.631404in}}%
\pgfpathlineto{\pgfqpoint{0.887226in}{3.040116in}}%
\pgfusepath{stroke}%
\end{pgfscope}%
\begin{pgfscope}%
\definecolor{textcolor}{rgb}{0.150000,0.150000,0.150000}%
\pgfsetstrokecolor{textcolor}%
\pgfsetfillcolor{textcolor}%
\pgftext[x=0.887226in,y=0.546682in,,top]{\color{textcolor}\rmfamily\fontsize{17.000000}{16.800000}\selectfont \(\displaystyle {0}\)}%
\end{pgfscope}%
\begin{pgfscope}%
\pgfpathrectangle{\pgfqpoint{0.703855in}{0.631404in}}{\pgfqpoint{4.034181in}{2.408712in}}%
\pgfusepath{clip}%
\pgfsetroundcap%
\pgfsetroundjoin%
\pgfsetlinewidth{1.003750pt}%
\definecolor{currentstroke}{rgb}{0.800000,0.800000,0.800000}%
\pgfsetstrokecolor{currentstroke}%
\pgfsetdash{}{0pt}%
\pgfpathmoveto{\pgfqpoint{2.117910in}{0.631404in}}%
\pgfpathlineto{\pgfqpoint{2.117910in}{3.040116in}}%
\pgfusepath{stroke}%
\end{pgfscope}%
\begin{pgfscope}%
\definecolor{textcolor}{rgb}{0.150000,0.150000,0.150000}%
\pgfsetstrokecolor{textcolor}%
\pgfsetfillcolor{textcolor}%
\pgftext[x=2.117910in,y=0.546682in,,top]{\color{textcolor}\rmfamily\fontsize{17.000000}{16.800000}\selectfont \(\displaystyle {50}\)}%
\end{pgfscope}%
\begin{pgfscope}%
\pgfpathrectangle{\pgfqpoint{0.703855in}{0.631404in}}{\pgfqpoint{4.034181in}{2.408712in}}%
\pgfusepath{clip}%
\pgfsetroundcap%
\pgfsetroundjoin%
\pgfsetlinewidth{1.003750pt}%
\definecolor{currentstroke}{rgb}{0.800000,0.800000,0.800000}%
\pgfsetstrokecolor{currentstroke}%
\pgfsetdash{}{0pt}%
\pgfpathmoveto{\pgfqpoint{3.348594in}{0.631404in}}%
\pgfpathlineto{\pgfqpoint{3.348594in}{3.040116in}}%
\pgfusepath{stroke}%
\end{pgfscope}%
\begin{pgfscope}%
\definecolor{textcolor}{rgb}{0.150000,0.150000,0.150000}%
\pgfsetstrokecolor{textcolor}%
\pgfsetfillcolor{textcolor}%
\pgftext[x=3.348594in,y=0.546682in,,top]{\color{textcolor}\rmfamily\fontsize{17.000000}{16.800000}\selectfont \(\displaystyle {100}\)}%
\end{pgfscope}%
\begin{pgfscope}%
\pgfpathrectangle{\pgfqpoint{0.703855in}{0.631404in}}{\pgfqpoint{4.034181in}{2.408712in}}%
\pgfusepath{clip}%
\pgfsetroundcap%
\pgfsetroundjoin%
\pgfsetlinewidth{1.003750pt}%
\definecolor{currentstroke}{rgb}{0.800000,0.800000,0.800000}%
\pgfsetstrokecolor{currentstroke}%
\pgfsetdash{}{0pt}%
\pgfpathmoveto{\pgfqpoint{4.579278in}{0.631404in}}%
\pgfpathlineto{\pgfqpoint{4.579278in}{3.040116in}}%
\pgfusepath{stroke}%
\end{pgfscope}%
\begin{pgfscope}%
\definecolor{textcolor}{rgb}{0.150000,0.150000,0.150000}%
\pgfsetstrokecolor{textcolor}%
\pgfsetfillcolor{textcolor}%
\pgftext[x=4.579278in,y=0.546682in,,top]{\color{textcolor}\rmfamily\fontsize{17.000000}{16.800000}\selectfont \(\displaystyle {150}\)}%
\end{pgfscope}%
\begin{pgfscope}%
\definecolor{textcolor}{rgb}{0.150000,0.150000,0.150000}%
\pgfsetstrokecolor{textcolor}%
\pgfsetfillcolor{textcolor}%
\pgftext[x=2.720945in,y=0.313349in,,top]{\color{textcolor}\rmfamily\fontsize{21.000000}{21.600000}\selectfont Epochs}%
\end{pgfscope}%
\begin{pgfscope}%
\pgfpathrectangle{\pgfqpoint{0.703855in}{0.631404in}}{\pgfqpoint{4.034181in}{2.408712in}}%
\pgfusepath{clip}%
\pgfsetroundcap%
\pgfsetroundjoin%
\pgfsetlinewidth{1.003750pt}%
\definecolor{currentstroke}{rgb}{0.800000,0.800000,0.800000}%
\pgfsetstrokecolor{currentstroke}%
\pgfsetdash{}{0pt}%
\pgfpathmoveto{\pgfqpoint{0.703855in}{0.745977in}}%
\pgfpathlineto{\pgfqpoint{4.738036in}{0.745977in}}%
\pgfusepath{stroke}%
\end{pgfscope}%
\begin{pgfscope}%
\definecolor{textcolor}{rgb}{0.150000,0.150000,0.150000}%
\pgfsetstrokecolor{textcolor}%
\pgfsetfillcolor{textcolor}%
\pgftext[x=0.368904in, y=0.676533in, left, base]{\color{textcolor}\rmfamily\fontsize{17.000000}{16.800000}\selectfont \(\displaystyle {0.0}\)}%
\end{pgfscope}%
\begin{pgfscope}%
\pgfpathrectangle{\pgfqpoint{0.703855in}{0.631404in}}{\pgfqpoint{4.034181in}{2.408712in}}%
\pgfusepath{clip}%
\pgfsetroundcap%
\pgfsetroundjoin%
\pgfsetlinewidth{1.003750pt}%
\definecolor{currentstroke}{rgb}{0.800000,0.800000,0.800000}%
\pgfsetstrokecolor{currentstroke}%
\pgfsetdash{}{0pt}%
\pgfpathmoveto{\pgfqpoint{0.703855in}{1.195808in}}%
\pgfpathlineto{\pgfqpoint{4.738036in}{1.195808in}}%
\pgfusepath{stroke}%
\end{pgfscope}%
\begin{pgfscope}%
\definecolor{textcolor}{rgb}{0.150000,0.150000,0.150000}%
\pgfsetstrokecolor{textcolor}%
\pgfsetfillcolor{textcolor}%
\pgftext[x=0.368904in, y=1.126364in, left, base]{\color{textcolor}\rmfamily\fontsize{17.000000}{16.800000}\selectfont \(\displaystyle {0.1}\)}%
\end{pgfscope}%
\begin{pgfscope}%
\pgfpathrectangle{\pgfqpoint{0.703855in}{0.631404in}}{\pgfqpoint{4.034181in}{2.408712in}}%
\pgfusepath{clip}%
\pgfsetroundcap%
\pgfsetroundjoin%
\pgfsetlinewidth{1.003750pt}%
\definecolor{currentstroke}{rgb}{0.800000,0.800000,0.800000}%
\pgfsetstrokecolor{currentstroke}%
\pgfsetdash{}{0pt}%
\pgfpathmoveto{\pgfqpoint{0.703855in}{1.645639in}}%
\pgfpathlineto{\pgfqpoint{4.738036in}{1.645639in}}%
\pgfusepath{stroke}%
\end{pgfscope}%
\begin{pgfscope}%
\definecolor{textcolor}{rgb}{0.150000,0.150000,0.150000}%
\pgfsetstrokecolor{textcolor}%
\pgfsetfillcolor{textcolor}%
\pgftext[x=0.368904in, y=1.576195in, left, base]{\color{textcolor}\rmfamily\fontsize{17.000000}{16.800000}\selectfont \(\displaystyle {0.2}\)}%
\end{pgfscope}%
\begin{pgfscope}%
\pgfpathrectangle{\pgfqpoint{0.703855in}{0.631404in}}{\pgfqpoint{4.034181in}{2.408712in}}%
\pgfusepath{clip}%
\pgfsetroundcap%
\pgfsetroundjoin%
\pgfsetlinewidth{1.003750pt}%
\definecolor{currentstroke}{rgb}{0.800000,0.800000,0.800000}%
\pgfsetstrokecolor{currentstroke}%
\pgfsetdash{}{0pt}%
\pgfpathmoveto{\pgfqpoint{0.703855in}{2.095470in}}%
\pgfpathlineto{\pgfqpoint{4.738036in}{2.095470in}}%
\pgfusepath{stroke}%
\end{pgfscope}%
\begin{pgfscope}%
\definecolor{textcolor}{rgb}{0.150000,0.150000,0.150000}%
\pgfsetstrokecolor{textcolor}%
\pgfsetfillcolor{textcolor}%
\pgftext[x=0.368904in, y=2.026026in, left, base]{\color{textcolor}\rmfamily\fontsize{17.000000}{16.800000}\selectfont \(\displaystyle {0.3}\)}%
\end{pgfscope}%
\begin{pgfscope}%
\pgfpathrectangle{\pgfqpoint{0.703855in}{0.631404in}}{\pgfqpoint{4.034181in}{2.408712in}}%
\pgfusepath{clip}%
\pgfsetroundcap%
\pgfsetroundjoin%
\pgfsetlinewidth{1.003750pt}%
\definecolor{currentstroke}{rgb}{0.800000,0.800000,0.800000}%
\pgfsetstrokecolor{currentstroke}%
\pgfsetdash{}{0pt}%
\pgfpathmoveto{\pgfqpoint{0.703855in}{2.545301in}}%
\pgfpathlineto{\pgfqpoint{4.738036in}{2.545301in}}%
\pgfusepath{stroke}%
\end{pgfscope}%
\begin{pgfscope}%
\definecolor{textcolor}{rgb}{0.150000,0.150000,0.150000}%
\pgfsetstrokecolor{textcolor}%
\pgfsetfillcolor{textcolor}%
\pgftext[x=0.368904in, y=2.475857in, left, base]{\color{textcolor}\rmfamily\fontsize{17.000000}{16.800000}\selectfont \(\displaystyle {0.4}\)}%
\end{pgfscope}%
\begin{pgfscope}%
\pgfpathrectangle{\pgfqpoint{0.703855in}{0.631404in}}{\pgfqpoint{4.034181in}{2.408712in}}%
\pgfusepath{clip}%
\pgfsetroundcap%
\pgfsetroundjoin%
\pgfsetlinewidth{1.003750pt}%
\definecolor{currentstroke}{rgb}{0.800000,0.800000,0.800000}%
\pgfsetstrokecolor{currentstroke}%
\pgfsetdash{}{0pt}%
\pgfpathmoveto{\pgfqpoint{0.703855in}{2.995133in}}%
\pgfpathlineto{\pgfqpoint{4.738036in}{2.995133in}}%
\pgfusepath{stroke}%
\end{pgfscope}%
\begin{pgfscope}%
\definecolor{textcolor}{rgb}{0.150000,0.150000,0.150000}%
\pgfsetstrokecolor{textcolor}%
\pgfsetfillcolor{textcolor}%
\pgftext[x=0.368904in, y=2.925688in, left, base]{\color{textcolor}\rmfamily\fontsize{17.000000}{16.800000}\selectfont \(\displaystyle {0.5}\)}%
\end{pgfscope}%
\begin{pgfscope}%
\definecolor{textcolor}{rgb}{0.150000,0.150000,0.150000}%
\pgfsetstrokecolor{textcolor}%
\pgfsetfillcolor{textcolor}%
\pgftext[x=0.313349in,y=1.835760in,,bottom,rotate=90.000000]{\color{textcolor}\rmfamily\fontsize{21.000000}{21.600000}\selectfont Loss \(\displaystyle \mathcal{L}\)}%
\end{pgfscope}%
\begin{pgfscope}%
\pgfpathrectangle{\pgfqpoint{0.703855in}{0.631404in}}{\pgfqpoint{4.034181in}{2.408712in}}%
\pgfusepath{clip}%
\pgfsetroundcap%
\pgfsetroundjoin%
\pgfsetlinewidth{3.011250pt}%
\definecolor{currentstroke}{rgb}{0.992157,0.384314,0.384314}%
\pgfsetstrokecolor{currentstroke}%
\pgfsetdash{}{0pt}%
\pgfpathmoveto{\pgfqpoint{0.703855in}{0.745977in}}%
\pgfpathlineto{\pgfqpoint{4.738036in}{0.745977in}}%
\pgfusepath{stroke}%
\end{pgfscope}%
\begin{pgfscope}%
\pgfpathrectangle{\pgfqpoint{0.703855in}{0.631404in}}{\pgfqpoint{4.034181in}{2.408712in}}%
\pgfusepath{clip}%
\pgfsetroundcap%
\pgfsetroundjoin%
\pgfsetlinewidth{2.208250pt}%
\definecolor{currentstroke}{rgb}{0.333333,0.556863,0.631373}%
\pgfsetstrokecolor{currentstroke}%
\pgfsetdash{}{0pt}%
\pgfpathmoveto{\pgfqpoint{0.887226in}{3.037441in}}%
\pgfpathlineto{\pgfqpoint{0.911840in}{2.968395in}}%
\pgfpathlineto{\pgfqpoint{0.936454in}{2.922397in}}%
\pgfpathlineto{\pgfqpoint{0.961067in}{2.912243in}}%
\pgfpathlineto{\pgfqpoint{0.985681in}{2.894755in}}%
\pgfpathlineto{\pgfqpoint{1.059522in}{2.850693in}}%
\pgfpathlineto{\pgfqpoint{1.084136in}{2.840261in}}%
\pgfpathlineto{\pgfqpoint{1.133363in}{2.830209in}}%
\pgfpathlineto{\pgfqpoint{1.157977in}{2.822909in}}%
\pgfpathlineto{\pgfqpoint{1.182591in}{2.814127in}}%
\pgfpathlineto{\pgfqpoint{1.207204in}{2.808216in}}%
\pgfpathlineto{\pgfqpoint{1.256432in}{2.799926in}}%
\pgfpathlineto{\pgfqpoint{1.305659in}{2.791531in}}%
\pgfpathlineto{\pgfqpoint{1.379500in}{2.779812in}}%
\pgfpathlineto{\pgfqpoint{1.428727in}{2.766565in}}%
\pgfpathlineto{\pgfqpoint{1.453341in}{2.763144in}}%
\pgfpathlineto{\pgfqpoint{1.527182in}{2.759677in}}%
\pgfpathlineto{\pgfqpoint{1.601023in}{2.751012in}}%
\pgfpathlineto{\pgfqpoint{1.724091in}{2.739718in}}%
\pgfpathlineto{\pgfqpoint{1.773319in}{2.736916in}}%
\pgfpathlineto{\pgfqpoint{1.847160in}{2.727904in}}%
\pgfpathlineto{\pgfqpoint{1.871773in}{2.723915in}}%
\pgfpathlineto{\pgfqpoint{1.896387in}{2.716516in}}%
\pgfpathlineto{\pgfqpoint{1.921001in}{2.700933in}}%
\pgfpathlineto{\pgfqpoint{1.945615in}{2.664064in}}%
\pgfpathlineto{\pgfqpoint{1.970228in}{2.576047in}}%
\pgfpathlineto{\pgfqpoint{2.019456in}{2.292507in}}%
\pgfpathlineto{\pgfqpoint{2.044069in}{2.228625in}}%
\pgfpathlineto{\pgfqpoint{2.068683in}{2.214322in}}%
\pgfpathlineto{\pgfqpoint{2.093297in}{2.193895in}}%
\pgfpathlineto{\pgfqpoint{2.117910in}{2.177866in}}%
\pgfpathlineto{\pgfqpoint{2.142524in}{2.151267in}}%
\pgfpathlineto{\pgfqpoint{2.167138in}{2.106483in}}%
\pgfpathlineto{\pgfqpoint{2.191751in}{2.075839in}}%
\pgfpathlineto{\pgfqpoint{2.216365in}{2.058332in}}%
\pgfpathlineto{\pgfqpoint{2.240979in}{2.030862in}}%
\pgfpathlineto{\pgfqpoint{2.265592in}{2.000019in}}%
\pgfpathlineto{\pgfqpoint{2.290206in}{1.976965in}}%
\pgfpathlineto{\pgfqpoint{2.314820in}{1.962087in}}%
\pgfpathlineto{\pgfqpoint{2.364047in}{1.942982in}}%
\pgfpathlineto{\pgfqpoint{2.413274in}{1.896442in}}%
\pgfpathlineto{\pgfqpoint{2.437888in}{1.884558in}}%
\pgfpathlineto{\pgfqpoint{2.487115in}{1.868909in}}%
\pgfpathlineto{\pgfqpoint{2.511729in}{1.864529in}}%
\pgfpathlineto{\pgfqpoint{2.536343in}{1.864975in}}%
\pgfpathlineto{\pgfqpoint{2.560956in}{1.858784in}}%
\pgfpathlineto{\pgfqpoint{2.585570in}{1.851288in}}%
\pgfpathlineto{\pgfqpoint{2.634797in}{1.848450in}}%
\pgfpathlineto{\pgfqpoint{2.684025in}{1.840284in}}%
\pgfpathlineto{\pgfqpoint{2.708638in}{1.834021in}}%
\pgfpathlineto{\pgfqpoint{2.733252in}{1.829013in}}%
\pgfpathlineto{\pgfqpoint{2.782479in}{1.827279in}}%
\pgfpathlineto{\pgfqpoint{2.831707in}{1.820918in}}%
\pgfpathlineto{\pgfqpoint{2.880934in}{1.816206in}}%
\pgfpathlineto{\pgfqpoint{2.930162in}{1.809863in}}%
\pgfpathlineto{\pgfqpoint{2.979389in}{1.804592in}}%
\pgfpathlineto{\pgfqpoint{3.028616in}{1.799199in}}%
\pgfpathlineto{\pgfqpoint{3.127071in}{1.788440in}}%
\pgfpathlineto{\pgfqpoint{3.176298in}{1.780572in}}%
\pgfpathlineto{\pgfqpoint{3.225526in}{1.769858in}}%
\pgfpathlineto{\pgfqpoint{3.250139in}{1.762886in}}%
\pgfpathlineto{\pgfqpoint{3.274753in}{1.754006in}}%
\pgfpathlineto{\pgfqpoint{3.299367in}{1.741333in}}%
\pgfpathlineto{\pgfqpoint{3.323980in}{1.722618in}}%
\pgfpathlineto{\pgfqpoint{3.348594in}{1.692758in}}%
\pgfpathlineto{\pgfqpoint{3.373208in}{1.638524in}}%
\pgfpathlineto{\pgfqpoint{3.397821in}{1.528954in}}%
\pgfpathlineto{\pgfqpoint{3.422435in}{1.331665in}}%
\pgfpathlineto{\pgfqpoint{3.447049in}{1.092605in}}%
\pgfpathlineto{\pgfqpoint{3.471662in}{0.899372in}}%
\pgfpathlineto{\pgfqpoint{3.496276in}{0.829205in}}%
\pgfpathlineto{\pgfqpoint{3.520890in}{0.853879in}}%
\pgfpathlineto{\pgfqpoint{3.545503in}{0.860424in}}%
\pgfpathlineto{\pgfqpoint{3.570117in}{0.834440in}}%
\pgfpathlineto{\pgfqpoint{3.594731in}{0.811085in}}%
\pgfpathlineto{\pgfqpoint{3.619344in}{0.799173in}}%
\pgfpathlineto{\pgfqpoint{3.668572in}{0.791054in}}%
\pgfpathlineto{\pgfqpoint{3.693185in}{0.791326in}}%
\pgfpathlineto{\pgfqpoint{3.717799in}{0.789115in}}%
\pgfpathlineto{\pgfqpoint{3.742413in}{0.783121in}}%
\pgfpathlineto{\pgfqpoint{3.767026in}{0.778397in}}%
\pgfpathlineto{\pgfqpoint{3.791640in}{0.770966in}}%
\pgfpathlineto{\pgfqpoint{3.816254in}{0.766293in}}%
\pgfpathlineto{\pgfqpoint{3.865481in}{0.763819in}}%
\pgfpathlineto{\pgfqpoint{3.988550in}{0.753536in}}%
\pgfpathlineto{\pgfqpoint{4.087004in}{0.750806in}}%
\pgfpathlineto{\pgfqpoint{4.136232in}{0.748592in}}%
\pgfpathlineto{\pgfqpoint{4.210073in}{0.747761in}}%
\pgfpathlineto{\pgfqpoint{4.333141in}{0.747008in}}%
\pgfpathlineto{\pgfqpoint{4.505437in}{0.746262in}}%
\pgfpathlineto{\pgfqpoint{4.554664in}{0.746252in}}%
\pgfpathlineto{\pgfqpoint{4.554664in}{0.746252in}}%
\pgfusepath{stroke}%
\end{pgfscope}%
\begin{pgfscope}%
\pgfsetrectcap%
\pgfsetmiterjoin%
\pgfsetlinewidth{1.254687pt}%
\definecolor{currentstroke}{rgb}{0.800000,0.800000,0.800000}%
\pgfsetstrokecolor{currentstroke}%
\pgfsetdash{}{0pt}%
\pgfpathmoveto{\pgfqpoint{0.703855in}{0.631404in}}%
\pgfpathlineto{\pgfqpoint{0.703855in}{3.040116in}}%
\pgfusepath{stroke}%
\end{pgfscope}%
\begin{pgfscope}%
\pgfsetrectcap%
\pgfsetmiterjoin%
\pgfsetlinewidth{1.254687pt}%
\definecolor{currentstroke}{rgb}{0.800000,0.800000,0.800000}%
\pgfsetstrokecolor{currentstroke}%
\pgfsetdash{}{0pt}%
\pgfpathmoveto{\pgfqpoint{4.738036in}{0.631404in}}%
\pgfpathlineto{\pgfqpoint{4.738036in}{3.040116in}}%
\pgfusepath{stroke}%
\end{pgfscope}%
\begin{pgfscope}%
\pgfsetrectcap%
\pgfsetmiterjoin%
\pgfsetlinewidth{1.254687pt}%
\definecolor{currentstroke}{rgb}{0.800000,0.800000,0.800000}%
\pgfsetstrokecolor{currentstroke}%
\pgfsetdash{}{0pt}%
\pgfpathmoveto{\pgfqpoint{0.703855in}{0.631404in}}%
\pgfpathlineto{\pgfqpoint{4.738036in}{0.631404in}}%
\pgfusepath{stroke}%
\end{pgfscope}%
\begin{pgfscope}%
\pgfsetrectcap%
\pgfsetmiterjoin%
\pgfsetlinewidth{1.254687pt}%
\definecolor{currentstroke}{rgb}{0.800000,0.800000,0.800000}%
\pgfsetstrokecolor{currentstroke}%
\pgfsetdash{}{0pt}%
\pgfpathmoveto{\pgfqpoint{0.703855in}{3.040116in}}%
\pgfpathlineto{\pgfqpoint{4.738036in}{3.040116in}}%
\pgfusepath{stroke}%
\end{pgfscope}%
\begin{pgfscope}%
\pgfsetbuttcap%
\pgfsetmiterjoin%
\pgfsetlinewidth{0.000000pt}%
\definecolor{currentstroke}{rgb}{0.000000,0.000000,0.000000}%
\pgfsetstrokecolor{currentstroke}%
\pgfsetstrokeopacity{0.000000}%
\pgfsetdash{}{0pt}%
\pgfpathmoveto{\pgfqpoint{5.567871in}{0.631404in}}%
\pgfpathlineto{\pgfqpoint{8.916242in}{0.631404in}}%
\pgfpathlineto{\pgfqpoint{8.916242in}{3.040116in}}%
\pgfpathlineto{\pgfqpoint{5.567871in}{3.040116in}}%
\pgfpathlineto{\pgfqpoint{5.567871in}{0.631404in}}%
\pgfpathclose%
\pgfusepath{}%
\end{pgfscope}%
\begin{pgfscope}%
\pgfsys@transformshift{5.570000in}{0.635208in}%
\pgftext[left,bottom]{\includegraphics[interpolate=true,width=3.350000in,height=2.410000in]{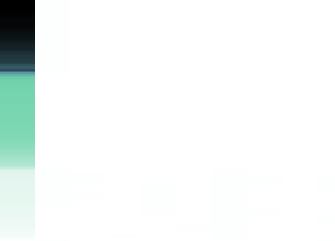}}%
\end{pgfscope}%
\begin{pgfscope}%
\definecolor{textcolor}{rgb}{0.150000,0.150000,0.150000}%
\pgfsetstrokecolor{textcolor}%
\pgfsetfillcolor{textcolor}%
\pgftext[x=6.096561in,y=0.546682in,,top]{\color{textcolor}\rmfamily\fontsize{17.000000}{16.800000}\selectfont \(\displaystyle {2}\)}%
\end{pgfscope}%
\begin{pgfscope}%
\definecolor{textcolor}{rgb}{0.150000,0.150000,0.150000}%
\pgfsetstrokecolor{textcolor}%
\pgfsetfillcolor{textcolor}%
\pgftext[x=6.801482in,y=0.546682in,,top]{\color{textcolor}\rmfamily\fontsize{17.000000}{16.800000}\selectfont \(\displaystyle {4}\)}%
\end{pgfscope}%
\begin{pgfscope}%
\definecolor{textcolor}{rgb}{0.150000,0.150000,0.150000}%
\pgfsetstrokecolor{textcolor}%
\pgfsetfillcolor{textcolor}%
\pgftext[x=7.506402in,y=0.546682in,,top]{\color{textcolor}\rmfamily\fontsize{17.000000}{16.800000}\selectfont \(\displaystyle {6}\)}%
\end{pgfscope}%
\begin{pgfscope}%
\definecolor{textcolor}{rgb}{0.150000,0.150000,0.150000}%
\pgfsetstrokecolor{textcolor}%
\pgfsetfillcolor{textcolor}%
\pgftext[x=8.211322in,y=0.546682in,,top]{\color{textcolor}\rmfamily\fontsize{17.000000}{16.800000}\selectfont \(\displaystyle {8}\)}%
\end{pgfscope}%
\begin{pgfscope}%
\definecolor{textcolor}{rgb}{0.150000,0.150000,0.150000}%
\pgfsetstrokecolor{textcolor}%
\pgfsetfillcolor{textcolor}%
\pgftext[x=8.916242in,y=0.546682in,,top]{\color{textcolor}\rmfamily\fontsize{17.000000}{16.800000}\selectfont \(\displaystyle {10}\)}%
\end{pgfscope}%
\begin{pgfscope}%
\pgfpathrectangle{\pgfqpoint{5.567871in}{0.631404in}}{\pgfqpoint{3.348371in}{2.408712in}}%
\pgfusepath{clip}%
\pgfsetroundcap%
\pgfsetroundjoin%
\pgfsetlinewidth{2.007500pt}%
\definecolor{currentstroke}{rgb}{0.827451,0.827451,0.827451}%
\pgfsetstrokecolor{currentstroke}%
\pgfsetdash{}{0pt}%
\pgfpathmoveto{\pgfqpoint{5.567871in}{0.631404in}}%
\pgfpathlineto{\pgfqpoint{5.567871in}{3.040116in}}%
\pgfusepath{stroke}%
\end{pgfscope}%
\begin{pgfscope}%
\pgfpathrectangle{\pgfqpoint{5.567871in}{0.631404in}}{\pgfqpoint{3.348371in}{2.408712in}}%
\pgfusepath{clip}%
\pgfsetroundcap%
\pgfsetroundjoin%
\pgfsetlinewidth{2.007500pt}%
\definecolor{currentstroke}{rgb}{0.827451,0.827451,0.827451}%
\pgfsetstrokecolor{currentstroke}%
\pgfsetdash{}{0pt}%
\pgfpathmoveto{\pgfqpoint{5.920331in}{0.631404in}}%
\pgfpathlineto{\pgfqpoint{5.920331in}{3.040116in}}%
\pgfusepath{stroke}%
\end{pgfscope}%
\begin{pgfscope}%
\pgfpathrectangle{\pgfqpoint{5.567871in}{0.631404in}}{\pgfqpoint{3.348371in}{2.408712in}}%
\pgfusepath{clip}%
\pgfsetroundcap%
\pgfsetroundjoin%
\pgfsetlinewidth{2.007500pt}%
\definecolor{currentstroke}{rgb}{0.827451,0.827451,0.827451}%
\pgfsetstrokecolor{currentstroke}%
\pgfsetdash{}{0pt}%
\pgfpathmoveto{\pgfqpoint{6.272791in}{0.631404in}}%
\pgfpathlineto{\pgfqpoint{6.272791in}{3.040116in}}%
\pgfusepath{stroke}%
\end{pgfscope}%
\begin{pgfscope}%
\pgfpathrectangle{\pgfqpoint{5.567871in}{0.631404in}}{\pgfqpoint{3.348371in}{2.408712in}}%
\pgfusepath{clip}%
\pgfsetroundcap%
\pgfsetroundjoin%
\pgfsetlinewidth{2.007500pt}%
\definecolor{currentstroke}{rgb}{0.827451,0.827451,0.827451}%
\pgfsetstrokecolor{currentstroke}%
\pgfsetdash{}{0pt}%
\pgfpathmoveto{\pgfqpoint{6.625252in}{0.631404in}}%
\pgfpathlineto{\pgfqpoint{6.625252in}{3.040116in}}%
\pgfusepath{stroke}%
\end{pgfscope}%
\begin{pgfscope}%
\pgfpathrectangle{\pgfqpoint{5.567871in}{0.631404in}}{\pgfqpoint{3.348371in}{2.408712in}}%
\pgfusepath{clip}%
\pgfsetroundcap%
\pgfsetroundjoin%
\pgfsetlinewidth{2.007500pt}%
\definecolor{currentstroke}{rgb}{0.827451,0.827451,0.827451}%
\pgfsetstrokecolor{currentstroke}%
\pgfsetdash{}{0pt}%
\pgfpathmoveto{\pgfqpoint{6.977712in}{0.631404in}}%
\pgfpathlineto{\pgfqpoint{6.977712in}{3.040116in}}%
\pgfusepath{stroke}%
\end{pgfscope}%
\begin{pgfscope}%
\pgfpathrectangle{\pgfqpoint{5.567871in}{0.631404in}}{\pgfqpoint{3.348371in}{2.408712in}}%
\pgfusepath{clip}%
\pgfsetroundcap%
\pgfsetroundjoin%
\pgfsetlinewidth{2.007500pt}%
\definecolor{currentstroke}{rgb}{0.827451,0.827451,0.827451}%
\pgfsetstrokecolor{currentstroke}%
\pgfsetdash{}{0pt}%
\pgfpathmoveto{\pgfqpoint{7.330172in}{0.631404in}}%
\pgfpathlineto{\pgfqpoint{7.330172in}{3.040116in}}%
\pgfusepath{stroke}%
\end{pgfscope}%
\begin{pgfscope}%
\pgfpathrectangle{\pgfqpoint{5.567871in}{0.631404in}}{\pgfqpoint{3.348371in}{2.408712in}}%
\pgfusepath{clip}%
\pgfsetroundcap%
\pgfsetroundjoin%
\pgfsetlinewidth{2.007500pt}%
\definecolor{currentstroke}{rgb}{0.827451,0.827451,0.827451}%
\pgfsetstrokecolor{currentstroke}%
\pgfsetdash{}{0pt}%
\pgfpathmoveto{\pgfqpoint{7.682632in}{0.631404in}}%
\pgfpathlineto{\pgfqpoint{7.682632in}{3.040116in}}%
\pgfusepath{stroke}%
\end{pgfscope}%
\begin{pgfscope}%
\pgfpathrectangle{\pgfqpoint{5.567871in}{0.631404in}}{\pgfqpoint{3.348371in}{2.408712in}}%
\pgfusepath{clip}%
\pgfsetroundcap%
\pgfsetroundjoin%
\pgfsetlinewidth{2.007500pt}%
\definecolor{currentstroke}{rgb}{0.827451,0.827451,0.827451}%
\pgfsetstrokecolor{currentstroke}%
\pgfsetdash{}{0pt}%
\pgfpathmoveto{\pgfqpoint{8.035092in}{0.631404in}}%
\pgfpathlineto{\pgfqpoint{8.035092in}{3.040116in}}%
\pgfusepath{stroke}%
\end{pgfscope}%
\begin{pgfscope}%
\pgfpathrectangle{\pgfqpoint{5.567871in}{0.631404in}}{\pgfqpoint{3.348371in}{2.408712in}}%
\pgfusepath{clip}%
\pgfsetroundcap%
\pgfsetroundjoin%
\pgfsetlinewidth{2.007500pt}%
\definecolor{currentstroke}{rgb}{0.827451,0.827451,0.827451}%
\pgfsetstrokecolor{currentstroke}%
\pgfsetdash{}{0pt}%
\pgfpathmoveto{\pgfqpoint{8.387552in}{0.631404in}}%
\pgfpathlineto{\pgfqpoint{8.387552in}{3.040116in}}%
\pgfusepath{stroke}%
\end{pgfscope}%
\begin{pgfscope}%
\pgfpathrectangle{\pgfqpoint{5.567871in}{0.631404in}}{\pgfqpoint{3.348371in}{2.408712in}}%
\pgfusepath{clip}%
\pgfsetroundcap%
\pgfsetroundjoin%
\pgfsetlinewidth{2.007500pt}%
\definecolor{currentstroke}{rgb}{0.827451,0.827451,0.827451}%
\pgfsetstrokecolor{currentstroke}%
\pgfsetdash{}{0pt}%
\pgfpathmoveto{\pgfqpoint{8.740012in}{0.631404in}}%
\pgfpathlineto{\pgfqpoint{8.740012in}{3.040116in}}%
\pgfusepath{stroke}%
\end{pgfscope}%
\begin{pgfscope}%
\definecolor{textcolor}{rgb}{0.150000,0.150000,0.150000}%
\pgfsetstrokecolor{textcolor}%
\pgfsetfillcolor{textcolor}%
\pgftext[x=7.242057in,y=0.313349in,,top]{\color{textcolor}\rmfamily\fontsize{21.000000}{21.600000}\selectfont \(\displaystyle \omega\)}%
\end{pgfscope}%
\begin{pgfscope}%
\definecolor{textcolor}{rgb}{0.150000,0.150000,0.150000}%
\pgfsetstrokecolor{textcolor}%
\pgfsetfillcolor{textcolor}%
\pgftext[x=5.385234in, y=0.569988in, left, base]{\color{textcolor}\rmfamily\fontsize{17.000000}{16.800000}\selectfont \(\displaystyle {0}\)}%
\end{pgfscope}%
\begin{pgfscope}%
\definecolor{textcolor}{rgb}{0.150000,0.150000,0.150000}%
\pgfsetstrokecolor{textcolor}%
\pgfsetfillcolor{textcolor}%
\pgftext[x=5.287318in, y=0.971440in, left, base]{\color{textcolor}\rmfamily\fontsize{17.000000}{16.800000}\selectfont \(\displaystyle {25}\)}%
\end{pgfscope}%
\begin{pgfscope}%
\definecolor{textcolor}{rgb}{0.150000,0.150000,0.150000}%
\pgfsetstrokecolor{textcolor}%
\pgfsetfillcolor{textcolor}%
\pgftext[x=5.287318in, y=1.372892in, left, base]{\color{textcolor}\rmfamily\fontsize{17.000000}{16.800000}\selectfont \(\displaystyle {50}\)}%
\end{pgfscope}%
\begin{pgfscope}%
\definecolor{textcolor}{rgb}{0.150000,0.150000,0.150000}%
\pgfsetstrokecolor{textcolor}%
\pgfsetfillcolor{textcolor}%
\pgftext[x=5.287318in, y=1.774344in, left, base]{\color{textcolor}\rmfamily\fontsize{17.000000}{16.800000}\selectfont \(\displaystyle {75}\)}%
\end{pgfscope}%
\begin{pgfscope}%
\definecolor{textcolor}{rgb}{0.150000,0.150000,0.150000}%
\pgfsetstrokecolor{textcolor}%
\pgfsetfillcolor{textcolor}%
\pgftext[x=5.189403in, y=2.175796in, left, base]{\color{textcolor}\rmfamily\fontsize{17.000000}{16.800000}\selectfont \(\displaystyle {100}\)}%
\end{pgfscope}%
\begin{pgfscope}%
\definecolor{textcolor}{rgb}{0.150000,0.150000,0.150000}%
\pgfsetstrokecolor{textcolor}%
\pgfsetfillcolor{textcolor}%
\pgftext[x=5.189403in, y=2.577248in, left, base]{\color{textcolor}\rmfamily\fontsize{17.000000}{16.800000}\selectfont \(\displaystyle {125}\)}%
\end{pgfscope}%
\begin{pgfscope}%
\definecolor{textcolor}{rgb}{0.150000,0.150000,0.150000}%
\pgfsetstrokecolor{textcolor}%
\pgfsetfillcolor{textcolor}%
\pgftext[x=5.133847in,y=1.835760in,,bottom,rotate=90.000000]{\color{textcolor}\rmfamily\fontsize{21.000000}{21.600000}\selectfont Epochs}%
\end{pgfscope}%
\begin{pgfscope}%
\pgfsetrectcap%
\pgfsetmiterjoin%
\pgfsetlinewidth{1.254687pt}%
\definecolor{currentstroke}{rgb}{0.800000,0.800000,0.800000}%
\pgfsetstrokecolor{currentstroke}%
\pgfsetdash{}{0pt}%
\pgfpathmoveto{\pgfqpoint{5.567871in}{0.631404in}}%
\pgfpathlineto{\pgfqpoint{5.567871in}{3.040116in}}%
\pgfusepath{stroke}%
\end{pgfscope}%
\begin{pgfscope}%
\pgfsetrectcap%
\pgfsetmiterjoin%
\pgfsetlinewidth{1.254687pt}%
\definecolor{currentstroke}{rgb}{0.800000,0.800000,0.800000}%
\pgfsetstrokecolor{currentstroke}%
\pgfsetdash{}{0pt}%
\pgfpathmoveto{\pgfqpoint{8.916242in}{0.631404in}}%
\pgfpathlineto{\pgfqpoint{8.916242in}{3.040116in}}%
\pgfusepath{stroke}%
\end{pgfscope}%
\begin{pgfscope}%
\pgfsetrectcap%
\pgfsetmiterjoin%
\pgfsetlinewidth{1.254687pt}%
\definecolor{currentstroke}{rgb}{0.800000,0.800000,0.800000}%
\pgfsetstrokecolor{currentstroke}%
\pgfsetdash{}{0pt}%
\pgfpathmoveto{\pgfqpoint{5.567871in}{0.631404in}}%
\pgfpathlineto{\pgfqpoint{8.916242in}{0.631404in}}%
\pgfusepath{stroke}%
\end{pgfscope}%
\begin{pgfscope}%
\pgfsetrectcap%
\pgfsetmiterjoin%
\pgfsetlinewidth{1.254687pt}%
\definecolor{currentstroke}{rgb}{0.800000,0.800000,0.800000}%
\pgfsetstrokecolor{currentstroke}%
\pgfsetdash{}{0pt}%
\pgfpathmoveto{\pgfqpoint{5.567871in}{3.040116in}}%
\pgfpathlineto{\pgfqpoint{8.916242in}{3.040116in}}%
\pgfusepath{stroke}%
\end{pgfscope}%
\begin{pgfscope}%
\definecolor{textcolor}{rgb}{0.150000,0.150000,0.150000}%
\pgfsetstrokecolor{textcolor}%
\pgfsetfillcolor{textcolor}%
\pgftext[x=5.532625in,y=3.096319in,left,base]{\color{textcolor}\rmfamily\fontsize{17.000000}{16.800000}\selectfont 9e-05}%
\end{pgfscope}%
\begin{pgfscope}%
\pgfsetbuttcap%
\pgfsetmiterjoin%
\pgfsetlinewidth{0.000000pt}%
\definecolor{currentstroke}{rgb}{0.000000,0.000000,0.000000}%
\pgfsetstrokecolor{currentstroke}%
\pgfsetstrokeopacity{0.000000}%
\pgfsetdash{}{0pt}%
\pgfpathmoveto{\pgfqpoint{8.996926in}{0.631404in}}%
\pgfpathlineto{\pgfqpoint{9.117361in}{0.631404in}}%
\pgfpathlineto{\pgfqpoint{9.117361in}{3.040116in}}%
\pgfpathlineto{\pgfqpoint{8.996926in}{3.040116in}}%
\pgfpathlineto{\pgfqpoint{8.996926in}{0.631404in}}%
\pgfpathclose%
\pgfusepath{}%
\end{pgfscope}%
\begin{pgfscope}%
\pgfpathrectangle{\pgfqpoint{8.996926in}{0.631404in}}{\pgfqpoint{0.120436in}{2.408712in}}%
\pgfusepath{clip}%
\pgfsetbuttcap%
\pgfsetmiterjoin%
\pgfsetlinewidth{0.000000pt}%
\definecolor{currentstroke}{rgb}{0.000000,0.000000,0.000000}%
\pgfsetstrokecolor{currentstroke}%
\pgfsetstrokeopacity{0.000000}%
\pgfsetdash{}{0pt}%
\pgfusepath{}%
\end{pgfscope}%
\begin{pgfscope}%
\pgfsetbuttcap%
\pgfsetroundjoin%
\definecolor{currentfill}{rgb}{0.150000,0.150000,0.150000}%
\pgfsetfillcolor{currentfill}%
\pgfsetlinewidth{1.254687pt}%
\definecolor{currentstroke}{rgb}{0.150000,0.150000,0.150000}%
\pgfsetstrokecolor{currentstroke}%
\pgfsetdash{}{0pt}%
\pgfsys@defobject{currentmarker}{\pgfqpoint{0.000000in}{0.000000in}}{\pgfqpoint{0.083333in}{0.000000in}}{%
\pgfpathmoveto{\pgfqpoint{0.000000in}{0.000000in}}%
\pgfpathlineto{\pgfqpoint{0.083333in}{0.000000in}}%
\pgfusepath{stroke,fill}%
}%
\begin{pgfscope}%
\pgfsys@transformshift{9.117361in}{1.141475in}%
\pgfsys@useobject{currentmarker}{}%
\end{pgfscope}%
\end{pgfscope}%
\begin{pgfscope}%
\definecolor{textcolor}{rgb}{0.150000,0.150000,0.150000}%
\pgfsetstrokecolor{textcolor}%
\pgfsetfillcolor{textcolor}%
\pgftext[x=9.249306in, y=1.072031in, left, base]{\color{textcolor}\rmfamily\fontsize{17.000000}{16.800000}\selectfont \(\displaystyle {2}\)}%
\end{pgfscope}%
\begin{pgfscope}%
\pgfsetbuttcap%
\pgfsetroundjoin%
\definecolor{currentfill}{rgb}{0.150000,0.150000,0.150000}%
\pgfsetfillcolor{currentfill}%
\pgfsetlinewidth{1.254687pt}%
\definecolor{currentstroke}{rgb}{0.150000,0.150000,0.150000}%
\pgfsetstrokecolor{currentstroke}%
\pgfsetdash{}{0pt}%
\pgfsys@defobject{currentmarker}{\pgfqpoint{0.000000in}{0.000000in}}{\pgfqpoint{0.083333in}{0.000000in}}{%
\pgfpathmoveto{\pgfqpoint{0.000000in}{0.000000in}}%
\pgfpathlineto{\pgfqpoint{0.083333in}{0.000000in}}%
\pgfusepath{stroke,fill}%
}%
\begin{pgfscope}%
\pgfsys@transformshift{9.117361in}{1.651549in}%
\pgfsys@useobject{currentmarker}{}%
\end{pgfscope}%
\end{pgfscope}%
\begin{pgfscope}%
\definecolor{textcolor}{rgb}{0.150000,0.150000,0.150000}%
\pgfsetstrokecolor{textcolor}%
\pgfsetfillcolor{textcolor}%
\pgftext[x=9.249306in, y=1.582104in, left, base]{\color{textcolor}\rmfamily\fontsize{17.000000}{16.800000}\selectfont \(\displaystyle {4}\)}%
\end{pgfscope}%
\begin{pgfscope}%
\pgfsetbuttcap%
\pgfsetroundjoin%
\definecolor{currentfill}{rgb}{0.150000,0.150000,0.150000}%
\pgfsetfillcolor{currentfill}%
\pgfsetlinewidth{1.254687pt}%
\definecolor{currentstroke}{rgb}{0.150000,0.150000,0.150000}%
\pgfsetstrokecolor{currentstroke}%
\pgfsetdash{}{0pt}%
\pgfsys@defobject{currentmarker}{\pgfqpoint{0.000000in}{0.000000in}}{\pgfqpoint{0.083333in}{0.000000in}}{%
\pgfpathmoveto{\pgfqpoint{0.000000in}{0.000000in}}%
\pgfpathlineto{\pgfqpoint{0.083333in}{0.000000in}}%
\pgfusepath{stroke,fill}%
}%
\begin{pgfscope}%
\pgfsys@transformshift{9.117361in}{2.161622in}%
\pgfsys@useobject{currentmarker}{}%
\end{pgfscope}%
\end{pgfscope}%
\begin{pgfscope}%
\definecolor{textcolor}{rgb}{0.150000,0.150000,0.150000}%
\pgfsetstrokecolor{textcolor}%
\pgfsetfillcolor{textcolor}%
\pgftext[x=9.249306in, y=2.092178in, left, base]{\color{textcolor}\rmfamily\fontsize{17.000000}{16.800000}\selectfont \(\displaystyle {6}\)}%
\end{pgfscope}%
\begin{pgfscope}%
\pgfsetbuttcap%
\pgfsetroundjoin%
\definecolor{currentfill}{rgb}{0.150000,0.150000,0.150000}%
\pgfsetfillcolor{currentfill}%
\pgfsetlinewidth{1.254687pt}%
\definecolor{currentstroke}{rgb}{0.150000,0.150000,0.150000}%
\pgfsetstrokecolor{currentstroke}%
\pgfsetdash{}{0pt}%
\pgfsys@defobject{currentmarker}{\pgfqpoint{0.000000in}{0.000000in}}{\pgfqpoint{0.083333in}{0.000000in}}{%
\pgfpathmoveto{\pgfqpoint{0.000000in}{0.000000in}}%
\pgfpathlineto{\pgfqpoint{0.083333in}{0.000000in}}%
\pgfusepath{stroke,fill}%
}%
\begin{pgfscope}%
\pgfsys@transformshift{9.117361in}{2.671696in}%
\pgfsys@useobject{currentmarker}{}%
\end{pgfscope}%
\end{pgfscope}%
\begin{pgfscope}%
\definecolor{textcolor}{rgb}{0.150000,0.150000,0.150000}%
\pgfsetstrokecolor{textcolor}%
\pgfsetfillcolor{textcolor}%
\pgftext[x=9.249306in, y=2.602251in, left, base]{\color{textcolor}\rmfamily\fontsize{17.000000}{16.800000}\selectfont \(\displaystyle {8}\)}%
\end{pgfscope}%
\begin{pgfscope}%
\definecolor{textcolor}{rgb}{0.150000,0.150000,0.150000}%
\pgfsetstrokecolor{textcolor}%
\pgfsetfillcolor{textcolor}%
\pgftext[x=9.402777in,y=1.835760in,,top,rotate=90.000000]{\color{textcolor}\rmfamily\fontsize{27.000000}{21.600000}\selectfont \(\displaystyle |c_{\omega}|\)}%
\end{pgfscope}%
\begin{pgfscope}%
\definecolor{textcolor}{rgb}{0.150000,0.150000,0.150000}%
\pgfsetstrokecolor{textcolor}%
\pgfsetfillcolor{textcolor}%
\pgftext[x=9.033056in,y=3.081782in,left,base]{\color{textcolor}\rmfamily\fontsize{17.000000}{16.800000}\selectfont \(\displaystyle \times{10^{\ensuremath{-}5}}{}\)}%
\end{pgfscope}%
\begin{pgfscope}%
\pgfsys@transformshift{9.000000in}{0.635208in}%
\pgftext[left,bottom]{\includegraphics[interpolate=true,width=0.120000in,height=2.410000in]{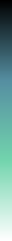}}%
\end{pgfscope}%
\begin{pgfscope}%
\pgfsetrectcap%
\pgfsetmiterjoin%
\pgfsetlinewidth{1.254687pt}%
\definecolor{currentstroke}{rgb}{0.800000,0.800000,0.800000}%
\pgfsetstrokecolor{currentstroke}%
\pgfsetdash{}{0pt}%
\pgfpathmoveto{\pgfqpoint{8.996926in}{0.631404in}}%
\pgfpathlineto{\pgfqpoint{9.057143in}{0.631404in}}%
\pgfpathlineto{\pgfqpoint{9.117361in}{0.631404in}}%
\pgfpathlineto{\pgfqpoint{9.117361in}{3.040116in}}%
\pgfpathlineto{\pgfqpoint{9.057143in}{3.040116in}}%
\pgfpathlineto{\pgfqpoint{8.996926in}{3.040116in}}%
\pgfpathlineto{\pgfqpoint{8.996926in}{0.631404in}}%
\pgfpathclose%
\pgfusepath{stroke}%
\end{pgfscope}%
\end{pgfpicture}%
\makeatother%
\endgroup%

%% file: Figures/figs_numerics/122_training.pgf
\begingroup%
\makeatletter%
\begin{pgfpicture}%
\pgfpathrectangle{\pgfpointorigin}{\pgfqpoint{9.748433in}{3.334937in}}%
\pgfusepath{use as bounding box, clip}%
\begin{pgfscope}%
\pgfsetbuttcap%
\pgfsetmiterjoin%
\pgfsetlinewidth{0.000000pt}%
\definecolor{currentstroke}{rgb}{1.000000,1.000000,1.000000}%
\pgfsetstrokecolor{currentstroke}%
\pgfsetdash{}{0pt}%
\pgfpathmoveto{\pgfqpoint{0.000000in}{0.000000in}}%
\pgfpathlineto{\pgfqpoint{9.748433in}{0.000000in}}%
\pgfpathlineto{\pgfqpoint{9.748433in}{3.334937in}}%
\pgfpathlineto{\pgfqpoint{0.000000in}{3.334937in}}%
\pgfpathlineto{\pgfqpoint{0.000000in}{0.000000in}}%
\pgfpathclose%
\pgfusepath{}%
\end{pgfscope}%
\begin{pgfscope}%
\pgfsetbuttcap%
\pgfsetmiterjoin%
\pgfsetlinewidth{0.000000pt}%
\definecolor{currentstroke}{rgb}{0.000000,0.000000,0.000000}%
\pgfsetstrokecolor{currentstroke}%
\pgfsetstrokeopacity{0.000000}%
\pgfsetdash{}{0pt}%
\pgfpathmoveto{\pgfqpoint{0.703855in}{0.631404in}}%
\pgfpathlineto{\pgfqpoint{4.743310in}{0.631404in}}%
\pgfpathlineto{\pgfqpoint{4.743310in}{3.038044in}}%
\pgfpathlineto{\pgfqpoint{0.703855in}{3.038044in}}%
\pgfpathlineto{\pgfqpoint{0.703855in}{0.631404in}}%
\pgfpathclose%
\pgfusepath{}%
\end{pgfscope}%
\begin{pgfscope}%
\pgfpathrectangle{\pgfqpoint{0.703855in}{0.631404in}}{\pgfqpoint{4.039455in}{2.406640in}}%
\pgfusepath{clip}%
\pgfsetroundcap%
\pgfsetroundjoin%
\pgfsetlinewidth{1.003750pt}%
\definecolor{currentstroke}{rgb}{0.800000,0.800000,0.800000}%
\pgfsetstrokecolor{currentstroke}%
\pgfsetdash{}{0pt}%
\pgfpathmoveto{\pgfqpoint{0.887466in}{0.631404in}}%
\pgfpathlineto{\pgfqpoint{0.887466in}{3.038044in}}%
\pgfusepath{stroke}%
\end{pgfscope}%
\begin{pgfscope}%
\definecolor{textcolor}{rgb}{0.150000,0.150000,0.150000}%
\pgfsetstrokecolor{textcolor}%
\pgfsetfillcolor{textcolor}%
\pgftext[x=0.887466in,y=0.546682in,,top]{\color{textcolor}\rmfamily\fontsize{17.000000}{16.800000}\selectfont \(\displaystyle {0}\)}%
\end{pgfscope}%
\begin{pgfscope}%
\pgfpathrectangle{\pgfqpoint{0.703855in}{0.631404in}}{\pgfqpoint{4.039455in}{2.406640in}}%
\pgfusepath{clip}%
\pgfsetroundcap%
\pgfsetroundjoin%
\pgfsetlinewidth{1.003750pt}%
\definecolor{currentstroke}{rgb}{0.800000,0.800000,0.800000}%
\pgfsetstrokecolor{currentstroke}%
\pgfsetdash{}{0pt}%
\pgfpathmoveto{\pgfqpoint{1.582307in}{0.631404in}}%
\pgfpathlineto{\pgfqpoint{1.582307in}{3.038044in}}%
\pgfusepath{stroke}%
\end{pgfscope}%
\begin{pgfscope}%
\definecolor{textcolor}{rgb}{0.150000,0.150000,0.150000}%
\pgfsetstrokecolor{textcolor}%
\pgfsetfillcolor{textcolor}%
\pgftext[x=1.582307in,y=0.546682in,,top]{\color{textcolor}\rmfamily\fontsize{17.000000}{16.800000}\selectfont \(\displaystyle {200}\)}%
\end{pgfscope}%
\begin{pgfscope}%
\pgfpathrectangle{\pgfqpoint{0.703855in}{0.631404in}}{\pgfqpoint{4.039455in}{2.406640in}}%
\pgfusepath{clip}%
\pgfsetroundcap%
\pgfsetroundjoin%
\pgfsetlinewidth{1.003750pt}%
\definecolor{currentstroke}{rgb}{0.800000,0.800000,0.800000}%
\pgfsetstrokecolor{currentstroke}%
\pgfsetdash{}{0pt}%
\pgfpathmoveto{\pgfqpoint{2.277147in}{0.631404in}}%
\pgfpathlineto{\pgfqpoint{2.277147in}{3.038044in}}%
\pgfusepath{stroke}%
\end{pgfscope}%
\begin{pgfscope}%
\definecolor{textcolor}{rgb}{0.150000,0.150000,0.150000}%
\pgfsetstrokecolor{textcolor}%
\pgfsetfillcolor{textcolor}%
\pgftext[x=2.277147in,y=0.546682in,,top]{\color{textcolor}\rmfamily\fontsize{17.000000}{16.800000}\selectfont \(\displaystyle {400}\)}%
\end{pgfscope}%
\begin{pgfscope}%
\pgfpathrectangle{\pgfqpoint{0.703855in}{0.631404in}}{\pgfqpoint{4.039455in}{2.406640in}}%
\pgfusepath{clip}%
\pgfsetroundcap%
\pgfsetroundjoin%
\pgfsetlinewidth{1.003750pt}%
\definecolor{currentstroke}{rgb}{0.800000,0.800000,0.800000}%
\pgfsetstrokecolor{currentstroke}%
\pgfsetdash{}{0pt}%
\pgfpathmoveto{\pgfqpoint{2.971988in}{0.631404in}}%
\pgfpathlineto{\pgfqpoint{2.971988in}{3.038044in}}%
\pgfusepath{stroke}%
\end{pgfscope}%
\begin{pgfscope}%
\definecolor{textcolor}{rgb}{0.150000,0.150000,0.150000}%
\pgfsetstrokecolor{textcolor}%
\pgfsetfillcolor{textcolor}%
\pgftext[x=2.971988in,y=0.546682in,,top]{\color{textcolor}\rmfamily\fontsize{17.000000}{16.800000}\selectfont \(\displaystyle {600}\)}%
\end{pgfscope}%
\begin{pgfscope}%
\pgfpathrectangle{\pgfqpoint{0.703855in}{0.631404in}}{\pgfqpoint{4.039455in}{2.406640in}}%
\pgfusepath{clip}%
\pgfsetroundcap%
\pgfsetroundjoin%
\pgfsetlinewidth{1.003750pt}%
\definecolor{currentstroke}{rgb}{0.800000,0.800000,0.800000}%
\pgfsetstrokecolor{currentstroke}%
\pgfsetdash{}{0pt}%
\pgfpathmoveto{\pgfqpoint{3.666828in}{0.631404in}}%
\pgfpathlineto{\pgfqpoint{3.666828in}{3.038044in}}%
\pgfusepath{stroke}%
\end{pgfscope}%
\begin{pgfscope}%
\definecolor{textcolor}{rgb}{0.150000,0.150000,0.150000}%
\pgfsetstrokecolor{textcolor}%
\pgfsetfillcolor{textcolor}%
\pgftext[x=3.666828in,y=0.546682in,,top]{\color{textcolor}\rmfamily\fontsize{17.000000}{16.800000}\selectfont \(\displaystyle {800}\)}%
\end{pgfscope}%
\begin{pgfscope}%
\pgfpathrectangle{\pgfqpoint{0.703855in}{0.631404in}}{\pgfqpoint{4.039455in}{2.406640in}}%
\pgfusepath{clip}%
\pgfsetroundcap%
\pgfsetroundjoin%
\pgfsetlinewidth{1.003750pt}%
\definecolor{currentstroke}{rgb}{0.800000,0.800000,0.800000}%
\pgfsetstrokecolor{currentstroke}%
\pgfsetdash{}{0pt}%
\pgfpathmoveto{\pgfqpoint{4.361669in}{0.631404in}}%
\pgfpathlineto{\pgfqpoint{4.361669in}{3.038044in}}%
\pgfusepath{stroke}%
\end{pgfscope}%
\begin{pgfscope}%
\definecolor{textcolor}{rgb}{0.150000,0.150000,0.150000}%
\pgfsetstrokecolor{textcolor}%
\pgfsetfillcolor{textcolor}%
\pgftext[x=4.361669in,y=0.546682in,,top]{\color{textcolor}\rmfamily\fontsize{17.000000}{16.800000}\selectfont \(\displaystyle {1000}\)}%
\end{pgfscope}%
\begin{pgfscope}%
\definecolor{textcolor}{rgb}{0.150000,0.150000,0.150000}%
\pgfsetstrokecolor{textcolor}%
\pgfsetfillcolor{textcolor}%
\pgftext[x=2.723582in,y=0.313349in,,top]{\color{textcolor}\rmfamily\fontsize{21.000000}{21.600000}\selectfont Epochs}%
\end{pgfscope}%
\begin{pgfscope}%
\pgfpathrectangle{\pgfqpoint{0.703855in}{0.631404in}}{\pgfqpoint{4.039455in}{2.406640in}}%
\pgfusepath{clip}%
\pgfsetroundcap%
\pgfsetroundjoin%
\pgfsetlinewidth{1.003750pt}%
\definecolor{currentstroke}{rgb}{0.800000,0.800000,0.800000}%
\pgfsetstrokecolor{currentstroke}%
\pgfsetdash{}{0pt}%
\pgfpathmoveto{\pgfqpoint{0.703855in}{0.765933in}}%
\pgfpathlineto{\pgfqpoint{4.743310in}{0.765933in}}%
\pgfusepath{stroke}%
\end{pgfscope}%
\begin{pgfscope}%
\definecolor{textcolor}{rgb}{0.150000,0.150000,0.150000}%
\pgfsetstrokecolor{textcolor}%
\pgfsetfillcolor{textcolor}%
\pgftext[x=0.368904in, y=0.696489in, left, base]{\color{textcolor}\rmfamily\fontsize{17.000000}{16.800000}\selectfont \(\displaystyle {0.0}\)}%
\end{pgfscope}%
\begin{pgfscope}%
\pgfpathrectangle{\pgfqpoint{0.703855in}{0.631404in}}{\pgfqpoint{4.039455in}{2.406640in}}%
\pgfusepath{clip}%
\pgfsetroundcap%
\pgfsetroundjoin%
\pgfsetlinewidth{1.003750pt}%
\definecolor{currentstroke}{rgb}{0.800000,0.800000,0.800000}%
\pgfsetstrokecolor{currentstroke}%
\pgfsetdash{}{0pt}%
\pgfpathmoveto{\pgfqpoint{0.703855in}{1.214364in}}%
\pgfpathlineto{\pgfqpoint{4.743310in}{1.214364in}}%
\pgfusepath{stroke}%
\end{pgfscope}%
\begin{pgfscope}%
\definecolor{textcolor}{rgb}{0.150000,0.150000,0.150000}%
\pgfsetstrokecolor{textcolor}%
\pgfsetfillcolor{textcolor}%
\pgftext[x=0.368904in, y=1.144920in, left, base]{\color{textcolor}\rmfamily\fontsize{17.000000}{16.800000}\selectfont \(\displaystyle {0.1}\)}%
\end{pgfscope}%
\begin{pgfscope}%
\pgfpathrectangle{\pgfqpoint{0.703855in}{0.631404in}}{\pgfqpoint{4.039455in}{2.406640in}}%
\pgfusepath{clip}%
\pgfsetroundcap%
\pgfsetroundjoin%
\pgfsetlinewidth{1.003750pt}%
\definecolor{currentstroke}{rgb}{0.800000,0.800000,0.800000}%
\pgfsetstrokecolor{currentstroke}%
\pgfsetdash{}{0pt}%
\pgfpathmoveto{\pgfqpoint{0.703855in}{1.662795in}}%
\pgfpathlineto{\pgfqpoint{4.743310in}{1.662795in}}%
\pgfusepath{stroke}%
\end{pgfscope}%
\begin{pgfscope}%
\definecolor{textcolor}{rgb}{0.150000,0.150000,0.150000}%
\pgfsetstrokecolor{textcolor}%
\pgfsetfillcolor{textcolor}%
\pgftext[x=0.368904in, y=1.593350in, left, base]{\color{textcolor}\rmfamily\fontsize{17.000000}{16.800000}\selectfont \(\displaystyle {0.2}\)}%
\end{pgfscope}%
\begin{pgfscope}%
\pgfpathrectangle{\pgfqpoint{0.703855in}{0.631404in}}{\pgfqpoint{4.039455in}{2.406640in}}%
\pgfusepath{clip}%
\pgfsetroundcap%
\pgfsetroundjoin%
\pgfsetlinewidth{1.003750pt}%
\definecolor{currentstroke}{rgb}{0.800000,0.800000,0.800000}%
\pgfsetstrokecolor{currentstroke}%
\pgfsetdash{}{0pt}%
\pgfpathmoveto{\pgfqpoint{0.703855in}{2.111226in}}%
\pgfpathlineto{\pgfqpoint{4.743310in}{2.111226in}}%
\pgfusepath{stroke}%
\end{pgfscope}%
\begin{pgfscope}%
\definecolor{textcolor}{rgb}{0.150000,0.150000,0.150000}%
\pgfsetstrokecolor{textcolor}%
\pgfsetfillcolor{textcolor}%
\pgftext[x=0.368904in, y=2.041781in, left, base]{\color{textcolor}\rmfamily\fontsize{17.000000}{16.800000}\selectfont \(\displaystyle {0.3}\)}%
\end{pgfscope}%
\begin{pgfscope}%
\pgfpathrectangle{\pgfqpoint{0.703855in}{0.631404in}}{\pgfqpoint{4.039455in}{2.406640in}}%
\pgfusepath{clip}%
\pgfsetroundcap%
\pgfsetroundjoin%
\pgfsetlinewidth{1.003750pt}%
\definecolor{currentstroke}{rgb}{0.800000,0.800000,0.800000}%
\pgfsetstrokecolor{currentstroke}%
\pgfsetdash{}{0pt}%
\pgfpathmoveto{\pgfqpoint{0.703855in}{2.559657in}}%
\pgfpathlineto{\pgfqpoint{4.743310in}{2.559657in}}%
\pgfusepath{stroke}%
\end{pgfscope}%
\begin{pgfscope}%
\definecolor{textcolor}{rgb}{0.150000,0.150000,0.150000}%
\pgfsetstrokecolor{textcolor}%
\pgfsetfillcolor{textcolor}%
\pgftext[x=0.368904in, y=2.490212in, left, base]{\color{textcolor}\rmfamily\fontsize{17.000000}{16.800000}\selectfont \(\displaystyle {0.4}\)}%
\end{pgfscope}%
\begin{pgfscope}%
\pgfpathrectangle{\pgfqpoint{0.703855in}{0.631404in}}{\pgfqpoint{4.039455in}{2.406640in}}%
\pgfusepath{clip}%
\pgfsetroundcap%
\pgfsetroundjoin%
\pgfsetlinewidth{1.003750pt}%
\definecolor{currentstroke}{rgb}{0.800000,0.800000,0.800000}%
\pgfsetstrokecolor{currentstroke}%
\pgfsetdash{}{0pt}%
\pgfpathmoveto{\pgfqpoint{0.703855in}{3.008087in}}%
\pgfpathlineto{\pgfqpoint{4.743310in}{3.008087in}}%
\pgfusepath{stroke}%
\end{pgfscope}%
\begin{pgfscope}%
\definecolor{textcolor}{rgb}{0.150000,0.150000,0.150000}%
\pgfsetstrokecolor{textcolor}%
\pgfsetfillcolor{textcolor}%
\pgftext[x=0.368904in, y=2.938643in, left, base]{\color{textcolor}\rmfamily\fontsize{17.000000}{16.800000}\selectfont \(\displaystyle {0.5}\)}%
\end{pgfscope}%
\begin{pgfscope}%
\definecolor{textcolor}{rgb}{0.150000,0.150000,0.150000}%
\pgfsetstrokecolor{textcolor}%
\pgfsetfillcolor{textcolor}%
\pgftext[x=0.313349in,y=1.834724in,,bottom,rotate=90.000000]{\color{textcolor}\rmfamily\fontsize{21.000000}{21.600000}\selectfont Loss \(\displaystyle \mathcal{L}\)}%
\end{pgfscope}%
\begin{pgfscope}%
\pgfpathrectangle{\pgfqpoint{0.703855in}{0.631404in}}{\pgfqpoint{4.039455in}{2.406640in}}%
\pgfusepath{clip}%
\pgfsetroundcap%
\pgfsetroundjoin%
\pgfsetlinewidth{2.208250pt}%
\definecolor{currentstroke}{rgb}{0.333333,0.556863,0.631373}%
\pgfsetstrokecolor{currentstroke}%
\pgfsetdash{}{0pt}%
\pgfpathmoveto{\pgfqpoint{0.887466in}{3.010279in}}%
\pgfpathlineto{\pgfqpoint{0.890940in}{2.997759in}}%
\pgfpathlineto{\pgfqpoint{0.894415in}{2.979478in}}%
\pgfpathlineto{\pgfqpoint{0.904837in}{2.957015in}}%
\pgfpathlineto{\pgfqpoint{0.911786in}{2.948313in}}%
\pgfpathlineto{\pgfqpoint{0.925682in}{2.934860in}}%
\pgfpathlineto{\pgfqpoint{0.943053in}{2.925608in}}%
\pgfpathlineto{\pgfqpoint{0.956950in}{2.917372in}}%
\pgfpathlineto{\pgfqpoint{0.967373in}{2.913897in}}%
\pgfpathlineto{\pgfqpoint{0.988218in}{2.903918in}}%
\pgfpathlineto{\pgfqpoint{1.002115in}{2.900283in}}%
\pgfpathlineto{\pgfqpoint{1.047279in}{2.894019in}}%
\pgfpathlineto{\pgfqpoint{1.082022in}{2.891248in}}%
\pgfpathlineto{\pgfqpoint{1.123712in}{2.888452in}}%
\pgfpathlineto{\pgfqpoint{1.165402in}{2.885784in}}%
\pgfpathlineto{\pgfqpoint{1.231412in}{2.884041in}}%
\pgfpathlineto{\pgfqpoint{1.460710in}{2.876816in}}%
\pgfpathlineto{\pgfqpoint{1.485029in}{2.873231in}}%
\pgfpathlineto{\pgfqpoint{1.491977in}{2.869578in}}%
\pgfpathlineto{\pgfqpoint{1.512823in}{2.851861in}}%
\pgfpathlineto{\pgfqpoint{1.516297in}{2.847169in}}%
\pgfpathlineto{\pgfqpoint{1.523245in}{2.826178in}}%
\pgfpathlineto{\pgfqpoint{1.533668in}{2.785887in}}%
\pgfpathlineto{\pgfqpoint{1.537142in}{2.760915in}}%
\pgfpathlineto{\pgfqpoint{1.540616in}{2.744586in}}%
\pgfpathlineto{\pgfqpoint{1.544090in}{2.736845in}}%
\pgfpathlineto{\pgfqpoint{1.547565in}{2.718706in}}%
\pgfpathlineto{\pgfqpoint{1.551039in}{2.706831in}}%
\pgfpathlineto{\pgfqpoint{1.557987in}{2.695063in}}%
\pgfpathlineto{\pgfqpoint{1.564936in}{2.668848in}}%
\pgfpathlineto{\pgfqpoint{1.571884in}{2.654573in}}%
\pgfpathlineto{\pgfqpoint{1.575358in}{2.650344in}}%
\pgfpathlineto{\pgfqpoint{1.582307in}{2.632442in}}%
\pgfpathlineto{\pgfqpoint{1.585781in}{2.630789in}}%
\pgfpathlineto{\pgfqpoint{1.589255in}{2.618068in}}%
\pgfpathlineto{\pgfqpoint{1.592729in}{2.616164in}}%
\pgfpathlineto{\pgfqpoint{1.603152in}{2.599186in}}%
\pgfpathlineto{\pgfqpoint{1.606626in}{2.596649in}}%
\pgfpathlineto{\pgfqpoint{1.617049in}{2.580219in}}%
\pgfpathlineto{\pgfqpoint{1.627471in}{2.567117in}}%
\pgfpathlineto{\pgfqpoint{1.634420in}{2.561540in}}%
\pgfpathlineto{\pgfqpoint{1.641368in}{2.553236in}}%
\pgfpathlineto{\pgfqpoint{1.655265in}{2.538819in}}%
\pgfpathlineto{\pgfqpoint{1.662213in}{2.536048in}}%
\pgfpathlineto{\pgfqpoint{1.676110in}{2.527551in}}%
\pgfpathlineto{\pgfqpoint{1.690007in}{2.522384in}}%
\pgfpathlineto{\pgfqpoint{1.703904in}{2.518818in}}%
\pgfpathlineto{\pgfqpoint{1.717801in}{2.516170in}}%
\pgfpathlineto{\pgfqpoint{1.728223in}{2.513200in}}%
\pgfpathlineto{\pgfqpoint{1.759491in}{2.506330in}}%
\pgfpathlineto{\pgfqpoint{1.773388in}{2.502820in}}%
\pgfpathlineto{\pgfqpoint{1.811604in}{2.499624in}}%
\pgfpathlineto{\pgfqpoint{1.995737in}{2.497112in}}%
\pgfpathlineto{\pgfqpoint{2.044376in}{2.494688in}}%
\pgfpathlineto{\pgfqpoint{2.072169in}{2.493105in}}%
\pgfpathlineto{\pgfqpoint{2.131231in}{2.490714in}}%
\pgfpathlineto{\pgfqpoint{2.172921in}{2.488936in}}%
\pgfpathlineto{\pgfqpoint{2.221560in}{2.486763in}}%
\pgfpathlineto{\pgfqpoint{2.277147in}{2.482456in}}%
\pgfpathlineto{\pgfqpoint{2.297992in}{2.478821in}}%
\pgfpathlineto{\pgfqpoint{2.304941in}{2.477676in}}%
\pgfpathlineto{\pgfqpoint{2.311889in}{2.478849in}}%
\pgfpathlineto{\pgfqpoint{2.315363in}{2.479989in}}%
\pgfpathlineto{\pgfqpoint{2.318838in}{2.479716in}}%
\pgfpathlineto{\pgfqpoint{2.329260in}{2.472698in}}%
\pgfpathlineto{\pgfqpoint{2.336209in}{2.472701in}}%
\pgfpathlineto{\pgfqpoint{2.343157in}{2.469946in}}%
\pgfpathlineto{\pgfqpoint{2.353580in}{2.468694in}}%
\pgfpathlineto{\pgfqpoint{2.360528in}{2.465469in}}%
\pgfpathlineto{\pgfqpoint{2.377899in}{2.464611in}}%
\pgfpathlineto{\pgfqpoint{2.384847in}{2.464424in}}%
\pgfpathlineto{\pgfqpoint{2.395270in}{2.463827in}}%
\pgfpathlineto{\pgfqpoint{2.419589in}{2.462530in}}%
\pgfpathlineto{\pgfqpoint{2.454332in}{2.461277in}}%
\pgfpathlineto{\pgfqpoint{2.489074in}{2.460810in}}%
\pgfpathlineto{\pgfqpoint{2.502970in}{2.461856in}}%
\pgfpathlineto{\pgfqpoint{2.516867in}{2.458628in}}%
\pgfpathlineto{\pgfqpoint{2.530764in}{2.459578in}}%
\pgfpathlineto{\pgfqpoint{2.544661in}{2.458337in}}%
\pgfpathlineto{\pgfqpoint{2.558558in}{2.458810in}}%
\pgfpathlineto{\pgfqpoint{2.575929in}{2.458532in}}%
\pgfpathlineto{\pgfqpoint{2.589825in}{2.458074in}}%
\pgfpathlineto{\pgfqpoint{2.669732in}{2.457454in}}%
\pgfpathlineto{\pgfqpoint{2.767010in}{2.457810in}}%
\pgfpathlineto{\pgfqpoint{2.777432in}{2.459775in}}%
\pgfpathlineto{\pgfqpoint{2.787855in}{2.462767in}}%
\pgfpathlineto{\pgfqpoint{2.794803in}{2.459603in}}%
\pgfpathlineto{\pgfqpoint{2.798278in}{2.457925in}}%
\pgfpathlineto{\pgfqpoint{2.801752in}{2.457644in}}%
\pgfpathlineto{\pgfqpoint{2.812174in}{2.459607in}}%
\pgfpathlineto{\pgfqpoint{2.822597in}{2.457745in}}%
\pgfpathlineto{\pgfqpoint{2.833020in}{2.458438in}}%
\pgfpathlineto{\pgfqpoint{2.846916in}{2.457793in}}%
\pgfpathlineto{\pgfqpoint{2.940720in}{2.457184in}}%
\pgfpathlineto{\pgfqpoint{3.093585in}{2.457925in}}%
\pgfpathlineto{\pgfqpoint{3.104007in}{2.460535in}}%
\pgfpathlineto{\pgfqpoint{3.110956in}{2.461621in}}%
\pgfpathlineto{\pgfqpoint{3.124853in}{2.457429in}}%
\pgfpathlineto{\pgfqpoint{3.135275in}{2.458741in}}%
\pgfpathlineto{\pgfqpoint{3.145698in}{2.457383in}}%
\pgfpathlineto{\pgfqpoint{3.156120in}{2.457847in}}%
\pgfpathlineto{\pgfqpoint{3.170017in}{2.457180in}}%
\pgfpathlineto{\pgfqpoint{3.180440in}{2.457161in}}%
\pgfpathlineto{\pgfqpoint{3.194337in}{2.457143in}}%
\pgfpathlineto{\pgfqpoint{3.246450in}{2.456804in}}%
\pgfpathlineto{\pgfqpoint{3.315934in}{2.456991in}}%
\pgfpathlineto{\pgfqpoint{3.336779in}{2.458222in}}%
\pgfpathlineto{\pgfqpoint{3.350676in}{2.458761in}}%
\pgfpathlineto{\pgfqpoint{3.374995in}{2.457023in}}%
\pgfpathlineto{\pgfqpoint{3.402789in}{2.456891in}}%
\pgfpathlineto{\pgfqpoint{3.437531in}{2.456856in}}%
\pgfpathlineto{\pgfqpoint{3.573025in}{2.456603in}}%
\pgfpathlineto{\pgfqpoint{3.607767in}{2.457476in}}%
\pgfpathlineto{\pgfqpoint{3.618189in}{2.459777in}}%
\pgfpathlineto{\pgfqpoint{3.628612in}{2.463495in}}%
\pgfpathlineto{\pgfqpoint{3.635560in}{2.461657in}}%
\pgfpathlineto{\pgfqpoint{3.642509in}{2.459373in}}%
\pgfpathlineto{\pgfqpoint{3.652931in}{2.459630in}}%
\pgfpathlineto{\pgfqpoint{3.663354in}{2.457909in}}%
\pgfpathlineto{\pgfqpoint{3.670302in}{2.458762in}}%
\pgfpathlineto{\pgfqpoint{3.684199in}{2.457306in}}%
\pgfpathlineto{\pgfqpoint{3.691148in}{2.457741in}}%
\pgfpathlineto{\pgfqpoint{3.705044in}{2.457001in}}%
\pgfpathlineto{\pgfqpoint{3.718941in}{2.456511in}}%
\pgfpathlineto{\pgfqpoint{3.764106in}{2.456095in}}%
\pgfpathlineto{\pgfqpoint{3.847487in}{2.455307in}}%
\pgfpathlineto{\pgfqpoint{3.916971in}{2.455849in}}%
\pgfpathlineto{\pgfqpoint{3.930868in}{2.458145in}}%
\pgfpathlineto{\pgfqpoint{3.941290in}{2.459983in}}%
\pgfpathlineto{\pgfqpoint{3.948239in}{2.457960in}}%
\pgfpathlineto{\pgfqpoint{3.955187in}{2.455618in}}%
\pgfpathlineto{\pgfqpoint{3.962135in}{2.456315in}}%
\pgfpathlineto{\pgfqpoint{3.969084in}{2.457207in}}%
\pgfpathlineto{\pgfqpoint{3.986455in}{2.455755in}}%
\pgfpathlineto{\pgfqpoint{3.996877in}{2.455857in}}%
\pgfpathlineto{\pgfqpoint{4.007300in}{2.455251in}}%
\pgfpathlineto{\pgfqpoint{4.024671in}{2.455346in}}%
\pgfpathlineto{\pgfqpoint{4.062887in}{2.455222in}}%
\pgfpathlineto{\pgfqpoint{4.160165in}{2.456527in}}%
\pgfpathlineto{\pgfqpoint{4.177536in}{2.458221in}}%
\pgfpathlineto{\pgfqpoint{4.201855in}{2.455493in}}%
\pgfpathlineto{\pgfqpoint{4.215752in}{2.455940in}}%
\pgfpathlineto{\pgfqpoint{4.240072in}{2.455278in}}%
\pgfpathlineto{\pgfqpoint{4.351246in}{2.455604in}}%
\pgfpathlineto{\pgfqpoint{4.368617in}{2.455058in}}%
\pgfpathlineto{\pgfqpoint{4.410308in}{2.455234in}}%
\pgfpathlineto{\pgfqpoint{4.445050in}{2.456697in}}%
\pgfpathlineto{\pgfqpoint{4.458946in}{2.458042in}}%
\pgfpathlineto{\pgfqpoint{4.469369in}{2.456540in}}%
\pgfpathlineto{\pgfqpoint{4.479792in}{2.455202in}}%
\pgfpathlineto{\pgfqpoint{4.514534in}{2.455063in}}%
\pgfpathlineto{\pgfqpoint{4.559698in}{2.455106in}}%
\pgfpathlineto{\pgfqpoint{4.559698in}{2.455106in}}%
\pgfusepath{stroke}%
\end{pgfscope}%
\begin{pgfscope}%
\pgfpathrectangle{\pgfqpoint{0.703855in}{0.631404in}}{\pgfqpoint{4.039455in}{2.406640in}}%
\pgfusepath{clip}%
\pgfsetroundcap%
\pgfsetroundjoin%
\pgfsetlinewidth{3.011250pt}%
\definecolor{currentstroke}{rgb}{1.000000,0.647059,0.000000}%
\pgfsetstrokecolor{currentstroke}%
\pgfsetdash{}{0pt}%
\pgfpathmoveto{\pgfqpoint{0.703855in}{0.765933in}}%
\pgfpathlineto{\pgfqpoint{4.743310in}{0.765933in}}%
\pgfusepath{stroke}%
\end{pgfscope}%
\begin{pgfscope}%
\pgfsetrectcap%
\pgfsetmiterjoin%
\pgfsetlinewidth{1.254687pt}%
\definecolor{currentstroke}{rgb}{0.800000,0.800000,0.800000}%
\pgfsetstrokecolor{currentstroke}%
\pgfsetdash{}{0pt}%
\pgfpathmoveto{\pgfqpoint{0.703855in}{0.631404in}}%
\pgfpathlineto{\pgfqpoint{0.703855in}{3.038044in}}%
\pgfusepath{stroke}%
\end{pgfscope}%
\begin{pgfscope}%
\pgfsetrectcap%
\pgfsetmiterjoin%
\pgfsetlinewidth{1.254687pt}%
\definecolor{currentstroke}{rgb}{0.800000,0.800000,0.800000}%
\pgfsetstrokecolor{currentstroke}%
\pgfsetdash{}{0pt}%
\pgfpathmoveto{\pgfqpoint{4.743310in}{0.631404in}}%
\pgfpathlineto{\pgfqpoint{4.743310in}{3.038044in}}%
\pgfusepath{stroke}%
\end{pgfscope}%
\begin{pgfscope}%
\pgfsetrectcap%
\pgfsetmiterjoin%
\pgfsetlinewidth{1.254687pt}%
\definecolor{currentstroke}{rgb}{0.800000,0.800000,0.800000}%
\pgfsetstrokecolor{currentstroke}%
\pgfsetdash{}{0pt}%
\pgfpathmoveto{\pgfqpoint{0.703855in}{0.631404in}}%
\pgfpathlineto{\pgfqpoint{4.743310in}{0.631404in}}%
\pgfusepath{stroke}%
\end{pgfscope}%
\begin{pgfscope}%
\pgfsetrectcap%
\pgfsetmiterjoin%
\pgfsetlinewidth{1.254687pt}%
\definecolor{currentstroke}{rgb}{0.800000,0.800000,0.800000}%
\pgfsetstrokecolor{currentstroke}%
\pgfsetdash{}{0pt}%
\pgfpathmoveto{\pgfqpoint{0.703855in}{3.038044in}}%
\pgfpathlineto{\pgfqpoint{4.743310in}{3.038044in}}%
\pgfusepath{stroke}%
\end{pgfscope}%
\begin{pgfscope}%
\pgfsetbuttcap%
\pgfsetmiterjoin%
\pgfsetlinewidth{0.000000pt}%
\definecolor{currentstroke}{rgb}{0.000000,0.000000,0.000000}%
\pgfsetstrokecolor{currentstroke}%
\pgfsetstrokeopacity{0.000000}%
\pgfsetdash{}{0pt}%
\pgfpathmoveto{\pgfqpoint{5.664356in}{0.631404in}}%
\pgfpathlineto{\pgfqpoint{9.017104in}{0.631404in}}%
\pgfpathlineto{\pgfqpoint{9.017104in}{3.038044in}}%
\pgfpathlineto{\pgfqpoint{5.664356in}{3.038044in}}%
\pgfpathlineto{\pgfqpoint{5.664356in}{0.631404in}}%
\pgfpathclose%
\pgfusepath{}%
\end{pgfscope}%
\begin{pgfscope}%
\pgfsys@transformshift{5.660000in}{0.634937in}%
\pgftext[left,bottom]{\includegraphics[interpolate=true,width=3.360000in,height=2.410000in]{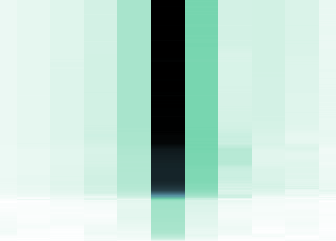}}%
\end{pgfscope}%
\begin{pgfscope}%
\definecolor{textcolor}{rgb}{0.150000,0.150000,0.150000}%
\pgfsetstrokecolor{textcolor}%
\pgfsetfillcolor{textcolor}%
\pgftext[x=5.999631in,y=0.546682in,,top]{\color{textcolor}\rmfamily\fontsize{17.000000}{16.800000}\selectfont \(\displaystyle {118}\)}%
\end{pgfscope}%
\begin{pgfscope}%
\definecolor{textcolor}{rgb}{0.150000,0.150000,0.150000}%
\pgfsetstrokecolor{textcolor}%
\pgfsetfillcolor{textcolor}%
\pgftext[x=6.670180in,y=0.546682in,,top]{\color{textcolor}\rmfamily\fontsize{17.000000}{16.800000}\selectfont \(\displaystyle {120}\)}%
\end{pgfscope}%
\begin{pgfscope}%
\definecolor{textcolor}{rgb}{0.150000,0.150000,0.150000}%
\pgfsetstrokecolor{textcolor}%
\pgfsetfillcolor{textcolor}%
\pgftext[x=7.340730in,y=0.546682in,,top]{\color{textcolor}\rmfamily\fontsize{17.000000}{16.800000}\selectfont \(\displaystyle {122}\)}%
\end{pgfscope}%
\begin{pgfscope}%
\definecolor{textcolor}{rgb}{0.150000,0.150000,0.150000}%
\pgfsetstrokecolor{textcolor}%
\pgfsetfillcolor{textcolor}%
\pgftext[x=8.011279in,y=0.546682in,,top]{\color{textcolor}\rmfamily\fontsize{17.000000}{16.800000}\selectfont \(\displaystyle {124}\)}%
\end{pgfscope}%
\begin{pgfscope}%
\definecolor{textcolor}{rgb}{0.150000,0.150000,0.150000}%
\pgfsetstrokecolor{textcolor}%
\pgfsetfillcolor{textcolor}%
\pgftext[x=8.681829in,y=0.546682in,,top]{\color{textcolor}\rmfamily\fontsize{17.000000}{16.800000}\selectfont \(\displaystyle {126}\)}%
\end{pgfscope}%
\begin{pgfscope}%
\pgfpathrectangle{\pgfqpoint{5.664356in}{0.631404in}}{\pgfqpoint{3.352748in}{2.406640in}}%
\pgfusepath{clip}%
\pgfsetroundcap%
\pgfsetroundjoin%
\pgfsetlinewidth{2.007500pt}%
\definecolor{currentstroke}{rgb}{0.827451,0.827451,0.827451}%
\pgfsetstrokecolor{currentstroke}%
\pgfsetdash{}{0pt}%
\pgfpathmoveto{\pgfqpoint{5.831993in}{0.631404in}}%
\pgfpathlineto{\pgfqpoint{5.831993in}{3.038044in}}%
\pgfusepath{stroke}%
\end{pgfscope}%
\begin{pgfscope}%
\pgfpathrectangle{\pgfqpoint{5.664356in}{0.631404in}}{\pgfqpoint{3.352748in}{2.406640in}}%
\pgfusepath{clip}%
\pgfsetroundcap%
\pgfsetroundjoin%
\pgfsetlinewidth{2.007500pt}%
\definecolor{currentstroke}{rgb}{0.827451,0.827451,0.827451}%
\pgfsetstrokecolor{currentstroke}%
\pgfsetdash{}{0pt}%
\pgfpathmoveto{\pgfqpoint{6.167268in}{0.631404in}}%
\pgfpathlineto{\pgfqpoint{6.167268in}{3.038044in}}%
\pgfusepath{stroke}%
\end{pgfscope}%
\begin{pgfscope}%
\pgfpathrectangle{\pgfqpoint{5.664356in}{0.631404in}}{\pgfqpoint{3.352748in}{2.406640in}}%
\pgfusepath{clip}%
\pgfsetroundcap%
\pgfsetroundjoin%
\pgfsetlinewidth{2.007500pt}%
\definecolor{currentstroke}{rgb}{0.827451,0.827451,0.827451}%
\pgfsetstrokecolor{currentstroke}%
\pgfsetdash{}{0pt}%
\pgfpathmoveto{\pgfqpoint{6.502543in}{0.631404in}}%
\pgfpathlineto{\pgfqpoint{6.502543in}{3.038044in}}%
\pgfusepath{stroke}%
\end{pgfscope}%
\begin{pgfscope}%
\pgfpathrectangle{\pgfqpoint{5.664356in}{0.631404in}}{\pgfqpoint{3.352748in}{2.406640in}}%
\pgfusepath{clip}%
\pgfsetroundcap%
\pgfsetroundjoin%
\pgfsetlinewidth{2.007500pt}%
\definecolor{currentstroke}{rgb}{0.827451,0.827451,0.827451}%
\pgfsetstrokecolor{currentstroke}%
\pgfsetdash{}{0pt}%
\pgfpathmoveto{\pgfqpoint{6.837818in}{0.631404in}}%
\pgfpathlineto{\pgfqpoint{6.837818in}{3.038044in}}%
\pgfusepath{stroke}%
\end{pgfscope}%
\begin{pgfscope}%
\pgfpathrectangle{\pgfqpoint{5.664356in}{0.631404in}}{\pgfqpoint{3.352748in}{2.406640in}}%
\pgfusepath{clip}%
\pgfsetroundcap%
\pgfsetroundjoin%
\pgfsetlinewidth{2.007500pt}%
\definecolor{currentstroke}{rgb}{0.827451,0.827451,0.827451}%
\pgfsetstrokecolor{currentstroke}%
\pgfsetdash{}{0pt}%
\pgfpathmoveto{\pgfqpoint{7.173092in}{0.631404in}}%
\pgfpathlineto{\pgfqpoint{7.173092in}{3.038044in}}%
\pgfusepath{stroke}%
\end{pgfscope}%
\begin{pgfscope}%
\pgfpathrectangle{\pgfqpoint{5.664356in}{0.631404in}}{\pgfqpoint{3.352748in}{2.406640in}}%
\pgfusepath{clip}%
\pgfsetroundcap%
\pgfsetroundjoin%
\pgfsetlinewidth{2.007500pt}%
\definecolor{currentstroke}{rgb}{0.827451,0.827451,0.827451}%
\pgfsetstrokecolor{currentstroke}%
\pgfsetdash{}{0pt}%
\pgfpathmoveto{\pgfqpoint{7.508367in}{0.631404in}}%
\pgfpathlineto{\pgfqpoint{7.508367in}{3.038044in}}%
\pgfusepath{stroke}%
\end{pgfscope}%
\begin{pgfscope}%
\pgfpathrectangle{\pgfqpoint{5.664356in}{0.631404in}}{\pgfqpoint{3.352748in}{2.406640in}}%
\pgfusepath{clip}%
\pgfsetroundcap%
\pgfsetroundjoin%
\pgfsetlinewidth{2.007500pt}%
\definecolor{currentstroke}{rgb}{0.827451,0.827451,0.827451}%
\pgfsetstrokecolor{currentstroke}%
\pgfsetdash{}{0pt}%
\pgfpathmoveto{\pgfqpoint{7.843642in}{0.631404in}}%
\pgfpathlineto{\pgfqpoint{7.843642in}{3.038044in}}%
\pgfusepath{stroke}%
\end{pgfscope}%
\begin{pgfscope}%
\pgfpathrectangle{\pgfqpoint{5.664356in}{0.631404in}}{\pgfqpoint{3.352748in}{2.406640in}}%
\pgfusepath{clip}%
\pgfsetroundcap%
\pgfsetroundjoin%
\pgfsetlinewidth{2.007500pt}%
\definecolor{currentstroke}{rgb}{0.827451,0.827451,0.827451}%
\pgfsetstrokecolor{currentstroke}%
\pgfsetdash{}{0pt}%
\pgfpathmoveto{\pgfqpoint{8.178917in}{0.631404in}}%
\pgfpathlineto{\pgfqpoint{8.178917in}{3.038044in}}%
\pgfusepath{stroke}%
\end{pgfscope}%
\begin{pgfscope}%
\pgfpathrectangle{\pgfqpoint{5.664356in}{0.631404in}}{\pgfqpoint{3.352748in}{2.406640in}}%
\pgfusepath{clip}%
\pgfsetroundcap%
\pgfsetroundjoin%
\pgfsetlinewidth{2.007500pt}%
\definecolor{currentstroke}{rgb}{0.827451,0.827451,0.827451}%
\pgfsetstrokecolor{currentstroke}%
\pgfsetdash{}{0pt}%
\pgfpathmoveto{\pgfqpoint{8.514192in}{0.631404in}}%
\pgfpathlineto{\pgfqpoint{8.514192in}{3.038044in}}%
\pgfusepath{stroke}%
\end{pgfscope}%
\begin{pgfscope}%
\pgfpathrectangle{\pgfqpoint{5.664356in}{0.631404in}}{\pgfqpoint{3.352748in}{2.406640in}}%
\pgfusepath{clip}%
\pgfsetroundcap%
\pgfsetroundjoin%
\pgfsetlinewidth{2.007500pt}%
\definecolor{currentstroke}{rgb}{0.827451,0.827451,0.827451}%
\pgfsetstrokecolor{currentstroke}%
\pgfsetdash{}{0pt}%
\pgfpathmoveto{\pgfqpoint{8.849466in}{0.631404in}}%
\pgfpathlineto{\pgfqpoint{8.849466in}{3.038044in}}%
\pgfusepath{stroke}%
\end{pgfscope}%
\begin{pgfscope}%
\definecolor{textcolor}{rgb}{0.150000,0.150000,0.150000}%
\pgfsetstrokecolor{textcolor}%
\pgfsetfillcolor{textcolor}%
\pgftext[x=7.340730in,y=0.313349in,,top]{\color{textcolor}\rmfamily\fontsize{21.000000}{21.600000}\selectfont \(\displaystyle \omega\)}%
\end{pgfscope}%
\begin{pgfscope}%
\definecolor{textcolor}{rgb}{0.150000,0.150000,0.150000}%
\pgfsetstrokecolor{textcolor}%
\pgfsetfillcolor{textcolor}%
\pgftext[x=5.481718in, y=0.563097in, left, base]{\color{textcolor}\rmfamily\fontsize{17.000000}{16.800000}\selectfont \(\displaystyle {0}\)}%
\end{pgfscope}%
\begin{pgfscope}%
\definecolor{textcolor}{rgb}{0.150000,0.150000,0.150000}%
\pgfsetstrokecolor{textcolor}%
\pgfsetfillcolor{textcolor}%
\pgftext[x=5.285887in, y=1.018038in, left, base]{\color{textcolor}\rmfamily\fontsize{17.000000}{16.800000}\selectfont \(\displaystyle {200}\)}%
\end{pgfscope}%
\begin{pgfscope}%
\definecolor{textcolor}{rgb}{0.150000,0.150000,0.150000}%
\pgfsetstrokecolor{textcolor}%
\pgfsetfillcolor{textcolor}%
\pgftext[x=5.285887in, y=1.472980in, left, base]{\color{textcolor}\rmfamily\fontsize{17.000000}{16.800000}\selectfont \(\displaystyle {400}\)}%
\end{pgfscope}%
\begin{pgfscope}%
\definecolor{textcolor}{rgb}{0.150000,0.150000,0.150000}%
\pgfsetstrokecolor{textcolor}%
\pgfsetfillcolor{textcolor}%
\pgftext[x=5.285887in, y=1.927921in, left, base]{\color{textcolor}\rmfamily\fontsize{17.000000}{16.800000}\selectfont \(\displaystyle {600}\)}%
\end{pgfscope}%
\begin{pgfscope}%
\definecolor{textcolor}{rgb}{0.150000,0.150000,0.150000}%
\pgfsetstrokecolor{textcolor}%
\pgfsetfillcolor{textcolor}%
\pgftext[x=5.285887in, y=2.382862in, left, base]{\color{textcolor}\rmfamily\fontsize{17.000000}{16.800000}\selectfont \(\displaystyle {800}\)}%
\end{pgfscope}%
\begin{pgfscope}%
\definecolor{textcolor}{rgb}{0.150000,0.150000,0.150000}%
\pgfsetstrokecolor{textcolor}%
\pgfsetfillcolor{textcolor}%
\pgftext[x=5.187972in, y=2.837804in, left, base]{\color{textcolor}\rmfamily\fontsize{17.000000}{16.800000}\selectfont \(\displaystyle {1000}\)}%
\end{pgfscope}%
\begin{pgfscope}%
\definecolor{textcolor}{rgb}{0.150000,0.150000,0.150000}%
\pgfsetstrokecolor{textcolor}%
\pgfsetfillcolor{textcolor}%
\pgftext[x=5.132416in,y=1.834724in,,bottom,rotate=90.000000]{\color{textcolor}\rmfamily\fontsize{21.000000}{21.600000}\selectfont Epochs}%
\end{pgfscope}%
\begin{pgfscope}%
\pgfsetrectcap%
\pgfsetmiterjoin%
\pgfsetlinewidth{1.254687pt}%
\definecolor{currentstroke}{rgb}{0.800000,0.800000,0.800000}%
\pgfsetstrokecolor{currentstroke}%
\pgfsetdash{}{0pt}%
\pgfpathmoveto{\pgfqpoint{5.664356in}{0.631404in}}%
\pgfpathlineto{\pgfqpoint{5.664356in}{3.038044in}}%
\pgfusepath{stroke}%
\end{pgfscope}%
\begin{pgfscope}%
\pgfsetrectcap%
\pgfsetmiterjoin%
\pgfsetlinewidth{1.254687pt}%
\definecolor{currentstroke}{rgb}{0.800000,0.800000,0.800000}%
\pgfsetstrokecolor{currentstroke}%
\pgfsetdash{}{0pt}%
\pgfpathmoveto{\pgfqpoint{9.017104in}{0.631404in}}%
\pgfpathlineto{\pgfqpoint{9.017104in}{3.038044in}}%
\pgfusepath{stroke}%
\end{pgfscope}%
\begin{pgfscope}%
\pgfsetrectcap%
\pgfsetmiterjoin%
\pgfsetlinewidth{1.254687pt}%
\definecolor{currentstroke}{rgb}{0.800000,0.800000,0.800000}%
\pgfsetstrokecolor{currentstroke}%
\pgfsetdash{}{0pt}%
\pgfpathmoveto{\pgfqpoint{5.664356in}{0.631404in}}%
\pgfpathlineto{\pgfqpoint{9.017104in}{0.631404in}}%
\pgfusepath{stroke}%
\end{pgfscope}%
\begin{pgfscope}%
\pgfsetrectcap%
\pgfsetmiterjoin%
\pgfsetlinewidth{1.254687pt}%
\definecolor{currentstroke}{rgb}{0.800000,0.800000,0.800000}%
\pgfsetstrokecolor{currentstroke}%
\pgfsetdash{}{0pt}%
\pgfpathmoveto{\pgfqpoint{5.664356in}{3.038044in}}%
\pgfpathlineto{\pgfqpoint{9.017104in}{3.038044in}}%
\pgfusepath{stroke}%
\end{pgfscope}%
\begin{pgfscope}%
\definecolor{textcolor}{rgb}{0.150000,0.150000,0.150000}%
\pgfsetstrokecolor{textcolor}%
\pgfsetfillcolor{textcolor}%
\pgftext[x=7.139565in,y=3.096049in,left,base]{\color{textcolor}\rmfamily\fontsize{17.000000}{16.800000}\selectfont 6e-05}%
\end{pgfscope}%
\begin{pgfscope}%
\pgfsetbuttcap%
\pgfsetmiterjoin%
\pgfsetlinewidth{0.000000pt}%
\definecolor{currentstroke}{rgb}{0.000000,0.000000,0.000000}%
\pgfsetstrokecolor{currentstroke}%
\pgfsetstrokeopacity{0.000000}%
\pgfsetdash{}{0pt}%
\pgfpathmoveto{\pgfqpoint{9.097893in}{0.631404in}}%
\pgfpathlineto{\pgfqpoint{9.218225in}{0.631404in}}%
\pgfpathlineto{\pgfqpoint{9.218225in}{3.038044in}}%
\pgfpathlineto{\pgfqpoint{9.097893in}{3.038044in}}%
\pgfpathlineto{\pgfqpoint{9.097893in}{0.631404in}}%
\pgfpathclose%
\pgfusepath{}%
\end{pgfscope}%
\begin{pgfscope}%
\pgfpathrectangle{\pgfqpoint{9.097893in}{0.631404in}}{\pgfqpoint{0.120332in}{2.406640in}}%
\pgfusepath{clip}%
\pgfsetbuttcap%
\pgfsetmiterjoin%
\pgfsetlinewidth{0.000000pt}%
\definecolor{currentstroke}{rgb}{0.000000,0.000000,0.000000}%
\pgfsetstrokecolor{currentstroke}%
\pgfsetstrokeopacity{0.000000}%
\pgfsetdash{}{0pt}%
\pgfusepath{}%
\end{pgfscope}%
\begin{pgfscope}%
\pgfsetbuttcap%
\pgfsetroundjoin%
\definecolor{currentfill}{rgb}{0.150000,0.150000,0.150000}%
\pgfsetfillcolor{currentfill}%
\pgfsetlinewidth{1.254687pt}%
\definecolor{currentstroke}{rgb}{0.150000,0.150000,0.150000}%
\pgfsetstrokecolor{currentstroke}%
\pgfsetdash{}{0pt}%
\pgfsys@defobject{currentmarker}{\pgfqpoint{0.000000in}{0.000000in}}{\pgfqpoint{0.083333in}{0.000000in}}{%
\pgfpathmoveto{\pgfqpoint{0.000000in}{0.000000in}}%
\pgfpathlineto{\pgfqpoint{0.083333in}{0.000000in}}%
\pgfusepath{stroke,fill}%
}%
\begin{pgfscope}%
\pgfsys@transformshift{9.218225in}{1.065214in}%
\pgfsys@useobject{currentmarker}{}%
\end{pgfscope}%
\end{pgfscope}%
\begin{pgfscope}%
\definecolor{textcolor}{rgb}{0.150000,0.150000,0.150000}%
\pgfsetstrokecolor{textcolor}%
\pgfsetfillcolor{textcolor}%
\pgftext[x=9.350169in, y=1.012407in, left, base]{\color{textcolor}\rmfamily\fontsize{17.000000}{13.200000}\selectfont \(\displaystyle {1}\)}%
\end{pgfscope}%
\begin{pgfscope}%
\pgfsetbuttcap%
\pgfsetroundjoin%
\definecolor{currentfill}{rgb}{0.150000,0.150000,0.150000}%
\pgfsetfillcolor{currentfill}%
\pgfsetlinewidth{1.254687pt}%
\definecolor{currentstroke}{rgb}{0.150000,0.150000,0.150000}%
\pgfsetstrokecolor{currentstroke}%
\pgfsetdash{}{0pt}%
\pgfsys@defobject{currentmarker}{\pgfqpoint{0.000000in}{0.000000in}}{\pgfqpoint{0.083333in}{0.000000in}}{%
\pgfpathmoveto{\pgfqpoint{0.000000in}{0.000000in}}%
\pgfpathlineto{\pgfqpoint{0.083333in}{0.000000in}}%
\pgfusepath{stroke,fill}%
}%
\begin{pgfscope}%
\pgfsys@transformshift{9.218225in}{1.499051in}%
\pgfsys@useobject{currentmarker}{}%
\end{pgfscope}%
\end{pgfscope}%
\begin{pgfscope}%
\definecolor{textcolor}{rgb}{0.150000,0.150000,0.150000}%
\pgfsetstrokecolor{textcolor}%
\pgfsetfillcolor{textcolor}%
\pgftext[x=9.350169in, y=1.446244in, left, base]{\color{textcolor}\rmfamily\fontsize{17.000000}{13.200000}\selectfont \(\displaystyle {2}\)}%
\end{pgfscope}%
\begin{pgfscope}%
\pgfsetbuttcap%
\pgfsetroundjoin%
\definecolor{currentfill}{rgb}{0.150000,0.150000,0.150000}%
\pgfsetfillcolor{currentfill}%
\pgfsetlinewidth{1.254687pt}%
\definecolor{currentstroke}{rgb}{0.150000,0.150000,0.150000}%
\pgfsetstrokecolor{currentstroke}%
\pgfsetdash{}{0pt}%
\pgfsys@defobject{currentmarker}{\pgfqpoint{0.000000in}{0.000000in}}{\pgfqpoint{0.083333in}{0.000000in}}{%
\pgfpathmoveto{\pgfqpoint{0.000000in}{0.000000in}}%
\pgfpathlineto{\pgfqpoint{0.083333in}{0.000000in}}%
\pgfusepath{stroke,fill}%
}%
\begin{pgfscope}%
\pgfsys@transformshift{9.218225in}{1.932888in}%
\pgfsys@useobject{currentmarker}{}%
\end{pgfscope}%
\end{pgfscope}%
\begin{pgfscope}%
\definecolor{textcolor}{rgb}{0.150000,0.150000,0.150000}%
\pgfsetstrokecolor{textcolor}%
\pgfsetfillcolor{textcolor}%
\pgftext[x=9.350169in, y=1.880082in, left, base]{\color{textcolor}\rmfamily\fontsize{17.000000}{13.200000}\selectfont \(\displaystyle {3}\)}%
\end{pgfscope}%
\begin{pgfscope}%
\pgfsetbuttcap%
\pgfsetroundjoin%
\definecolor{currentfill}{rgb}{0.150000,0.150000,0.150000}%
\pgfsetfillcolor{currentfill}%
\pgfsetlinewidth{1.254687pt}%
\definecolor{currentstroke}{rgb}{0.150000,0.150000,0.150000}%
\pgfsetstrokecolor{currentstroke}%
\pgfsetdash{}{0pt}%
\pgfsys@defobject{currentmarker}{\pgfqpoint{0.000000in}{0.000000in}}{\pgfqpoint{0.083333in}{0.000000in}}{%
\pgfpathmoveto{\pgfqpoint{0.000000in}{0.000000in}}%
\pgfpathlineto{\pgfqpoint{0.083333in}{0.000000in}}%
\pgfusepath{stroke,fill}%
}%
\begin{pgfscope}%
\pgfsys@transformshift{9.218225in}{2.366726in}%
\pgfsys@useobject{currentmarker}{}%
\end{pgfscope}%
\end{pgfscope}%
\begin{pgfscope}%
\definecolor{textcolor}{rgb}{0.150000,0.150000,0.150000}%
\pgfsetstrokecolor{textcolor}%
\pgfsetfillcolor{textcolor}%
\pgftext[x=9.350169in, y=2.313919in, left, base]{\color{textcolor}\rmfamily\fontsize{17.000000}{13.200000}\selectfont \(\displaystyle {4}\)}%
\end{pgfscope}%
\begin{pgfscope}%
\pgfsetbuttcap%
\pgfsetroundjoin%
\definecolor{currentfill}{rgb}{0.150000,0.150000,0.150000}%
\pgfsetfillcolor{currentfill}%
\pgfsetlinewidth{1.254687pt}%
\definecolor{currentstroke}{rgb}{0.150000,0.150000,0.150000}%
\pgfsetstrokecolor{currentstroke}%
\pgfsetdash{}{0pt}%
\pgfsys@defobject{currentmarker}{\pgfqpoint{0.000000in}{0.000000in}}{\pgfqpoint{0.083333in}{0.000000in}}{%
\pgfpathmoveto{\pgfqpoint{0.000000in}{0.000000in}}%
\pgfpathlineto{\pgfqpoint{0.083333in}{0.000000in}}%
\pgfusepath{stroke,fill}%
}%
\begin{pgfscope}%
\pgfsys@transformshift{9.218225in}{2.800563in}%
\pgfsys@useobject{currentmarker}{}%
\end{pgfscope}%
\end{pgfscope}%
\begin{pgfscope}%
\definecolor{textcolor}{rgb}{0.150000,0.150000,0.150000}%
\pgfsetstrokecolor{textcolor}%
\pgfsetfillcolor{textcolor}%
\pgftext[x=9.350169in, y=2.747757in, left, base]{\color{textcolor}\rmfamily\fontsize{17.000000}{13.200000}\selectfont \(\displaystyle {5}\)}%
\end{pgfscope}%
\begin{pgfscope}%
\definecolor{textcolor}{rgb}{0.150000,0.150000,0.150000}%
\pgfsetstrokecolor{textcolor}%
\pgfsetfillcolor{textcolor}%
\pgftext[x=9.391767in,y=1.934724in,,top,rotate=90.000000]{\color{textcolor}\rmfamily\fontsize{27.000000}{14.400000}\selectfont \(\displaystyle |c_{\omega}|\)}%
\end{pgfscope}%
\begin{pgfscope}%
\definecolor{textcolor}{rgb}{0.150000,0.150000,0.150000}%
\pgfsetstrokecolor{textcolor}%
\pgfsetfillcolor{textcolor}%
\pgftext[x=9.218225in,y=3.079711in,right,base]{\color{textcolor}\rmfamily\fontsize{17.000000}{13.200000}\selectfont \(\displaystyle \times{10^{\ensuremath{-}5}}{}\)}%
\end{pgfscope}%
\begin{pgfscope}%
\pgfsys@transformshift{9.100000in}{0.634937in}%
\pgftext[left,bottom]{\includegraphics[interpolate=true,width=0.120000in,height=2.410000in]{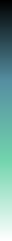}}%
\end{pgfscope}%
\begin{pgfscope}%
\pgfsetrectcap%
\pgfsetmiterjoin%
\pgfsetlinewidth{1.254687pt}%
\definecolor{currentstroke}{rgb}{0.800000,0.800000,0.800000}%
\pgfsetstrokecolor{currentstroke}%
\pgfsetdash{}{0pt}%
\pgfpathmoveto{\pgfqpoint{9.097893in}{0.631404in}}%
\pgfpathlineto{\pgfqpoint{9.158059in}{0.631404in}}%
\pgfpathlineto{\pgfqpoint{9.218225in}{0.631404in}}%
\pgfpathlineto{\pgfqpoint{9.218225in}{3.038044in}}%
\pgfpathlineto{\pgfqpoint{9.158059in}{3.038044in}}%
\pgfpathlineto{\pgfqpoint{9.097893in}{3.038044in}}%
\pgfpathlineto{\pgfqpoint{9.097893in}{0.631404in}}%
\pgfpathclose%
\pgfusepath{stroke}%
\end{pgfscope}%
\end{pgfpicture}%
\makeatother%
\endgroup%

%% file: Figures/figs_numerics/variance_redundancies.tikz
\resizebox{\textwidth}{!}{
\begin{tikzpicture}[remember picture]
    
    \node[inner sep=0pt] (plot) {\input{Figures/figs_numerics/figA.pgf}};
    \draw [draw=grayplot, fill=grayplot!50] (-3.95,-.01) rectangle (-3.41,1.345);
    \begin{scope}[overlay]
        \node at (-3.7,.625) { 
            \begin{adjustbox}{height=.7cm}
            \begin{quantikz}[row sep = 1mm,column sep =2mm]
            &  \gate[1]{e^{ixH}} & \\
            &   \gate[1]{e^{ixH}} &    \\
            &  \wireoverride{n}\raisebox{6pt}{\vdots}  \\
            &   \gate[1]{e^{ixH}} &    \\
        \end{quantikz}
        \end{adjustbox}
        };
    \draw [draw=grayplot, fill=grayplot!50] (4.22,-.01) rectangle (4.82,1.345);
        \node at (4.5,.625) { 
            \begin{adjustbox}{height=.7cm}
            \begin{quantikz}[row sep = 1mm,column sep =2mm]
            &  \gate[1]{e^{ixH_1}}  & \\
            &   \gate[1]{e^{ixH_2}} &    \\\vspace{-20mm}
            &  \wireoverride{n}\raisebox{6pt}{\vdots}  \\
            &   \gate[1]{e^{ixH_5}} &    \\
        \end{quantikz}
        \end{adjustbox}
        };
    \end{scope}
\end{tikzpicture}
}

%% file: Figures/figs_numerics/figC_quad_wCircs.tikz
\resizebox{\textwidth}{!}{
\begin{tikzpicture}[remember picture]
    \node[inner sep=0pt] (plot) {\input{Figures/figs_numerics/figC_quad.pgf}};
    \begin{scope}[overlay]
        \node at (-5,-.9) { 
            \begin{adjustbox}{height=.8cm}
            \begin{quantikz}
            & \gate[1,style={fill=gray!60}]{V} & \gate[4]{S(x)} & \gate[1,style={fill=gray!60}]{V} & \meter[4]{}\\
            & \gate[1,style={fill=gray!60}]{V} & & \gate[1,style={fill=gray!60}]{V} &  \\
            & \gate[1,style={fill=gray!60}]{V} & & \gate[1,style={fill=gray!60}]{V} &  \\
            & \gate[1,style={fill=gray!60}]{V} & & \gate[1,style={fill=gray!60}]{V} &  \\
        \end{quantikz}
        \end{adjustbox}
        };
        \node at (-0.415,-.9) { 
            \begin{adjustbox}{height=.8cm}
            \begin{quantikz}
            & \gate[2,style={fill=gray!60}]{V} & \gate[4]{S(x)} & \gate[2,style={fill=gray!60}]{V} & \meter[4]{}\\
            &  & &  &  \\
            & \gate[2,style={fill=gray!60}]{V} & & \gate[2,style={fill=gray!60}]{V} &  \\
            &  & &  & \\
        \end{quantikz}
        \end{adjustbox}
        };
        \node at (4.17,-.85) { 
            \begin{adjustbox}{height=.75cm}
            \begin{quantikz}
            & \gate[4,style={fill=gray!60}]{V} & \gate[4]{S(x)} & \gate[4,style={fill=gray!60}]{V} & \meter[4]{}\\ [.5cm]
            &  & &  & \\[.7cm]
            &  & &  & \\[.5cm]
            &  & &  &  
        \end{quantikz}
        \end{adjustbox}
        };
    \end{scope}
\end{tikzpicture}
}

%% file: Figures/figs_numerics/diff_encoding.tikz
\resizebox{\textwidth}{!}{
\begin{tikzpicture}[remember picture]
    \node[inner sep=0pt] (plot) {\input{Figures/figs_numerics/diff_encoding.pgf}};
    \begin{scope}[overlay]
        \node at (-1.3,.67) { 
            \begin{adjustbox}{height=.8cm}
            \begin{quantikz}
            & \gate[2,style={fill=gray!60}]{V} & \gate[4]{S(x)} & \gate[2,style={fill=gray!60}]{V} & \meter[4]{}\\
            &  & &  &  \\
            & \gate[2,style={fill=gray!60}]{V} & & \gate[2,style={fill=gray!60}]{V} &  \\
            &  & &  & \\
        \end{quantikz}
        \end{adjustbox}
        };
        \node at (6.28,.67) { 
            \begin{adjustbox}{height=.8cm}
            \begin{quantikz}
            & \gate[2,style={fill=gray!60}]{V} & \gate[4]{S(x)} & \gate[2,style={fill=gray!60}]{V} & \meter[4]{}\\
            &  & &  &  \\
            & \gate[2,style={fill=gray!60}]{V} & & \gate[2,style={fill=gray!60}]{V} &  \\
            &  & &  & \\
        \end{quantikz}
        \end{adjustbox}
        };
    \end{scope}
\end{tikzpicture}
}

%% file: Figures/figs_numerics/diff_encoding_local.tikz
\resizebox{\textwidth}{!}{
\begin{tikzpicture}[remember picture]
    \node[inner sep=0pt] (plot) {\input{Figures/figs_numerics/diff_encoding_local.pgf}};
    \begin{scope}[overlay]
        \draw [dashed, fill=grayplot] (-.75,.23) rectangle (-.45,1.41);
        \node at (-1.2,.76) { 
            \begin{adjustbox}{height=.7cm}
            \begin{quantikz}
            & \gate[2,style={fill=gray!60}]{V} & \gate[4]{S(x)} & \gate[2,style={fill=gray!60}]{V} & \meter{}\\
            &  & &  &\meter{}  \\
            & \gate[2,style={fill=gray!60}]{V} & & \gate[2,style={fill=gray!60}]{V} & \meter{} \\
            &  & &  & \meter{}\\
        \end{quantikz}
        \end{adjustbox}
        };
        \draw [dashed, fill=grayplot] (6.73,.23) rectangle (7.03,1.41);
        \node at (6.28,.76) { 
            \begin{adjustbox}{height=.7cm}
            \begin{quantikz}
            & \gate[2,style={fill=gray!60}]{V} & \gate[4]{S(x)} & \gate[2,style={fill=gray!60}]{V} & \meter{} \\
            &  & &  & \meter{} \\
            & \gate[2,style={fill=gray!60}]{V} & & \gate[2,style={fill=gray!60}]{V} &  \meter{} \\
            &  & &  & \meter{}\\
        \end{quantikz}
        \end{adjustbox}
        };
    \end{scope}
\end{tikzpicture}
}

%% file: Figures/figs_numerics/lightcone_Bound_global_wCircs.tikz
\resizebox{\textwidth}{!}{
\begin{tikzpicture}[remember picture]
    \node[inner sep=0pt] (plot) {\input{Figures/figs_numerics/lightcone_Bound_global.pgf}};
    \begin{scope}[overlay]
        \node at (-3.1,.82) { 
            \begin{adjustbox}{height=.6cm}
            \begin{quantikz}[row sep= 2pt]
            & \gate[2,style={fill=gray!60}]{V} &  &  & & & &\\
            &  & \gate[2,style={fill=gray!60}]{V} & \gate[6]{S(x)} & \gate[2,style={fill=gray!60}]{V}  & & &\\
            & \gate[2,style={fill=gray!60}]{V}  &  &  & & \gate[2,style={fill=gray!60}]{V}  & &\\
            &  & \gate[2,style={fill=gray!60}]{V}  & & \gate[2,style={fill=gray!60}]{V}  &  & \gate[2,style={fill=gray!60}]{V} & \meter[2]{}\\
            & \gate[2,style={fill=gray!60}]{V} &  &  & & \gate[2,style={fill=gray!60}]{V} & &\\
            &  & \gate[2,style={fill=gray!60}]{V}  & & \gate[2,style={fill=gray!60}]{V} & & &\\
            & \gate[2,style={fill=gray!60}]{V}  & & & & & & \\
            & & & & & & & \\
        \end{quantikz}
        \end{adjustbox}
        };
       
        \node at (1.48,.82) { 
            \begin{adjustbox}{height=.6cm}
            \begin{quantikz}[row sep= 2pt]
            & \gate[2,style={fill=gray!60}]{V} &  &  & & & &\\
            &  & \gate[2,style={fill=gray!60}]{V} &  &  & & &\\
            & \gate[2,style={fill=gray!60}]{V}  &  & \gate[2,style={fill=gray!60}]{V} & \gate[4]{S(x)} &  \gate[2,style={fill=gray!60}]{V}  & &\\
            &  & \gate[2,style={fill=gray!60}]{V}  &  &  & & \gate[2,style={fill=gray!60}]{V} & \meter[2]{}\\
            & \gate[2,style={fill=gray!60}]{V} &  & \gate[2,style={fill=gray!60}]{V}   & & \gate[2,style={fill=gray!60}]{V} & &\\
            &  & \gate[2,style={fill=gray!60}]{V}  & & & & &\\
            & \gate[2,style={fill=gray!60}]{V}  & & & & & & \\
            & & & & & & & \\
        \end{quantikz}
        \end{adjustbox}
        };
        
        \node at (6.05,.82) { 
        \begin{adjustbox}{height=.6cm}
            \begin{quantikz}[row sep= 2pt]
            & \gate[2,style={fill=gray!60}]{V} &  &  & & & &\\
            &  & \gate[2,style={fill=gray!60}]{V} &  &  & & &\\
            & \gate[2,style={fill=gray!60}]{V}  &  & \gate[2,style={fill=gray!60}]{V} &  & & &\\
            &  & \gate[2,style={fill=gray!60}]{V}  &  &  \gate[2,style={fill=gray!60}]{V}  &\gate[2]{S(x)} & \gate[2,style={fill=gray!60}]{V} & \meter[2]{}\\
            & \gate[2,style={fill=gray!60}]{V} &  & \gate[2,style={fill=gray!60}]{V}   & &  & &\\
            &  & \gate[2,style={fill=gray!60}]{V}  & & & & &\\
            & \gate[2,style={fill=gray!60}]{V}  & & & & & & \\
            & & & & & & & \\
        \end{quantikz}
        \end{adjustbox}
        };
    \end{scope}
\end{tikzpicture}
}

%% file: Figures/figs_numerics/lightcone_Bound_global_exp_gol.tikz
\resizebox{\textwidth}{!}{
\begin{tikzpicture}[remember picture]
    
    \node[inner sep=0pt] (plot) {\input{Figures/figs_numerics/lightcone_Bound_global_exp_gol.pgf}};
    \begin{scope}[overlay]
        
        \node at (-.85,.95) { 
            \begin{adjustbox}{height=.5cm}
            \begin{quantikz}[row sep= 3pt]
            & \gate[2,style={fill=gray!60}]{V} &  &  & & & &\\
            &  & \gate[2,style={fill=gray!60}]{V} &  &  & & &\\
            & \gate[2,style={fill=gray!60}]{V}  &  & \gate[2,style={fill=gray!60}]{V} & \gate[4]{S(x)} &  \gate[2,style={fill=gray!60}]{V}  & &\\
            &  & \gate[2,style={fill=gray!60}]{V}  &  &  & & \gate[2,style={fill=gray!60}]{V} & \meter[2]{}\\
            & \gate[2,style={fill=gray!60}]{V} &  & \gate[2,style={fill=gray!60}]{V}   & & \gate[2,style={fill=gray!60}]{V} & &\\
            &  & \gate[2,style={fill=gray!60}]{V}  & & & & &\\
            & \gate[2,style={fill=gray!60}]{V}  & & & & & & \\
            & & & & & & & \\
        \end{quantikz}
        \end{adjustbox}
        };
        \node at (6,.82) { 
            \begin{adjustbox}{height=.6cm}
            \begin{quantikz}[row sep= 3pt]
            & \gate[2,style={fill=gray!60}]{V} &  &  & & & &\\
            &  & \gate[2,style={fill=gray!60}]{V} &  &  & & &\\
            & \gate[2,style={fill=gray!60}]{V}  &  & \gate[2,style={fill=gray!60}]{V} & \gate[4]{S(x)} &  \gate[2,style={fill=gray!60}]{V}  & &\\
            &  & \gate[2,style={fill=gray!60}]{V}  &  &  & & \gate[2,style={fill=gray!60}]{V} & \meter[2]{}\\
            & \gate[2,style={fill=gray!60}]{V} &  & \gate[2,style={fill=gray!60}]{V}   & & \gate[2,style={fill=gray!60}]{V} & &\\
            &  & \gate[2,style={fill=gray!60}]{V}  & & & & &\\
            & \gate[2,style={fill=gray!60}]{V}  & & & & & & \\
            & & & & & & & \\
        \end{quantikz}
        \end{adjustbox}
        };
        
    \end{scope}
\end{tikzpicture}
}

%% file: A0_Preliminaries.tex
\newpage
\onecolumn
\section{Appendix: Preliminaries}

\input{Proofs/01_pqcs_fourier_coeffs}

%% file: Proofs/01_pqcs_fourier_coeffs.tex
\subsection{Quantum Models with Hamiltonian encoding as Large Fourier series}\label{appendix:VQC_fourier}

For completeness, we revisit the proof in \cite{schuld_effect_2021}, showing that a Quantum model with Hamiltonian encoding can be expressed as a truncated Fourier series of the form in Eq.\eqref{Eq:quantum_Model}, with a spectrum constructed from the encoding Hamiltonians eigenvalues defined in Eq.\eqref{Eq:Spectrum_def}.

We start by detailing the proof for the one-dimensional input case ($x \in \mathbb{R}$) and generalize it afterwards to the $D$-dimensional input case ($x \in \mathbb{R}^D$).

To do so, we consider an $L$ layered ansatz of the following form $U(x;\theta)= W^{(L+1)}(\theta)S^{(L)}(x) \dots S^{(1)}(x)W^{(1)}(\theta)$ as described in Eq.\eqref{eq:circuit_ansatz}. In what follows, we drop the explicit dependence of the trainable layers $W^{(l)}(\theta)$ on the parameter vector $\theta$ for simplicity and recall that the trainable and encoding unitaries are $d$-dimensional matrices acting on $n$ qubits ($d=2^n$).

We consider the Hamiltonian encoding scheme where the classical input $x \in \mathbb{R}$ is encoded as the time evolution of some Hamiltonian. Thus, the encoding unitary in each layer $l$ is of the form $S^{l}(x)=e^{-ixH_l} \quad \forall l \in \{1,\dots,L\}$. If one considers $H_l=P_lD_lP_l^{\dagger}$ and injects $P_l$ and $P_l^{\dagger}$ in the parameterized unitaries  $W^{l+1}$ and $W^{l}$ respectively, then $S^{l}(x)$ can be rewritten in the following form $S^{l}(x) = \text{diag}(e^{-ix\lambda_1^{l}},\dots,e^{-ix\lambda_d^{l}})$ where $\lambda_i^l$s are eigenvalues of the underlying $d$-dimensional encoding Hamiltonian $H_l$.

We start by applying the first layer $S^{(1)}(x)W^{(1)}$ on the $\ket{0}$ computational basis state and we iterate through the remaining layers to obtain $\ket{\psi(x;\theta)} = U(x;\theta)\ket{0}$.

\begin{equation*} 
\begin{split}
S^{(1)}(x)W^{(1)}\ket{0} & = \sum_{j_1=1}^d W^{(1)}_{j_11} e^{-ix\lambda_{j_1}^1} \ket{j_1} \\
S^{(2)}(x)W^{(2)}S^{(1)}(x)W^{(1)}\ket{0} & = \sum_{j_2=1}^d\sum_{j_1=1}^d W^{(2)}_{j_2j_1}W^{(1)}_{j_11} e^{-ix(\lambda_{j_1}^1 + \lambda_{j_2}^2)} \ket{j_2} 
\\ & \qquad \qquad \qquad \quad \vdots  \\
W^{(L+1)}\prod_{l=1}^L S^{(l)}(x)W^{(l)} \ket{0} & = \sum_{k=1}^d\sum_{\boldsymbol{J}\in \left[d\right]^L} W^{(L+1)}_{kj_L}\dots W^{(1)}_{j_11} e^{-ix\Lambda_{\boldsymbol{J}}}\ket{k}
\end{split}
\end{equation*}
where $\Lambda_J= \sum_{j_l  \in J} \lambda_{j_l}$ and $J=(j_1,\dots,j_L) \in [|1,d|]^L$.

The full quantum model can be written as a truncated Fourier series:
\begin{equation*}
    \begin{split}
           f(x;\theta)  = \langle 0|U(x; \theta)^\dagger O U(x; \theta) |0\rangle  &= \sum_{k,k'=1}^d\sum_{\boldsymbol{J,J'}\in \left[d\right]^L} (W^{(L+1)}_{k'j'_L}\dots W^{(1)}_{j'_11})^*  
           \langle k'|O|k \rangle
           (W^{(L+1)}_{kj_L}\dots W^{(1)}_{j_11}) e^{-ix(\Lambda_{\boldsymbol{J}}-\Lambda_{\boldsymbol{J'}})} \\
        & = \sum_{\omega \in \Omega}\sum_{\boldsymbol{J},\boldsymbol{J'}\in R(\omega)} c_{\omega} e^{-ix\omega}
    \end{split}
\end{equation*}

Hence, the expression of the Fourier coefficient is obtained by grouping multi-indices $(J,J')$ that lead to the same frequency $\omega$ (i.e. $\Lambda_J - \Lambda_{J'} = \omega$), i.e. in the frequency generator $R(\omega)$:
\begin{equation}\label{eq:Fourier_coeff_appendix}
c_{\omega} = \sum_{J,J' \in R(\omega)} \sum_{k,k'} W^{(1)*}_{j'_1 0}  W^{(2)*}_{j'_2 j'_1} \dots W^{(L+1)*}_{k'j'_L} O_{k'k} W^{(L+1)}_{kj_L}\dots W^{(2)}_{j_2 j_1} W^{(1)}_{j_1 0}
\end{equation}
with
\begin{equation}\label{eq:Omega}
    \begin{aligned}
         \Omega &= \left\{ \sum_{l \in [L]} (\lambda^l_{j_l} -  \lambda^{l}_{j'_{l}} )\middle| (j_1, \dots, j_L), (j'_1, \dots, j'_L)  \in \llbracket 1,d \rrbracket^L  \right\} \\
         R(\omega) &= \left\{ (J,J') \in \llbracket 1,d \rrbracket^L \times \llbracket 1,d \rrbracket^L \middle| \sum_{l \in [L]} (\lambda^l_{j_l} -  \lambda^{l}_{j'_{l}} ) = \omega  \right\}
    \end{aligned}
\end{equation}

Here we recall that the trainable unitaries $W^{(l)}$ implicitly contain the projector $P_l$ formed by the eigenvectors of the encoding Hamiltonian $H_l$. Thus, in the following section, whenever an assumption on the distribution of the trainable unitaries is made, we should take into account their multiplication by projectors $P_l$.

Moreover, we note that using encoding Hamiltonians that act non trivially only on a subset of the $n$ qubits will increase the eigenvalues degeneracy. Hence the redundancies $|R(\omega)|$  of all the frequencies $\omega \in \Omega$ according to Definition \ref{def:Redundancy} will be multiplied by the same factor.

\color{black}
\subsubsection{Quantum Fourier model with high-dimensional input}\label{app:high_dim}
We can now generalize the Quantum model Fourier Series representation to the $D$-dimensional input setting. Assume that for $x=(x_1, \dots, x_D) \subset \mathbb{R}^D$, each component $x_k$ is encoded as the time evolution of some Hamiltonian $H_{l}^{(k)}$ in the $l^{th}$ uploading layer, such that $S^l(x)=\prod_{k=1}^D S^l(x_k) = \prod_{k=1}^D e^{-ix_k H_l^{(k)}}$.

In general, writing down the expression of a Fourier coefficient with high dimensional input data is intricate.  For simplicity, we give the expression of the a Fourier coefficient $c_\omega$ for the frequency vector $\omega = (\omega_1,\dots,\omega_D)$ under the assumption that the encoding unitaries of the different components commute. Precisely, for each layer $l$, $\{S^{l}(x_k)\}_{k=1}^D$ mutually commute (i.e. equivalently  $\{H^{(k)}_l\}_{k=1}^D$ mutually commute). Using this assumption, the global encoding unitary in the $l^{th}$ layer $S^l(x)$ can be diagonalized as 
\begin{equation}
    S^l(x) = P_l \prod_{k=1}^D e^{-ix_k D_l^{(k)}} P_l^\dagger = P_l  e^{-i \sum_{k=1}^D x_k D_l^{(k)}} P_l^\dagger = P_l \left(\sum_{i=1}^d  \sum_{k=1}^D x_k \lambda_{i}^{k,l} \ketbra{i}{i} \right)P_l^\dagger\;,
\end{equation}
where $P_l$ is an orthogonal projector and each $D_l^{(k)}:= diag(\lambda_{1}^{k,l},\dots,\lambda_{d}^{k,l})$ is the diagonal matrix of eigenvalues of the Hamiltonian $H_l^{(k)}$.
Similarly to the analysis done for a one-dimensional input vector, the quantum spectrum in the $D$-dimensional setting will be of the form
\begin{equation}
    \Omega= \{ \omega= (\omega_k)_{k=1}^D \;, \omega_k = \lambda^k_{i}-\lambda^k_{j} \;,\forall i,j \in [|1,2^n|]\}
\end{equation}
Here, we see that the maximal size of the spectrum is independent of the dimensionality $D$, i.e. $|\Omega| \leq 2^{2n}$. This independence is mainly due to the assumption that the encoding Hamiltonians commute. However, in the most general case, the $D-$dimensional spectrum can scale up to $\mathcal{O}(\prod_{k=1}^D |\Omega_k|)$ where $\Omega_k$ is the spectrum constructed solely from the eigenvalues of Hamiltonians $H^l_k$ as defined in Eq.\eqref{eq:Omega}.

Consequently, it can be easily seen that all of our theoretical results can generalize to high-dimensional input vectors when the encoding Hamiltonians commute within a single uploading layer. 
Precisely, there is a one to one correspondence between frequency vectors $\omega \in \Omega$ and computational basis states. Hence, the associated Fourier coefficient can be simply expressed in a very similar way to the one dimensional case in Eq.\eqref{eq:Fourier_coeff_appendix}. Consequently, the proofs of our main theorems trivially apply to this setting.
However, our techniques no longer hold if we consider non commuting encoding unitaries.
\color{black}

\subsection{Spectrum distribution}\label{appendix:spec_distribution}
From the above spectrum construction, one can make several observations regarding the spectrum structure and its size by considering different Hamiltonian encoding schemes. We further comment on how the spectrum distribution interacts with the Fourier coefficient expression in Eq.\eqref{eq:Fourier_coeff_appendix}.

In what follows, we will focus on the one-dimensional setting ($x \in \mathbb{R}$).

\begin{lemma}\label{parallel_sequence_encoding}(Sequential and parallel encoding)
Consider the spectrum $\Omega_{\text{parallel}}$ obtained from a re-uploading circuit with $L$ layers and encoding Hamiltonians $H_1,\dots,H_L$ acting each on $n$-qubits, such that $H_l = P_l D_l P_l^{\dagger}, \quad \forall l \in [L]$. The spectrum $\Omega_{\text{parallel}}$
is the same as the spectrum $\Omega_{\text{sequential}}$  obtained by considering a single layered circuit acting on $nL$-qubits with an encoding Hamiltonian of the form $H=P\sum_{l=1}^L D_l^{(nl,n(l+1))} P^{\dagger}$ where $D_l^{(nl,n(l+1))}$ acts non trivially on the subset of the total $nL$ qubits indexed from $nl$ to $n(l+1)$ and $P= P_1 \otimes P_2 \otimes \dots \otimes P_L$.    
\end{lemma}

\begin{proof}
We consider a single layer reuploading circuit of the form in Eq.\eqref{eq:circuit_ansatz} ($L=1$) with the following encoding layer unitary acting on $nL$ qubits:
    \begin{equation*}
    \begin{aligned}
        S(x)&:= e^{-ixH_1}\otimes e^{-ixH_2}\otimes \dots e^{-ixH_L}\\
        &=P_1e^{-ixD_1} P_1^{\dagger} \otimes P_2 e^{-ixD_2} P_2^{\dagger}\otimes P_L \dots e^{-ixD_L} P_L^{\dagger}\\
        &=P e^{-ix\sum_{l=1}^L D_l^{(nl,n(l+1))}} P^{\dagger}
    \end{aligned}
    \end{equation*}
    where  $H_l = P_l D_l P_l^{\dagger}, D_l= \text{diag}(\lambda_1^l,\dots,\lambda_d^l)$ contains the eigenvalues of the Hamiltonian $H_l$ and $P= P_1 \otimes P_2 \otimes \dots \otimes P_L$.
    
\end{proof}

This result was already proven in \cite{schuld_effect_2021} to convey that we can use the encoding gates in sequence or in parallel (reupoading model) to get the same spectrum. In practice, we will have a combination of the \textit{sequential} and the \textit{parallel} encodings since each encoding layer in the reuploading model is usually made of one-qubit or two-qubit gates.
Hence a more general form of the spectrum is the following:
{\scriptsize{
\begin{equation}\label{eq:mixed_spectrum}
     \Omega = \left\{ \sum_{l \in [L]} \sum_{k \in [K_l]}(\lambda^{l,k}_{j_{l,k}} - \lambda^{l,k}_{j'_{l,k}}) \middle| (j_{1,1}, \dots,j_{1,K},\dots j_{L,1}, \dots,j_{L,K}), (j'_{1,1}, \dots,j'_{1,K},\dots j'_{L,1}, \dots,j'_{L,K})  \in \llbracket 1,2^{n/K} \rrbracket^{KL}  \right\}
\end{equation}}}
where $K_l$ is the number of blocks of the $l^{th}$ encoding layer and $\lambda^{l,k}_{j_{l,k}}$ is the eigenvalue of the local Hamiltonian $H_{l,k}$ acting non trivially on the $k^{th}$ subsystem in layer $l$.

From the above construction, one can easily see that the spectrum $\Omega$ is conserved under permutation of the encoding layers and even under permutation of tensor product unitaries within the layers.
However, the permutation invariance of the spectrum does not imply that we get the same quantum model under encoding blocks permutation since the expression of the Fourier coefficients in Eq.\eqref{eq:Fourier_coeff_appendix} is not invariant under permutation of the indices in $(J,J')\in R(\omega)$.

We introduce the \textit{partial} frequency generator as it will serve to derive Theorem \ref{thm:formal_2design_reup} and to make comments on the spectrum distribution in general.

\begin{definition}[Partial Frequency Generator]\label{def:Partial_Redundancy}
    Considering an $L$-layered quantum Fourier model as described in Eqs.(\ref{Eq:quantum_Model}-\ref{Eq:Spectrum_def}), we denote by $R_h^l(\omega)$ the ensemble of eigenvalue indices giving rise to the frequency $\omega$ by considering only encoding layers from $h$ to $l$. The partial redundancy $|R^l_h(\omega)|$ is the size of this set:
    \begin{equation}
        R^l_h(\omega) = \left\{ ((j_h,\dots,j_l),(j'_h,\dots,j'_l)) \in \llbracket 1,d \rrbracket^{l-h+1} \times \llbracket 1,d \rrbracket^{l-h+1} \middle| \; \sum_{k=h}^l (\lambda_{j_k}^k - \lambda_{j'_k}^k) = \omega  \right\}
    \end{equation}
    For simplicity, we denote the frequency generator  $R_1^L(\omega)$ by $R(\omega)$ as defined in Definition \ref{def:Redundancy} and the frequency generator from a single layer $R_l^l(\omega)$ by $R_l(\omega)$.
\end{definition}

\begin{lemma}[Recursive spectrum construction] \label{lemma:recursive_spec} 
We recall that $R_h^L(\omega) = \{(J:=(j_h,\dots,j_L),J':=(j'_h,\dots,j'_L)) \in \left[d\right]^{L-h+1} \times \left[d\right]^{L-h+1}\ ; \Lambda_J-\Lambda_J'   = \omega\}$. We denote by $\Omega^l$ the spectrum obtained from the $l^{th}$ encoding layer and its corresponding frequency generator $R_l(\omega)$ as defined in Definition \ref{def:Partial_Redundancy}.
Consider a Quantum Fourier model on $n$-qubits with $L$ layers and a spectrum $\Omega$.
For a fixed frequency $\omega \in \Omega$, we have the following recursive relation between $R_h^L(\omega)$ and $R_{h+1}^L(\omega)$: $$R^L_h(\omega) = \bigcup_{\substack{j_h, j'_h = 1 \\  k_h= \lambda_{j_h} - \lambda_{j'_h}}}^d \{R_{h+1}^L(\omega-k_h) \times \{(j_h, j'_h)\}\} $$
Consequently, $|R^L_1(\omega)| = \sum_{k_L,\dots, k_2, k_1} \beta_{k_L} \dots \beta_{k_1} \delta_{\sum_{l=1}^L k_l}^{\omega}$
 with $k_l \in \Omega^l_{distinct}, \forall l \in [L]$, $\Omega^l$ is the spectrum generated by the $l^{th}$ encoding layer and $\beta_{k_l}:=|R_l(k_l)|$ is the redundancy of the frequency $k_l$ in $\Omega^l$. 
\end{lemma}

\begin{proof}
    Let $(J,J') \in R_h^L(\omega)$, 
$$\omega = \Lambda_J-\Lambda_J' = \sum_{l=h}^L (\lambda_{j_l} - \lambda_{j'_l}) = \sum_{l=h+1}^{L} (\lambda_{j_l} - \lambda_{j_l'}) +  (\lambda_{j_h} - \lambda_{j_h'})$$
$$R^L_h(\omega) = \bigcup_{\substack{j_h, j'_h = 1 \\  k_h= \lambda_{j_h} - \lambda_{j'_h}}}^d \{R_{h+1}^L(\omega-k_h) \times \{(j_h, j'_h)\}\} $$
    $$|R_h^L(\omega)|= \sum_{\substack{j_h, j'_h = 1 \\  k_h= \lambda_{j_h} - \lambda_{j'_h}}}^d |R_{h+1}^L(\omega-k_h)| = \sum_{k_h \in \Omega^h_{distinct}} |R_h(k_h)| |R_{h+1}^L(\omega-k_h)| = \sum_{k_h \in \Omega^h_{distinct}} \beta_{k_h} |R_{h+1}^L(\omega-k_h)|$$

    Therefore we obtain : $|R^L_1(\omega)| = \sum_{k_L, \dots, k_2} \beta_{k_L} \dots \beta_{k_2} |R_1(\omega-\sum_{l=2}^L k_l)| =\sum_{k_L,\dots, k_2, k_1} \beta_{k_L} \dots \beta_{k_1} \delta_{\sum_{l=1}^L k_l}^{\omega}$
 with each $k_l \in \Omega^l_{distinct}$, $\Omega^l$ is the spectrum generated by the $l^{th}$ encoding layer and $\beta_{k_l}$ is the redundancy of the frequency $k_l$ in $\Omega^l$. 
\end{proof}

Whether we are considering sequential encoding, parallel encoding or both, the result of Lemma \ref{lemma:recursive_spec} enables us to track the evolution of the paths in the quantum tree (See Fig.\ref{fig:quantum_Spectrum_Trees}) leading to a certain frequency and mainly to characterize the size of the spectrum $|\Omega|$ and the frequency redundancies $|R(\omega)|$.

First, one can easily see that $\forall l \in [L], |R_l(0)| \geq 2^n$.
Hence the recursive construction of the spectrum implies that all frequencies that can be generated from a subset of $\kappa<L$ layers have an exponential redundancy in the number of qubits and in the number of the remaining $(L-\kappa)$ layers. Specifically, for these frequencies we have $|R(\omega)| \geq 2^{n(L-\kappa)}$. 

Consequently, adopting the reuploading scheme leads inevitably to exponential redundancies in $n$ as mentioned in \cite{barthe_gradients_2023}. 
Moreover, if one considers tensor product encoding unitaries within a single encoding layer (i.e. Pauli rotations), then the frequencies already generated by few blocks will have exponential redundancies in $n$. 

We give some examples of encoding Hamiltonians and explicit the obtained spectrum, its size and the redundancies scaling in each case. For more details, we refer the reader to a  discussion of the degeneracy of a Quantum Fourier model spectrum in \cite{peters_generalization_2023} (See Appendix C).

\subsubsection{Pauli encoding}\label{sec:Appendix_Pauli_Encoding}
The Pauli encoding scheme consists in using Pauli matrices $\{\sigma_x,\sigma_y,\sigma_z\}$ as encoding Hamiltonians, such that a single encoding layer is a tensor product of Pauli rotations parameterized by $x$: $S^l(x):= \otimes_{i=1}^n e^{-ix \sigma/2}, \sigma \in \{\sigma_x,\sigma_y,\sigma_z\}$.  In this case, all eigenvalues are $\lambda = \pm 1/2$. Hence by using Eq.\ref{eq:mixed_spectrum}, the set of frequencies we obtain by applying Pauli rotations on all $n$-qubits in each of the $L$ layers  are consecutive integer frequencies from $-nL$ to $nL$. Hence the size of the spectrum $|\Omega|=2nL$ and $|R(k)|= \binom{2nL}{nL-|k|} \forall k \in \llbracket -nL,nL\rrbracket$. The redundancy value can be obtained by using Lemma \ref{lemma:recursive_spec}.

\subsubsection{Exponential encoding}\label{sec:Appendix_Exponential_Encoding}
If we consider single qubit Pauli rotations for encoding with a scaling coefficient, the largest spectrum we can get is 
of size $|\Omega|=3^{nL}$ since the spectrum of a scaled Pauli matrix is of the form $\{-\lambda,\lambda\}$. One spectrum that reaches this limit is the set of consecutive integer frequencies from $-\frac{3^{nL}}{2}$ to $\frac{3^{nL}}{2}$ with $|\Omega|=3^{nL}$.
This spectrum is obtained by using the scaling coefficient $3^{jl}$ in the Pauli rotation gate acting on qubit $j$ in layer $l$. This scheme was introduced in \cite{shin_exponential_2022} under the name of \emph{exponential encoding}, which refers to the exponential size of the spectrum in the number of qubits/layers. We note that $|R(\omega=k)| = 2^{nL-T(|k|)}$ where $T(|k|)$ takes values in $\llbracket 0,nL\rrbracket$. 
More details about the proof are provided in \cite{peters_generalization_2023,shin_exponential_2022}.

\subsubsection{Non degenerate encoding (Golomb encoding)}\label{sec:Appendix_Golomb_Encoding}

As mentioned in Lemma \ref{lemma:recursive_spec}, with the reuploading scheme we will certainly have frequencies with redundancies exponential in the number of layers. However to surpass the limit $|\Omega^l|=3^n$ per layer $l$, one needs to consider non separable encoding unitaries in a single layer. A perfectly non degenerate spectrum for a single encoding layer $(|R(\omega)|=1 \forall \omega \in \Omega\backslash \{0\})$ was proposed in \cite{peters_generalization_2023} by setting the diagonal of the data-encoding Hamiltonian to
the elements of a Golomb ruler. However, we stress again that using this scheme in the reuploading model will give rise to redundancies that are still exponential in $nL$ for frequencies generated from few layers since the null frequency in a single layer has redundancy $|R(0)|=2^n$.





%% file: A1_proof.tex
\section{Appendix: Proofs}

\input{Proofs/02_global_2design}
\input{Proofs/03_local_2design}
\input{Proofs/Local2design_proof}

%% file: Proofs/02_global_2design.tex
\subsection{Fourier coefficients variance with 2-design trainable unitaries }\label{Proof:Global_2design}

We present the formal version of Theorem \ref{thm:single_layer_gloabl_2design} showcasing an exact expression for the expectation and variance of Fourier coefficients corresponding to a single layer model.

\begin{theorem}\label{thm:formal_2design}
 Consider a quantum model of the form in Eq.\eqref{Eq:quantum_Model} and a parametrized circuit of the form in Eq.\eqref{eq:circuit_ansatz} with $L=1$ layers and fixed encoding Hamiltonians resulting in a spectrum $\Omega$. We assume that each of the  trainable layers $W^l(\theta)\;,l \in \{1,2\}$ form independently a 2-design. The expectation and variance of each Fourier coefficient $c_{\omega}(\theta)$ for the frequencies $ \omega \in \Omega$ appearing in the model Fourier decomposition in Eq.\eqref{Eq:quantum_Fourier_Model} are given by
\begin{equation}\label{eq:coeff_variance_2design_single_app}
    \begin{aligned}
        \E_{\theta}[c_{\omega}(\theta)]& = \quad\frac{Tr(O)}{d}\delta_{\omega}^0\;,\\
        \Var_{\theta}[c_{\omega}(\theta)] &=	\frac{d||O||_2^2-Tr(O)^2}{d(d^2-1)} \frac{|R(\omega)|}{d(d+1)}  +   \frac{Tr(O)^2-d||O||^2}{d^2(d^2-1)}\delta_{\omega}^0\;,
    \end{aligned}
\end{equation}
where we recall that $d=2^n$.
\end{theorem}

The following Theorem provides a generalization of  Theorem~\ref{thm:formal_2design} to the case of reuploading models with $L\geq 1$ layers. Indeed, for $L=1$, we retrieve the results in Theorem \ref{thm:formal_2design}. Hence, the proof of Theorem~\ref{thm:formal_2design_reup} covers also the proof of Theorem~\ref{thm:formal_2design}.
 \begin{theorem}\label{thm:formal_2design_reup}
Consider a quantum model of the form in Eq.\eqref{Eq:quantum_Model} and a parametrized circuit of the form in Eq.\eqref{eq:circuit_ansatz} with $L\geq 1$ layers and fixed encoding Hamiltonians resulting in a spectrum $\Omega$. We assume that each of the  trainable layers $W^l(\theta)\;,l \in \{1,\dots,L\}$ form independently a 2-design. The expectation and variance of each Fourier coefficient $c_{\omega}(\theta)$ for the frequencies $ \omega \in \Omega$ appearing in the model Fourier decomposition in Eq.\eqref{Eq:quantum_Fourier_Model} are given by
\begin{equation}\label{eq:coeff_variance_2design_reuploading}
    \begin{aligned}
\E_{\theta}[c_{\omega}(\theta)]\quad&= \quad\frac{Tr(O)}{d}\delta_{\omega}^0\;,\\
        \Var_{\theta}[c_{\omega}(\theta)] &\simeq	 \frac{d||O||_2^2-Tr(O)^2}{d(d^2-1)}\left[\frac{|R_1^L(\omega)|- |R_2^{L}(\omega)|}{d(d+1)(d^2-1)^{L-1}}  + \sum_{j=3}^{L} \frac{|R_j^L(\omega)|}{d(d^2-1)^{L-j+2}} \right] +   \frac{Tr(O)^2-d||O||^2}{d^2(d^2-1)}\delta_{\omega}^0\;.
        \end{aligned}
        \end{equation}

Here $R_j^l(\omega)$ denotes the partial spectrum formed from the $j^{th}$ up to the $l^{th}$ encoding layers as defined in Definition \ref{def:Partial_Redundancy}. 
\end{theorem}

\begin{proof}

    We begin recall the expression of the Fourier coefficient derived in Appendix \ref{appendix:VQC_fourier}.
    \begin{align}
c_{\omega} &= \sum_{J,J' \in R(\omega)} \sum_{k,k'} W^{(1)*}_{j'_1 0}  W^{(2)*}_{j'_2 j'_1} \dots W^{(L+1)*}_{k'j'_L} O_{k'k} W^{(L+1)}_{kj_L}\dots W^{(2)}_{j_2 j_1} W^{(1)}_{j_1 0}
\end{align}
    We start by proving the expression of the expectation value of a Fourier coefficient and note that it is sufficient to establish this result under the 1-design hypothesis. To do so, we apply the Weingarten formula for the first moment \cite{mele_introduction_2023} and obtain

    \begin{equation}
        \begin{split}
           \E_{W^{(1)},\dots,W^{(L+1)} \sim U(N)} \left[c_{\omega}\right] & =  \smashoperator{\sum_{\substack{k,k' \\ J,J' \in R(\omega)}}}  \frac{\delta_{j_1}^{j'_1} \delta_{j_2}^{j'_2} \ldots \delta_{j_L}^{j'_L} \delta_{k}^{k'} O_{k'k}}{d^{L+1}}  = \smashoperator{\sum_{\substack{k \\ J,J' \in R(\omega)}}} \frac{\delta_{j_1}^{j'_1} \delta_{j_2}^{j'_2} \dots \delta_{j_L}^{j'_L}  O_{kk}}{d^{L+1}}  = \sum_{J,J' \in R(\omega)} \delta_J^{J'} \frac{\Tr(O)}{d^{L+1}}  \\& = \frac{\Tr(O)}{d} \delta_{\omega}^0
        \end{split}
    \end{equation}

The variance of a Fourier coefficient for a reuploading VQC (i.e $L>1$) is obtained recursively starting from the variance of a single-layered circuit and using the recursive relation between the partial redundancies $R_h^L(\omega)$ established in Lemma \ref{lemma:recursive_spec}.

From Eq.\eqref{eq:Fourier_coeff_appendix}, we give the expression of $\E_{W^{(1)},\dots,W^{(L+1)} \sim U(N)} \left[|c_{\omega}|^2\right] $

 \begin{equation}\label{eq:var_coeff_expanded}
      \E_{W^{(1)},\dots,W^{(L+1)} \sim U(N)}\left[|c_{\omega}|^2\right]  =  \sum_{\substack{I,I' \in R(\omega) \\ J,J' \in R(\omega)}}  \prod_{l=1}^{L+1} X_l(I,I',J,J') 
 \end{equation}
 where we introduce the shorthands
 \begin{align}
     X_l(I,I',J,J') = \begin{cases}
         \E \left[  
                W^{(1)}_{j_1 1} W^{(1)}_{i'_1 1} W^{(1)*}_{j'_1 1} W^{(1)*}_{i_1 1} \right] & \text{if } l=1\;,\\
                \E \left[  
                W^{(l)}_{j_l j_{l-1}} W^{(l)}_{i'_l i'_{l-1}} W^{(l)*}_{j'_l j'_{l-1}} W^{(l)*}_{i_l i_{l-1}} \right] & \text{if } 2\leq l \leq L\;,\\
                \sum_{\substack{k,k'\\ h,h'}} \E \left[  
                W^{(L+1)}_{k j_L} W^{(L+1)}_{h' i'_L} W^{(L+1)*}_{k' j'_L } W_{h i_L }^{(L+1)*} \right] O_{k'k}O_{hh'} & \text{if } l=L+1\;.
     \end{cases}
 \end{align}

Using the Weingarten calculus for first and second moment monomials \cite{mele_introduction_2023}, we compute independently the terms $  X_l(I,I',J,J')$ appearing in the sum of Eq.(\ref{eq:var_coeff_expanded}).
     \begin{itemize}

        \item The first term depends on the  parameterized part of the first layer $W^{(1)}$.
        \begin{equation}
             X_1(I,I',J,J') = \frac{1}{d(d+1)}\left(\delta_{j_1}^{j'_1} \delta_{i_1}^{i'_1}+\delta_{i_1}^{j_1} \delta_{i'_1}^{j'_1}\right)
        \end{equation}

        \item The last term depends on the  parameterized part  $W^{(L+1)}$ and the observable $O$.
        \begin{equation}
            \begin{aligned}
                 X_{L+1}(I,I',J,J') &=
                \delta_{j_L}^{j'_L} \delta_{i_L}^{i'_L} \left(\frac{d\Tr(O)^2-||O||^2}{d(d^2-1)}\right) +  \delta_{i_L}^{j_L} \delta_{i'_L}^{j'_L} \left(\frac{d||O||^2-\Tr(O)^2}{d(d^2-1)}\right)\\
                &= C_1 \delta_{j_L}^{j'_L} \delta_{i_L}^{i'_L} + C_2 \delta_{i_L}^{j_L} \delta_{i'_L}^{j'_L}
            \end{aligned}
        \end{equation}

        where $C_1 = \frac{d\Tr(O)^2-||O||^2}{d(d^2-1)}$ and $C_2=\frac{d||O||^2-\Tr(O)^2}{d(d^2-1)}$.
        
        \item For the remaining layers, we get $ \forall l \notin \{1,L+1\} $
        \begin{align*}
        \begin{split}
                  X_l(I,I',J,J')&= \frac{\delta_{j_l}^{j'_l} \delta_{i_l}^{i'_l} \delta_{j_{l-1}}^{j'_{l-1}} \delta_{i_{l-1}}^{i'_{l-1}} + \delta_{i_l}^{j_l} \delta_{i'_l}^{j'_l} \delta_{i_{l-1}}^{j_{l-1}} \delta_{i'_{l-1}}^{j'_{l-1}}}{d^2-1}  - \frac{\delta_{j_l}^{j'_l} \delta_{i_l}^{i'_l} \delta_{i_{l-1}}^{j_{l-1}} \delta_{i'_{l-1}}^{j'_{l-1}} + \delta_{i_l}^{j_l} \delta_{i'_l}^{j'_l} \delta_{j_{l-1}}^{j'_{l-1}} \delta_{i_{l-1}}^{i'_{l-1}} }{d(d^2-1)}
        \end{split}
        \end{align*}
    \end{itemize}

We denote by $c_{\omega}^L$ the Fourier coefficient if we consider the full circuit up to the $L$ layer and we establish in what follows a recursive relation between $\E[|c_{\omega}^L|^2]$ and $\E[|c_{\omega}^{L-1}|^2]$ where $c_{\omega}^{L-1}$ is a Fourier coefficient of the quantum model generated by the first $L-1$ layers.

We begin by recalling that the expectation value of the modulus squared of  the Fourier coefficients $c_\omega^L$ and $c_\omega^{L-1}$  can be written as 
\begin{align}
            \E_{W^{(1)},\dots,W^{(L+1)} \sim U(N)} \left[|c_{\omega}^{L}|^2\right] & =  \sum_{\substack{J,J' \in R_1^{L}(\omega) \\ I,I' \in R_1^{L}(\omega)}}  
            \left(\prod_{l=1}^{L-1} X_l(J,J',I,I')\right) X_L(J,J',I,I') \left( C_1 \delta_{j_L}^{j'_L} \delta_{i_L}^{i'_L} + C_2 \delta_{i_L}^{j_L} \delta_{i'_L}^{j'_L}\right)\label{eq:cL}\\
        \E_{W^{(1)},\dots,W^{L} \sim U(N)} \left[|c_{\omega}^{L-1}|^2\right] &= \sum_{\substack{J,J' \in R_1^{L-1}(\omega) \\ I,I' \in R_1^{L-1}(\omega)}}  \left(\prod_{l=1}^{L-1} X_l(J,J',I,I')\right) (C_1 \delta_{j_{L-1}}^{j'_{L-1}} \delta_{i_{L-1}}^{i'_{L-1}} + C_2 \delta_{i_{L-1}}^{j_{L-1}} \delta_{i'_{L-1}}^{j'_{L-1}})
    \end{align}

    Now, we exploit the recursive relation between the frequency generators up to layer $L-1$ and layer $L$ derived in Lemma \ref{lemma:recursive_spec} and expand the term $X_L(J,J',I,I') ( C_1 \delta_{j_L}^{j'_L} \delta_{i_L}^{i'_L} + C_2 \delta_{i_L}^{j_L} \delta_{i'_L}^{j'_L})$ in Eq.\eqref{eq:cL} to make appear the expression of $\E_{W^{(1)},\dots,W^{(L)} \sim U(N)} |c_{\omega}^{L-1}|^2$. For ease of notation, we introduce the shorthand 
    $\widetilde{X}_{L-1}:= \prod_{l=1}^{L-1} X_l(J,J',I,I')$ and drop the explicit dependence on the multi-indices $J,J',I,I'$. Moreover, we introduce the shorthands $A_1^L(\omega) $ and $A_2^L(\omega)$ obtained as follows.

        \begin{align}
            \E  \left[|c_{\omega}^{L}|^2\right]  &=  \sum_{\substack{J,J' \in R_1^{L}(\omega) \\ I,I' \in R_1^{L}(\omega)}}  
            \widetilde{X}_{L-1} \E \left[  
                W^{(L)}_{j_L j_{L-1}} W^{(L)}_{i'_L i'_{L-1}} W^{(L)*}_{j'_L j'_{L-1}} W^{(L)*}_{i_L i_{L-1}} \right] 
                (C_1 \delta_{j_L}^{j'_L} \delta_{i_L}^{i'_L} + C_2 \delta_{i_L}^{j_L} \delta_{i'_L}^{j'_L}) \\
                 & = C_1 A_1^L(\omega) + C_2 A_2^L(\omega)
                 \end{align}
                 where we define
                 \begin{align}
        A_1^L(\omega) &=  \sum_{\substack{J,J' \in R_1^{L}(\omega) \\ I,I' \in R_1^{L}(\omega)}} \widetilde{X}_{L-1} \delta_{j_L}^{j'_L} \delta_{i_L}^{i'_L}  \E \left[  
                W^{(L)}_{j_L j_{L-1}} W^{(L)}_{i'_L i'_{L-1}} W^{(L)*}_{j'_L j'_{L-1}} W^{(L)*}_{i_L i_{L-1}} \right] \\
        A_2^L(\omega) &=  \sum_{\substack{J,J' \in R_1^{L}(\omega) \\ I,I' \in R_1^{L}(\omega)}} \widetilde{X}_{L-1} \delta_{i_L}^{j_L} \delta_{i'_L}^{j'_L}  \E \left[  
                    W^{(L)}_{j_L j_{L-1}} W^{(L)}_{i'_L i'_{L-1}} W^{(L)*}_{j'_L j'_{L-1}} W^{(L)*}_{i_L i_{L-1}} \right]
    \end{align}

    In what follows, we focus on deriving recursive relations for $A_1^L(\omega) $ and $A_2^L(\omega)$ separately. Hence, we start with the first step of the recursion and compute both quantities for $L=1$.

We first show that $A_1^1(\omega) = \delta_{\omega}^0$.
     \begin{align}
         A_1^1(\omega) & =  \sum_{\substack{j_1,j'_1 \in R_1(\omega) \\ i_1,i'_1 \in R_1(\omega)}}  \E \left[  
                W^{(1)}_{j_1 1} W^{(1)}_{i'_1 1} W^{(1)*}_{j'_1 1} W^{(1)*}_{i_1 1} \right] \delta_{j_1}^{j'_1} \delta_{i_1}^{i'_1}\\
                &= \sum_{\substack{j_1,j'_1 \in R_1(\omega) \\ i_1,i'_1 \in R_1(\omega)}}  \left(\frac{ \delta_{j_{1}}^{j'_{1}} \delta_{i_{1}}^{i'_{1}} +  \delta_{i_{1}}^{j_{1}} \delta_{i'_{1}}^{j'_{1}}}{d(d+1)}\right) \delta_{j_1}^{j'_1} \delta_{i_1}^{i'_1}\\
                & = \sum_{\substack{j_1,j'_1 \in R_1(\omega) \\ i_1,i'_1 \in R_1(\omega)}} \frac{ \delta_{j_{1}}^{j'_{1}} \delta_{i_{1}}^{i'_{1}}}{d(d+1)} + \sum_{\substack{j_1,j'_1 \in R_1(\omega) \\ i_1,i'_1 \in R_1(\omega)}}  \frac{ \delta_{j_{1}}^{j'_{1}} \delta_{i_{1}}^{i'_{1}}  \delta_{i_{1}}^{j_{1}} \delta_{i'_{1}}^{j'_{1}}}{d(d+1)}\\
                &= \sum_{\substack{j_1,j_1 \in R_1(\omega) \\ i_1,i_1 \in R_1(\omega)}} \frac{ 1}{d(d+1)} + \sum_{\substack{j_1,j_1 \in R_1(\omega)}}  \frac{1}{d(d+1)}\\
                & = \frac{d^2}{d(d+1)} \delta_{\omega}^0 + \frac{d}{d(d+1)} \delta_{\omega}^0 = \delta_{\omega}^0 
                \end{align}
For $A_2^1(\omega)$, we obtain
        \begin{align}
           A_2^1(\omega)  =  \sum_{\substack{j_1,j'_1 \in R_1(\omega) \\ i_1,i'_1 \in R_1(\omega)}}  \E \left[ 
                W^{(1)}_{j_1 1} W^{(1)}_{i'_1 1} W^{(1)*}_{j'_1 1} W^{(1)*}_{i_1 1} \right] \delta_{i_1}^{j_1} \delta_{i'_1}^{j'_1}
                = \sum_{\substack{j_1,j'_1 \in R_1(\omega) \\ i_1,i'_1 \in R_1(\omega)}}  \frac{ \delta_{j_{1}}^{j'_{1}} \delta_{i_{1}}^{i'_{1}} +  \delta_{i_{1}}^{j_{1}} \delta_{i'_{1}}^{j'_{1}}}{d(d+1)} \delta_{i_1}^{j_1} \delta_{i'_1}^{j'_1}  = \frac{\delta_{\omega}^0}{d+1}  + \frac{|R_1(\omega)|}{d(d+1)}
     \end{align}

The recursive relation between $A_1^L(\omega) $ and $A_1^{L-1}(\omega) $ is obtained as follows.
\begin{equation}
     A_1^L(\omega)  =  \sum_{\substack{J,J' \in R_1^{L}(\omega) \\ I,I' \in R_1^{L}(\omega)}} \widetilde{X}_{L-1} \delta_{j_L}^{j'_L} \delta_{i_L}^{i'_L}  \E \left[  
                W^{(L)}_{j_L j_{L-1}} W^{(L)}_{i'_L i'_{L-1}} W^{(L)*}_{j'_L j'_{L-1}} W^{(L)*}_{i_L i_{L-1}} \right]
\end{equation}
Now, we invoke the recursive relation between $R_1^L$ and $R_1^{L-1}$ established in Lemma \ref{lemma:recursive_spec}.
 \begin{equation}
     \begin{aligned}
         A_1^L(\omega) 
        & = \smashoperator{\sum_{\substack{j_L, j'_L = 1 \\ i_L, i'_L = 1 \\ J,J' \in R_1^{L-1}(\omega-k_L) \times (j_L, j'_L) \\ I,I' \in R_1^{L-1}(\omega-k_L) \times (i_L, i'_L)}}^d} \widetilde{X}_{L-1} \delta_{j_L}^{j'_L} \delta_{i_L}^{i'_L}
             \left(\frac{\delta_{j_l}^{j'_l} \delta_{i_l}^{i'_l} \delta_{j_{l-1}}^{j'_{l-1}} \delta_{i_{l-1}}^{i'_{l-1}} + \delta_{i_l}^{j_l} \delta_{i'_l}^{j'_l} \delta_{i_{l-1}}^{j_{l-1}} \delta_{i'_{l-1}}^{j'_{l-1}}}{d^2-1}
             - \frac{\delta_{j_l}^{j'_l} \delta_{i_l}^{i'_l} \delta_{i_{l-1}}^{j_{l-1}} \delta_{i'_{l-1}}^{j'_{l-1}} + \delta_{i_l}^{j_l} \delta_{i'_l}^{j'_l} \delta_{j_{l-1}}^{j'_{l-1}} \delta_{i_{l-1}}^{i'_{l-1}} }{d(d^2-1)}\right)
        \\
        & = \smashoperator{\sum_{\substack{j_L =1\\i_L=1}}^d \sum_{\substack{J,J' \in R_1^{L-1}(\omega-0) \times (j_L, j_L) \\ I,I' \in R_1^{L-1}(\omega-0) \times (i_L, i_L)}}}  \widetilde{X}_{L-1}\left(\frac{ \delta_{j_{l-1}}^{j'_{l-1}} \delta_{i_{l-1}}^{i'_{l-1}} + \delta_{i_l}^{j_l} \delta_{i'_l}^{j'_l} \delta_{i_{l-1}}^{j_{l-1}} \delta_{i'_{l-1}}^{j'_{l-1}}}{d^2-1}  - \frac{ \delta_{i_{l-1}}^{j_{l-1}} \delta_{i'_{l-1}}^{j'_{l-1}} + \delta_{i_l}^{j_l} \delta_{i'_l}^{j'_l} \delta_{j_{l-1}}^{j'_{l-1}} \delta_{i_{l-1}}^{i'_{l-1}} }{d(d^2-1)}\right) \\
        & =  \sum_{j_L =1}^d \sum_{i_L=j_L}\sum_{\substack{J,J' \in R_1^{L-1}(\omega-0) \times (j_L, j_L) \\ I,I' \in R_1^{L-1}(\omega-0) \times (j_L, j_L)}}  \widetilde{X}_{L-1}\left(\frac{ \delta_{j_{l-1}}^{j'_{l-1}} \delta_{i_{l-1}}^{i'_{l-1}} +  \delta_{i_{l-1}}^{j_{l-1}} \delta_{i'_{l-1}}^{j'_{l-1}}}{d^2-1}  - \frac{ \delta_{i_{l-1}}^{j_{l-1}} \delta_{i'_{l-1}}^{j'_{l-1}} +  \delta_{j_{l-1}}^{j'_{l-1}} \delta_{i_{l-1}}^{i'_{l-1}} }{d(d^2-1)}\right) \\
        & \quad + \sum_{j_L =1}^d \sum_{i_L \neq j_L}\sum_{\substack{J,J' \in R_1^{L-1}(\omega-0) \times (j_L, j_L) \\ I,I' \in R_1^{L-1}(\omega-0) \times (i_L, i_L)}}  \widetilde{X}_{L-1}\left(\frac{ \delta_{j_{l-1}}^{j'_{l-1}} \delta_{i_{l-1}}^{i'_{l-1}} }{d^2-1}  - \frac{ \delta_{i_{l-1}}^{j_{l-1}} \delta_{i'_{l-1}}^{j'_{l-1}} }{d(d^2-1)}\right) \\
        & =  \sum_{j_L =1}^d \sum_{\substack{J,J' \in R_1^{L-1}(\omega-0) \times (j_L, j_L) \\ I,I' \in R_1^{L-1}(\omega-0) \times (j_L, j_L)}}  \widetilde{X}_{L-1}\left(\frac{ \delta_{j_{l-1}}^{j'_{l-1}} \delta_{i_{l-1}}^{i'_{l-1}} +  \delta_{i_{l-1}}^{j_{l-1}} \delta_{i'_{l-1}}^{j'_{l-1}}}{d(d+1)} \right)\\
        & \hspace{10pt}  +  \sum_{j_L =1}^d \sum_{i_L \neq j_L}\sum_{\substack{J,J' \in R_1^{L-1}(\omega-0) \times (j_L, j_L) \\ I,I' \in R_1^{L-1}(\omega-0) \times (i_L, i_L)}}  \widetilde{X}_{L-1}\left(\frac{ \delta_{j_{l-1}}^{j'_{l-1}} \delta_{i_{l-1}}^{i'_{l-1}} }{d^2-1}  - \frac{ \delta_{i_{l-1}}^{j_{l-1}} \delta_{i'_{l-1}}^{j'_{l-1}} }{d(d^2-1)}\right) \\
        & = d \sum_{\substack{J,J' \in R_1^{L-1}(\omega-0)  \\ I,I' \in R_1^{L-1}(\omega-0)}} \widetilde{X}_{L-1}\left(\frac{ \delta_{j_{l-1}}^{j'_{l-1}} \delta_{i_{l-1}}^{i'_{l-1}} +  \delta_{i_{l-1}}^{j_{l-1}} \delta_{i'_{l-1}}^{j'_{l-1}}}{d(d+1)} \right) + d(d-1)\sum_{\substack{J,J' \in R_1^{L-1}(\omega-0)  \\ I,I' \in R_1^{L-1}(\omega-0) }}  \widetilde{X}_{L-1}\left(\frac{ \delta_{j_{l-1}}^{j'_{l-1}} \delta_{i_{l-1}}^{i'_{l-1}} }{d^2-1}  - \frac{ \delta_{i_{l-1}}^{j_{l-1}} \delta_{i'_{l-1}}^{j'_{l-1}} }{d(d^2-1)}\right)\\
        & = \frac{1}{d+1} \left[A_1^{L-1}(\omega)+ A_2^{L-1}(\omega)\right] + \frac{d}{d+1} A_1^{L-1}(\omega)-\frac{1}{d+1}A_2^{L-1}(\omega) \\
        & = A_1^{L-1}(\omega)
     \end{aligned}
 \end{equation}
 
 We showed that $\forall \omega \in \Omega , \forall L \geq 2 , A_1^{L}(\omega) = A_1^{L-1}(\omega)$. Moreover, we have $A_1^1(\omega) = \delta_{\omega}^0$. Hence, we obtain
 \begin{equation}\label{eq:final_A1L}
     \forall \omega \in \Omega,  \forall L \geq 1 , A_1^L(\omega) = A_1^1(\omega) = \delta_{\omega}^0\;.
 \end{equation}
 This implies that apart from the frequency zero, the term which contributes to the variance expression of a Fourier coefficient is $ A_2^{L}(\omega)$.
 
 For the recursive relation between $A_2^L(\omega)$ and $A_2^{L-1}(\omega)$, we get the following:
 \begin{equation}
     \begin{split}
         A_2^L(\omega)& =   \sum_{\substack{J,J' \in R_1^{L}(\omega) \\ I,I' \in R_1^{L}(\omega)}} \widetilde{X}_{L-1} \delta_{i_L}^{j_L} \delta_{i'_L}^{j'_L}  \E \left[  
                W^{(L)}_{j_L j_{L-1}} W^{(L)}_{i'_L i'_{L-1}} W^{(L)*}_{j'_L j'_{L-1}} W^{(L)*}_{i_L i_{L-1}} \right]\\
                &=\smashoperator{\sum_{\substack{i_L, i'_L = 1 \\ j_L, j'_L = 1\\I,I' \in R_1^{L-1}(\omega-k_L) \times (i_L, i'_L) \\ J,J' \in R_1^{L-1}(\omega-k_L) \times (j_L, j'_L)}}}  \widetilde{X}_{L-1} \delta_{i_L}^{j_L} \delta_{i'_L}^{j'_L}
        \left(
            \frac{\delta_{j_l}^{j'_l} \delta_{i_l}^{i'_l} \delta_{j_{l-1}}^{j'_{l-1}} \delta_{i_{l-1}}^{i'_{l-1}} + \delta_{i_l}^{j_l} \delta_{i'_l}^{j'_l} \delta_{i_{l-1}}^{j_{l-1}} \delta_{i'_{l-1}}^{j'_{l-1}}}{d^2-1} 
             - \frac{\delta_{j_l}^{j'_l} \delta_{i_l}^{i'_l} \delta_{i_{l-1}}^{j_{l-1}} \delta_{i'_{l-1}}^{j'_{l-1}} + \delta_{i_l}^{j_l} \delta_{i'_l}^{j'_l} \delta_{j_{l-1}}^{j'_{l-1}} \delta_{i_{l-1}}^{i'_{l-1}} }{d(d^2-1)}\right)
        \\
        & = \sum_{j_L, j'_L =1}^d \sum_{\substack{J,J' \in R_1^{L-1}(\omega-k_L) \times (j_L, j'_L)}}  \widetilde{X}_{L-1}\left(\frac{ \delta_{j_l}^{j'_l} \delta_{j_{l-1}}^{j'_{l-1}} \delta_{i_{l-1}}^{i'_{l-1}} +  \delta_{i_{l-1}}^{j_{l-1}} \delta_{i'_{l-1}}^{j'_{l-1}}}{d^2-1}  - \frac{  \delta_{j_l}^{j'_l} \delta_{i_{l-1}}^{j_{l-1}} \delta_{i'_{l-1}}^{j'_{l-1}} +  \delta_{j_{l-1}}^{j'_{l-1}} \delta_{i_{l-1}}^{i'_{l-1}} }{d(d^2-1)}\right) \\
        & =  \sum_{j_L =1}^d \sum_{j'_L=j_L}\sum_{\substack{J,J' \in R_1^{L-1}(\omega-k_L) \times (j_L, j_L)}}  \widetilde{X}_{L-1}\left(\frac{ \delta_{j_{l-1}}^{j'_{l-1}} \delta_{i_{l-1}}^{i'_{l-1}} +  \delta_{i_{l-1}}^{j_{l-1}} \delta_{i'_{l-1}}^{j'_{l-1}}}{d^2-1}  - \frac{ \delta_{i_{l-1}}^{j_{l-1}} \delta_{i'_{l-1}}^{j'_{l-1}} +  \delta_{j_{l-1}}^{j'_{l-1}} \delta_{i_{l-1}}^{i'_{l-1}} }{d(d^2-1)}\right) \\
        & \quad +  \sum_{j_L =1}^d \sum_{j'_L \neq j_L}\sum_{\substack{J,J' \in R_1^{L-1}(\omega-k_L) \times (j_L, j'_L) }}  \widetilde{X}_{L-1}\left(\frac{ \delta_{i_{l-1}}^{j_{l-1}} \delta_{i'_{l-1}}^{j'_{l-1}} }{d^2-1}  - \frac{ \delta_{j_{l-1}}^{j'_{l-1}} \delta_{i_{l-1}}^{i'_{l-1}} }{d(d^2-1)}\right) \\
        & = \sum_{j_L =1}^d \sum_{j'_L=j_L}\sum_{\substack{J,J' \in R_1^{L-1}(\omega-0) \times (j_L, j_L)}}  \widetilde{X}_{L-1}\left(\frac{ \delta_{j_{l-1}}^{j'_{l-1}} \delta_{i_{l-1}}^{i'_{l-1}} +  \delta_{i_{l-1}}^{j_{l-1}} \delta_{i'_{l-1}}^{j'_{l-1}}}{d(d+1)} \right) \\
        & \quad +  \sum_{j_L =1}^d \sum_{j'_L \neq j_L}\sum_{\substack{J,J' \in R_1^{L-1}(\omega-k_L) \times (j_L, j'_L) }}  \widetilde{X}_{L-1}\left(\frac{ \delta_{i_{l-1}}^{j_{l-1}} \delta_{i'_{l-1}}^{j'_{l-1}} }{d^2-1}  - \frac{ \delta_{j_{l-1}}^{j'_{l-1}} \delta_{i_{l-1}}^{i'_{l-1}} }{d(d^2-1)}\right) \\
        & = d \sum_{\substack{J,J' \in R_1^{L-1}(\omega-0) }}\widetilde{X}_{L-1}\left(\frac{ \delta_{j_{l-1}}^{j'_{l-1}} \delta_{i_{l-1}}^{i'_{l-1}} +  \delta_{i_{l-1}}^{j_{l-1}} \delta_{i'_{l-1}}^{j'_{l-1}}}{d(d+1)} \right) +  \sum_{\substack{ {j_L,j'_L} \\ {j'_L \neq j_L}}} \sum_{\substack{J,J' \in R_1^{L-1}(\omega-k_L)  }}  \widetilde{X}_{L-1}\left(\frac{ \delta_{i_{l-1}}^{j_{l-1}} \delta_{i'_{l-1}}^{j'_{l-1}} }{d^2-1}  - \frac{ \delta_{j_{l-1}}^{j'_{l-1}} \delta_{i_{l-1}}^{i'_{l-1}} }{d(d^2-1)}\right)\\
        & = \frac{1}{d+1} \left[A_1^{L-1}(\omega)+ A_2^{L-1}(\omega)\right] + \sum_{\substack{ {j_L,j'_L} \\ {j'_L \neq j_L}}} \left[\frac{1}{d^2-1} A_2^{L-1}(\omega - k_L) - \frac{1}{d(d^2-1)} A_1^{L-1}(\omega - k_L)\right]\\
        & = \frac{1}{d+1} \left[\delta_{\omega}^0+ A_2^{L-1}(\omega)\right] + \sum_{\substack{ {j_L,j'_L} \\ {j'_L \neq j_L}}} \frac{A_2^{L-1}(\omega - k_L)}{d^2-1}  - \frac{|R_L(\omega)|-d \delta_{\omega}^0}{d(d^2-1)}\\
        &= \frac{d-1}{d^2-1} A_2^{L-1}(\omega) + \sum_{\substack{ {j_L,j'_L} \\ {j'_L \neq j_L}}} \frac{A_2^{L-1}(\omega - k_L)}{d^2-1} + \frac{d \delta_{\omega}^0}{d^2-1} - \frac{|R_L(\omega)|}{d(d^2-1)}  \\
        & = \sum_{j_L, j'_L} \frac{A_2^{L-1}(\omega - k_L)}{d^2-1} -\frac{A_2^{L-1}(\omega)}{d^2-1} + \frac{d \delta_{\omega}^0}{d^2-1} - \frac{|R_L(\omega)|}{d(d^2-1)} \\
        & = \sum_{k_L \in \Omega^L_{distinct}} \frac{|R_L(k_L)|- \delta_{k_L}^0}{d^2-1}  A_2^{L-1}(\omega - k_L)  + \frac{d \delta_{\omega}^0}{d^2-1} - \frac{|R_L(\omega)|}{d(d^2-1)}
     \end{split}
 \end{equation}

 Thus, the final recursion formula for $A_2^L(\omega) $ is the following:
 \begin{align*}
     \forall \omega , \forall L \geq 2 , A_2^L(\omega) &= \sum_{k_L \in \Omega^L_{distinct}} \frac{|R_L(k_L)|- \delta_{k_L}^0}{d^2-1}  A_2^{L-1}(\omega - k_L)  + \frac{d \delta_{\omega}^0}{d^2-1} - \frac{|R_L(\omega)|}{d(d^2-1)}\\
     &= \sum_{k_L \in \Omega^L_{distinct}} \alpha_{k_L} A_2^{L-1}(\omega - k_L)  + \frac{d \delta_{\omega}^0}{d^2-1} - \frac{|R_L(\omega)|}{d(d^2-1)}
 \end{align*}

 We iterate through $L$ to find the expression of $A_2^L(\omega)$ :
\begin{equation}\label{eq:A2L}
\begin{aligned}
    A_2^L(\omega) &=  \sum_{k_L, \dots, k_2} \alpha_{k_L}\dots \alpha_{k_2} \left[\frac{\delta_{\omega - \sum_{l=2}^L k_l}^{0}}{d+1} + \frac{|R_1(\omega - \sum_{l=2}^L k_l)|}{d(d+1)}\right]\\
    &+ \sum_{k_L, \dots, k_3} \alpha_{k_L}\dots \alpha_{k_3} \left[\frac{d \delta_{\omega - \sum_{l=3}^L k_l}^0}{d^2-1} - \frac{|R_2(\omega - \sum_{l=3}^L k_l)|} {d(d^2-1)}\right]\\
    &+ \dots + \sum_{k_L, \dots, k_h} \alpha_{k_L}\dots \alpha_{k_h} \left[\frac{d \delta_{\omega - \sum_{l=h}^L k_l}^0}{d^2-1} - \frac{|R_{h-1}(\omega - \sum_{l=h}^L k_l)|} {d(d^2-1)}\right] + \dots\\
    &+ \sum_{k_L} \alpha_{k_L} \left[\frac{d \delta_{\omega - k_L}^0}{d^2-1} - \frac{|R_{L-1}(\omega - k_L)|}{d(d^2-1)}\right]
    + \frac{d \delta_{\omega}^0}{d^2-1} - \frac{|R_L(\omega)|}{d(d^2-1)}
    \end{aligned}
\end{equation}
We have $\alpha_{k_l} = \frac{(|R_l(k_l)|- \delta_{k_l}^0)}{d^2-1} =  \frac{(\beta_{k_l}- \delta_{k_l}^0)}{d^2-1} \simeq \frac{\beta_{k_l}}{d^2-1} $ so we can replace $\alpha_{k_l}$ with $\frac{\beta_{k_l}}{d^2-1}$ in the above expression since $\beta_0 \geq d$.

In this step, we use the expression of the recursive redundancy relation in Lemma \ref{lemma:recursive_spec} to simplify the generic term in the summation of Eq.\eqref{eq:A2L}:
\begin{equation}\label{eq:redundancy_A2L}
\begin{aligned}
    \frac{1}{(d^2-1)^{L-h+1}}   \sum_{k_L, \dots, k_h} \beta_{k_L} & \dots \beta_{k_h} \left[\frac{d \delta_{\omega - \sum_{l=h}^L k_l}^0}{d^2-1} - \frac{|R_{h-1}(\omega - \sum_{l=h}^L k_l)|}{d(d^2-1)}\right] \\ & = \frac{d}{(d^2-1)^{L-h+2}} |R_h^L(\omega)| - \frac{1}{d(d^2-1)^{L-h+2}} |R_{h-1}^L(\omega)|
\end{aligned}
\end{equation}

Finally, we inject Eq.\eqref{eq:redundancy_A2L} in Eq.\eqref{eq:A2L} and get:
\begin{equation}\label{eq:final_A2L}
\begin{aligned}
     A_2^L(\omega) &= \frac{1}{(d^2-1)^{L-1}}  \sum_{k_L, \dots, k_2} \beta_{k_L}\dots \beta_{k_2} \left[\frac{\delta_{\omega - \sum_{l=2}^L k_l}^{0}}{d+1} + \frac{|R_1(\omega - \sum_{h=2}^L k_l)|}{d(d+1)}\right]\\
     &+ \sum_{h=3}^L \left[\frac{d}{(d^2-1)^{L-h+2}} |R_h^L(\omega)| - \frac{1}{d(d^2-1)^{L-h+2}} |R_{h-1}^L(\omega)|\right]\\
    &+ \frac{d \delta_{\omega}^0}{d^2-1} - \frac{|R_L(\omega)|}{d(d^2-1)}\\
    &= \frac{1}{d(d+1)(d^2-1)^{L-1}} |R_1^L(\omega)| + \left(\frac{1}{(d+1)(d^2-1)^{L-1}} - \frac{1}{d(d^2-1)^{L-1}}\right) |R_2^L(\omega)|\\
    &+ \sum_{h=3}^{L-1} \left(\frac{d}{(d^2-1)^{L-h+2}}-\frac{1}{d(d^2-1)^{L-h+1}}\right) |R_h^L(\omega)| + \left(\frac{d}{(d^2-1)^2} - \frac{1}{d(d^2-1)}\right) |R_L(\omega)| + \frac{d \delta_{\omega}^0}{d^2-1}\\
    &= \frac{1}{d(d+1)(d^2-1)^{L-1}} |R_1^L(\omega)| - \frac{1}{d(d+1)(d^2-1)^{L-1}} |R_2^{L}(\omega)|+ \sum_{h=3}^{L} \frac{1}{d(d^2-1)^{L-h+2}} |R_h^L(\omega)| +  \frac{d \delta_{\omega}^0}{d^2-1}
    \end{aligned}
\end{equation}

By combining Eq.\eqref{eq:final_A1L} and Eq.\eqref{eq:final_A2L}, we retrieve the result from Theorem  \ref{thm:formal_2design_reup}:
$$\E[|c_{\omega}^L|^2]=C_1 A_1^L(\omega) + C_2 A_2^L(\omega) = C_1 \delta_{\omega}^0 + C_2 A_2^L(\omega)$$

\end{proof}

%% file: Proofs/03_local_2design.tex
\subsection{Fourier coefficients variance with approximate 2-design trainable unitaries}\label{Proof:approx_2design}

We present the formal version of Theorem \ref{thm:bound_approx_2design_informal}.

    \begin{theorem}[Fourier coefficients variance with approximate 2-design trainable unitaries, Formal]\label{thm:bound_approx_2design}
       Consider a quantum model of the form in Eq.\eqref{Eq:quantum_Model} and a parametrized circuit of the form in Eq.\eqref{eq:circuit_ansatz} with $L=1$ layers and fixed encoding Hamiltonians resulting in a spectrum $\Omega$. We assume that each of the  trainable layers $W^l(\theta)\;,l \in \{1,2\}$ form independently an $\varepsilon$-approximate 2-design. The  variance of each Fourier coefficient $c_{\omega}(\theta)$ for the frequencies $ \omega \in \Omega$ appearing in the model Fourier decomposition in Eq.\eqref{Eq:quantum_Fourier_Model} is upper bounded as
    \begin{equation}
     \Var_{\theta}[c_{\omega}(\theta)] \leq \Var_{\text{2-design}}[c_{\omega}(\theta)]+Q( \varepsilon,|R(\omega)|)
   \end{equation}
   with $\Var_{\text{2-design}}[c_{\omega}]$  the variance of a Fourier coefficient under the 2-design assumption given in Theorem \ref{thm:single_layer_gloabl_2design} and $Q$ is a polynomial in $\varepsilon$ and the frequency redundancy $|R(\omega)|$. 
   Moreover, the polynomial $Q$ is defined as follows for different measures of approximate 2-design

   \begin{enumerate}
       \item For diamond norm-based approximate 2-design, we have
       \begin{equation}\label{eq:diamondnorm}
       \begin{aligned}
    Q(\varepsilon_{\diamond},|R(\omega)|)&= C_1 \varepsilon_{\diamond} + \|O\|_1^2 \varepsilon_{\diamond}^2\\
    &\quad + \frac{\|O\|_{\infty}^2}{d(d+1)} |R(\omega)| \varepsilon_{\diamond}
    \end{aligned}
       \end{equation}
       \item For monomial-based approximate 2-design, we have
       \begin{equation}\label{eq:monomialNorm}
       \begin{aligned}
           Q(\varepsilon_{M},|R(\omega)|)&= \left( \frac{C_1 }{d^2} + \frac{C_2}{d(d+1)}\right)\varepsilon_M|R(\omega)|\\
           &\quad + \frac{C_2}{d^2}(\varepsilon_M|R(\omega)|)^2
           \end{aligned}
       \end{equation}
       \item For spectral norm-based approximate 2-design, we have
       \begin{equation}\label{eq:infinityNorm}
       \begin{aligned}
           Q(\varepsilon_{\infty},|R(\omega)|)&= \left( C_1+ \frac{\|O\|_2^2}{d(d+1)}\right)\varepsilon_{\infty}\sqrt{|R(\omega)|}\\
           &\quad + \|O\|_2^2 \varepsilon_{\infty}^2|R(\omega)|
           \end{aligned}
       \end{equation}
   \end{enumerate}
   

\noindent Here, we use the shorthand $\varepsilon_{\diamond}:= \|\mathcal{A}^{(2)}\|_{\diamond}$ for the diamond norm defined in Eq.\eqref{eq:diamond_norm}, $\varepsilon_{\infty} :=\norm{O}_{\infty}$ for the spectral norm and $\varepsilon_{M}:= d^2 max_{i,j} |\mathcal{A}^{(2)}|_{i,j}$.  The constants $C_1$ and $C_2$ are defined as follows $C_1 = \frac{d||O||^2-Tr(O)^2}{d(d^2-1)},C_2=\sum_{l,k} \frac{|[O^{\otimes{2}}]_{l,k}|}{d^2} $.
\end{theorem}

Here we provide an upper bound on the variance of each Fourier coefficient in the setting of a single layered QFM when each of the two parameterized layers form an $\varepsilon$-approximate 2 design according to the diamond norm, spectral norm and the monomial definition introduced in Definition \ref{def:monomial_norm}.

We note that if we assume that the parameterized layers form each a 1-design, then the expectation value of each Fourier coefficient is zero except for the null frequency as mentioned in \ref{Proof:Global_2design}.

The main takeaway from this result is that the bound is polynomial in the frequency redundancy. This shows that the degree to which a Fourier coefficient concentrates around its mean (which is 0 under the 1-design hypothesis for non null frequencies) is restricted by its frequency in the spectrum construction, thus introducing an encoding dependent \emph{inductive bias}. The proof of this Theorem follows a similar proof to the one given in \cite{holmes_connecting_2022} where the authors used the same techniques to compute a bound on the variance of the quantum model's gradient.
\begin{proof}   
Let us first recall the Fourier coefficient expression for a single encoding layer ($L=1$ in Eq.\eqref{eq:Fourier_coeff_appendix}):

\begin{equation}\label{eq:Fourier_coeff_L1}
\begin{aligned}
    c_{\omega} &= \sum_{j_1,j'_1 \in R(\omega)} \sum_{k,k'} W^{(1)*}_{j'_1 0}  W^{(2)*}_{j'_2 j'_1}  O_{k'k}  W^{(2)}_{j_2 j_1} W^{(1)}_{j_1 0}\\
    &=  \sum_{j_1,j'_1 \in R(\omega)} \left( W^{(1)} \ket{0}\bra{0}  W^{(1) \dagger}\right)_{j_1,j'_1} \left( W^{(2) \dagger} O W^{(2)}\right)_{j'_1,j_1}\\
    &= \sum_{j_1,j'_1 \in R(\omega)} Tr \left[W^{(1)} \ket{0}\bra{0}  W^{(1) \dagger} \ket{j'_1}\bra{j_1}\right] Tr \left[ W^{(2) \dagger} O W^{(2)} \ket{j_1}\bra{j'_1}\right]
\end{aligned}
\end{equation}

Then, the expectation of the modulus squared of the coefficient $c_{\omega}$ is given by:

\begin{equation}\label{eq:exp_squared_fourier_L1}
\begin{aligned}
    \mathbb{E}\left[|c_\omega|^2\right]  
    &= \sumrw \Tr\left[\mathbb{E}_{W^{(1)}}\left[W^{(1)\otimes 2} \ket{00}\bra{00} W^{(1)\dagger \otimes 2} \ket{j'_1 i_1}\bra{j_1 i'_1}\right] \right] \Tr\left[\mathbb{E}_{W^{(2)}}\left[ W^{(2)\dagger\otimes2}O^{\otimes 2} W^{(2)\otimes2}\ket{j_1 i'_1}\bra{j'_1 i_1}\right]\right]
\end{aligned}
\end{equation}

where we use in the second equality the property $Tr[A] \times Tr[B]=Tr[A \otimes B]$.



We recall here that in Theorem \ref{thm:bound_approx_2design} we consider that both parameterized unitaries have the same $\varepsilon$-distance to a 2-design. However the following proof holds if each of the parameterized unitaries have different distributions with different distances to a 2-design.

Expectation terms in Eq.\eqref{eq:exp_squared_fourier_L1} can be written using the superoperator $\mathcal{A}_{\mathbb{W}}^{(2)}(\cdot):=  \int_{\text{Haar}} d \mu(W) W^{\otimes 2}(\cdot)\left(W^{\dagger}\right)^{\otimes 2}  -\int_{\mathbb{W}} d W W^{\otimes 2}(\cdot)\left(W^{\dagger}\right)^{\otimes 2}$. Moreover, we use the property $\int_{\mathbb{W}} d W W^{\otimes 2}(\cdot)\left(W^{\dagger}\right)^{\otimes 2} = \int_{\mathbb{W}} d W \left(W^{\dagger}\right)^{\otimes 2}(\cdot)W^{\otimes 2}$ \cite{mele_introduction_2023} and obtain:
\begin{equation}\label{eq:terms_forbound}
\begin{aligned}
    \Tr\left[\mathbb{E}_{W^{(1)}}\left[W^{(1)\otimes 2} \ket{00}\bra{00} W^{(1)\dagger \otimes 2} \ket{j'_1 i_1}\bra{j_1 i'_1}\right] \right] &= \mathbb{E}_{\text{Haar}} \left[ \Tr\left[W^{(1)\otimes 2} \ket{00}\bra{00} W^{(1)\dagger \otimes 2} \ket{j'_1 i_1}\bra{j_1 i'_1} \right] \right]\\& \quad - \Tr[\mathcal{A}_{\mathbb{W}}^{(2)}(\ket{00}\bra{00})\ket{j'_1i_1}\bra{j_1 i'_1}]\\
    &= \frac{\delta_{j_1,j'_1}\delta_{i_1,i'_1}(\delta_{\omega,0})+\delta_{i_1,j_1}\delta_{i'_1,j'_1}}{d(d+1)} - \Tr[\mathcal{A}_{\mathbb{W}}^{(2)}(\ket{00}\bra{00})\ket{j'_1i_1}\bra{j_1 i'_1}]\\
    \Tr[\mathbb{E}_{W^{(2)}}\left[W^{(2)\dagger\otimes2}O^{\otimes 2} W^{(2)\otimes2}\ket{j_1 i'_1}\bra{j'_1 i_1}\right]] &= 
    \mathbb{E}_{\text{Haar}} \left[\Tr[W^{(2)\dagger\otimes2}O^{\otimes 2} W^{(2)\otimes2}\ket{j_1 i'_1}\bra{j'_1 i_1}]\right]\\& \quad - \Tr\left[ \mathcal{A}_{\mathbb{W}}^{(2)}\left(O^{\otimes 2}\right)\ket{j_1,i'_1}\bra{j'_1,i_1}\right]\\
    &= K_1\delta_{j_1,j'_1}\delta_{i_1,i'_1} + K_2\delta_{i_1,j_1}\delta_{i'_1,j'_1} - \Tr\left[ \mathcal{A}_{\mathbb{W}}^{(2)}\left(O^{\otimes 2}\right)\ket{j_1,i'_1}\bra{j'_1,i_1}\right]\\
    &= K_1\delta_{j_1,j'_1}\delta_{i_1,i'_1} + K_2\delta_{i_1,j_1}\delta_{i'_1,j'_1} - \Tr\left[ \mathcal{A}_{\mathbb{W}}^{(2)}\left(\ket{j_1,i'_1}\bra{j'_1,i_1}\right)O^{\otimes 2}\right]
\end{aligned}
\end{equation}

The expectation with respect to the Haar measure in both terms is simply computed using the Weingarten calculus formula for the second moment monomials \cite{mele_introduction_2023} where

\begin{equation}\label{eq:constants_K1_K2}
    K_1 = \frac{d \Tr (O)^2-\|O\|_2^2}{d(d^2-1)} \quad \text{ and } \quad K_2 = \frac{d\|O\|^2-\Tr(O)^2}{d(d^2-1)}
\end{equation}

By substituting the terms of Eq.\eqref{eq:terms_forbound} in Eq.\eqref{eq:exp_squared_fourier_L1}, we get the following expression:
\begin{align}\label{eq:var_coeff_4terms}
    \begin{split}
\E \left[|c_\omega|^2\right] &= \E_{\text{Haar}} \left[|c_\omega|^2\right] - \sumrw  \left(\frac{\delta_{j_1,j'_1}\delta_{i_1,i'_1}(\delta_{\omega,0})+\delta_{i_1,j_1}\delta_{i'_1,j'_1}}{d(d+1)}\right) \Tr\left[\mathcal{A}^{\otimes2}(\ket{j_1,i'_1}\bra{j'_1,i_1})O^{\otimes2}\right]  \\
&\qquad- \sumrw \left(K_1\delta_{j_1,j'_1}\delta_{i_1,i'_1}(\delta_{\omega,0}) + K_2\delta_{i_1,j_1}\delta_{i'_1,j'_1} \right) \Tr \left[\mathcal{A}_{\mathbb{W}}^{(2)}(\ket{00}\bra{00})\ket{j'_1,i_1}\bra{j_1,i'_1}\right]  \\ 
&\qquad+ \sumrw \Tr \left[\mathcal{A}_{\mathbb{W}}^{(2)}(\ket{00}\bra{00})\ket{j'_1,i_1}\bra{j_1,i'_1}\right] \Tr\left[\mathcal{A}_{\mathbb{W}}^{(2)\otimes2}(\ket{j_1,i'_1}\bra{j'_1,i_1})O^{\otimes2}\right] 
    \end{split}
\end{align}

$\forall \omega \in \Omega \backslash \{0\}$, we can further simplify the expression in Eq.\eqref{eq:var_coeff_4terms},

\begin{align}\label{eq:var_4terms}
    \begin{split}
\E \left[|c_\omega|^2\right] &= \E_{\text{Haar}} \left[|c_\omega|^2\right] -   \frac{1}{d(d+1)} \Tr\left[\mathcal{A}_{\mathbb{W}}^{(2)}\left(\sum_{j_1,j'_1\in R(\omega)}\ket{j_1,j'_1}\bra{j'_1,j_1}\right)O^{\otimes2}\right]   \\
&\qquad-  K_2 \Tr \left[\mathcal{A}_{\mathbb{W}}^{(2)}\left(\ket{00}\bra{00}\right)\sum_{j_1,j'_1\in R(\omega)}\ket{j'_1,j_1}\bra{j_1,j'_1}\right] \\ 
&\qquad+   \Tr \left[\left(\mathcal{A}_{\mathbb{W}}^{(2)}(\ket{00}\bra{00}) \otimes \mathcal{A}_{\mathbb{W}}^{(2)}(O^{\otimes2}) \right) \left( \sumrw \ket{j'_1,i_1}\bra{j_1,i'_1} \otimes \ket{j_1,i'_1}\bra{j'_1,i_1}\right)\right] 
    \end{split}
\end{align}

Different norms of $\mathcal{A}_{\mathbb{W}}^{(2)}(X^{\otimes2})$ can be used to further obtain an upper bound on $\E\left[|c_\omega|^2\right]$. The norm choice leading to the tightest upper bound is highly dependent on the observable norm and the frequency redundancy. We list here the different operator norms and observable norms leading to similar upper bounds on the Fourier coefficients variance but with different scaling depending on the setting.

All the operators norms used in the following proof are Schatten p-norms. For completeness, we recall that the scahtten p-norm of an operator $\Omega$ is defined as follows: $\norm{\Omega}_p = \left[Tr\left(|\Omega|^p\right)\right]^{1/p} = \left[\sum_i \sigma_i^p\right]^{1/p}$ where $|\Omega| = \sqrt{\Omega^{\dagger}\Omega}$ and $\sigma_i$s are the singular values of $\Omega$.
We also recall that the  diamond norm of a Hermiticity preserving linear map $\mathcal{S}_A$ is defined as
\begin{equation}\label{eq:diamond_norm}
    \|\mathcal{S}_A\|_{\diamond} = \sup_n \sum_{\Omega_{AB} \neq 0} \frac{\|(\mathcal{S}_A \otimes \mathcal{I}_B^{(n)})(\Omega_{AB})\|}{\|\Omega_{AB}\|_1}
\end{equation}
where $\Omega_{AB} \in \mathcal{L}(\mathcal{H}_A \otimes \mathcal{H}_B)$. The diamond norm distance $\norm{\mathcal{M} - \mathcal{N}}_{\diamond}$ is a measure of the distinguishability of two quantum operations $\mathcal{M}$ and $\mathcal{N}$.

\paragraph{Diamond norm bound:}

We start by applying the triangular inequality on all terms in Eq.\eqref{eq:var_4terms} and obtain:
\begin{equation}
    \begin{aligned}
        \E \left[|c_\omega|^2\right] &\leq \E_{\text{Haar}} \left[|c_\omega|^2\right] +\frac{||O^{\otimes 2}||_{\infty}} {d(d+1)} \| \sum_{j_1,j'_1\in R(\omega)}\ket{j_1,j'_1}\bra{j'_1,j_1} \|_1 \varepsilon_{\diamond} \\
        & \qquad + K_2 ||\sum_{j_1,j'_1\in R(\omega)}\ket{j'_1,j_1}\bra{j_1,j'_1}||_{\infty} \varepsilon_{\diamond}\\
        & \qquad + ||\mathcal{A}_{\mathbb{W}}^{(2)}(\ket{00}\bra{00}) \otimes \mathcal{A}_{\mathbb{W}}^{(2)}(O^{\otimes2})||_1 ||\sumrw \ket{j'_1,i_1}\bra{j_1,i'_1} \otimes \ket{j_1,i'_1}\bra{j'_1,i_1}||_{\infty}\\
        &\leq \E_{\text{Haar}} \left[|c_\omega|^2\right]  +  \frac{||O||^2_{\infty}} {d(d+1)} |R(\omega)| \varepsilon_{\diamond} + K_2 \varepsilon_{\diamond} + ||O||_1^2 \varepsilon_{\diamond}^2 
    \end{aligned}
\end{equation}

where in the first inequality we used hölder's inequality $|Tr[A B]| \leq \|A\|_1 \|B\|_{\infty} $ and the property $\|\mathcal{A}_{\mathbb{W}}^{(2)}(X)\|_1 \leq \|\mathcal{A}_{\mathbb{W}}^{(2)}\|_{\diamond} \|X\|_1$ with the shorthand $\varepsilon_{\diamond} := \|\mathcal{A}_{\mathbb{W}}^{(2)}\|_{\diamond}$ and in the second inequality we used the property $\|A \otimes B\|= \|A\| \|B\|$.
We also used the following elementary calculations to show that the singular values of matrices of the form $\sum_{j_1,j'_1\in R(\omega)}\ket{j_1,j'_1}\bra{j'_1,j_1}$ are zeros and ones with 1s having multiplicity $|R(\omega)|$:
\begin{align}
    \left(\sum_{j_1,j'_1\in R(\omega)}\ket{j_1,j'_1}\bra{j'_1,j_1}\right)^{\dagger} \left(\sum_{i_1,i'_1\in R(\omega)}\ket{i_1,i'_1}\bra{i'_1,i_1}\right)  &= \sumrw \ket{j'_1,j_1}\bra{j_1,j'_1}  \ket{i_1,i'_1}\bra{i'_1,i_1} \\
    &= \sum_{j_1,j'_1\in R(\omega)} \ket{j'_1,j_1}\bra{j'_1,j_1}
\end{align}

We note that the bound on the last term can be slightly modified and its tightness will depend on the observable norm and the frequency redundancy. Specifically, we can obtain:
\begin{equation}
\begin{aligned}
    \E \left[|c_\omega|^2\right] &\leq \E_{\text{Haar}} \left[|c_\omega|^2\right] + \frac{||O||^2_{\infty}} {d(d+1)} |R(\omega)| \varepsilon_{\diamond} + K_2 \varepsilon_\diamond+   ||O||^2_{1}  \varepsilon_{\diamond}^2\\
    & \leq \E_{\text{Haar}} \left[|c_\omega|^2\right] + \frac{||O||^2_{\infty}} {d^2} |R(\omega)| \varepsilon_{\diamond} + \frac{\norm{O}_2^2}{d^2} \varepsilon_\diamond+   ||O||^2_{1}  \varepsilon_{\diamond}^2
    \end{aligned}
\end{equation}

\paragraph{Spectral norm bound:}
\begin{equation}
    \begin{aligned}
        \E \left[|c_\omega|^2\right] &\leq \E_{\text{Haar}} \left[|c_\omega|^2\right] +\frac{||O^{\otimes 2}||_2}{d(d+1)} \norm{ \sum_{j_1,j'_1\in R(\omega)}\ket{j_1,j'_1}\bra{j'_1,j_1}}_2 \varepsilon_{\infty} \\
        & \qquad + K_2 \norm{\sum_{j_1,j'_1\in R(\omega)}\ket{j'_1,j_1}\bra{j_1,j'_1}}_2\varepsilon_{\infty}\\
        & \qquad + \norm{\mathcal{A}_{\mathbb{W}}^{(2)}(\ket{00}\bra{00}) \otimes \mathcal{A}_{\mathbb{W}}^{(2)}(O^{\otimes2})}_2 \norm{\sumrw \ket{j'_1,i_1}\bra{j_1,i'_1} \otimes \ket{j_1,i'_1}\bra{j'_1,i_1}}_2\\
        &\leq \E_{\text{Haar}} \left[|c_\omega|^2\right]  +  \frac{||O||^2_{2}} {d(d+1)} \sqrt{|R(\omega)|} \varepsilon_{\infty} + K_2 \sqrt{|R(\omega)|} \varepsilon_{\infty} + ||O||_2^2 |R(\omega)| \varepsilon_{\infty}^2 
    \end{aligned}
\end{equation}

where in the first inequality we use $|Tr[A B]| \leq \|A\|_2 \|B\|_2 $ and the property $\|\mathcal{A}_{\mathbb{W}}^{(2)}(X)\|_2 \leq \|\mathcal{A}_{\mathbb{W}}^{(2)}\|_{\infty} \|X\|_2$ with the shorthand $\varepsilon_{\infty} := \|\mathcal{A}_{\mathbb{W}}^{(2)}\|_{\infty}$.
The above upper bound can be further simplified. Precisely, we obtain 
\begin{equation}
     \Var_{\theta}[c_{\omega}(\theta)] \leq \Var_{\text{Haar}}[c_{\omega}(\theta)]+ 2\frac{ \|O\|_2^2}{d}\sqrt{|\widetilde{R}(\omega)|}\varepsilon_{\infty} +d^2\|O\|_2^2 |\widetilde{R}(\omega)|\varepsilon_{\infty}^2
\end{equation}
where we introduced the normalized frequency redundancy $|\widetilde{R}(\omega)|:= |R(\omega)|/d^2$ and used the inequality $K_2 \leq \norm{O}^2/d^2$.
\paragraph{Monomial norm bound:}

\begin{definition}\label{def:monomial_norm}[Monomial definition of $\varepsilon$-approximate 2-design]
    An ansatz $U(\Theta)$ forms a monomial $\varepsilon$-approximate 2-design if:
    \begin{equation}
        \max_{p,q,r,s \in [d]} \; |(\mathcal{A}^{(2)}_{U(\Theta)})_{p,q,r,s}| \leq \frac{\varepsilon}{d^2}
    \end{equation}
    where $(\mathcal{A}^{(2)}_{U(\Theta)})_{p,q,r,s}$ is a coefficient of the $d^4$-dimensional matrix $\mathcal{A}^{(2)}_{U(\Theta)}$.
\end{definition}

 The monomial definition \ref{def:monomial_norm} of approximate 2-design  gives an upper bound on the superoperator coefficients in the computational basis. Hence, we take   
$$\varepsilon=\max_{\substack{i_1,i'_1,j_1,j'_1 \in [d] \\ k,k' \in [d^{4}]}} d^2 \left|\mathcal{A}_{\mathbb{W}}^{(2)}(\ket{j_1,i'_1}\bra{j'_1,i_1})_{k,k'}\right| $$
Consequently, we get
 $$|\Tr\left[\mathcal{A}_{\mathbb{W}}^{(2)}(\ket{00}\bra{00})\ket{j'_1,i_1}\bra{j_1,i'_1}\right]| \leq \frac{\epsilon}{d^{2}}$$

 and

 $$
 \Tr\left[ \mathcal{A}_{\mathbb{W}}^{(2)}(\ket{j'_1,i_1}\bra{j_1,i'_1})O^{\otimes2} \right] = \sum_{k,k'\in \llbracket d \rrbracket} |\left(\mathcal{A}_{\mathbb{W}}^{(2)}\ket{j_1,i'_1}\bra{j'_1,i_1}\right)_{k,k'}||O^{\otimes2}_{kk'}| \leq \frac{\epsilon}{d^{2}} \sum_{k,k'\in \llbracket d \rrbracket} |O^{\otimes2}_{kk'}|
 $$

By applying the triangular inequality in Eq.\eqref{eq:var_coeff_4terms}, one arrives to the upper bound obtained in Theorem \ref{thm:bound_approx_2design}:

 \begin{equation}\label{eq:monom_f}
    \E\left[|c_\omega|^2\right] \leq \E_{\text{Haar}}\left[|c_\omega|^2\right] + \frac{|R(\omega)|}{d(d+1)}\varepsilon \sum_{kk'}\frac{|[O^{\otimes 2}]_{kk'}|}{d^2} + K_2\frac{|R(\omega)|}{d^2}\varepsilon + \frac{|R(\omega)|^2\varepsilon^2}{d^2}\sum_{kk'}\frac{|[O^{\otimes2}]_{kk'}|}{d^2}
 \end{equation}
 The term $\sum_{kk'}[O^{\otimes2}]_{kk'}$ can be related to the observable 2-norm as follows. Indeed, by definition the observable schatten 2-norm can be written as
 \begin{align}
     \norm{O}_2^2 = Tr[O^\dagger O] = \sum_{k,k'=1}^d |O_{k,k'}|^2\;.
 \end{align}
 Hence, we obtain
\begin{align}
    \sum_{k,k'=1}^{d^2} |[O^{\otimes2}]_{kk'}| &\leq \sqrt{d^4 \sum_{k,k'=1}^{d^2} |[O^{\otimes2}]_{kk'}|^2}\\
    &= d^2 \norm{O^{\otimes 2}}_2\\
    &= d^2 \norm{O}_2^2
\end{align}
 Consequently, Eq.\eqref{eq:monom_f} becomes 
 \begin{equation}
 \begin{aligned}
     \Var_\theta[c_\omega(\theta)] &\leq \Var_{\text{Haar}} [c_\omega(\theta)] + (\norm{O}_2^2 + \frac{\norm{O}_2^2}{d^2}) |\widetilde{R}(\omega)| \varepsilon + d^2 \norm{O}_2^2 |\widetilde{R}(\omega)|^2 \varepsilon^2\\
     &\leq \Var_{\text{Haar}} [c_\omega(\theta)] + 2\norm{O}_2^2 |\widetilde{R}(\omega)| \varepsilon + d^2 \norm{O}_2^2 |\widetilde{R}(\omega)|^2 \varepsilon^2
     \end{aligned}
 \end{equation}

\end{proof}

\subsection{Model variance with approximate 2-design trainable unitaries}\label{Proof:approx_2design_f}

We present a formal version of Corollary \ref{cor:bound_model_approx_2design_informal} in the following Corollary.

\begin{corollary}\label{cor:bound_model_approx_2design_formal}
      Consider a quantum model  $f(x,\theta)$ of the form in Eq.\eqref{Eq:quantum_Model} and a parametrized circuit of the form in Eq.\eqref{eq:circuit_ansatz} with $L=1$ layers and fixed encoding Hamiltonians. We assume that each of the  trainable layer $W^l(\theta)\;,l \in \{1,2\}$ form independently an $\varepsilon_M$-approximate 2-design according to the monomial definition introduced in Definition \ref{def:monomial_norm}. 
     For a fixed $x \in \mathcal{X}$, the variance of the model $f(x,\theta)$ is upper bounded for any $x \in \mathbb{R}$ as

\begin{equation}
  \Var_\theta[f(x,\theta)]\leq \frac{\norm{O}_2^2}{d^2} + \norm{O}_2^2 \varepsilon\;.
\end{equation}

\end{corollary}

\begin{proof}
We recall that $f(x,\theta)$ for a fixed data point $x \in \mathbb{R}$ is a real valued function. Hence, $\forall x \in \mathbb{R}$, its variance is given by $ \Var_{\theta}[f(x,\theta)]= \E_{\theta}[f^2(x,\theta)]-\E_{\theta}[f(x,\theta)]^2$.

We also recall that the model $f$ is given by 
\begin{equation}
    f(x,\theta) = Tr[W^{(1) \dagger} S^\dagger(x) W^{(2) \dagger} O W^{(2)} S(x) W^{(1)} \ketbra{0}{0}]
\end{equation}
where the dependence on trainable parameters $\theta$ is hidden in the trainable unitaries $W^{(1)}$ and $W^{(2)}$.

Hence the model's second moment w.r.t to the distributions over $W^{(1)}$ and $W^{(2)}$ can be expressed as 
\begin{align}
    \E_{\theta}[f^2(x,\theta)] &:=  \E_{W^{(1)}\sim \mathbb{W},W^{(2)} \sim \mathbb{W}}[f^2(x)]\\
    &= \E_{W^{(1)}\sim \mathbb{W},W^{(2)} \sim \mathbb{W}}[f^2(x)] - \E_{W^{(1)}\sim \text{Haar},W^{(2)} \sim \mathbb{W}}[f^2(x)] + \E_{W^{(1)}\sim \text{Haar},W^{(2)} \sim \mathbb{W}}[f^2(x)]\\
    &= \E_{W^{(2)} \sim \mathbb{W}} \left[ \E_{W^{(1)}\sim  \mathbb{W}}[f^2(x)] - \E_{W^{(1)}\sim  \text{Haar}}[f^2(x)]\right] + \E_{W^{(1)}\sim \text{Haar},W^{(2)} \sim \mathbb{W}}[f^2(x)]
\end{align}

Using the invariance property of the Haar measure, one can easily show that 
 \begin{equation}
     (\E_{W^{(1)}\sim \text{Haar},W^{(2)} \sim \text{Haar}}[f(x)])^2 =  (\E_{W^{(1)} \sim \text{Haar}}[Tr[ O W^{(1)}  \ketbra{0}{0} W^{(1) \dagger}]])^2
 \end{equation}
 Similarly, we have 
 \begin{equation}
     \E_{W^{(1)}\sim \text{Haar},W^{(2)} \sim \mathbb{W}}[f^2(x)] = \E_{W^{(1)} \sim \text{Haar}}[Tr[ O W^{(1)}  \ketbra{0}{0} W^{(1) \dagger}]^2]
 \end{equation}

 Hence, the model's variance becomes 
 \begin{align}
     \Var[f(x)] &= \E_{W^{(2)} \sim \mathbb{W}} \left[ \E_{W^{(1)}\sim  \mathbb{W}}[f^2(x)] - \E_{W^{(1)}\sim  \text{Haar}}[f^2(x)]\right] + \Var_{W^{(1)} \sim \text{Haar}}[Tr[ O W^{(1)}  \ketbra{0}{0} W^{(1) \dagger}]]\\
     &=  \E_{W^{(2)} \sim \mathbb{W}} \left[\Tr\left[ S(x)^{\dagger\otimes 2}W^{(2)\dagger \otimes 2} O^{\otimes 2}W^{(2)\otimes 2} S(x)^{\otimes 2} (\E_{W^{(1)} \sim \mathbb{W}}-\E_{W^{(1)} \sim \text{Haar}}) [W^{(1)\otimes 2}\ketbra{00}{00} W^{(1)\dagger\otimes 2}]  \right]\right]  \\
     &\quad + \Var_{W^{(1)} \sim \text{Haar}}[Tr[ O W^{(1)}  \ketbra{0}{0} W^{(1) \dagger}]]\\
     &= \E_{W^{(2)} \sim \mathbb{W}} \left[\Tr\left[ S(x)^{\dagger\otimes 2}W^{(2)\dagger \otimes 2} O^{\otimes 2}W^{(2)\otimes 2} S(x)^{\otimes 2} \mathcal{A}_{\mathbb{W}}(\ketbra{00}{00})   \right]\right]  \\
     &\quad + \Var_{W^{(1)} \sim \text{Haar}}[Tr[ O W^{(1)}  \ketbra{0}{0} W^{(1) \dagger}]]
 \end{align}
 In what follows we focus on bounding the first term in the variance expression above.
Using Holder's inequality, we obtain
\begin{align}
    \Tr\left[ S(x)^{\dagger\otimes 2}W^{(2)\dagger \otimes 2} O^{\otimes 2}W^{(2)\otimes 2} S(x)^{\otimes 2} \mathcal{A}_{\mathbb{W}}(\ketbra{00}{00})   \right] &\leq \norm{ S(x)^{\dagger\otimes 2}W^{(2)\dagger \otimes 2} O^{\otimes 2}W^{(2)\otimes 2} S(x)^{\otimes 2}}_2 \norm{\mathcal{A}_{\mathbb{W}}(\ketbra{00}{00})}_2\\
    &= \norm{  O^{\otimes 2}}_2 \norm{\mathcal{A}_{\mathbb{W}}(\ketbra{00}{00})}_2 \\
    &= \norm{O}_2^2 \sqrt{\sum_{k,k'=1}^{d^2} |\mathcal{A}_{\mathbb{W}}(\ketbra{00}{00})|^2_{k,k'}}\\
    & \leq  \norm{O}_2^2 \sqrt{\sum_{k,k'=1}^{d^2} \left(\frac{\varepsilon}{d^2}\right)^2}\\
    &= \norm{O}_2^2 \varepsilon
\end{align}

Now, we go back to the term $\Var_{W^{(1)} \sim \text{Haar}}[Tr[ O W^{(1)}  \ketbra{0}{0} W^{(1) \dagger}]]$ which can be directly computed using Weingarten calculus \cite{mele_introduction_2023}.
\begin{align}
    \Var_{W^{(1)} \sim \text{Haar}}[Tr[ O W^{(1)}  \ketbra{0}{0} W^{(1) \dagger}]] &= \E_{W^{(1)} \sim \text{Haar}}[Tr[ O W^{(1)}  \ketbra{0}{0} W^{(1) \dagger}]^2] - \E_{W^{(1)} \sim \text{Haar}}[Tr[ O W^{(1)}  \ketbra{0}{0} W^{(1) \dagger}]]^2\\
    &= \frac{Tr[O^2] +Tr[O]^2}{d(d+1)}  -\frac{Tr[O]^2}{d^2}\\
    &\leq \frac{\norm{O}_2^2}{d^2}
\end{align}

which concludes the proof.
\end{proof}

%% file: Proofs/Local2design_proof.tex
\subsection{Variance under the local 2-design setting}\label{app:local2design}

\subsubsection{Preliminaries}\label{proof:premliminaries}
In this section, we present properties of Haar integration over Haar random unitaries that we will use in the main proof of Theorem \ref{thm:bound_local2design}.

\begin{restatable}[]{lem}{Lemma1}\label{lemma:TN_first_layer}
    Let $\mathcal{H}= \mathcal{H}_w \otimes \mathcal{H}_{\overline{w}} $ be a bipartite Hilbert space and let $\{W(\theta)\}_{\theta \in \Theta}$ be a parameterized unitary acting on $\mathcal{H}_w $ and forming a 2-design when considering a uniform distribution over the variational parameters $\Theta$. Then, for any arbitrary linear operators $A,B,C,D : \mathcal{H} \rightarrow \mathcal{H}$ such that $A=A_w \otimes A_{\overline{w}} \text{ and } C=C_w \otimes C_{\overline{w}}$, we have:
     \begin{equation}\label{Eq:block_int_2design}
    \begin{aligned}
    &\int_{\Theta} Tr_w[A(W \otimes \mathbb{1}_{\overline{w}})B(W^{\dagger} \otimes \mathbb{1}_{\overline{w}})] Tr_w[C(W \otimes \mathbb{1}_{\overline{w}})D(W^{\dagger} \otimes \mathbb{1}_{\overline{w}})] dW(\Theta)\\
     &=\int_W Tr_w[A(W \otimes \mathbb{1}_{\overline{w}})B(W^{\dagger} \otimes \mathbb{1}_{\overline{w}})] Tr_w[C(W \otimes \mathbb{1}_{\overline{w}})D(W^{\dagger} \otimes \mathbb{1}_{\overline{w}})] dW\\
     &= \frac{Tr_w[B] Tr_w[D]}{2^{2m}-1} A_{\overline{w}} C_{\overline{w}} \left( Tr[A_w]Tr[C_w]-\frac{Tr[A_w C_w]}{2^m} \right) +
     \frac{Tr_w[B(C_{\overline{w}} \otimes \mathbb{1}_{w})D]}{2^{2m}-1}  \left( Tr[A_w C_w] - \frac{Tr[A_w]Tr[C_w]}{2^m} \right)
 \end{aligned} 
 \end{equation}
\end{restatable}

\begin{restatable}[]{lem}{LemmaPartial}\label{lemma:partial}
    Let $\mathcal{H}= \mathcal{H}_{\omega} \otimes \mathcal{H}_{\overline{\omega}}$ be a bipartite Hilbert space. Then, for any arbitrary linear operators  $A_{\omega}, B_{\omega} : \mathcal{H_{\omega}} \rightarrow \mathcal{H_{\omega}}, S: \mathcal{H} \rightarrow \mathcal{H} $, we have:
    $$Tr_{\overline{\omega}}[|p\rangle \langle q| (A_{\omega} \otimes \mathbb{1}_{\overline{\omega}}) S (B_{\omega} \otimes \mathbb{1}_{\overline{\omega}})]=|p\rangle \langle q|_{\omega} A_{\omega} Tr_{\overline{\omega}}[(\mathbb{1}_{\omega} \otimes |p\rangle \langle q|_{\overline{\omega}}) S]
            B_{\omega}
        $$
\end{restatable}

\begin{restatable}[]{lem}{LemmaTauAInteg}\label{lemma:tau_a_integ}
    Let $\mathcal{H}$ be a composite Hilbert space such that  
    $\mathcal{H}=\mathcal{H}_{\tau_a}\otimes \mathcal{H}_{\tau_b} \otimes \mathcal{H}_{z} \otimes \mathcal{H}_{y}$ and $\mathcal{H}_x$  a subspace of $\mathcal{H}_{\tau_a}\otimes \mathcal{H}_{\tau_b} \otimes \mathcal{H}_{z}$. Then, for any unitary X acting on $\mathcal{H}$
    ,
    $W_{\tau}$ acting non trivially on subspace $\mathcal{H}_\tau =\mathcal{H}_{\tau_a}\otimes \mathcal{H}_{\tau_b}$ and $Y_i=(W_{\tau}^{\dagger} \otimes \mathbb{1}_{\overline{\tau}}) X_i (W_{\tau}\otimes \mathbb{1}_{\overline{\tau}}) \forall i \in \{1,2\}$, we get:
    
    \begin{equation}
        E_{\tau}[Tr[Tr_x[Y_1]Tr_x[Y_2]]]  =\left\{
  \begin{array}{@{}ll@{}}
  \frac{2^{m/2}}{2^m+1}Tr\left[ Tr_{x,\tau}[X_1]Tr_{x,\tau}[X_2] + Tr_{x \backslash \tau}[X_1]Tr_{x \backslash \tau}[X_2]\right], & \text{if}\ x \cap \tau \in \{\tau_a, \tau_b\} \\
    Tr[Tr_x[X_1]Tr_x[X_2]], & \text{otherwise} 
  \end{array}\right.
    \end{equation}
\end{restatable}

\begin{proof}
For ease of notation, we denote the Hilbert space by its associated subsytem index.

    \begin{equation*}
            Tr_x[Y]
        =\left\{
  \begin{array}{@{}ll@{}}
  Tr_{\tau}[(W_{\tau}^{\dagger} \otimes \mathbb{1}_{y}) X (W_{\tau}\otimes \mathbb{1}_{y})], & \text{if}\ x = \tau \\
    (W_{\tau}^{\dagger} \otimes \mathbb{1}_{y}) Tr_x[X] (W_{\tau}\otimes \mathbb{1}_{y}), & \text{if}\ x \cap \tau = \varnothing \\
    Tr_{\tau}[(W_{\tau}^{\dagger} \otimes \mathbb{1}_{y}) Tr_{x \backslash \tau}[X] (W_{\tau}\otimes \mathbb{1}_{y})], & \text{if}\ \tau \subset x\\
    Tr_{\tau_a}[(W_{\tau}^{\dagger} \otimes \mathbb{1}_{y}) Tr_{x \backslash \tau}[X] (W_{\tau}\otimes \mathbb{1}_{y})], & \text{if}\ x \cap \tau = \tau_a\\
    Tr_{\tau_b}[(W_{\tau}^{\dagger} \otimes \mathbb{1}_{y}) Tr_{x \backslash \tau}[X] (W_{\tau}\otimes \mathbb{1}_{y})], & \text{if}\ x \cap \tau = \tau_b
  \end{array}\right.
    \end{equation*}

    For the case where $x \cap \tau = \varnothing$,

    \begin{equation*}
        \begin{aligned}
            E_{\tau}[Tr[Tr_x[Y_1]Tr_x[Y_2]]]&= E_{\tau}[Tr[(W_{\tau}^{\dagger} \otimes \mathbb{1}_{y}) Tr_x[X_1] Tr_x[X_2] (W_{\tau} \otimes \mathbb{1}_{y})]]\\
            &= E_{\tau}[Tr[ Tr_x[X_1] Tr_x[X_2] ]]\\
            &=Tr[ Tr_x[X_1] Tr_x[X_2] ]
        \end{aligned}
    \end{equation*}

In both cases $x=\tau \text{ and } \tau \subset x$, we get $Tr_x[Y]= Tr_{\tau}[(W_{\tau}^{\dagger} \otimes \mathbb{1}_{y}) Tr_z[X] (W_{\tau} \otimes \mathbb{1}_{y})]$ where $z= \varnothing , x \backslash \tau$ respectively.
Hence, to compute $E_{\tau}[Tr[Tr_x[Y_1]Tr_x[Y_2]]]$, we apply \ref{lemma:partial} for $\omega= \tau, A=C=\mathbb{1}, B=Tr_z[X_1] \text{ and 
 } D=Tr_z[X_2] $:

\begin{equation*}
    \begin{aligned}
         E_{\tau}[Tr[Tr_x[Y_1]Tr_x[Y_2]]]&= Tr[Tr_{\tau,z}[X_1]Tr_{\tau,z}[X_2]]\\
         &= Tr[Tr_x[X_1]Tr_x[X_2]]
    \end{aligned}
\end{equation*}

For the last 2 cases where the intersection between $x \text{ and } \tau$ is a singleton (i.e. $x \cap \tau \in \{\tau_a,\tau_b\}$), we use the \textit{Random Tensor Network Integrator} (RTNI) package \cite{fukuda_rtni_2019} and get the following:
$$E_{\tau}[Tr[Tr_x[Y_1]Tr_x[Y_2]]] = \frac{2^{m/2}}{2^m+1}\left[ Tr[Tr_{x,\tau}[X_1]Tr_{x,\tau}[X_2]] + Tr[Tr_{x \backslash \tau}[X_1]Tr_{x \backslash \tau}[X_2]]\right]$$
 
\end{proof}

\subsubsection{Integrate over a brickwise circuit of local 2-design blocks}\label{proof:local2design_moment}

In this section, we provide a general method to compute the second moment superoperator of a brickwise circuit made of parameterized local 2-design blocks (no encoding part is evolved).
The methods used and the expression of the second moment superoperator will be useful to prove Theorem \ref{thm:bound_local2design}
and to give a better intuition about the monomial $\varepsilon$-distance to a 2-design as defined in Definition \ref{def:monomial_norm}, for this type of circuits.

We therefore consider a circuit with a brickwise structure as described in Fig.\ref{fig:Figure_Framework_Local2design} acting on $n=m \times A$ qubits with depth $L$, $A$ blocks per layer and circular connectivity. Each block acts on $m$ qubits and forms a 2-design on the corresponding $m$-qubit subsystem. We denote by $U(\boldsymbol{\theta})$ the parameterized unitary generated by this circuit and by $U_{j,l}(\theta)$ the unitary corresponding to the $j^{th}$ block in the $l^{th}$ layer.

For layers with an odd index $l$, we denote the set of subsystems on which its blocks act non-trivially by $\{s_h\}_{h=1}^{A}$ and for even layers, we use the notation $\{\tau_h\}_{h=1}^{A}$ for the corresponding subsystems. Using this notation, we can decompose each subsystem $s_h$ into $s_h = (s_h)_1 \cup (s_h)_2$ and similarly  each subsystem $\tau_h$ into $\tau_h = (\tau_h)_1 \cup (\tau_h)_2$ where each subsystem $(s_h)_i, (\tau_h)_i$ acts on $m/2$ qubits.

We note that in the following proof, we will consider that each layer $l$ contains $A_l$ blocks in order to account for brickwise circuits with slightly different configurations (mainly the brickwise light cone). However, to compute the second moment superoperator, we simply consider that $A_l=A \; \forall l \in [L]$.


We recall here the expression of a single coefficient of the second moment superoperator corresponding to the circuit unitary $U(\boldsymbol{\theta})$ indexed by multi-indices $\boldsymbol{qp}:=(q_1,q_2,p_1,p_2)\in [2^{4n}] $ and $\boldsymbol{rs}:=(r_1,r_2,s_1,s_2)\in [2^{4n}] $ :

\begin{equation}\label{eq:secondmoment_def}
 \begin{aligned}
    (M_{U(\boldsymbol{\theta})}^{(2)})_{\textbf{qp,rs}}
    &= \int_{\boldsymbol{\theta}} dU(\theta) (U(\theta)^{\otimes 2} \otimes (U(\theta)^*)^{\otimes 2})_{\textbf{qp,rs}} \\
    &:=\int_{\boldsymbol{\theta}} U_{q_1,r_1}U_{q_2,r_2} U^{*}_{p_1,s_1} U^{*}_{p_2,s_2} dU(\theta)
    \end{aligned}
\end{equation}

For convenience, we rewrite the second moment superoperator as follows:

\begin{equation}
    \begin{aligned}       (M_{U(\boldsymbol{\theta})}^{(2)})_{\textbf{p,q,r,s}} &=\int_{\boldsymbol{\theta}} U_{q_1,r_1}U_{q_2,r_2} U^{*}_{p_1,s_1} U^{*}_{p_2,s_2} dU(\theta)\\
    &= \int_{\boldsymbol{\mathbb{U}}}
             Tr[ |p_1\rangle \langle q_1| U Y_1 U^{\dagger}]Tr[ |p_2\rangle \langle q_2| U Y_2 U^{\dagger}] dU\\
             &:= E_{\mathbb{U}}[Tr[ |p_1\rangle \langle q_1| U Y_1 U^{\dagger}]Tr[ |p_2\rangle \langle q_2| U Y_2 U^{\dagger}]]
    \end{aligned}
\end{equation}
where $Y_1=|r_1\rangle \langle s_1|$ and $Y_2=|r_2\rangle \langle s_2|$.

In order to compute the above expectation value, one needs to integrate over all the local 2-design blocks forming the circuit unitary $U(\boldsymbol{\theta})$. Under the assumption that the blocks are independent, we can perform the integration iteratively. 
To do so, we first start by integrating over the blocks of the first layer and then we integrate over the remaining blocks (blocks from the $2^{nd}$ up to the $L^{th}$ layer).

Considering that the $h^{th}$ block of the first layer acts non trivially on subsystem $s_h$ (with the corresponding unitary matrix $U_{h,1}(\theta)$), we introduce the following notations where we drop $\theta$ for simplicity:
\begin{equation}
    X_h^{pq}:= Tr_{h^+}[(\mathbb{1}_{h^{-}} \otimes |p\rangle \langle q|_{h^+} ) \Tilde{U}^{(h,1)}] \quad \forall h \in \{1, \dots, A_1\}
\end{equation}
\begin{equation}\label{eq:def_f_h_local2design}
    f_h(x):= Tr[Tr_x[X_h^{p_1q_1}]Tr_x[X_h^{p_2q_2}]] \quad \forall h< A_1, \forall x \subseteq h^{-}
\end{equation}

where $h^+ := \{s_{h+1} \cup s_{h+2} \cup \dots, s_{A_1}\}$, $h^{-}:=\{s_1\cup s_2 \cup \dots, s_{h-1} \cup s_{h}\}$, $\Tilde{U}^{(h,1)} := U^{(h,1)}  Y U^{(h,1)\dagger}$ and $U^{(h,1)}$ is the circuit unitary $U$ after removing the first $h$ blocks from the first layer.
We notice here that $X_h^{pq}$ acts on subsystem $h^{-}$ and that it does no longer depend on the first $h$ blocks of the first layer. We note that $X_0^{pq}:= Tr[|p\rangle \langle q|UYU^{\dagger}]$ and $X_{A}^{pq}:= \Tilde{U}^{(A_1,1)} $.

We give the following recursive relation between $X_h^{pq}$ and $X_{h+1}^{pq}$ using Lemma \ref{lemma:partial} and for ease of notation, we shortly use $h+1$ to denote subsystem $s_{h+1}$:
\begin{align*}
    X_h^{pq} &= Tr_{h+1}[Tr_{(h+1)^+}[(\mathbb{1}_{h^{-}} \otimes |p\rangle \langle q|_{h^+} ) (\mathbb{1}_{h^{-}} \otimes U_{h+1,1} \otimes \mathbb{1}_{(h+1)^+}  ) \Tilde{U}^{(h+1)}(\mathbb{1}_{h^{-}} \otimes U_{h+1,1}^{\dagger}   \otimes \mathbb{1}_{(h+1)^+})]]\\
    &= Tr_{h+1}[(\mathbb{1}_{h^{-}} \otimes |p\rangle \langle q|_{h+1} ) (\mathbb{1}_{h^{-}} \otimes U_{h+1,1}) X_{h+1}^{pq} (\mathbb{1}_{h^{-}} \otimes U_{h+1,1}^{\dagger}   )]
\end{align*}

We then apply Lemma \ref{lemma:TN_first_layer} and get the following $\forall h< A_1, \forall x \subseteq h^{-}$, 
\begin{equation}\label{Eq:Recursion_int_Xh}
   \begin{aligned}
   E_{U_{h+1,1}}[f_h(x)] &= 
    E_{U_{h+1,1}}[Tr[Tr_x[X_h^{p_1q_1}]Tr_x[X_h^{p_2q_2}]]]\\
    &= \alpha_{h+1} Tr[Tr_{h+1}[Tr_x[X_{h+1}^{p_1q_1}]]Tr_{h+1}[Tr_x[X_{h+1}^{p_2q_2}]]] + \beta_{h+1} Tr[Tr_x[X_{h+1}^{p_1q_1}]Tr_x[X_{h+1}^{p_2q_2}]]\\
    &= \alpha_{h+1} f_{h+1}(x \cup (h+1)) + \beta_{h+1} f_{h+1}(x) 
\end{aligned} 
\end{equation}

with the coefficients $\alpha_{h+1}$ and $\beta_{h+1}$ defined as follows and explicitly dependent on indices $(p_1,q_1,p_2,q_2)$: \begin{align*}
    \alpha_{h+1} (p_1,q_1,p_2,q_2)&:= \frac{1}{2^{2m}-1}(\delta_{(p_1,q_1)_{h+1}} \delta_{(p_2,q_2)_{h+1}} - \frac{\delta_{(q_1,p_2)_{h+1}} \delta_{(p_1,q_2)_{h+1}}}{2^m})\\
    &= \frac{1}{2^{2m}-1} \left(C_{h+1}(p_1,q_1,p_2,q_2) - \frac{D_{h+1}(p_1,q_1,p_2,q_2)}{2^m} \right)\\
    \beta_{h+1} (p_1,q_1,p_2,q_2)&:= \frac{1}{2^{2m}-1}(\delta_{(q_1,p_2)_{h+1}} \delta_{(p_1,q_2)_{h+1}} - \frac{\delta_{(p_1,q_1)_{h+1}} \delta_{(p_2,q_2)_{h+1}}}{2^m})\\&= \frac{1}{2^{2m}-1} \left( D_{h+1}(p_1,q_1,p_2,q_2) - \frac{C_{h+1}(p_1,q_1,p_2,q_2)}{2^m} \right)
\end{align*}

Using Eq.\eqref{Eq:Recursion_int_Xh}, we integrate iteratively over the $A_1$ blocks of the first layer as follows:
\begin{equation}\label{eq:int_first}
    \begin{aligned}
        &\prod_{j=1}^{A_1} E_{U_{j1}}[Tr[ |p_1\rangle \langle q_1| W Y_1 U^{\dagger} ]Tr[|p_2\rangle \langle q_2| W Y_2 U^{\dagger}]]\\
        &=  \prod_{j=1}^{A_1}  E_{U_{j1}}[f_0(\varnothing)]\\
        &= \prod_{j=2}^{A_1} E_{U_{j1}}[\alpha_1 f_1(1) + \beta_1 f_1(\varnothing)]\\
        &= \prod_{j=3}^{A_1} E_{U_{j1}}[\alpha_1 \left( \alpha_2 f_2(12) + \beta_2 f_2(1) \right) + \beta_1 (\alpha_2 f_2(2) + \beta_2 f_2(\varnothing))]\\
        & \vdots\\
        &= \sum_{x_1 \in P_1} a_{x_1}(p_1,q_1,p_2,q_2) f_{A_1}(x_1), \quad a_{x_1}(p_1,q_1,p_2,q_2):= \prod_{h \in x_1} \alpha_h(p_1,q_1,p_2,q_2) \prod_{h \in \overline{x_1}} \beta_h(p_1,q_1,p_2,q_2) 
    \end{aligned}
\end{equation}
Here $P_1$ is the power set of the ensemble $\{1,2,\dots,A_1\}$, hence containing $2^{A_1}$ subsystems. We also recall from Eq.\eqref{eq:def_f_h_local2design} that:
\begin{equation*}
    f_{A_1}(x_1):= Tr[Tr_{x_1}[U^{(A_1,1)}Y_1 U^{(A_1,1)\dagger}]Tr_{x_1}[U^{(A_1,1)}Y_2 U^{(A_1,1)\dagger}]]     
\end{equation*}
where $U^{(A_1,1)}$ is the unitary representing the circuit from the $2^{nd}$ up to the $L^{th}$ layer (i.e. the $A_1$ blocks of the first layer are removed).

We notice that in the expression of $f_{A_1}(x_1)$, we no longer have the dependence on the indices $\boldsymbol{qp}$, which contribution is now contained in the coefficients $a_{x_1}$.

 To integrate over the remaining blocks, we introduce the following notations $ \forall t \in \{1,2\},  \forall l \in \{2, \dots, L\} \text{ and } \forall h \in \{1,\dots,A_l\}$ 
 \begin{equation*}
     \begin{aligned}
         Z_{h,l}^t &:= U^{(h,l)} Y_t U^{(h,l)\dagger} \\
         g_{h,l}(x)&:= Tr[Tr_x[Z_{h,l}^1]Tr_x[Z_{h,l}^2]] 
     \end{aligned}
 \end{equation*}

where $U^{(h,l)}$ is the circuit unitary after removing the layers $i \in \{1,\dots,l-1\}$ and the first $h$ blocks of the $l^{th}$ layer. We also note that $g_{0,l}(x) = g_{A_{l-1}, l-1}(x) \; \forall l \in \{3,\dots,L\}$, $g_{0,2}(x) := f_{A_1}(x)$ and $g_{A_L,L}(x):= Tr[Tr_x[Y_1]Tr_x[Y_2]]$.
 
By applying Lemma \ref{lemma:tau_a_integ},
we obtain the following formula allowing to integrate recursively over the remaining blocks. We note that depending on the parity of the layer index $l$, we use  $s_h$ or $\tau_h$ (here we use $\tau_h$ but we get the same formula for $s_h$) to denote the subsystem on which the block $h$ is acting non trivially:

\begin{equation}\label{eq:intgerate_intersection}
       E_{U_{{h+1},l}}[g_{h,l}(x)]=\left\{
  \begin{array}{@{}ll@{}}
    g_{h+1,l}(x), & \text{if}\ \tau_{h+1} \cap x \in \{ \varnothing, \tau_{h+1} \} \\
    \frac{2^{m/2}}{2^m+1} \left( g_{h+1,l}(x \cup \tau_{h+1}) + g_{h+1,l}(x \backslash \tau_{h+1}) \right), & \text{otherwise}
  \end{array}\right.
\end{equation}

According to the recursive formula above, one can easily notice that for a fixed layer $l$, integrating $g_{0,l}(x)$ over its $A_l$ blocks amounts to determining all the \emph{non trivial} intersections between the current subsystem $x$ and the layer blocks subsystems $\{\tau_h\}_{h=1}^{A_l}$ (if $l$ is even otherwise $\{s_h\}_{h=1}^{A_l}$). In other terms, the set $I_l(x):=\{h | x\cap \tau_{h} \in \{(\tau_{h})_1 , (\tau_{h+1})_2\} \forall h \in [A_l] \}$  will contain the blocks over which we are going to effectively integrate by adding the multiplicative factor $\frac{2^{m/2}}{2^m+1}$ and \emph{branching} subsystem $x$ into the two new subsystems $x \cup \tau_{h+1}$ and $x \backslash \tau_{h+1}$.

Formally, we get $\;\forall \; 2 \leq l \leq L$:
\begin{equation}\label{eq:int_second}
     \prod_{j=1}^{A_l} E_{U_{jl}}[g_{0,l}(x_{l-1})]=\sum_{x_{l} \in P_{l}(x_{l-1})} (\frac{2^{m/2}}{2^m+1})^{d_{l}(x_{l-1})} g_{A_l,l}(x_{l})
\end{equation}

where $d_{l}(x_{l-1}):= |I_l(x_{l-1})|$  and $P_l(x_{l-1})$ is the set of subsystems generated after each \emph{branching} of $x_{l-1}$ obtained by following the rule in Eq.\eqref{eq:intgerate_intersection}.



For ease of notation we drop the explicit dependence of $a_{x_1}$ on $(p_1,q_1,p_2,q_2)$. By combining Eq.\eqref{eq:int_first} and Eq.\eqref{eq:int_second}, we get the final expression of the second moment coefficient:
\begin{equation}\label{eq:sencond_moment_haarblock}
    \begin{aligned}  
 (M_{U(\boldsymbol{\theta})}^{(2)})_{\textbf{p,q,r,s}}&= \prod_{i=1}^L \prod_{j=1}^{A_i} E_{U_{ji}}[f_0(\varnothing)]\\
    &= \prod_{i=3}^L \prod_{j=1}^{A_i} E_{U_{ji}}\left[\sum_{x_1 \in P_1} a_{x_1} \prod_{j=1}^{A_2} E_{U_{j2}}[f_{A_1}(x_1)] \right]\\
    &= \prod_{i=3}^L \prod_{j=1}^{A_i} E_{U_{ji}}\left[\sum_{x_1 \in P_1} a_{x_1} \prod_{j=1}^{A_2} E_{U_{j2}}[g_{0,2}(x_1)] \right]\\
    &= \prod_{i=3}^L \prod_{j=1}^{A_i} E_{U_{ji}}\left[\sum_{x_1 \in P_1} a_{x_1}  \sum_{x_2 \in P_2(x_1)} (\frac{2^{m/2}}{2^m+1})^{d_2(x_1)} g_{0,3}(x_2) \right]\\
    &=\sum_{x_1 \in P_1} a_{x_1} \sum_{x_2 \in P_2(x_1)} \sum_{x_3 \in P_3(x_2)} \dots \sum_{x_L \in P_L(x_{L-1})}(\frac{2^{m/2}}{2^m+1})^{\sum_{i=2}^L d_i(x_{i-1})} g_{A_L,L}(x_L)\\
    &= \sum_{x_1 \in P_1} a_{x_1} \sum_{x_2 \in P_2(x_1)} \sum_{x_3 \in P_3(x_2)} \dots \sum_{x_L \in P_L(x_{L-1})}(\frac{2^{m/2}}{2^m+1})^{\sum_{i=2}^L d_i(x_{i-1})} Tr[Tr_{x_L}[Y_1]Tr_{x_L}[Y_2]]\\
    &= \sum_{x_1 \in P_1} a_{x_1} \sum_{x_L \in P_f(x_1) } (\frac{2^{m/2}}{2^m+1})^{d(x_L)} Tr[Tr_{x_L}[Y_1]Tr_{x_L}[Y_2]]\\
    &=  \sum_{x_1 \in P_1} a_{x_1} \sum_{x_L \in P_f(x_1) } (\frac{2^{m/2}}{2^m+1})^{d(x_L)} \delta(r_1,s_1)_{x_L} \delta(r_1,s_2)_{\overline{x_L}}
    \delta(r_2,s_2)_{x_L} \delta(r_2,s_1)_{\overline{x_L}}
    \end{aligned}
\end{equation}

Where $P_f(x_1)$ is the set of all subsystems obtained by applying the branching rule in Eq.\eqref{eq:intgerate_intersection} starting from $x_1$ and iterating through all the blocks in the circuits.
We note as well that $d(x_L):=\sum_{i=2}^L d_i(x_{i-1})$ is the total number of branchings of $x_1$ leading to the final subsystem $x_L \in P_f(x_1)$.

One can remark from the above expression of the second moment superoperator coefficient that the coefficient $a_{x_1}$ may scale in the number of blocks of the first layer $A_1 := A$ and that the sum $\sum_{x_L \in P_f(x_1) } (\frac{2^{m/2}}{2^m+1})^{d(x_L)}$ will eventually scale in $L$, the circuit depth.

Moreover, we note that the expression holds for a light cone made of local 2-design blocks and acting on $n = m \times A_1$ qubits where $A_l=L-l+1, \forall l \in [L]$. In this setting, we have exactly the same expression as in Eq.\eqref{eq:sencond_moment_haarblock}, with $P_f(x_1)$ the set of subsystems obtained by applying the branching rule from Eq.\ref{eq:intgerate_intersection} on the light cone blocks.
Therefore, what differs between considering a full circuit made of local 2-design blocks and a light cone of the same circuit is the set of final subsystems  $P_f(x_1)$ and its corresponding branching count $d(x_L) \forall x_L \in P_f(x_1)$.

However, in both cases, the construction of $P_f(x_1) \; \forall x_1 \in P_1$ and hence the calculation of $d(x_L)$ is quiet cumbersome. Therefore, we derive the following upper bound on this sum.

To do so, we know that
\begin{equation}
    f_A(x_1)= Tr[Tr_{x_1}[U^{(A)}Y_1U^{(A)\dagger}]Tr_{x_1}[U^{(A)}Y_2U^{(A)\dagger}]] =  \sum_{x_L \in P_f(x_1) } (\frac{2^{m/2}}{2^m+1})^{d(x_L)} Tr[Tr_{x_L}[Y_1]Tr_{x_L}[Y_2]]
\end{equation}

 If $Y_1,Y_2$ are projectors (i.e. $r_i = s_i, i \in \{1,2\}$),
 then using the sub multiplicative property of positive semi definite matrices, we get $f_A(x_1) \leq Tr[Y_1] Tr[Y_2] = 1$

 For $Y_1=Y_2=|0\rangle \langle 0|$, we get $f_A(x_1) = \sum_{x \in P_f(x_1)} (\frac{2^{m/2}}{2^m+1})^{d(x)} \leq 1$.
 Therefore for arbitrary $Y_1$ and $Y_2$, we get:
 \begin{equation}\label{eq:bound1}
     f_A(x_1) \leq Tr[Tr_{x_L}[Y_1]Tr_{x_L}[Y_2]]
 \end{equation}

We note that this upper bound holds for a full brickwise circuit or a lightcone over the same circuit, which can be observed in Section \ref{sec:Simulations} (See Fig.\ref{fig:lightcone_global_meas_exp_gol}).

Indeed, as will be detailed in the next section, Eq.\eqref{eq:sencond_moment_haarblock} and Eq.\eqref{eq:bound1} will be key in obtaining the expression of the Fourier coefficients variance over a light cone, leading to the upper bound in theorem \ref{thm:bound_local2design}.





 \subsubsection{Proof of Theorem \ref{thm:bound_local2design}}\label{proof:local2design_coeff}

We consider the brick-wise circuit architecture made of $m$-local encoding blocks and $m$-local 2-design parameterized blocks as illustrated in Fig.\ref{fig:Figure_Framework_Local2design}.
We recall Theorem \ref{thm:bound_local2design} in this setting.

\begin{theorem}
 Consider a quantum model of the form in Eq.\eqref{Eq:quantum_Model} and a parametrized circuit of the form in Eq.\eqref{eq:circuit_ansatz} using a brickwise architecture with $L=1$ layers and observable $O= \hat{O}_{s_k} \otimes \mathbb{1}_{\overline{s_k}} $ acting non trivially on the $m$-qubit subsystem $s_k$. Assume that each trainable $m$-qubit unitary forms a local 2-design. The  variance of each Fourier coefficient $c_{\omega}(\theta)$ for the frequencies $ \omega \in \Omega$ appearing in the model Fourier decomposition in Eq.\eqref{Eq:quantum_Fourier_Model} is upper bounded as

\begin{enumerate}
    \item If $||\hat{O}_{s_k}||_2^2 \leq 2^m$ , we have
\begin{equation}
    \Var[c_{\omega}] \leq \left(  \frac{2^{m+1}}{2^{2m}-1}\right)^{2L_2} |R_{E_k}(\omega)|^2.
\end{equation}
\item If $\hat{O}_{s_k}$ is a projector of rank $r$ , we have
\begin{equation}
    \Var[c_{\omega}] \leq \left(  \frac{2^{m+1}}{2^{2m}-1}\right)^{2L_2} \left(\frac{r}{2^m}\right)^2 |R_{E_k}(\omega)|^2.
\end{equation}
\end{enumerate}
Here $R_{E_k}(\omega)$ is the frequency generator obtained from the encoding blocks inside the observable backward light cone $\mathcal{L}_k$ (acting non trivially on $\mathcal{S}_{E_k}$) and $L_2$ is the depth of the post-encoding parameterized block.
\end{theorem}

\begin{proof}
In the following proof, using similar techniques as presented in \ref{proof:local2design_moment}, we give an expression of the variance of the Fourier coefficients under the local 2-design setting described in Section \ref{subsec:local_2-design}. However, as the obtained expression involves cumbersome calculations, we give in Theorem \ref{thm:bound_local2design} an upper bound on the Fourier coefficients variance that depends on the circuit depth and the frequency redundancy.

In what follows, we use the same notations from the previous section \ref{proof:local2design_moment} and we consider a circuit with a first parameterized layer $W^{(1)}$ of depth $L_1$, a layer of encoding blocks $S(x)$ and a final parameterized layer $W^{(2)}$ of depth $L_2$ as depicted in Fig.\ref{fig:Figure_Framework_Local2design} acting initially on $n= m \times A$ qubits. 
We recall here the expression of the expectation of the modulus squared of a Fourier coefficient $c_{\omega}$ (See Eq.\eqref{eq:exp_squared_fourier_L1} for details):

\begin{equation}\label{eq:var_Fourier_L1}
  \begin{aligned}
\mathbb{E}\left[|c_\omega|^2\right] =  \sumrwb &\mathbb{E}_{W^{(1)}} \left[ \Tr[\ket{J'}\bra{J}W^{(1)}|0 \rangle \langle 0| W^{(1)\dagger}]\Tr[\ket{I}\bra{I'}W^{(1)}|0 \rangle \langle 0| W^{(1)\dagger}] \right] \\
& \times \mathbb{E}_{W^{(2)}} \left[ \Tr[\ket{J}\bra{J'}W^{(2)\dagger}O W^{(2)}]\Tr[\ket{I'}\bra{I}W^{(2)\dagger}O W^{(2)}] \right]\\
&= \sumrwb \Lambda(J,J',I,I') 
\Gamma(J,J',I,I') 
\end{aligned}   
\end{equation}

Here $J,J' \in R(\omega)$ are multi indices of length $A$, the number of blocks in a single layer  and each component $(j_k,j'_k)$ corresponds to the indices of 2 eigenvalues of the $2^m$-dimensional encoding unitary acting on subsystem $s_k, \forall k \in \{1,\dots,A\}$.
For ease of notation, we introduce the shorthand $\Lambda(J,J',I,I')$ for the expectation over the pre-encoding unitary $W^{(1)}$ and $\Gamma(J,J',I,I')$ for the expectation over the post-encoding unitary $W^{(2)}$ in Eq.\eqref{eq:var_Fourier_L1}.

We then consider an $m$-local observable of the form $O_k= \hat{O}_{s_k} \otimes \mathbb{1}_{\overline{s_k}}$ acting non trivially on subsystem $s_k$ and we denote by $\mathcal{L}_k$ the backward light cone associated to this observable. Precisely, $\mathcal{L}_k$ is the subcircuit containing all blocks with at least one qubit causally connected to the input qubits of $\hat{O}_{s_k}$. We also denote the subsystem on which $\mathcal{L}_k$ acts non trivially by $\mathcal{S}_{\mathcal{L}_k}$.

Due to the brick-wise circuit architecture and the locality of the observable, the quantum model is reduced to an effective model obtained by considering the restricted action of the circuit unitary on subsystem $\mathcal{S}_{\mathcal{L}_k}$:
\begin{equation}
    f_{\mathcal{L}_k}(x) := Tr[\ket{0}\bra{0}_{\mathcal{L}_k}U(x,\theta)^\dagger_{\mathcal{L}_k} O_{\mathcal{L}_k} U(x,\theta)_{\mathcal{L}_k}]
\end{equation}
where $U(x,\theta)_{\mathcal{L}_k}^{\dagger}=W^{(1)\dagger}_{\mathcal{L}_k} S(x)_{\mathcal{L}_k}^{\dagger} W^{(2)\dagger}_{\mathcal{L}_k}$ is the adjoint circuit unitary restricted to the light cone acting on subsystem $\mathcal{S}_{\mathcal{L}_k}$.

Hence, Eq.\eqref{eq:var_Fourier_L1} can be rewritten using the following: 

\begin{align*}
    \Lambda(J,J',I,I') &= \mathbb{E}_{W^{(1)}_{\mathcal{L}_k}} \left[ \Tr[\ket{J'}\bra{J}_{\mathcal{L}_k} W^{(1)}_{\mathcal{L}_k}|0 \rangle \langle 0|_{\mathcal{L}_k} 
W^{(1)\dagger}_{\mathcal{L}_k} ]\Tr[\ket{I}\bra{I'}_{\mathcal{L}_k} W^{(1)}_{\mathcal{L}_k} |0 \rangle \langle 0|_{\mathcal{L}_k}  W^{(1)\dagger}_{\mathcal{L}_k} ] \right]\\
\Gamma(J,J',I,I')  &= \mathbb{E}_{W^{(2)}_{\mathcal{L}_k}} \left[ \Tr[\ket{J}\bra{J'}_{\mathcal{L}_k}W^{(2)\dagger}_{\mathcal{L}_k}O_{\mathcal{L}_k} W^{(2)}_{\mathcal{L}_k}]\Tr[\ket{I'}\bra{I}_{\mathcal{L}_k}W^{(2)\dagger}_{\mathcal{L}_k}O_{\mathcal{L}_k} W^{(2)}_{\mathcal{L}_k}] \right]
\end{align*}


Another consequence of restricting calculations to the light cone $\mathcal{L}_k$ is spectrum reduction where only encoding blocks acting on subsystem $\mathcal{S}_{\mathcal{L}_k}$ are involved. Hence, in what follows $R(\omega)$ refers to the frequency generator made only of encoding blocks acting non trivially on subsystem $\mathcal{S}_{\mathcal{L}_k}$.  Let us denote by $\mathcal{S}_{E_k}$ the subsystem on which $W^{(2)}_{\mathcal{L}_k}$ (and similarly $S(x)$) acts non trivially and $\mathcal{S}_{\overline{{E_k}}}:= \mathcal{S}_{E_k} \backslash \mathcal{S}_{\overline{{E_k}}}$ as depicted in Fig.\ref{fig:Figure_Framework_Local2design}. Consequently, we get 

\begin{equation}\label{eq:var_lightcone}
\begin{aligned}
    \Gamma(J,J',I,I') &= \delta_{(J,J')_{\overline{{E_k}}} } 
    \delta_{(I,I')_{\overline{{E_k}}} }\mathbb{E}_{W^{(2)}_{{E_k}}} \left[ \Tr[\ket{J}\bra{J'}_{{E_k}}W^{(2)\dagger}_{{E_k}}O_{{E_k}} W^{(2)}_{{E_k}}]\Tr[\ket{I'}\bra{I}_{{E_k}}W^{(2)\dagger}_{{E_k}}O_{{E_k}} W^{(2)}_{{E_k}}] \right]\\
    &= \delta_{(J,J')_{\overline{{E_k}}} } 
    \delta_{(I,I')_{\overline{{E_k}}} } \Gamma_2(J,J',I,I')
    \end{aligned}
\end{equation}
In what follows whenever $\Gamma(J,J',I,I')$ is used, it actually refers to $ \Gamma(J,J',I,I')_2$. Therefore, Eq.\eqref{eq:var_Fourier_L1} becomes:
\begin{equation}\label{eq:coeff_var_LC}
    \mathbb{E}\left[|c_\omega|^2\right] = \sumrwb \delta_{(J,J')_{\overline{E_k}} } 
    \delta_{(I,I')_{\overline{E_k}} } \Lambda(J,J',I,I') 
\Gamma(J,J',I,I') 
\end{equation}

This implies that the above sum over $R(\omega)$ will effectively contain just the pairs $(J,J') \in R(\omega)$ such that $\forall s_l \in \mathcal{S}_{\overline{E_k}}, \; \lambda_{j_l} - \lambda_{j'_l} = 0$. 
Therefore the frequency generator becomes $R(\omega):= \{(J,J') \in [|2^{m}|]^{L_2} \times [|2^{m}|]^{L_2} | \sum_{l \in \mathcal{S}_{E_k}} (\lambda_{j_l} - \lambda_{j'_l})  =\omega \} \times \{(J,J) \in [|2^m|]^{L_1-1}\}$. In other terms, the effective spectrum is the one generated by encoding blocks acting non trivially on subsystems of $\mathcal{S}_{E_k}$ and that is made redundant by adding null contributions from the encoding blocks outside the light cone.

In order to calculate $\Gamma(J,J',I,I')$ and $\Lambda(J,J',I,I')$, we use the same calculations from Eq.\eqref{eq:sencond_moment_haarblock} with a minor change of notations where $A$ will take the values $L_1+L_2-1$ and $L_2$ respectively. Therefore, we get: 

\begin{equation}\label{eq:gamma_lambda_calcul}
    \begin{aligned}
        \Lambda(J,J',I',I) &= \sum_{x_1 \in P_1^1} a_{x_1}(J,J',I',I) f^1(x_1,|0 \rangle \langle 0|_{\mathcal{L}_k})\\ 
        &= \sum_{x_1 \in P_1^1} a_{x_1}(J,J',I',I) \sum_{x_L \in P_f^1(x_1) } (\frac{2^{m/2}}{2^m+1})^{d(x_L)} Tr[Tr_{x_L}[|0 \rangle \langle 0|_{\mathcal{L}_k}]^2]\\  \Gamma(J,J',I',I) &= \sum_{x_1 \in P_1^2} a_{x_1}(J,J',I,I') f^2 (x_1,O_{E_k})\\
        &= \sum_{x_1 \in P_1^2} a_{x_1}(J,J',I',I) \sum_{x_L \in P_f^2(x_1) } (\frac{2^{m/2}}{2^m+1})^{d(x_L)} Tr[Tr_{x_L}[O_{E_k}]^2]
    \end{aligned}
\end{equation}

where $P_1^1$ and $P_1^2$ are the power sets of  $\mathcal{S}_{\mathcal{L}_k}$ and $\mathcal{S}_{E_k}$ respectively, $P_f^1$ and $P_f^2$ are the sets of final subsystems after branching over blocks of $W^{(1)}_{\mathcal{L}_k}$ and $W^{(2)}_{E_k}$ respectively, according to the construction detailed in Eq.\eqref{eq:sencond_moment_haarblock} and $f^i(x_1,H)$ is a shorthand for $\sum_{x_L \in P_f^i(x_1) } (\frac{2^{m/2}}{2^m+1})^{d(x_L)} Tr[Tr_{x_L}[H]^2], \; \forall i \in \{1,2\}$.

By substituting Eq.\eqref{eq:gamma_lambda_calcul} in Eq.\eqref{eq:coeff_var_LC}, we get the following expression of $ \mathbb{E}\left[|c_\omega|^2\right]$:
\begin{equation}\label{eq:sum_Ebar}
    \begin{aligned}
        \mathbb{E}\left[|c_\omega|^2\right] &= \sumrwb \delta(J,J')_{\overline{ E_k}} \delta(I,I')_{\overline{ E_k}}\Lambda(J,J',I',I) 
        \Gamma(J,J',I',I) \\
        &= \sumrwb \delta(J,J')_{\overline{ E_k}} \delta(I,I')_{\overline{ E_k}} \left(\sum_{x_1 \in P_1^1} a_{x_1}(J,J',I',I) f^1(x_1,|0 \rangle \langle 0|_{\mathcal{L}_k})\right) \left(\sum_{y_1 \in P_1^2} a_{y_1}(J,J',I',I) f^2(y_1,O_{E_k})\right)\\
        &= \sum_{\substack{J,J' \in R_{E_k}(\omega)\\ J,J'  \in R_{E_k}(\omega)}} 
        \left(\sum_{y_1 \in P_1^2} a_{y_1} f^2(y_1,O_{E_k})\right)
        \sum_{\substack{(J,J')_{\overline{E_k}}\\ (I,I')_{\overline{E_k}}}}
        \delta(J,J')_{\overline{ E_k}} \delta(I,I')_{\overline{ E_k}} 
        \left(\sum_{x_1 \in P_1^1} a_{x_1} f^1(x_1,|0 \rangle \langle 0|_{\mathcal{L}_k})\right)\\ &= \left( \frac{1}{2^{2m}-1} \right)^{L_1+2L_2-1} \sum_{\substack{J,J' \in R_{E_k}(\omega)\\ J,J'  \in R_{E_k}(\omega)}} 
        \left(\sum_{y_1 \in P_1^2} \tilde{a}_{y_1} f^2(y_1,O_{E_k})\right)
        \sum_{x_1 \in P_1^1}  f^1(x_1,|0 \rangle \langle 0|_{\mathcal{L}_k})\sum_{\substack{(JJ')_{\Bar{E_k}}\\ (I,I')_{\Bar{E_k}}}}
        \delta(J,J')_{\Bar{ E_k}} \delta(I,I')_{\Bar{ E_k}} 
         \tilde{a}_{x_1}
    \end{aligned}
\end{equation}

where $a_{x_1} := \left( \frac{1}{2^{2m}-1} \right)^{L_1+L_2-1}  \tilde{a}_{x_1}$ and $a_{y_1} := \left( \frac{1}{2^{2m}-1} \right)^{L_2}  \tilde{a}_{y_1}$.

Summing over indices in $\overline{ E_k}$ gives the following:

\begin{equation*}
    \begin{aligned}
        \sum_{\substack{(JJ')_{\overline{E_k}}\\ (I,I')_{\overline{E_k}}}}
        \delta(J,J')_{\overline{ E_k}} \delta(I,I')_{\overline{ E_k}} 
         \tilde{a}_{x_1} &= \sum_{J_{\overline{E_k}},I_{\overline{E_k}}} \tilde{a}_{x_1}  \delta(J,J')_{\overline{ E_k}} \delta(I,I')_{\overline{ E_k}} \\
         &= \sum_{J_{\overline{E_k}},I_{\overline{E_k}}} \prod_{l \in x_1} (C_l - \frac{D_l}{2^m}) \prod_{l \in \overline{x_1}} (D_l - \frac{C_l}{2^m}) (\delta(J,J')_{\overline{ E_k}} \delta(I,I')_{\overline{ E_k}}) \\
         &= \prod_{l \in x_1 \cap E_k} (C_l - \frac{D_l}{2^m}) \prod_{l \in \overline{x_1} \cap E_k} (D_l - \frac{C_l}{2^m})\sum_{J_{\overline{E_k}},I_{\overline{E_k}}} \prod_{l \in x_1 \cap \overline{E_k}} (1 - \frac{\delta_{j_l,i_l}}{2^m}) \prod_{l \in \overline{x_1} \cap \overline{E_k}} (\delta_{j_l,i_l} - \frac{1}{2^m})\\
         &= \prod_{l \in x_1 \cap E_k} (C_l - \frac{D_l}{2^m}) \prod_{l \in \overline{x_1} \cap E_k} (D_l - \frac{C_l}{2^m})   \prod_{l \in x_1 \cap \overline{E_k}} \sum_{j_l,i_l} (1 - \frac{\delta_{j_l,i_l}}{2^m}) \prod_{l \in \overline{x_1} \cap \overline{E_k}} \sum_{j_l,i_l}(\delta_{j_l,i_l} - \frac{1}{2^m})\\ &=  \prod_{l \in x_1 \cap E_k} (C_l - \frac{D_l}{2^m}) \prod_{l \in \overline{x_1} \cap E_k} (D_l - \frac{C_l}{2^m}) \prod_{l \in x_1 \cap \overline{E_k}} (2^{2m}-1)
         \prod_{l \in \overline{x_1} \cap \overline{E_k}} (2^m-2^m)\\
         &= \prod_{l \in x_1 \cap E_k} (C_l - \frac{D_l}{2^m}) \prod_{l \in \overline{x_1} \cap E_k} (D_l - \frac{C_l}{2^m})  (2^{2m}-1)^{|x_1 \cap \overline{E_k}|} 0^{|\overline{x_1} \cap \overline{E_k}|}\\
         &= \left\{
  \begin{array}{@{}ll@{}}
  \prod_{l \in x_1 \cap E_k} (C_l - \frac{D_l}{2^m}) \prod_{l \in \overline{x_1} \cap E_k} (D_l - \frac{C_l}{2^m})  (2^{2m}-1)^{| \overline{E_k}|}, & \text{if}\ \overline{E_k} \subset x_1 \\
    0, & \text{otherwise} 
  \end{array}\right.\\
  &= \left\{
  \begin{array}{@{}ll@{}}
  \prod_{l \in x_1 \cap E_k} (C_l - \frac{D_l}{2^m}) \prod_{l \in \overline{x_1} \cap E_k} (D_l - \frac{C_l}{2^m})  (2^{2m}-1)^{L_1-1}, & \text{if}\ \overline{E_k} \subset x_1 \\
    0, & \text{otherwise} 
  \end{array}\right.
    \end{aligned}
\end{equation*}

Thus, by substituting this above sum in Eq.\eqref{eq:sum_Ebar}, we obtain: 

\begin{equation}
    \begin{aligned}        \mathbb{E}\left[|c_\omega|^2\right] &=(\frac{1}{2^{2m}-1})^{2L_2} \sum_{\substack{J,J' \in R_{E_k}(\omega)\\ I,I'  \in R_{E_k}(\omega)}} \sum_{y_1 \in P_1^2} \tilde{a}_{y_1} f^2(y_1,O_{E_k})  \sum_{\substack{x_1 \in P_1^1\\ \overline{E_k} \subset x_1}}  \prod_{l \in x_1 \cap E_k} (C_l - \frac{D_l}{2^m}) \prod_{l \in \overline{x_1} \cap E_k} (D_l - \frac{C_l}{2^m})   f^1(x_1,|0 \rangle \langle 0|_{\mathcal{L}_k})\\
        &= (\frac{1}{2^{2m}-1})^{2L_2} \sum_{\substack{J,J' \in R_{E_k}(\omega)\\ I,I'  \in R_{E_k}(\omega)}} \sum_{y_1 \in P_1^2} \tilde{a}_{y_1} f^2(y_1,O_{E_k})  \sum_{x_1 \in P_1^2 }  \Tilde{a}_{x_1}   f^1(x_1 \cup \overline{E_k},|0 \rangle \langle 0|_{\mathcal{L}_k}) 
    \end{aligned}
\end{equation}

In the above expression of $\mathbb{E}\left[|c_\omega|^2\right]$, 
we have $|\Tilde{a}_x|, \leq 1 \; \forall x \in P_1^2$ and from Eq.\eqref{eq:bound1} we have: 
\begin{equation*}
    \begin{aligned}
         f^1(x_1 \cup \overline{E_k},|0 \rangle \langle 0|_{\mathcal{L}_k}) &= \sum_{x \in P_f^1(x_1\cup \overline{E_k}) } (\frac{2^{m/2}}{2^m+1})^{d(x)} Tr[Tr_{x_L}[|0 \rangle \langle 0|_{\mathcal{L}_k}]^2]\\
         & \leq  Tr[Tr_{x_L}[|0 \rangle \langle 0|_{\mathcal{L}_k}]^2] = 1\\
         f^2(y_1,O_{E_k}) &= \sum_{x \in P_f^2(y_1) } (\frac{2^{m/2}}{2^m+1})^{d(x)} Tr[Tr_{x_L}[O_{E_k}]^2]\\
         & \leq Tr[Tr_{x_L}[O_{E_k}]^2]\\
         &\leq 2^{2mL_2} 
    \end{aligned}
\end{equation*}
where we use in the last inequality the hypothesis that $Tr[\hat{O}_{s_k}^2]\leq 2^m$.

Finally, by combining all the previous steps, we retrieve the upper bound in Theorem \ref{thm:bound_local2design}:

\begin{equation*}
    \begin{aligned}
    \mathbb{E}\left[|c_\omega|^2\right] \leq \left( \frac{2^{m+1}}{2^{2m}-1} \right)^{2L_2} |R_{E_k}(\omega)|^2
    \end{aligned}
\end{equation*}
\end{proof}
\subsubsection{Fourier coefficients variance in a light cone forming a 2-design }
 In this section, we consider the same settings and notations from the previous section \ref{proof:local2design_coeff} but we assume that $W^{(1)}_{\mathcal{L}_k}$ and $W^{(2)}_{E_k}$ form each a 2-design on subsystems $\mathcal{S}_{\mathcal{L}_k}$ and $\mathcal{S}_{E_k}$  respectively.
 We follow the same steps as in the previous proof until we get to Eq.\eqref{eq:coeff_var_LC}.
 Then, we compute $\Gamma(J,J',I,I')$ and $\Lambda(J,J',I,I')$ under the 2-design assumption.
 Hence, by applying Weingarten calculus expression of the second moment \cite{mele_introduction_2023},  we obtain: 
 \begin{equation}\label{eq:gamma_lambda_2design}
     \begin{aligned}
         \Lambda(J,J',I,I') &= \frac{1}{2^{2m(L_1+L_2-1)}-1} \left(1-\frac{1}{2^{m(L_1+L_2-1)}}\right) \left[ \delta_{J,J'} \delta_{I,I'} + \delta_{J,I} \delta_{J',I'} \right] \\ \Gamma(J,J',I,I') &= \frac{1}{2^{2m L_2}-1} \left[ \delta_{(J,J')_{E_k}} \delta_{(I,I')_{E_k}} 
         \left(Tr[O_{E_k}]^2  - \frac{Tr[O_{E_k}^2]}{2^{mL_2}} \right) + 
         \delta_{(J,I)_{E_k}} \delta_{(J',I')_{E_k}}
         \left( Tr[O_{E_k}^2] - \frac{Tr[O_{E_k}]^2}{2^{mL_2}}  \right)
          \right]   
     \end{aligned}
 \end{equation}

Therefore, we substitute $\Gamma(J,J',I,I')$ and $\Lambda(J,J',I,I')$ in Eq.\eqref{eq:coeff_var_LC} using the new expressions in Eq.\eqref{eq:gamma_lambda_2design} and we get $\forall \omega \in \Omega \backslash \{0\}$:
\begin{equation}\label{eq:2design_lightcone}
    \begin{aligned}
    \mathbb{E}\left[|c_\omega|^2\right] &= \frac{1}{2^{m(L_1+L2-1)}(2^{m(L_1+L2-1)}+1)(2^{2mL_2}-1)} \sumrwb  \delta_{(J,J')_{\overline{E_k}} } 
    \delta_{(I,I')_{\overline{E_k}} }
     \delta_{J,I} \delta_{J',I'} 
         \left( Tr[O_{E_k}^2] - \frac{Tr[O_{E_k}]^2}{2^{mL_2}}  \right) \\
    &= \frac{ \left( Tr[O_{E_k}^2] - \frac{Tr[O_{E_k}]^2}{2^{mL_2}}  \right)}{2^{m(L_1+L2-1)}(2^{m(L_1+L2-1)}+1)(2^{2mL_2}-1)} \sum_{\substack{(I,I')_{\overline{E_k}} \\ (J,J')_{\overline{E_k}} }} \delta_{(J,J')_{\overline{E_k}} } 
    \delta_{(I,I')_{\overline{E_k}} }
    \delta_{(J,I)_{\overline{E_k}} } 
    \delta_{(J',I')_{\overline{E_k}} }\\
    & \hspace{8cm} \times
    \sum_{\substack{I,I' \in R_{E_k}(\omega)\\J,J' \in R_{E_k}(\omega)}} \delta_{(J,I)_{{E_k}} } 
    \delta_{(J',I')_{{E_k}} }\\
    &= \frac{2^{m(L_2-1)}\left( Tr[\hat{O}_k^2] - \frac{Tr[\hat{O}_k]^2}{2^m}  \right)}{2^{m(L_1+L2-1)}(2^{m(L_1+L2-1)}+1)(2^{2mL_2}-1)} 2^{m(L_1-1)} |R_{E_k}(\omega)|\\
    &= \frac{1}{2^m (2^{m(L_1+L2-1)}+1)(2^{2mL_2}-1)} \left( Tr[\hat{O}_k^2] - \frac{Tr[\hat{O}_k]^2}{2^m}  \right) |R_{E_k}(\omega)|
         \end{aligned}
\end{equation}

\subsection{Proof of theorem \ref{thm:Mass_Conservation}}
\label{app:proof_mass_conservation}

We recall Theorem \ref{thm:Mass_Conservation}.

\begin{theorem}[Fourier Norm Bound]
    Consider a quantum model  $f(x,\theta)$ of the form in Eq.\eqref{Eq:quantum_Model} using an observable $O$ and a parametrized circuit of the form in Eq.\eqref{eq:circuit_ansatz} with $L\geq 1 $ layers. Also assume that the encoding Hamiltonians are fixed, giving rise to a spectrum $\Omega$.
     Then,
    \begin{equation}
        \forall x \in \mathbb{R}^d, \forall \theta \in \Theta, |f(x,\theta)|^2 \leq ||O||^2_{\infty}
    \end{equation}
    \begin{equation}
        \forall \theta \in \Theta, \sum_{\omega \in \Omega}|c_\omega(\theta)|^2 \leq ||O||^2_{\infty}
    \end{equation}
\end{theorem}
\begin{proof}
    The first point can be proven by remarking that $\langle\psi|O|\psi\rangle$ can be maximized by taking $|\psi\rangle$ as the eigenvector associated to the largest eigenvalue of O.

    For the second point, one can write by considering the half spectrum $\Omega_+$
    \begin{align}
        |f(x)|^2 &= \bigg|\sum_{\omega \in \Omega_+} c_\omega e^{-i\omega^\top x} + c_\omega^* e^{i\omega^\top x}\bigg|^2\\
        &=\sum_{\omega \in \Omega_+} (c_\omega e^{-i\omega^\top x} + c_\omega^* e^{i\omega^\top x})(c_\omega^* e^{i\omega^\top x} + c_\omega e^{-i\omega^\top x})\\
        &+2\sum_{\omega_1 \neq \omega_2} (c_{\omega_1} e^{-i\omega_1^\top x} + c_{\omega_1}^* e^{i\omega_1^\top x})(c_{\omega_2} e^{-i\omega_2^\top x} + c_{\omega_2}^* e^{i\omega_2^\top x})\\
        &= \sum_{\omega \in \Omega_+} 2|c_\omega|^2+c_\omega^2e^{-2i\omega^\top x} + c_\omega^{2*}e^{2i\omega^\top x}\\
        &+2\sum_{\omega_1 \neq \omega_2} c_{\omega_1}c_{\omega_2}^* e^{-i(\omega_1 - \omega_2)^\top x} + c_{\omega_1}c_{\omega_2}e^{-i(\omega_1 + \omega_2)^\top x} + c_{\omega_1}^* c_{\omega_2}^* e^{i(\omega_1 + \omega_2)^\top x}+
        c_{\omega_1}^* c_{\omega_2}e^{i(\omega_1 - \omega_2)^\top x}\\
        &= 2\sum_{\omega \in \Omega_+} |c_\omega|^2 + g(x)
    \end{align}
    We finish the proof by finding $x_0$ such that $g(x_0)$ = 0. This can be done with the lemma \ref{lemma:cancellation}. We first reduce the problem to one variable and we introduce $h(t) = \sum_{\omega \in \Omega} a_\omega cos(\omega_1 t + \omega_{2:d}^\top x_0' + \phi_\omega)$ where $x'_0$ is an arbitrary vector of size $d-1$.
\end{proof}

\begin{lemma}
\label{lemma:cancellation}
    Let $h(t) = \sum_{\omega \in \Omega} a_\omega cos(\omega t + \phi_\omega)$, with $\Omega$ being a discrete subset of $\mathbb{R}^d$, and $\phi_\omega \in \mathbb{R}$.
    Then it exists $t_0$ such that $g(t_0) = 0$.
\end{lemma}
\begin{proof}
Let us suppose that $h(t)$ is of constant sign, we can assume it is positive on $\mathbb{R}$. If $h$ is not of constant sign, by continuity it means that it exists $t_0$ such that $h(t_0) = 0$ and the proof is finished. We will show that $\forall \varepsilon >0$ it exists $T$ such that $\big|\int_0^T h(t)dt\big| < \varepsilon$. Then it means that $h$ is equal to 0 over all the interval $[0, T]$, which proves the result.

We will now prove that $\forall \varepsilon >0$, there exists a real $T$ and integers $q_\omega$s such that $\forall \omega \in \Omega, |2 \pi T - q_\omega 2\pi/\omega| \leq \varepsilon$ which proves that $\big|\int_0^{2 \pi T} h(t)dt\big| < \varepsilon$ since for each frequency we integrate over an integer number of periods.

We will show how to construct $T$ and $q_\omega$s for three values $\omega_1$, $\omega_2$ and $\omega_3$.

Let $R_2 = \{k\frac{\omega_2}{\omega_1} - \lfloor k\frac{\omega_2}{\omega_1} \rfloor, k \in \mathbb{N}, k\in [0, N^2]\}$

If one divides the interval $[0, 1]$ into $N$ equal subintervals, then there are at least $N+1$ elements of $R_2$ that are in the same subinterval. Let $\{k_1,\dots k_{N+1}\}$ the integers corresponding to these elements.

Then let $R_3 = \{k\frac{\omega_3}{\omega_1} - \lfloor k\frac{\omega_3}{\omega_1} \rfloor, k \in \mathbb{N}, k\in \{k_1,\dots k_{N+1}\}\}$.
If one divides again the interval $[0, 1]$ into $N$ equal subintervals, then there are at least 2 elements of $R_3$ that are in the same subinterval.

Then there exists $k, k'$ such that 
\begin{gather*}
    \bigg|k\frac{\omega_3}{\omega_1} - \lfloor k\frac{\omega_3}{\omega_1} \rfloor - (k'\frac{\omega_3}{\omega_1} - \lfloor k'\frac{\omega_3}{\omega_1} \rfloor)\bigg| \leq \frac{1}{N}\\
    \bigg|(k-k')\frac{\omega_3}{\omega_1} - \big(\lfloor k\frac{\omega_3}{\omega_1} \rfloor  - \lfloor k'\frac{\omega_3}{\omega_1}\big) \rfloor)\bigg| \leq \frac{1}{N}\\
    \bigg|(k-k')\frac{1}{\omega_1} - \frac{1}{\omega_3}\big(\lfloor k\frac{\omega_3}{\omega_1} \rfloor  - \lfloor k'\frac{\omega_3}{\omega_1}\big) \rfloor)\bigg| \leq \frac{1}{N\omega_3}
\end{gather*}
And since $k$ and $k'$ are in $\{k_1,\dots k_{N+1}\}$, we also have
\begin{gather*}
    \bigg|(k-k')\frac{1}{\omega_1} - \frac{1}{\omega_2}\big(\lfloor k\frac{\omega_2}{\omega_1} \rfloor  - \lfloor k'\frac{\omega_2}{\omega_1}\big) \rfloor)\bigg| \leq \frac{1}{N\omega_2}
\end{gather*}
By taking $T = (k-k')\frac{1}{\omega_1}$ and $1/N = \varepsilon$, we proved the result for 3 numbers. One can apply the same construction for $|\Omega|$ number of frequencies by taking $N^{|\Omega|}$ integers at the beginning. 

\end{proof}

%% file: A2_additional_numerics.tex
\section{Additional Numerics}
\label{app:additional_numerics}

\subsection{Fourier coefficients variance in the 2-design setting for a reuploading model}

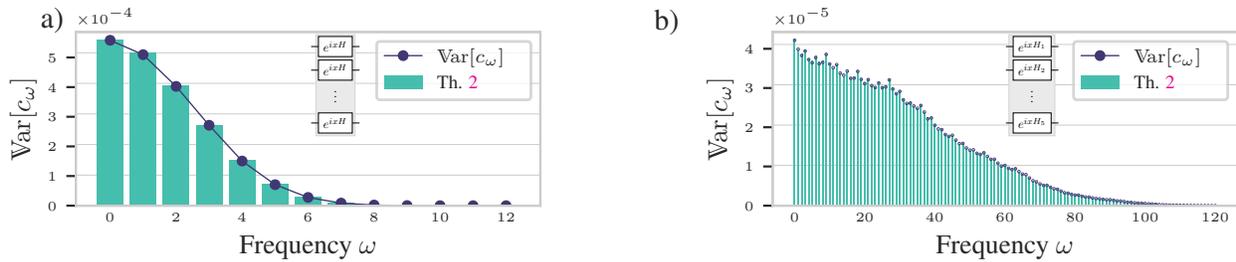
\begin{figure*}[ht]        \hspace{-0.4cm}\resizebox{1.05\textwidth}{!}{\input{Figures/figs_numerics/variance_redundancies_bis.tikz}}
    \caption{For $n=4$ qubits, two circuit layers ($L=2$) with the same encoding gates in each of the two encoding layers, five repetitions of the strongly entangling ansatz per trainable layer (see Appendix \ref{ansatze}); relation between the variance of each Fourier coefficient $\Var\left[c_\omega\right]$ (blue dots) and its corresponding partial redundancies given by theorem \ref{thm:formal_2design_reup} (green bars). Values are given for two different encoding strategies, a) Pauli encoding and b) exponential encoding.}
    \label{fig:thm2_sim}
    \end{figure*}

We consider the same settings of Theorem \ref{thm:formal_2design_reup} where we take a reuploading circuit with $L = 2$ circuit layers acting on $n=4$ qubits. In Fig.\ref{fig:thm2_sim}, we see that the simulated variance of the Fourier coefficients match the theoretical variance values predicted by Theorem \ref{thm:formal_2design_reup} for the Pauli and exponential encoding strategies. The explicit dependence on the redundancies is harder to visualize as was done in Fig.\ref{fig:variance_redundancies} because the variance expression from Theorem \ref{thm:formal_2design_reup} include the partial redundancies. Nonetheless, we clearly observe in the case of Pauli encoding that we have a gaussian distribution (negative frequencies are not plotted) with half of the frequencies suppressed. For the exponential encoding strategy, we see a gaussian distribution but with 
a higher variance than the Pauli case which is due to redundancies caused by the reuploading scheme and the fact that we are using the same encoding layer twice as explained in Appendix \ref{appendix:spec_distribution}.

\subsection{Fourier coefficients variance in the approximate 2-design setting for a reuploading model}\label{app:reuploading_approx}
\begin{figure*}[ht]
\centering
\begin{minipage}{0.7\textwidth}
\input{Figures/figs_numerics/var_vs_redundancies_wCircs.tikz}
\end{minipage}
\caption{Fitting Fourier coefficients variance $\Var[c_{\omega}]$ to a second degree polynomial in the redundancies $|R(\omega)|$ for four different circuit architecture (connectivity).
As the locality of the trainable block decreases $V$, $\varepsilon$ gets bigger and the trainable layers farther away from being a 2-design. 
 R2 assesses the goodness of the , the closer to one the better. }
\label{fig:simulation_approx_2desgin_L>1}
\end{figure*}
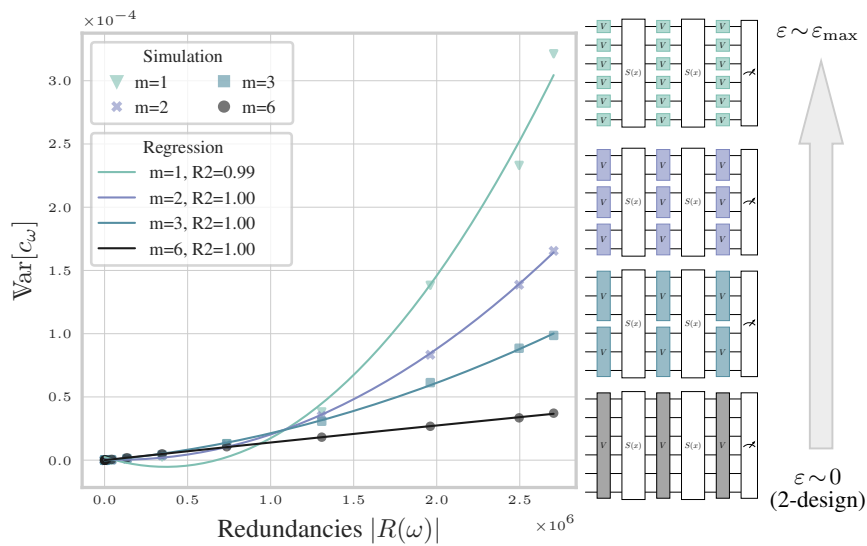

In Section \ref{subsec:approx_2-design}, we showed in Theorem \ref{thm:bound_approx_2design} an upper bound on $\Var(c_\omega)$ in the case of a model with a single circuir layer ($L=1$), being a second degree polynomial in $|R(\omega)|$. 
It was left as an open question to demonstrate a similar bound for $L>1$. 
Fig.\ref{fig:simulation_approx_2desgin_L>1} shows a simulation in the case of $L=2$ circuit layers. We tried to fit a second degree polynomial in order to express the relation between $\Var(c_\omega)$ and $|R(\omega)|$ directly. 
Even though one cannot simulate an upper bound but just a direct correlation, seeing this second-degree polynomial fitting well could indicate that the same kind of bound could be expected.

%% file: Figures/figs_numerics/variance_redundancies_bis.tikz
\resizebox{\textwidth}{!}{
\begin{tikzpicture}[remember picture]
    \node[inner sep=0pt] (plot) {\input{Figures/figs_numerics/figA_bis.pgf}};
    \begin{scope}[overlay]
       
        \draw [draw=grayplot, fill=grayplot!50] (-4.03,-.07) rectangle (-3.52,1.25);
        \node at (-3.8,.55) { 
            \begin{adjustbox}{height=.7cm}
            \begin{quantikz}[row sep = 1mm,column sep =2mm]
            &  \gate[1]{e^{ixH}} &  \\
            &   \gate[1]{e^{ixH}} &    \\
            &  \wireoverride{n}\raisebox{6pt}{\vdots}  \\
            &   \gate[1]{e^{ixH}} &    \\
        \end{quantikz}
        \end{adjustbox}
        };
\draw [draw=grayplot, fill=grayplot!50] (5.02,-.07) rectangle (5.62,1.25);
        \node at (5.3,.55) { 
            \begin{adjustbox}{height=.7cm}
            \begin{quantikz}[row sep = 1mm,column sep =2mm]
            &  \gate[1]{e^{ixH_1}} &  \\
            &   \gate[1]{e^{ixH_2}} &    \\\vspace{-20mm}
            &  \wireoverride{n}\raisebox{6pt}{\vdots}  \\
            &   \gate[1]{e^{ixH_5}} &    \\
        \end{quantikz}
        \end{adjustbox}
        };
    \end{scope}
\end{tikzpicture}
}

%% file: Figures/figs_numerics/var_vs_redundancies_wCircs.tikz
\resizebox{.7\textwidth}{!}{
\begin{tikzpicture}[remember picture]
    \node[inner sep=0pt] (plot) at (-1.1,0) {\input{Figures/figs_numerics/var_vs_redundancies.pgf}};
    \begin{scope}[overlay]
        \node at (4cm,2.7cm) { 
            \begin{adjustbox}{height=.9cm}
            \begin{quantikz}[row sep={0.8cm,between origins}]
            & \gate[1,style={draw=blue3,fill=blue3!60}]{V} & \gate[6]{S(x)} & \gate[1,style={draw=blue3,fill=blue3!60}]{V} & \gate[6]{S(x)} & \gate[1,style={draw=blue3,fill=blue3!60}]{V} & \meter[6]{}\\
            & \gate[1,style={draw=blue3,fill=blue3!60}]{ V} & & \gate[1,style={draw=blue3,fill=blue3!60}]{V} & & \gate[1,style={draw=blue3,fill=blue3!60}]{V} & \\
            & \gate[1,style={draw=blue3,fill=blue3!60}]{V} & & \gate[1,style={draw=blue3,fill=blue3!60}]{V} & & \gate[1,style={draw=blue3,fill=blue3!60}]{V} & \\
            & \gate[1,style={draw=blue3,fill=blue3!60}]{V} & & \gate[1,style={draw=blue3,fill=blue3!60}]{V} & & \gate[1,style={draw=blue3,fill=blue3!60}]{V} & \\
            & \gate[1,style={draw=blue3,fill=blue3!60}]{V} & & \gate[1,style={draw=blue3,fill=blue3!60}]{V} & & \gate[1,style={draw=blue3,fill=blue3!60}]{V} & \\
            & \gate[1,style={draw=blue3,fill=blue3!60}]{V} & & \gate[1,style={draw=blue3,fill=blue3!60}]{ V} & & \gate[1,style={draw=blue3,fill=blue3!60}]{V} & \\
        \end{quantikz}
        \end{adjustbox}
        };
        \node at (4cm,2.6cm-1.65cm) { 
            \begin{adjustbox}{height=.9cm}
            \begin{quantikz}[row sep={0.8cm,between origins}]
            & \gate[2,style={draw=blue2,fill=blue2!60}]{V} & \gate[6]{S(x)} & \gate[2,style={draw=blue2,fill=blue2!60}]{V} & \gate[6]{S(x)} & \gate[2,style={draw=blue2,fill=blue2!60}]{V} & \meter[6]{}\\
            & & & & & & \\
            & \gate[2,style={draw=blue2,fill=blue2!60}]{V} & & \gate[2,style={draw=blue2,fill=blue2!60}]{V} & & \gate[2,style={draw=blue2,fill=blue2!60}]{V} & \\
            & & & & & & \\
            & \gate[2,style={draw=blue2,fill=blue2!60}]{V} & & \gate[2,style={draw=blue2,fill=blue2!60}]{V} & & \gate[2,style={draw=blue2,fill=blue2!60}]{V} & \\
            & & & & & & \\
        \end{quantikz}
        \end{adjustbox}
        };
        \node at (4cm,2.6cm-2*1.65cm) { 
            \begin{adjustbox}{height=.9cm}
            \begin{quantikz}[row sep={0.8cm,between origins}]
            & \gate[3,style={draw=blue1,fill=blue1!60}]{V} & \gate[6]{S(x)} & \gate[3,style={draw=blue1,fill=blue1!60}]{V} & \gate[6]{S(x)} & \gate[3,style={draw=blue1,fill=blue1!60}]{V} & \meter[6]{}\\
            & & & & & & \\
            &  & &  & &  & \\
            & \gate[3,style={draw=blue1,fill=blue1!60}]{V} & & \gate[3,style={draw=blue1,fill=blue1!60}]{V} & & \gate[3,style={draw=blue1,fill=blue1!60}]{V} & \\
            & & &  & &  & \\
            & & & & & & \\
        \end{quantikz}
        \end{adjustbox}
        };
        \node at (4cm,2.6cm-3*1.65cm) { 
            \begin{adjustbox}{height=.9cm}
            \begin{quantikz}[row sep={0.8cm,between origins}]
            & \gate[6,style={fill=black!35}]{V} & \gate[6]{\small S(x)} & \gate[6,style={fill=black!35}]{V}& \gate[6]{S(x)} & \gate[6,style={fill=black!35}]{V} & \meter[6]{}\\
            & & & & & & \\
            &  & &  & &  & \\
            & & & & & & \\
            & & &  & &  & \\
            & & & & & & \\
        \end{quantikz}
        \end{adjustbox}
        };
        \node at (6,.3) {\input{Figures/figs_numerics/arrow.tikz}};
        \node at (6,-2.7) {\footnotesize $\varepsilon\!\sim\!0$};
        \node at (6,3.3) {\footnotesize $\varepsilon\!\sim\!\varepsilon_{\max}$};
        \node at (6,-3.05) {\footnotesize{(2-design)}};

    \end{scope}
\end{tikzpicture}}

%% file: Figures/figs_numerics/var_vs_redundancies.pgf
\begingroup%
\makeatletter%
\begin{pgfpicture}%
\pgfpathrectangle{\pgfpointorigin}{\pgfqpoint{3.211352in}{3.044343in}}%
\pgfusepath{use as bounding box, clip}%
\begin{pgfscope}%
\pgfsetbuttcap%
\pgfsetmiterjoin%
\definecolor{currentfill}{rgb}{1.000000,1.000000,1.000000}%
\pgfsetfillcolor{currentfill}%
\pgfsetlinewidth{0.000000pt}%
\definecolor{currentstroke}{rgb}{1.000000,1.000000,1.000000}%
\pgfsetstrokecolor{currentstroke}%
\pgfsetdash{}{0pt}%
\pgfpathmoveto{\pgfqpoint{0.000000in}{0.000000in}}%
\pgfpathlineto{\pgfqpoint{3.211352in}{0.000000in}}%
\pgfpathlineto{\pgfqpoint{3.211352in}{3.044343in}}%
\pgfpathlineto{\pgfqpoint{0.000000in}{3.044343in}}%
\pgfpathlineto{\pgfqpoint{0.000000in}{0.000000in}}%
\pgfpathclose%
\pgfusepath{fill}%
\end{pgfscope}%
\begin{pgfscope}%
\pgfsetbuttcap%
\pgfsetmiterjoin%
\definecolor{currentfill}{rgb}{1.000000,1.000000,1.000000}%
\pgfsetfillcolor{currentfill}%
\pgfsetlinewidth{0.000000pt}%
\definecolor{currentstroke}{rgb}{0.000000,0.000000,0.000000}%
\pgfsetstrokecolor{currentstroke}%
\pgfsetstrokeopacity{0.000000}%
\pgfsetdash{}{0pt}%
\pgfpathmoveto{\pgfqpoint{0.476352in}{0.397839in}}%
\pgfpathlineto{\pgfqpoint{3.111352in}{0.397839in}}%
\pgfpathlineto{\pgfqpoint{3.111352in}{2.823339in}}%
\pgfpathlineto{\pgfqpoint{0.476352in}{2.823339in}}%
\pgfpathlineto{\pgfqpoint{0.476352in}{0.397839in}}%
\pgfpathclose%
\pgfusepath{fill}%
\end{pgfscope}%
\begin{pgfscope}%
\pgfpathrectangle{\pgfqpoint{0.476352in}{0.397839in}}{\pgfqpoint{2.635000in}{2.425500in}}%
\pgfusepath{clip}%
\pgfsetroundcap%
\pgfsetroundjoin%
\pgfsetlinewidth{0.301125pt}%
\definecolor{currentstroke}{rgb}{0.800000,0.800000,0.800000}%
\pgfsetstrokecolor{currentstroke}%
\pgfsetdash{}{0pt}%
\pgfpathmoveto{\pgfqpoint{0.596124in}{0.397839in}}%
\pgfpathlineto{\pgfqpoint{0.596124in}{2.823339in}}%
\pgfusepath{stroke}%
\end{pgfscope}%
\begin{pgfscope}%
\definecolor{textcolor}{rgb}{0.150000,0.150000,0.150000}%
\pgfsetstrokecolor{textcolor}%
\pgfsetfillcolor{textcolor}%
\pgftext[x=0.596124in,y=0.356173in,,top]{\color{textcolor}\rmfamily\fontsize{4.000000}{4.800000}\selectfont \(\displaystyle {0.0}\)}%
\end{pgfscope}%
\begin{pgfscope}%
\pgfpathrectangle{\pgfqpoint{0.476352in}{0.397839in}}{\pgfqpoint{2.635000in}{2.425500in}}%
\pgfusepath{clip}%
\pgfsetroundcap%
\pgfsetroundjoin%
\pgfsetlinewidth{0.301125pt}%
\definecolor{currentstroke}{rgb}{0.800000,0.800000,0.800000}%
\pgfsetstrokecolor{currentstroke}%
\pgfsetdash{}{0pt}%
\pgfpathmoveto{\pgfqpoint{1.039045in}{0.397839in}}%
\pgfpathlineto{\pgfqpoint{1.039045in}{2.823339in}}%
\pgfusepath{stroke}%
\end{pgfscope}%
\begin{pgfscope}%
\definecolor{textcolor}{rgb}{0.150000,0.150000,0.150000}%
\pgfsetstrokecolor{textcolor}%
\pgfsetfillcolor{textcolor}%
\pgftext[x=1.039045in,y=0.356173in,,top]{\color{textcolor}\rmfamily\fontsize{4.000000}{4.800000}\selectfont \(\displaystyle {0.5}\)}%
\end{pgfscope}%
\begin{pgfscope}%
\pgfpathrectangle{\pgfqpoint{0.476352in}{0.397839in}}{\pgfqpoint{2.635000in}{2.425500in}}%
\pgfusepath{clip}%
\pgfsetroundcap%
\pgfsetroundjoin%
\pgfsetlinewidth{0.301125pt}%
\definecolor{currentstroke}{rgb}{0.800000,0.800000,0.800000}%
\pgfsetstrokecolor{currentstroke}%
\pgfsetdash{}{0pt}%
\pgfpathmoveto{\pgfqpoint{1.481966in}{0.397839in}}%
\pgfpathlineto{\pgfqpoint{1.481966in}{2.823339in}}%
\pgfusepath{stroke}%
\end{pgfscope}%
\begin{pgfscope}%
\definecolor{textcolor}{rgb}{0.150000,0.150000,0.150000}%
\pgfsetstrokecolor{textcolor}%
\pgfsetfillcolor{textcolor}%
\pgftext[x=1.481966in,y=0.356173in,,top]{\color{textcolor}\rmfamily\fontsize{4.000000}{4.800000}\selectfont \(\displaystyle {1.0}\)}%
\end{pgfscope}%
\begin{pgfscope}%
\pgfpathrectangle{\pgfqpoint{0.476352in}{0.397839in}}{\pgfqpoint{2.635000in}{2.425500in}}%
\pgfusepath{clip}%
\pgfsetroundcap%
\pgfsetroundjoin%
\pgfsetlinewidth{0.301125pt}%
\definecolor{currentstroke}{rgb}{0.800000,0.800000,0.800000}%
\pgfsetstrokecolor{currentstroke}%
\pgfsetdash{}{0pt}%
\pgfpathmoveto{\pgfqpoint{1.924887in}{0.397839in}}%
\pgfpathlineto{\pgfqpoint{1.924887in}{2.823339in}}%
\pgfusepath{stroke}%
\end{pgfscope}%
\begin{pgfscope}%
\definecolor{textcolor}{rgb}{0.150000,0.150000,0.150000}%
\pgfsetstrokecolor{textcolor}%
\pgfsetfillcolor{textcolor}%
\pgftext[x=1.924887in,y=0.356173in,,top]{\color{textcolor}\rmfamily\fontsize{4.000000}{4.800000}\selectfont \(\displaystyle {1.5}\)}%
\end{pgfscope}%
\begin{pgfscope}%
\pgfpathrectangle{\pgfqpoint{0.476352in}{0.397839in}}{\pgfqpoint{2.635000in}{2.425500in}}%
\pgfusepath{clip}%
\pgfsetroundcap%
\pgfsetroundjoin%
\pgfsetlinewidth{0.301125pt}%
\definecolor{currentstroke}{rgb}{0.800000,0.800000,0.800000}%
\pgfsetstrokecolor{currentstroke}%
\pgfsetdash{}{0pt}%
\pgfpathmoveto{\pgfqpoint{2.367808in}{0.397839in}}%
\pgfpathlineto{\pgfqpoint{2.367808in}{2.823339in}}%
\pgfusepath{stroke}%
\end{pgfscope}%
\begin{pgfscope}%
\definecolor{textcolor}{rgb}{0.150000,0.150000,0.150000}%
\pgfsetstrokecolor{textcolor}%
\pgfsetfillcolor{textcolor}%
\pgftext[x=2.367808in,y=0.356173in,,top]{\color{textcolor}\rmfamily\fontsize{4.000000}{4.800000}\selectfont \(\displaystyle {2.0}\)}%
\end{pgfscope}%
\begin{pgfscope}%
\pgfpathrectangle{\pgfqpoint{0.476352in}{0.397839in}}{\pgfqpoint{2.635000in}{2.425500in}}%
\pgfusepath{clip}%
\pgfsetroundcap%
\pgfsetroundjoin%
\pgfsetlinewidth{0.301125pt}%
\definecolor{currentstroke}{rgb}{0.800000,0.800000,0.800000}%
\pgfsetstrokecolor{currentstroke}%
\pgfsetdash{}{0pt}%
\pgfpathmoveto{\pgfqpoint{2.810729in}{0.397839in}}%
\pgfpathlineto{\pgfqpoint{2.810729in}{2.823339in}}%
\pgfusepath{stroke}%
\end{pgfscope}%
\begin{pgfscope}%
\definecolor{textcolor}{rgb}{0.150000,0.150000,0.150000}%
\pgfsetstrokecolor{textcolor}%
\pgfsetfillcolor{textcolor}%
\pgftext[x=2.810729in,y=0.356173in,,top]{\color{textcolor}\rmfamily\fontsize{4.000000}{4.800000}\selectfont \(\displaystyle {2.5}\)}%
\end{pgfscope}%
\begin{pgfscope}%
\definecolor{textcolor}{rgb}{0.150000,0.150000,0.150000}%
\pgfsetstrokecolor{textcolor}%
\pgfsetfillcolor{textcolor}%
\pgftext[x=1.793852in,y=0.238889in,,top]{\color{textcolor}\rmfamily\fontsize{10.000000}{12.000000}\selectfont Redundancies \(\displaystyle |R(\omega)|\)}%
\end{pgfscope}%
\begin{pgfscope}%
\definecolor{textcolor}{rgb}{0.150000,0.150000,0.150000}%
\pgfsetstrokecolor{textcolor}%
\pgfsetfillcolor{textcolor}%
\pgftext[x=3.111352in,y=0.252778in,right,top]{\color{textcolor}\rmfamily\fontsize{4.000000}{4.800000}\selectfont \(\displaystyle \times{10^{6}}{}\)}%
\end{pgfscope}%
\begin{pgfscope}%
\pgfpathrectangle{\pgfqpoint{0.476352in}{0.397839in}}{\pgfqpoint{2.635000in}{2.425500in}}%
\pgfusepath{clip}%
\pgfsetroundcap%
\pgfsetroundjoin%
\pgfsetlinewidth{0.301125pt}%
\definecolor{currentstroke}{rgb}{0.800000,0.800000,0.800000}%
\pgfsetstrokecolor{currentstroke}%
\pgfsetdash{}{0pt}%
\pgfpathmoveto{\pgfqpoint{0.476352in}{0.543811in}}%
\pgfpathlineto{\pgfqpoint{3.111352in}{0.543811in}}%
\pgfusepath{stroke}%
\end{pgfscope}%
\begin{pgfscope}%
\definecolor{textcolor}{rgb}{0.150000,0.150000,0.150000}%
\pgfsetstrokecolor{textcolor}%
\pgfsetfillcolor{textcolor}%
\pgftext[x=0.294444in, y=0.519698in, left, base]{\color{textcolor}\rmfamily\fontsize{4.000000}{4.800000}\selectfont \(\displaystyle {0.0}\)}%
\end{pgfscope}%
\begin{pgfscope}%
\pgfpathrectangle{\pgfqpoint{0.476352in}{0.397839in}}{\pgfqpoint{2.635000in}{2.425500in}}%
\pgfusepath{clip}%
\pgfsetroundcap%
\pgfsetroundjoin%
\pgfsetlinewidth{0.301125pt}%
\definecolor{currentstroke}{rgb}{0.800000,0.800000,0.800000}%
\pgfsetstrokecolor{currentstroke}%
\pgfsetdash{}{0pt}%
\pgfpathmoveto{\pgfqpoint{0.476352in}{0.881387in}}%
\pgfpathlineto{\pgfqpoint{3.111352in}{0.881387in}}%
\pgfusepath{stroke}%
\end{pgfscope}%
\begin{pgfscope}%
\definecolor{textcolor}{rgb}{0.150000,0.150000,0.150000}%
\pgfsetstrokecolor{textcolor}%
\pgfsetfillcolor{textcolor}%
\pgftext[x=0.294444in, y=0.857275in, left, base]{\color{textcolor}\rmfamily\fontsize{4.000000}{4.800000}\selectfont \(\displaystyle {0.5}\)}%
\end{pgfscope}%
\begin{pgfscope}%
\pgfpathrectangle{\pgfqpoint{0.476352in}{0.397839in}}{\pgfqpoint{2.635000in}{2.425500in}}%
\pgfusepath{clip}%
\pgfsetroundcap%
\pgfsetroundjoin%
\pgfsetlinewidth{0.301125pt}%
\definecolor{currentstroke}{rgb}{0.800000,0.800000,0.800000}%
\pgfsetstrokecolor{currentstroke}%
\pgfsetdash{}{0pt}%
\pgfpathmoveto{\pgfqpoint{0.476352in}{1.218963in}}%
\pgfpathlineto{\pgfqpoint{3.111352in}{1.218963in}}%
\pgfusepath{stroke}%
\end{pgfscope}%
\begin{pgfscope}%
\definecolor{textcolor}{rgb}{0.150000,0.150000,0.150000}%
\pgfsetstrokecolor{textcolor}%
\pgfsetfillcolor{textcolor}%
\pgftext[x=0.294444in, y=1.194851in, left, base]{\color{textcolor}\rmfamily\fontsize{4.000000}{4.800000}\selectfont \(\displaystyle {1.0}\)}%
\end{pgfscope}%
\begin{pgfscope}%
\pgfpathrectangle{\pgfqpoint{0.476352in}{0.397839in}}{\pgfqpoint{2.635000in}{2.425500in}}%
\pgfusepath{clip}%
\pgfsetroundcap%
\pgfsetroundjoin%
\pgfsetlinewidth{0.301125pt}%
\definecolor{currentstroke}{rgb}{0.800000,0.800000,0.800000}%
\pgfsetstrokecolor{currentstroke}%
\pgfsetdash{}{0pt}%
\pgfpathmoveto{\pgfqpoint{0.476352in}{1.556540in}}%
\pgfpathlineto{\pgfqpoint{3.111352in}{1.556540in}}%
\pgfusepath{stroke}%
\end{pgfscope}%
\begin{pgfscope}%
\definecolor{textcolor}{rgb}{0.150000,0.150000,0.150000}%
\pgfsetstrokecolor{textcolor}%
\pgfsetfillcolor{textcolor}%
\pgftext[x=0.294444in, y=1.532427in, left, base]{\color{textcolor}\rmfamily\fontsize{4.000000}{4.800000}\selectfont \(\displaystyle {1.5}\)}%
\end{pgfscope}%
\begin{pgfscope}%
\pgfpathrectangle{\pgfqpoint{0.476352in}{0.397839in}}{\pgfqpoint{2.635000in}{2.425500in}}%
\pgfusepath{clip}%
\pgfsetroundcap%
\pgfsetroundjoin%
\pgfsetlinewidth{0.301125pt}%
\definecolor{currentstroke}{rgb}{0.800000,0.800000,0.800000}%
\pgfsetstrokecolor{currentstroke}%
\pgfsetdash{}{0pt}%
\pgfpathmoveto{\pgfqpoint{0.476352in}{1.894116in}}%
\pgfpathlineto{\pgfqpoint{3.111352in}{1.894116in}}%
\pgfusepath{stroke}%
\end{pgfscope}%
\begin{pgfscope}%
\definecolor{textcolor}{rgb}{0.150000,0.150000,0.150000}%
\pgfsetstrokecolor{textcolor}%
\pgfsetfillcolor{textcolor}%
\pgftext[x=0.294444in, y=1.870003in, left, base]{\color{textcolor}\rmfamily\fontsize{4.000000}{4.800000}\selectfont \(\displaystyle {2.0}\)}%
\end{pgfscope}%
\begin{pgfscope}%
\pgfpathrectangle{\pgfqpoint{0.476352in}{0.397839in}}{\pgfqpoint{2.635000in}{2.425500in}}%
\pgfusepath{clip}%
\pgfsetroundcap%
\pgfsetroundjoin%
\pgfsetlinewidth{0.301125pt}%
\definecolor{currentstroke}{rgb}{0.800000,0.800000,0.800000}%
\pgfsetstrokecolor{currentstroke}%
\pgfsetdash{}{0pt}%
\pgfpathmoveto{\pgfqpoint{0.476352in}{2.231692in}}%
\pgfpathlineto{\pgfqpoint{3.111352in}{2.231692in}}%
\pgfusepath{stroke}%
\end{pgfscope}%
\begin{pgfscope}%
\definecolor{textcolor}{rgb}{0.150000,0.150000,0.150000}%
\pgfsetstrokecolor{textcolor}%
\pgfsetfillcolor{textcolor}%
\pgftext[x=0.294444in, y=2.207580in, left, base]{\color{textcolor}\rmfamily\fontsize{4.000000}{4.800000}\selectfont \(\displaystyle {2.5}\)}%
\end{pgfscope}%
\begin{pgfscope}%
\pgfpathrectangle{\pgfqpoint{0.476352in}{0.397839in}}{\pgfqpoint{2.635000in}{2.425500in}}%
\pgfusepath{clip}%
\pgfsetroundcap%
\pgfsetroundjoin%
\pgfsetlinewidth{0.301125pt}%
\definecolor{currentstroke}{rgb}{0.800000,0.800000,0.800000}%
\pgfsetstrokecolor{currentstroke}%
\pgfsetdash{}{0pt}%
\pgfpathmoveto{\pgfqpoint{0.476352in}{2.569268in}}%
\pgfpathlineto{\pgfqpoint{3.111352in}{2.569268in}}%
\pgfusepath{stroke}%
\end{pgfscope}%
\begin{pgfscope}%
\definecolor{textcolor}{rgb}{0.150000,0.150000,0.150000}%
\pgfsetstrokecolor{textcolor}%
\pgfsetfillcolor{textcolor}%
\pgftext[x=0.294444in, y=2.545156in, left, base]{\color{textcolor}\rmfamily\fontsize{4.000000}{4.800000}\selectfont \(\displaystyle {3.0}\)}%
\end{pgfscope}%
\begin{pgfscope}%
\definecolor{textcolor}{rgb}{0.150000,0.150000,0.150000}%
\pgfsetstrokecolor{textcolor}%
\pgfsetfillcolor{textcolor}%
\pgftext[x=0.238889in,y=1.610589in,,bottom,rotate=90.000000]{\color{textcolor}\rmfamily\fontsize{10.000000}{12.000000}\selectfont \(\displaystyle \Var[c_\omega]\)}%
\end{pgfscope}%
\begin{pgfscope}%
\definecolor{textcolor}{rgb}{0.150000,0.150000,0.150000}%
\pgfsetstrokecolor{textcolor}%
\pgfsetfillcolor{textcolor}%
\pgftext[x=0.476352in,y=2.865006in,left,base]{\color{textcolor}\rmfamily\fontsize{4.000000}{4.800000}\selectfont \(\displaystyle \times{10^{\ensuremath{-}4}}{}\)}%
\end{pgfscope}%
\begin{pgfscope}%
\pgfpathrectangle{\pgfqpoint{0.476352in}{0.397839in}}{\pgfqpoint{2.635000in}{2.425500in}}%
\pgfusepath{clip}%
\pgfsetbuttcap%
\pgfsetroundjoin%
\definecolor{currentfill}{rgb}{0.498039,0.749020,0.698039}%
\pgfsetfillcolor{currentfill}%
\pgfsetfillopacity{0.600000}%
\pgfsetlinewidth{1.003750pt}%
\definecolor{currentstroke}{rgb}{0.498039,0.749020,0.698039}%
\pgfsetstrokecolor{currentstroke}%
\pgfsetstrokeopacity{0.600000}%
\pgfsetdash{}{0pt}%
\pgfsys@defobject{currentmarker}{\pgfqpoint{-0.018373in}{-0.018373in}}{\pgfqpoint{0.018373in}{0.018373in}}{%
\pgfpathmoveto{\pgfqpoint{-0.000000in}{-0.018373in}}%
\pgfpathlineto{\pgfqpoint{0.018373in}{0.018373in}}%
\pgfpathlineto{\pgfqpoint{-0.018373in}{0.018373in}}%
\pgfpathlineto{\pgfqpoint{-0.000000in}{-0.018373in}}%
\pgfpathclose%
\pgfusepath{stroke,fill}%
}%
\begin{pgfscope}%
\pgfsys@transformshift{2.991579in}{2.713089in}%
\pgfsys@useobject{currentmarker}{}%
\end{pgfscope}%
\begin{pgfscope}%
\pgfsys@transformshift{2.807314in}{2.117179in}%
\pgfsys@useobject{currentmarker}{}%
\end{pgfscope}%
\begin{pgfscope}%
\pgfsys@transformshift{2.333487in}{1.477190in}%
\pgfsys@useobject{currentmarker}{}%
\end{pgfscope}%
\begin{pgfscope}%
\pgfsys@transformshift{1.754366in}{0.801306in}%
\pgfsys@useobject{currentmarker}{}%
\end{pgfscope}%
\begin{pgfscope}%
\pgfsys@transformshift{1.247635in}{0.631742in}%
\pgfsys@useobject{currentmarker}{}%
\end{pgfscope}%
\begin{pgfscope}%
\pgfsys@transformshift{0.902717in}{0.560789in}%
\pgfsys@useobject{currentmarker}{}%
\end{pgfscope}%
\begin{pgfscope}%
\pgfsys@transformshift{0.715355in}{0.547645in}%
\pgfsys@useobject{currentmarker}{}%
\end{pgfscope}%
\begin{pgfscope}%
\pgfsys@transformshift{0.633776in}{0.544336in}%
\pgfsys@useobject{currentmarker}{}%
\end{pgfscope}%
\begin{pgfscope}%
\pgfsys@transformshift{0.605537in}{0.543896in}%
\pgfsys@useobject{currentmarker}{}%
\end{pgfscope}%
\begin{pgfscope}%
\pgfsys@transformshift{0.597917in}{0.543819in}%
\pgfsys@useobject{currentmarker}{}%
\end{pgfscope}%
\begin{pgfscope}%
\pgfsys@transformshift{0.596368in}{0.543812in}%
\pgfsys@useobject{currentmarker}{}%
\end{pgfscope}%
\begin{pgfscope}%
\pgfsys@transformshift{0.596145in}{0.543811in}%
\pgfsys@useobject{currentmarker}{}%
\end{pgfscope}%
\begin{pgfscope}%
\pgfsys@transformshift{0.596125in}{0.543811in}%
\pgfsys@useobject{currentmarker}{}%
\end{pgfscope}%
\end{pgfscope}%
\begin{pgfscope}%
\pgfpathrectangle{\pgfqpoint{0.476352in}{0.397839in}}{\pgfqpoint{2.635000in}{2.425500in}}%
\pgfusepath{clip}%
\pgfsetbuttcap%
\pgfsetroundjoin%
\definecolor{currentfill}{rgb}{0.498039,0.533333,0.749020}%
\pgfsetfillcolor{currentfill}%
\pgfsetfillopacity{0.600000}%
\pgfsetlinewidth{1.505625pt}%
\definecolor{currentstroke}{rgb}{0.498039,0.533333,0.749020}%
\pgfsetstrokecolor{currentstroke}%
\pgfsetstrokeopacity{0.600000}%
\pgfsetdash{}{0pt}%
\pgfsys@defobject{currentmarker}{\pgfqpoint{-0.020127in}{-0.020127in}}{\pgfqpoint{0.020127in}{0.020127in}}{%
\pgfpathmoveto{\pgfqpoint{-0.020127in}{-0.020127in}}%
\pgfpathlineto{\pgfqpoint{0.020127in}{0.020127in}}%
\pgfpathmoveto{\pgfqpoint{-0.020127in}{0.020127in}}%
\pgfpathlineto{\pgfqpoint{0.020127in}{-0.020127in}}%
\pgfusepath{stroke,fill}%
}%
\begin{pgfscope}%
\pgfsys@transformshift{2.991579in}{1.661052in}%
\pgfsys@useobject{currentmarker}{}%
\end{pgfscope}%
\begin{pgfscope}%
\pgfsys@transformshift{2.807314in}{1.480157in}%
\pgfsys@useobject{currentmarker}{}%
\end{pgfscope}%
\begin{pgfscope}%
\pgfsys@transformshift{2.333487in}{1.106355in}%
\pgfsys@useobject{currentmarker}{}%
\end{pgfscope}%
\begin{pgfscope}%
\pgfsys@transformshift{1.754366in}{0.780440in}%
\pgfsys@useobject{currentmarker}{}%
\end{pgfscope}%
\begin{pgfscope}%
\pgfsys@transformshift{1.247635in}{0.624126in}%
\pgfsys@useobject{currentmarker}{}%
\end{pgfscope}%
\begin{pgfscope}%
\pgfsys@transformshift{0.902717in}{0.568783in}%
\pgfsys@useobject{currentmarker}{}%
\end{pgfscope}%
\begin{pgfscope}%
\pgfsys@transformshift{0.715355in}{0.550814in}%
\pgfsys@useobject{currentmarker}{}%
\end{pgfscope}%
\begin{pgfscope}%
\pgfsys@transformshift{0.633776in}{0.545362in}%
\pgfsys@useobject{currentmarker}{}%
\end{pgfscope}%
\begin{pgfscope}%
\pgfsys@transformshift{0.605537in}{0.544109in}%
\pgfsys@useobject{currentmarker}{}%
\end{pgfscope}%
\begin{pgfscope}%
\pgfsys@transformshift{0.597917in}{0.543866in}%
\pgfsys@useobject{currentmarker}{}%
\end{pgfscope}%
\begin{pgfscope}%
\pgfsys@transformshift{0.596368in}{0.543819in}%
\pgfsys@useobject{currentmarker}{}%
\end{pgfscope}%
\begin{pgfscope}%
\pgfsys@transformshift{0.596145in}{0.543812in}%
\pgfsys@useobject{currentmarker}{}%
\end{pgfscope}%
\begin{pgfscope}%
\pgfsys@transformshift{0.596125in}{0.543811in}%
\pgfsys@useobject{currentmarker}{}%
\end{pgfscope}%
\end{pgfscope}%
\begin{pgfscope}%
\pgfpathrectangle{\pgfqpoint{0.476352in}{0.397839in}}{\pgfqpoint{2.635000in}{2.425500in}}%
\pgfusepath{clip}%
\pgfsetbuttcap%
\pgfsetroundjoin%
\definecolor{currentfill}{rgb}{0.329412,0.556863,0.631373}%
\pgfsetfillcolor{currentfill}%
\pgfsetfillopacity{0.600000}%
\pgfsetlinewidth{1.003750pt}%
\definecolor{currentstroke}{rgb}{0.329412,0.556863,0.631373}%
\pgfsetstrokecolor{currentstroke}%
\pgfsetstrokeopacity{0.600000}%
\pgfsetdash{}{0pt}%
\pgfsys@defobject{currentmarker}{\pgfqpoint{-0.018373in}{-0.018373in}}{\pgfqpoint{0.018373in}{0.018373in}}{%
\pgfpathmoveto{\pgfqpoint{-0.018373in}{-0.018373in}}%
\pgfpathlineto{\pgfqpoint{0.018373in}{-0.018373in}}%
\pgfpathlineto{\pgfqpoint{0.018373in}{0.018373in}}%
\pgfpathlineto{\pgfqpoint{-0.018373in}{0.018373in}}%
\pgfpathlineto{\pgfqpoint{-0.018373in}{-0.018373in}}%
\pgfpathclose%
\pgfusepath{stroke,fill}%
}%
\begin{pgfscope}%
\pgfsys@transformshift{2.991579in}{1.209785in}%
\pgfsys@useobject{currentmarker}{}%
\end{pgfscope}%
\begin{pgfscope}%
\pgfsys@transformshift{2.807314in}{1.141244in}%
\pgfsys@useobject{currentmarker}{}%
\end{pgfscope}%
\begin{pgfscope}%
\pgfsys@transformshift{2.333487in}{0.957704in}%
\pgfsys@useobject{currentmarker}{}%
\end{pgfscope}%
\begin{pgfscope}%
\pgfsys@transformshift{1.754366in}{0.751626in}%
\pgfsys@useobject{currentmarker}{}%
\end{pgfscope}%
\begin{pgfscope}%
\pgfsys@transformshift{1.247635in}{0.631629in}%
\pgfsys@useobject{currentmarker}{}%
\end{pgfscope}%
\begin{pgfscope}%
\pgfsys@transformshift{0.902717in}{0.572223in}%
\pgfsys@useobject{currentmarker}{}%
\end{pgfscope}%
\begin{pgfscope}%
\pgfsys@transformshift{0.715355in}{0.552504in}%
\pgfsys@useobject{currentmarker}{}%
\end{pgfscope}%
\begin{pgfscope}%
\pgfsys@transformshift{0.633776in}{0.546448in}%
\pgfsys@useobject{currentmarker}{}%
\end{pgfscope}%
\begin{pgfscope}%
\pgfsys@transformshift{0.605537in}{0.544463in}%
\pgfsys@useobject{currentmarker}{}%
\end{pgfscope}%
\begin{pgfscope}%
\pgfsys@transformshift{0.597917in}{0.543930in}%
\pgfsys@useobject{currentmarker}{}%
\end{pgfscope}%
\begin{pgfscope}%
\pgfsys@transformshift{0.596368in}{0.543827in}%
\pgfsys@useobject{currentmarker}{}%
\end{pgfscope}%
\begin{pgfscope}%
\pgfsys@transformshift{0.596145in}{0.543812in}%
\pgfsys@useobject{currentmarker}{}%
\end{pgfscope}%
\begin{pgfscope}%
\pgfsys@transformshift{0.596125in}{0.543811in}%
\pgfsys@useobject{currentmarker}{}%
\end{pgfscope}%
\end{pgfscope}%
\begin{pgfscope}%
\pgfpathrectangle{\pgfqpoint{0.476352in}{0.397839in}}{\pgfqpoint{2.635000in}{2.425500in}}%
\pgfusepath{clip}%
\pgfsetbuttcap%
\pgfsetroundjoin%
\definecolor{currentfill}{rgb}{0.100000,0.100000,0.100000}%
\pgfsetfillcolor{currentfill}%
\pgfsetfillopacity{0.600000}%
\pgfsetlinewidth{1.003750pt}%
\definecolor{currentstroke}{rgb}{0.100000,0.100000,0.100000}%
\pgfsetstrokecolor{currentstroke}%
\pgfsetstrokeopacity{0.600000}%
\pgfsetdash{}{0pt}%
\pgfsys@defobject{currentmarker}{\pgfqpoint{-0.018373in}{-0.018373in}}{\pgfqpoint{0.018373in}{0.018373in}}{%
\pgfpathmoveto{\pgfqpoint{0.000000in}{-0.018373in}}%
\pgfpathcurveto{\pgfqpoint{0.004873in}{-0.018373in}}{\pgfqpoint{0.009546in}{-0.016437in}}{\pgfqpoint{0.012992in}{-0.012992in}}%
\pgfpathcurveto{\pgfqpoint{0.016437in}{-0.009546in}}{\pgfqpoint{0.018373in}{-0.004873in}}{\pgfqpoint{0.018373in}{0.000000in}}%
\pgfpathcurveto{\pgfqpoint{0.018373in}{0.004873in}}{\pgfqpoint{0.016437in}{0.009546in}}{\pgfqpoint{0.012992in}{0.012992in}}%
\pgfpathcurveto{\pgfqpoint{0.009546in}{0.016437in}}{\pgfqpoint{0.004873in}{0.018373in}}{\pgfqpoint{0.000000in}{0.018373in}}%
\pgfpathcurveto{\pgfqpoint{-0.004873in}{0.018373in}}{\pgfqpoint{-0.009546in}{0.016437in}}{\pgfqpoint{-0.012992in}{0.012992in}}%
\pgfpathcurveto{\pgfqpoint{-0.016437in}{0.009546in}}{\pgfqpoint{-0.018373in}{0.004873in}}{\pgfqpoint{-0.018373in}{0.000000in}}%
\pgfpathcurveto{\pgfqpoint{-0.018373in}{-0.004873in}}{\pgfqpoint{-0.016437in}{-0.009546in}}{\pgfqpoint{-0.012992in}{-0.012992in}}%
\pgfpathcurveto{\pgfqpoint{-0.009546in}{-0.016437in}}{\pgfqpoint{-0.004873in}{-0.018373in}}{\pgfqpoint{0.000000in}{-0.018373in}}%
\pgfpathlineto{\pgfqpoint{0.000000in}{-0.018373in}}%
\pgfpathclose%
\pgfusepath{stroke,fill}%
}%
\begin{pgfscope}%
\pgfsys@transformshift{2.991579in}{0.794612in}%
\pgfsys@useobject{currentmarker}{}%
\end{pgfscope}%
\begin{pgfscope}%
\pgfsys@transformshift{2.807314in}{0.769682in}%
\pgfsys@useobject{currentmarker}{}%
\end{pgfscope}%
\begin{pgfscope}%
\pgfsys@transformshift{2.333487in}{0.724625in}%
\pgfsys@useobject{currentmarker}{}%
\end{pgfscope}%
\begin{pgfscope}%
\pgfsys@transformshift{1.754366in}{0.666515in}%
\pgfsys@useobject{currentmarker}{}%
\end{pgfscope}%
\begin{pgfscope}%
\pgfsys@transformshift{1.247635in}{0.615413in}%
\pgfsys@useobject{currentmarker}{}%
\end{pgfscope}%
\begin{pgfscope}%
\pgfsys@transformshift{0.902717in}{0.577853in}%
\pgfsys@useobject{currentmarker}{}%
\end{pgfscope}%
\begin{pgfscope}%
\pgfsys@transformshift{0.715355in}{0.556887in}%
\pgfsys@useobject{currentmarker}{}%
\end{pgfscope}%
\begin{pgfscope}%
\pgfsys@transformshift{0.633776in}{0.547876in}%
\pgfsys@useobject{currentmarker}{}%
\end{pgfscope}%
\begin{pgfscope}%
\pgfsys@transformshift{0.605537in}{0.544809in}%
\pgfsys@useobject{currentmarker}{}%
\end{pgfscope}%
\begin{pgfscope}%
\pgfsys@transformshift{0.597917in}{0.544000in}%
\pgfsys@useobject{currentmarker}{}%
\end{pgfscope}%
\begin{pgfscope}%
\pgfsys@transformshift{0.596368in}{0.543837in}%
\pgfsys@useobject{currentmarker}{}%
\end{pgfscope}%
\begin{pgfscope}%
\pgfsys@transformshift{0.596145in}{0.543814in}%
\pgfsys@useobject{currentmarker}{}%
\end{pgfscope}%
\begin{pgfscope}%
\pgfsys@transformshift{0.596125in}{0.543811in}%
\pgfsys@useobject{currentmarker}{}%
\end{pgfscope}%
\end{pgfscope}%
\begin{pgfscope}%
\pgfpathrectangle{\pgfqpoint{0.476352in}{0.397839in}}{\pgfqpoint{2.635000in}{2.425500in}}%
\pgfusepath{clip}%
\pgfsetroundcap%
\pgfsetroundjoin%
\pgfsetlinewidth{0.803000pt}%
\definecolor{currentstroke}{rgb}{0.498039,0.749020,0.698039}%
\pgfsetstrokecolor{currentstroke}%
\pgfsetdash{}{0pt}%
\pgfpathmoveto{\pgfqpoint{0.596125in}{0.560099in}}%
\pgfpathlineto{\pgfqpoint{0.644987in}{0.545689in}}%
\pgfpathlineto{\pgfqpoint{0.693534in}{0.533682in}}%
\pgfpathlineto{\pgfqpoint{0.741813in}{0.524024in}}%
\pgfpathlineto{\pgfqpoint{0.789874in}{0.516672in}}%
\pgfpathlineto{\pgfqpoint{0.837766in}{0.511591in}}%
\pgfpathlineto{\pgfqpoint{0.885542in}{0.508754in}}%
\pgfpathlineto{\pgfqpoint{0.933251in}{0.508146in}}%
\pgfpathlineto{\pgfqpoint{0.980946in}{0.509762in}}%
\pgfpathlineto{\pgfqpoint{1.028679in}{0.513603in}}%
\pgfpathlineto{\pgfqpoint{1.076501in}{0.519684in}}%
\pgfpathlineto{\pgfqpoint{1.124463in}{0.528027in}}%
\pgfpathlineto{\pgfqpoint{1.172614in}{0.538663in}}%
\pgfpathlineto{\pgfqpoint{1.221005in}{0.551634in}}%
\pgfpathlineto{\pgfqpoint{1.269683in}{0.566990in}}%
\pgfpathlineto{\pgfqpoint{1.318693in}{0.584789in}}%
\pgfpathlineto{\pgfqpoint{1.368082in}{0.605100in}}%
\pgfpathlineto{\pgfqpoint{1.417891in}{0.627997in}}%
\pgfpathlineto{\pgfqpoint{1.468162in}{0.653564in}}%
\pgfpathlineto{\pgfqpoint{1.518934in}{0.681892in}}%
\pgfpathlineto{\pgfqpoint{1.570244in}{0.713078in}}%
\pgfpathlineto{\pgfqpoint{1.622128in}{0.747229in}}%
\pgfpathlineto{\pgfqpoint{1.674620in}{0.784456in}}%
\pgfpathlineto{\pgfqpoint{1.727750in}{0.824876in}}%
\pgfpathlineto{\pgfqpoint{1.781549in}{0.868615in}}%
\pgfpathlineto{\pgfqpoint{1.836044in}{0.915803in}}%
\pgfpathlineto{\pgfqpoint{1.891261in}{0.966576in}}%
\pgfpathlineto{\pgfqpoint{1.947225in}{1.021075in}}%
\pgfpathlineto{\pgfqpoint{2.003960in}{1.079446in}}%
\pgfpathlineto{\pgfqpoint{2.061485in}{1.141843in}}%
\pgfpathlineto{\pgfqpoint{2.119822in}{1.208420in}}%
\pgfpathlineto{\pgfqpoint{2.178988in}{1.279341in}}%
\pgfpathlineto{\pgfqpoint{2.239001in}{1.354770in}}%
\pgfpathlineto{\pgfqpoint{2.299878in}{1.434879in}}%
\pgfpathlineto{\pgfqpoint{2.361633in}{1.519843in}}%
\pgfpathlineto{\pgfqpoint{2.424280in}{1.609842in}}%
\pgfpathlineto{\pgfqpoint{2.487832in}{1.705058in}}%
\pgfpathlineto{\pgfqpoint{2.552302in}{1.805681in}}%
\pgfpathlineto{\pgfqpoint{2.617701in}{1.911902in}}%
\pgfpathlineto{\pgfqpoint{2.684039in}{2.023918in}}%
\pgfpathlineto{\pgfqpoint{2.751326in}{2.141928in}}%
\pgfpathlineto{\pgfqpoint{2.819572in}{2.266138in}}%
\pgfpathlineto{\pgfqpoint{2.888785in}{2.396754in}}%
\pgfpathlineto{\pgfqpoint{2.958972in}{2.533989in}}%
\pgfpathlineto{\pgfqpoint{2.991579in}{2.599382in}}%
\pgfpathlineto{\pgfqpoint{2.991579in}{2.599382in}}%
\pgfusepath{stroke}%
\end{pgfscope}%
\begin{pgfscope}%
\pgfpathrectangle{\pgfqpoint{0.476352in}{0.397839in}}{\pgfqpoint{2.635000in}{2.425500in}}%
\pgfusepath{clip}%
\pgfsetroundcap%
\pgfsetroundjoin%
\pgfsetlinewidth{0.803000pt}%
\definecolor{currentstroke}{rgb}{0.498039,0.533333,0.749020}%
\pgfsetstrokecolor{currentstroke}%
\pgfsetdash{}{0pt}%
\pgfpathmoveto{\pgfqpoint{0.596125in}{0.547013in}}%
\pgfpathlineto{\pgfqpoint{0.669653in}{0.545856in}}%
\pgfpathlineto{\pgfqpoint{0.743163in}{0.546922in}}%
\pgfpathlineto{\pgfqpoint{0.816690in}{0.550211in}}%
\pgfpathlineto{\pgfqpoint{0.890266in}{0.555728in}}%
\pgfpathlineto{\pgfqpoint{0.963925in}{0.563481in}}%
\pgfpathlineto{\pgfqpoint{1.037700in}{0.573483in}}%
\pgfpathlineto{\pgfqpoint{1.111624in}{0.585750in}}%
\pgfpathlineto{\pgfqpoint{1.185731in}{0.600303in}}%
\pgfpathlineto{\pgfqpoint{1.260050in}{0.617166in}}%
\pgfpathlineto{\pgfqpoint{1.334616in}{0.636368in}}%
\pgfpathlineto{\pgfqpoint{1.409457in}{0.657939in}}%
\pgfpathlineto{\pgfqpoint{1.484606in}{0.681917in}}%
\pgfpathlineto{\pgfqpoint{1.560092in}{0.708341in}}%
\pgfpathlineto{\pgfqpoint{1.635944in}{0.737253in}}%
\pgfpathlineto{\pgfqpoint{1.712191in}{0.768700in}}%
\pgfpathlineto{\pgfqpoint{1.788860in}{0.802732in}}%
\pgfpathlineto{\pgfqpoint{1.865979in}{0.839402in}}%
\pgfpathlineto{\pgfqpoint{1.943573in}{0.878767in}}%
\pgfpathlineto{\pgfqpoint{2.021668in}{0.920886in}}%
\pgfpathlineto{\pgfqpoint{2.100289in}{0.965821in}}%
\pgfpathlineto{\pgfqpoint{2.179459in}{1.013640in}}%
\pgfpathlineto{\pgfqpoint{2.259202in}{1.064409in}}%
\pgfpathlineto{\pgfqpoint{2.339539in}{1.118202in}}%
\pgfpathlineto{\pgfqpoint{2.420492in}{1.175092in}}%
\pgfpathlineto{\pgfqpoint{2.502081in}{1.235155in}}%
\pgfpathlineto{\pgfqpoint{2.584326in}{1.298472in}}%
\pgfpathlineto{\pgfqpoint{2.667246in}{1.365125in}}%
\pgfpathlineto{\pgfqpoint{2.750860in}{1.435198in}}%
\pgfpathlineto{\pgfqpoint{2.835184in}{1.508779in}}%
\pgfpathlineto{\pgfqpoint{2.920235in}{1.585956in}}%
\pgfpathlineto{\pgfqpoint{2.991579in}{1.652989in}}%
\pgfpathlineto{\pgfqpoint{2.991579in}{1.652989in}}%
\pgfusepath{stroke}%
\end{pgfscope}%
\begin{pgfscope}%
\pgfpathrectangle{\pgfqpoint{0.476352in}{0.397839in}}{\pgfqpoint{2.635000in}{2.425500in}}%
\pgfusepath{clip}%
\pgfsetroundcap%
\pgfsetroundjoin%
\pgfsetlinewidth{0.803000pt}%
\definecolor{currentstroke}{rgb}{0.329412,0.556863,0.631373}%
\pgfsetstrokecolor{currentstroke}%
\pgfsetdash{}{0pt}%
\pgfpathmoveto{\pgfqpoint{0.596125in}{0.541989in}}%
\pgfpathlineto{\pgfqpoint{0.715210in}{0.554361in}}%
\pgfpathlineto{\pgfqpoint{0.834411in}{0.568981in}}%
\pgfpathlineto{\pgfqpoint{0.953746in}{0.585857in}}%
\pgfpathlineto{\pgfqpoint{1.073237in}{0.605001in}}%
\pgfpathlineto{\pgfqpoint{1.192904in}{0.626426in}}%
\pgfpathlineto{\pgfqpoint{1.312766in}{0.650145in}}%
\pgfpathlineto{\pgfqpoint{1.432844in}{0.676175in}}%
\pgfpathlineto{\pgfqpoint{1.553156in}{0.704531in}}%
\pgfpathlineto{\pgfqpoint{1.673724in}{0.735233in}}%
\pgfpathlineto{\pgfqpoint{1.794565in}{0.768301in}}%
\pgfpathlineto{\pgfqpoint{1.915699in}{0.803755in}}%
\pgfpathlineto{\pgfqpoint{2.037145in}{0.841620in}}%
\pgfpathlineto{\pgfqpoint{2.158922in}{0.881918in}}%
\pgfpathlineto{\pgfqpoint{2.281046in}{0.924676in}}%
\pgfpathlineto{\pgfqpoint{2.403538in}{0.969920in}}%
\pgfpathlineto{\pgfqpoint{2.526413in}{1.017679in}}%
\pgfpathlineto{\pgfqpoint{2.649691in}{1.067982in}}%
\pgfpathlineto{\pgfqpoint{2.773388in}{1.120861in}}%
\pgfpathlineto{\pgfqpoint{2.897520in}{1.176347in}}%
\pgfpathlineto{\pgfqpoint{2.991579in}{1.220006in}}%
\pgfpathlineto{\pgfqpoint{2.991579in}{1.220006in}}%
\pgfusepath{stroke}%
\end{pgfscope}%
\begin{pgfscope}%
\pgfpathrectangle{\pgfqpoint{0.476352in}{0.397839in}}{\pgfqpoint{2.635000in}{2.425500in}}%
\pgfusepath{clip}%
\pgfsetroundcap%
\pgfsetroundjoin%
\pgfsetlinewidth{0.803000pt}%
\definecolor{currentstroke}{rgb}{0.100000,0.100000,0.100000}%
\pgfsetstrokecolor{currentstroke}%
\pgfsetdash{}{0pt}%
\pgfpathmoveto{\pgfqpoint{0.596125in}{0.544058in}}%
\pgfpathlineto{\pgfqpoint{1.339844in}{0.623300in}}%
\pgfpathlineto{\pgfqpoint{2.083445in}{0.700295in}}%
\pgfpathlineto{\pgfqpoint{2.826932in}{0.775044in}}%
\pgfpathlineto{\pgfqpoint{2.991579in}{0.791295in}}%
\pgfpathlineto{\pgfqpoint{2.991579in}{0.791295in}}%
\pgfusepath{stroke}%
\end{pgfscope}%
\begin{pgfscope}%
\pgfsetrectcap%
\pgfsetmiterjoin%
\pgfsetlinewidth{1.254687pt}%
\definecolor{currentstroke}{rgb}{0.800000,0.800000,0.800000}%
\pgfsetstrokecolor{currentstroke}%
\pgfsetdash{}{0pt}%
\pgfpathmoveto{\pgfqpoint{0.476352in}{0.397839in}}%
\pgfpathlineto{\pgfqpoint{0.476352in}{2.823339in}}%
\pgfusepath{stroke}%
\end{pgfscope}%
\begin{pgfscope}%
\pgfsetrectcap%
\pgfsetmiterjoin%
\pgfsetlinewidth{1.254687pt}%
\definecolor{currentstroke}{rgb}{0.800000,0.800000,0.800000}%
\pgfsetstrokecolor{currentstroke}%
\pgfsetdash{}{0pt}%
\pgfpathmoveto{\pgfqpoint{3.111352in}{0.397839in}}%
\pgfpathlineto{\pgfqpoint{3.111352in}{2.823339in}}%
\pgfusepath{stroke}%
\end{pgfscope}%
\begin{pgfscope}%
\pgfsetrectcap%
\pgfsetmiterjoin%
\pgfsetlinewidth{1.254687pt}%
\definecolor{currentstroke}{rgb}{0.800000,0.800000,0.800000}%
\pgfsetstrokecolor{currentstroke}%
\pgfsetdash{}{0pt}%
\pgfpathmoveto{\pgfqpoint{0.476352in}{0.397839in}}%
\pgfpathlineto{\pgfqpoint{3.111352in}{0.397839in}}%
\pgfusepath{stroke}%
\end{pgfscope}%
\begin{pgfscope}%
\pgfsetrectcap%
\pgfsetmiterjoin%
\pgfsetlinewidth{1.254687pt}%
\definecolor{currentstroke}{rgb}{0.800000,0.800000,0.800000}%
\pgfsetstrokecolor{currentstroke}%
\pgfsetdash{}{0pt}%
\pgfpathmoveto{\pgfqpoint{0.476352in}{2.823339in}}%
\pgfpathlineto{\pgfqpoint{3.111352in}{2.823339in}}%
\pgfusepath{stroke}%
\end{pgfscope}%
\begin{pgfscope}%
\pgfsetbuttcap%
\pgfsetmiterjoin%
\definecolor{currentfill}{rgb}{1.000000,1.000000,1.000000}%
\pgfsetfillcolor{currentfill}%
\pgfsetfillopacity{0.800000}%
\pgfsetlinewidth{1.003750pt}%
\definecolor{currentstroke}{rgb}{0.800000,0.800000,0.800000}%
\pgfsetstrokecolor{currentstroke}%
\pgfsetstrokeopacity{0.800000}%
\pgfsetdash{}{0pt}%
\pgfpathmoveto{\pgfqpoint{0.544408in}{2.339389in}}%
\pgfpathlineto{\pgfqpoint{1.581755in}{2.339389in}}%
\pgfpathquadraticcurveto{\pgfqpoint{1.601200in}{2.339389in}}{\pgfqpoint{1.601200in}{2.358834in}}%
\pgfpathlineto{\pgfqpoint{1.601200in}{2.755284in}}%
\pgfpathquadraticcurveto{\pgfqpoint{1.601200in}{2.774728in}}{\pgfqpoint{1.581755in}{2.774728in}}%
\pgfpathlineto{\pgfqpoint{0.544408in}{2.774728in}}%
\pgfpathquadraticcurveto{\pgfqpoint{0.524963in}{2.774728in}}{\pgfqpoint{0.524963in}{2.755284in}}%
\pgfpathlineto{\pgfqpoint{0.524963in}{2.358834in}}%
\pgfpathquadraticcurveto{\pgfqpoint{0.524963in}{2.339389in}}{\pgfqpoint{0.544408in}{2.339389in}}%
\pgfpathlineto{\pgfqpoint{0.544408in}{2.339389in}}%
\pgfpathclose%
\pgfusepath{stroke,fill}%
\end{pgfscope}%
\begin{pgfscope}%
\definecolor{textcolor}{rgb}{0.150000,0.150000,0.150000}%
\pgfsetstrokecolor{textcolor}%
\pgfsetfillcolor{textcolor}%
\pgftext[x=0.802712in,y=2.668324in,left,base]{\color{textcolor}\rmfamily\fontsize{7.000000}{8.400000}\selectfont Simulation}%
\end{pgfscope}%
\begin{pgfscope}%
\pgfsetbuttcap%
\pgfsetmiterjoin%
\definecolor{currentfill}{rgb}{0.498039,0.749020,0.698039}%
\pgfsetfillcolor{currentfill}%
\pgfsetfillopacity{0.600000}%
\pgfsetlinewidth{1.505625pt}%
\definecolor{currentstroke}{rgb}{0.498039,0.749020,0.698039}%
\pgfsetstrokecolor{currentstroke}%
\pgfsetstrokeopacity{0.600000}%
\pgfsetdash{}{0pt}%
\pgfsys@defobject{currentmarker}{\pgfqpoint{-0.022222in}{-0.022222in}}{\pgfqpoint{0.022222in}{0.022222in}}{%
\pgfpathmoveto{\pgfqpoint{-0.000000in}{-0.022222in}}%
\pgfpathlineto{\pgfqpoint{0.022222in}{0.022222in}}%
\pgfpathlineto{\pgfqpoint{-0.022222in}{0.022222in}}%
\pgfpathlineto{\pgfqpoint{-0.000000in}{-0.022222in}}%
\pgfpathclose%
\pgfusepath{stroke,fill}%
}%
\begin{pgfscope}%
\pgfsys@transformshift{0.661074in}{2.566781in}%
\pgfsys@useobject{currentmarker}{}%
\end{pgfscope}%
\end{pgfscope}%
\begin{pgfscope}%
\definecolor{textcolor}{rgb}{0.150000,0.150000,0.150000}%
\pgfsetstrokecolor{textcolor}%
\pgfsetfillcolor{textcolor}%
\pgftext[x=0.758297in,y=2.532753in,left,base]{\color{textcolor}\rmfamily\fontsize{7.000000}{8.400000}\selectfont m=1}%
\end{pgfscope}%
\begin{pgfscope}%
\pgfsetbuttcap%
\pgfsetroundjoin%
\definecolor{currentfill}{rgb}{0.498039,0.533333,0.749020}%
\pgfsetfillcolor{currentfill}%
\pgfsetfillopacity{0.600000}%
\pgfsetlinewidth{1.505625pt}%
\definecolor{currentstroke}{rgb}{0.498039,0.533333,0.749020}%
\pgfsetstrokecolor{currentstroke}%
\pgfsetstrokeopacity{0.600000}%
\pgfsetdash{}{0pt}%
\pgfsys@defobject{currentmarker}{\pgfqpoint{-0.022222in}{-0.022222in}}{\pgfqpoint{0.022222in}{0.022222in}}{%
\pgfpathmoveto{\pgfqpoint{-0.022222in}{-0.022222in}}%
\pgfpathlineto{\pgfqpoint{0.022222in}{0.022222in}}%
\pgfpathmoveto{\pgfqpoint{-0.022222in}{0.022222in}}%
\pgfpathlineto{\pgfqpoint{0.022222in}{-0.022222in}}%
\pgfusepath{stroke,fill}%
}%
\begin{pgfscope}%
\pgfsys@transformshift{0.661074in}{2.431210in}%
\pgfsys@useobject{currentmarker}{}%
\end{pgfscope}%
\end{pgfscope}%
\begin{pgfscope}%
\definecolor{textcolor}{rgb}{0.150000,0.150000,0.150000}%
\pgfsetstrokecolor{textcolor}%
\pgfsetfillcolor{textcolor}%
\pgftext[x=0.758297in,y=2.397182in,left,base]{\color{textcolor}\rmfamily\fontsize{7.000000}{8.400000}\selectfont m=2}%
\end{pgfscope}%
\begin{pgfscope}%
\pgfsetbuttcap%
\pgfsetmiterjoin%
\definecolor{currentfill}{rgb}{0.329412,0.556863,0.631373}%
\pgfsetfillcolor{currentfill}%
\pgfsetfillopacity{0.600000}%
\pgfsetlinewidth{1.505625pt}%
\definecolor{currentstroke}{rgb}{0.329412,0.556863,0.631373}%
\pgfsetstrokecolor{currentstroke}%
\pgfsetstrokeopacity{0.600000}%
\pgfsetdash{}{0pt}%
\pgfsys@defobject{currentmarker}{\pgfqpoint{-0.022222in}{-0.022222in}}{\pgfqpoint{0.022222in}{0.022222in}}{%
\pgfpathmoveto{\pgfqpoint{-0.022222in}{-0.022222in}}%
\pgfpathlineto{\pgfqpoint{0.022222in}{-0.022222in}}%
\pgfpathlineto{\pgfqpoint{0.022222in}{0.022222in}}%
\pgfpathlineto{\pgfqpoint{-0.022222in}{0.022222in}}%
\pgfpathlineto{\pgfqpoint{-0.022222in}{-0.022222in}}%
\pgfpathclose%
\pgfusepath{stroke,fill}%
}%
\begin{pgfscope}%
\pgfsys@transformshift{1.233220in}{2.566781in}%
\pgfsys@useobject{currentmarker}{}%
\end{pgfscope}%
\end{pgfscope}%
\begin{pgfscope}%
\definecolor{textcolor}{rgb}{0.150000,0.150000,0.150000}%
\pgfsetstrokecolor{textcolor}%
\pgfsetfillcolor{textcolor}%
\pgftext[x=1.330443in,y=2.532753in,left,base]{\color{textcolor}\rmfamily\fontsize{7.000000}{8.400000}\selectfont m=3}%
\end{pgfscope}%
\begin{pgfscope}%
\pgfsetbuttcap%
\pgfsetroundjoin%
\definecolor{currentfill}{rgb}{0.100000,0.100000,0.100000}%
\pgfsetfillcolor{currentfill}%
\pgfsetfillopacity{0.600000}%
\pgfsetlinewidth{1.505625pt}%
\definecolor{currentstroke}{rgb}{0.100000,0.100000,0.100000}%
\pgfsetstrokecolor{currentstroke}%
\pgfsetstrokeopacity{0.600000}%
\pgfsetdash{}{0pt}%
\pgfsys@defobject{currentmarker}{\pgfqpoint{-0.022222in}{-0.022222in}}{\pgfqpoint{0.022222in}{0.022222in}}{%
\pgfpathmoveto{\pgfqpoint{0.000000in}{-0.022222in}}%
\pgfpathcurveto{\pgfqpoint{0.005893in}{-0.022222in}}{\pgfqpoint{0.011546in}{-0.019881in}}{\pgfqpoint{0.015713in}{-0.015713in}}%
\pgfpathcurveto{\pgfqpoint{0.019881in}{-0.011546in}}{\pgfqpoint{0.022222in}{-0.005893in}}{\pgfqpoint{0.022222in}{0.000000in}}%
\pgfpathcurveto{\pgfqpoint{0.022222in}{0.005893in}}{\pgfqpoint{0.019881in}{0.011546in}}{\pgfqpoint{0.015713in}{0.015713in}}%
\pgfpathcurveto{\pgfqpoint{0.011546in}{0.019881in}}{\pgfqpoint{0.005893in}{0.022222in}}{\pgfqpoint{0.000000in}{0.022222in}}%
\pgfpathcurveto{\pgfqpoint{-0.005893in}{0.022222in}}{\pgfqpoint{-0.011546in}{0.019881in}}{\pgfqpoint{-0.015713in}{0.015713in}}%
\pgfpathcurveto{\pgfqpoint{-0.019881in}{0.011546in}}{\pgfqpoint{-0.022222in}{0.005893in}}{\pgfqpoint{-0.022222in}{0.000000in}}%
\pgfpathcurveto{\pgfqpoint{-0.022222in}{-0.005893in}}{\pgfqpoint{-0.019881in}{-0.011546in}}{\pgfqpoint{-0.015713in}{-0.015713in}}%
\pgfpathcurveto{\pgfqpoint{-0.011546in}{-0.019881in}}{\pgfqpoint{-0.005893in}{-0.022222in}}{\pgfqpoint{0.000000in}{-0.022222in}}%
\pgfpathlineto{\pgfqpoint{0.000000in}{-0.022222in}}%
\pgfpathclose%
\pgfusepath{stroke,fill}%
}%
\begin{pgfscope}%
\pgfsys@transformshift{1.233220in}{2.431210in}%
\pgfsys@useobject{currentmarker}{}%
\end{pgfscope}%
\end{pgfscope}%
\begin{pgfscope}%
\definecolor{textcolor}{rgb}{0.150000,0.150000,0.150000}%
\pgfsetstrokecolor{textcolor}%
\pgfsetfillcolor{textcolor}%
\pgftext[x=1.330443in,y=2.397182in,left,base]{\color{textcolor}\rmfamily\fontsize{7.000000}{8.400000}\selectfont m=6}%
\end{pgfscope}%
\begin{pgfscope}%
\pgfsetbuttcap%
\pgfsetmiterjoin%
\definecolor{currentfill}{rgb}{1.000000,1.000000,1.000000}%
\pgfsetfillcolor{currentfill}%
\pgfsetfillopacity{0.800000}%
\pgfsetlinewidth{1.003750pt}%
\definecolor{currentstroke}{rgb}{0.800000,0.800000,0.800000}%
\pgfsetstrokecolor{currentstroke}%
\pgfsetstrokeopacity{0.800000}%
\pgfsetdash{}{0pt}%
\pgfpathmoveto{\pgfqpoint{0.544408in}{1.583148in}}%
\pgfpathlineto{\pgfqpoint{1.575044in}{1.583148in}}%
\pgfpathquadraticcurveto{\pgfqpoint{1.594488in}{1.583148in}}{\pgfqpoint{1.594488in}{1.602592in}}%
\pgfpathlineto{\pgfqpoint{1.594488in}{2.270184in}}%
\pgfpathquadraticcurveto{\pgfqpoint{1.594488in}{2.289628in}}{\pgfqpoint{1.575044in}{2.289628in}}%
\pgfpathlineto{\pgfqpoint{0.544408in}{2.289628in}}%
\pgfpathquadraticcurveto{\pgfqpoint{0.524963in}{2.289628in}}{\pgfqpoint{0.524963in}{2.270184in}}%
\pgfpathlineto{\pgfqpoint{0.524963in}{1.602592in}}%
\pgfpathquadraticcurveto{\pgfqpoint{0.524963in}{1.583148in}}{\pgfqpoint{0.544408in}{1.583148in}}%
\pgfpathlineto{\pgfqpoint{0.544408in}{1.583148in}}%
\pgfpathclose%
\pgfusepath{stroke,fill}%
\end{pgfscope}%
\begin{pgfscope}%
\definecolor{textcolor}{rgb}{0.150000,0.150000,0.150000}%
\pgfsetstrokecolor{textcolor}%
\pgfsetfillcolor{textcolor}%
\pgftext[x=0.802568in,y=2.183224in,left,base]{\color{textcolor}\rmfamily\fontsize{7.000000}{8.400000}\selectfont Regression}%
\end{pgfscope}%
\begin{pgfscope}%
\pgfsetroundcap%
\pgfsetroundjoin%
\pgfsetlinewidth{0.803000pt}%
\definecolor{currentstroke}{rgb}{0.498039,0.749020,0.698039}%
\pgfsetstrokecolor{currentstroke}%
\pgfsetdash{}{0pt}%
\pgfpathmoveto{\pgfqpoint{0.563852in}{2.081681in}}%
\pgfpathlineto{\pgfqpoint{0.661074in}{2.081681in}}%
\pgfpathlineto{\pgfqpoint{0.758297in}{2.081681in}}%
\pgfusepath{stroke}%
\end{pgfscope}%
\begin{pgfscope}%
\definecolor{textcolor}{rgb}{0.150000,0.150000,0.150000}%
\pgfsetstrokecolor{textcolor}%
\pgfsetfillcolor{textcolor}%
\pgftext[x=0.836074in,y=2.047653in,left,base]{\color{textcolor}\rmfamily\fontsize{7.000000}{8.400000}\selectfont m=1, R2=0.99}%
\end{pgfscope}%
\begin{pgfscope}%
\pgfsetroundcap%
\pgfsetroundjoin%
\pgfsetlinewidth{0.803000pt}%
\definecolor{currentstroke}{rgb}{0.498039,0.533333,0.749020}%
\pgfsetstrokecolor{currentstroke}%
\pgfsetdash{}{0pt}%
\pgfpathmoveto{\pgfqpoint{0.563852in}{1.946110in}}%
\pgfpathlineto{\pgfqpoint{0.661074in}{1.946110in}}%
\pgfpathlineto{\pgfqpoint{0.758297in}{1.946110in}}%
\pgfusepath{stroke}%
\end{pgfscope}%
\begin{pgfscope}%
\definecolor{textcolor}{rgb}{0.150000,0.150000,0.150000}%
\pgfsetstrokecolor{textcolor}%
\pgfsetfillcolor{textcolor}%
\pgftext[x=0.836074in,y=1.912082in,left,base]{\color{textcolor}\rmfamily\fontsize{7.000000}{8.400000}\selectfont m=2, R2=1.00}%
\end{pgfscope}%
\begin{pgfscope}%
\pgfsetroundcap%
\pgfsetroundjoin%
\pgfsetlinewidth{0.803000pt}%
\definecolor{currentstroke}{rgb}{0.329412,0.556863,0.631373}%
\pgfsetstrokecolor{currentstroke}%
\pgfsetdash{}{0pt}%
\pgfpathmoveto{\pgfqpoint{0.563852in}{1.810539in}}%
\pgfpathlineto{\pgfqpoint{0.661074in}{1.810539in}}%
\pgfpathlineto{\pgfqpoint{0.758297in}{1.810539in}}%
\pgfusepath{stroke}%
\end{pgfscope}%
\begin{pgfscope}%
\definecolor{textcolor}{rgb}{0.150000,0.150000,0.150000}%
\pgfsetstrokecolor{textcolor}%
\pgfsetfillcolor{textcolor}%
\pgftext[x=0.836074in,y=1.776511in,left,base]{\color{textcolor}\rmfamily\fontsize{7.000000}{8.400000}\selectfont m=3, R2=1.00}%
\end{pgfscope}%
\begin{pgfscope}%
\pgfsetroundcap%
\pgfsetroundjoin%
\pgfsetlinewidth{0.803000pt}%
\definecolor{currentstroke}{rgb}{0.100000,0.100000,0.100000}%
\pgfsetstrokecolor{currentstroke}%
\pgfsetdash{}{0pt}%
\pgfpathmoveto{\pgfqpoint{0.563852in}{1.674968in}}%
\pgfpathlineto{\pgfqpoint{0.661074in}{1.674968in}}%
\pgfpathlineto{\pgfqpoint{0.758297in}{1.674968in}}%
\pgfusepath{stroke}%
\end{pgfscope}%
\begin{pgfscope}%
\definecolor{textcolor}{rgb}{0.150000,0.150000,0.150000}%
\pgfsetstrokecolor{textcolor}%
\pgfsetfillcolor{textcolor}%
\pgftext[x=0.836074in,y=1.640941in,left,base]{\color{textcolor}\rmfamily\fontsize{7.000000}{8.400000}\selectfont m=6, R2=1.00}%
\end{pgfscope}%
\end{pgfpicture}%
\makeatother%
\endgroup%

%% file: Figures/figs_numerics/arrow.tikz
\tikzset{every picture/.style={line width=0.75pt}} 

\begin{tikzpicture}[x=0.75pt,y=0.75pt,yscale=-.9,xscale=.4]

\draw  [color={rgb, 255:red, 204; green, 204; blue, 204 }  ,draw opacity=1 ][fill={rgb, 255:red, 237; green, 237; blue, 237 }  ,fill opacity=1 ] (2,52.17) -- (29.91,5) -- (57.82,52.17) -- (44.41,52.17) -- (44.41,226.38) -- (15.42,226.38) -- (15.42,52.17) -- cycle ;

\end{tikzpicture}

%% file: A3_ansatze.tex
\section{Ansätze for trainable layers}\label{ansatze}

This section serves as a reference to the different circuit architectures that we considered in Section \ref{sec:Simulations}, explaining how a particular ansatz structure can be scaled in the number of qubits.

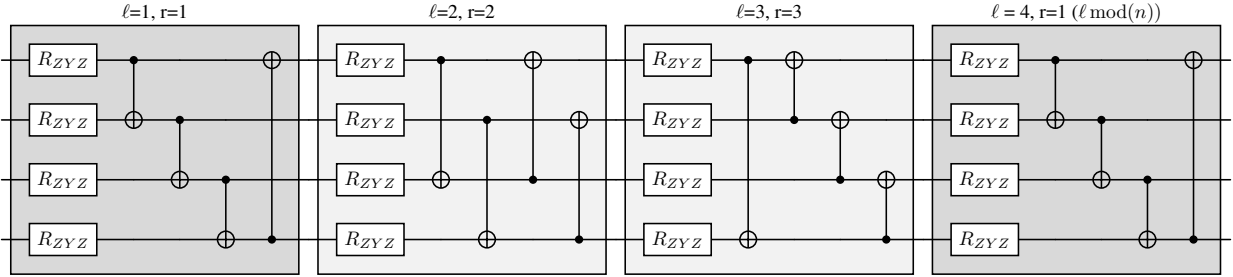
\begin{figure*}
\centering
\begin{adjustbox}{width=\textwidth}
    \begin{quantikz}
    & \gate{R_{ZYZ}} \gategroup[4,steps=5,style={inner sep=6pt,fill=darkgray!20},background]{$\ell$=1, r=1}& \ctrl{1} & & & \targ{} & & \gate{R_{ZYZ}}\gategroup[4,steps=5,style={inner sep=6pt,fill=lightgray!20},background]{$\ell$=2, r=2} & \ctrl{2} & & \targ{} & & & \gate{R_{ZYZ}}\gategroup[4,steps=5,style={inner sep=6pt,fill=lightgray!20},background]{$\ell$=3, r=3} & \ctrl{3} & \targ{}& & & &\gate{R_{ZYZ}} \gategroup[4,steps=5,style={inner sep=6pt,fill=darkgray!20},background]{$\ell$ = 4, r=1 ($\ell \operatorname{mod}(n))$}& \ctrl{1} & & & \targ{} & \\
    & \gate{R_{ZYZ}} & \targ{} & \ctrl{1} & & & & \gate{R_{ZYZ}} & & \ctrl{2} & & \targ{} & & \gate{R_{ZYZ}} & &\ctrl{-1}&\targ{}& & &\gate{R_{ZYZ}} & \targ{} & \ctrl{1} & & & \\
    & \gate{R_{ZYZ}} & & \targ{} & \ctrl{1} & & & \gate{R_{ZYZ}} & \targ{} & & \ctrl{-2} & & & \gate{R_{ZYZ}} &&&\ctrl{-1}&\targ{} & &\gate{R_{ZYZ}} & & \targ{} & \ctrl{1} & & \\
    & \gate{R_{ZYZ}} & & & \targ{} & \ctrl{-3} & & \gate{R_{ZYZ}} & & \targ{} & & \ctrl{-2} & & \gate{R_{ZYZ}} & \targ{} &&&\ctrl{-1} &  & \gate{R_{ZYZ}} & & & \targ{} & \ctrl{-3} & 
    \end{quantikz}
\end{adjustbox}
\caption{\emph{Strongly Entangling Ansatz} \cite{schuld_circuit-centric_2020}: The Ansatz depth varies with the number of qubits. To have all possible entanglers between qubits, one needs to have the number of blocks $\ell$ to be $\ell\geq n-1$, with $n$ the number of qubits and $r$ is the range of the control gates  given by $r = \ell \operatorname{mod}(n)$.}
\label{circ:strongly_entangling}
\end{figure*}

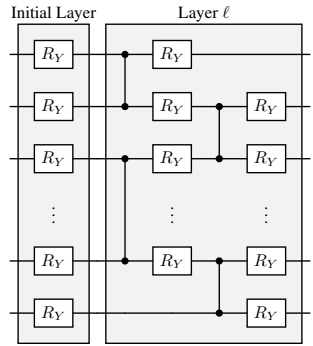
\begin{figure*}[ht]
    \centering
    \begin{adjustbox}{height=2.5cm}
    \begin{quantikz}
        & \gate{R_Y} \gategroup[6,steps=1,style={inner sep=6pt,fill=lightgray!20},background]{Initial Layer}  &&\ctrl{1} \gategroup[6,steps=4,style={inner sep=6pt,fill=lightgray!20},background]{Layer $\ell$}& \gate{R_Y}& &&\\
        & \gate{R_Y} && \control{} & \gate{R_Y} &\ctrl{1}& \gate{R_Y} &\\
        & \gate{R_Y} && \ctrl{2} & \gate{R_Y} &\control{}& \gate{R_Y} &\\
        &\wireoverride{n}\vdots&\wireoverride{n}&\wireoverride{n}&\wireoverride{n}\vdots& \wireoverride{n} & \wireoverride{n} \vdots\\
        & \gate{R_Y} && \control{} & \gate{R_Y} &\ctrl{1}& \gate{R_Y} &\\
        & \gate{R_Y} && &&\control{}& \gate{R_Y} &
    \end{quantikz}
\end{adjustbox}
\caption{\emph{Simplified Two Design Ansatz} \cite{cerezo_cost_2021}: An initial layer of Pauli-Y rotations and controlled-Z entanglers. Note that this ansatz has the same approximately the same amount of entangler gates as for the basic entangling ansatz layer, but double the number of parameters. It is composed of an initial layer and then the periodic layer is the one that will be repeated.}
\label{SimplifiedTwoDesign}
\end{figure*}